%% file: structure_manuscript.tex
\documentclass[english,american,nofootinbib]{revtex4-1}
\usepackage[T1]{fontenc}
\usepackage[latin9]{inputenc}
\usepackage{xcolor}
\usepackage{babel}
\usepackage{array}
\usepackage{prettyref}
\usepackage{booktabs}
\usepackage{mathtools}
\usepackage{amsmath}
\usepackage{amssymb}
\usepackage{stmaryrd}
\usepackage{graphicx}
\usepackage{feyn}
\usepackage[unicode=true,pdfusetitle,
 bookmarks=true,bookmarksnumbered=false,bookmarksopen=false,
 breaklinks=false,pdfborder={0 0 1},backref=false,colorlinks=false]
 {hyperref}
\hypersetup{
 colorlinks=true,linkcolor=blue,citecolor=blue,urlcolor=blue,filecolor=blue}

\makeatletter

\providecommand{\tabularnewline}{\\}

\usepackage{mathtools}
\usepackage{feynmp-auto}
\usepackage{bbm}
\usepackage{MnSymbol}

\newrefformat{cap}{\hyperref[#1]{Figure~\ref{#1}}}
\newrefformat{fig}{\hyperref[#1]{Figure~\ref{#1}}}
\newrefformat{tab}{\hyperref[#1]{Table ~\ref{#1}}}
\newrefformat{sec}{\hyperref[#1]{Section~\ref{#1}}}
\newrefformat{subsec}{\hyperref[#1]{Section~\ref{#1}}}
\newrefformat{sub}{\hyperref[#1]{Section~\ref{#1}}}
\newrefformat{cha}{\hyperref[#1]{Chapter~\ref{#1}}}
\newrefformat{app}{\hyperref[#1]{Appendix~\ref{#1}}}

\makeatother

\begin{document}
\include{macros}

\title{Self-consistent formulations for stochastic nonlinear neuronal dynamics}
\author{Jonas Stapmanns$^{1,2}$} \thanks{J. Stapmanns and T. K\"{u}hn contributed equally to this work \\ \-\hspace{-3.5mm} $^{\dagger}$ Current address: Laboratoire de Physique de l'ENS, 24, Rue Lhomond, 75231 Paris Cedex 05, France} \author{Tobias K\"{u}hn$^{1,2,\dagger}$} \thanks{J. Stapmanns and T. K\"{u}hn contributed equally to this work \\ \-\hspace{-3.5mm} $^{\dagger}$ Current address: Laboratoire de Physique de l'ENS, 24, Rue Lhomond, 75231 Paris Cedex 05, France} \author{David Dahmen$^{1}$} \author{Thomas Luu$^{3}$} \author{Carsten Honerkamp$^{2,4}$} \author{Moritz Helias$^{1,2}$}

\affiliation{$^{1}$Institute of Neuroscience and Medicine (INM-6) and Institute for Advanced Simulation (IAS-6) and JARA BRAIN Institute I, Jülich Research Centre, Jülich, Germany}

\affiliation{$^{2}$Institute for Theoretical Solid State Physics, RWTH Aachen University, 52074 Aachen, Germany}

\affiliation{$^{3}$Institut für Kernphysik (IKP-3), Institute for Advanced Simulation (IAS-4) and Jülich Center for Hadron Physics, Jülich Research Centre, Jülich, Germany}

\affiliation{$^{4}$JARA-FIT, Jülich Aachen Research Alliance - Fundamentals of Future Information Technology, Germany}
\date{\today}
\begin{abstract}
Neural dynamics is often investigated with tools from bifurcation
theory. However, many neuron models are stochastic, mimicking fluctuations
in the input from unknown parts of the brain or the spiking nature
of signals. Noise changes the dynamics with respect to the deterministic
model; in particular classical bifurcation theory cannot be applied.
We formulate the stochastic neuron dynamics in the Martin-Siggia-Rose
de Dominicis-Janssen (MSRDJ) formalism and present the fluctuation
expansion of the effective action and the functional renormalization
group (fRG) as two systematic ways to incorporate corrections to the
mean dynamics and time-dependent statistics due to fluctuations in
the presence of nonlinear neuronal gain. To formulate self-consistency
equations, we derive a fundamental link between the effective action
in the Onsager-Machlup (OM) formalism, which allows the study of phase
transitions, and the MSRDJ effective action, which is computationally
advantageous. These results in particular allow the derivation of
an OM effective action for systems with non-Gaussian noise. This approach
naturally leads to effective deterministic equations for the first
moment of the stochastic system; they explain how nonlinearities and
noise cooperate to produce memory effects. Moreover, the MSRDJ formulation
yields an effective linear system that has identical power spectra
and linear response. Starting from the better known loopwise approximation,
we then discuss the use of the fRG as a method to obtain self-consistency
beyond the mean. We present a new efficient truncation scheme for
the hierarchy of flow equations for the vertex functions by adapting
the Blaizot, Méndez and Wschebor (BMW) approximation from the derivative
expansion to the vertex expansion. The methods are presented by means
of the simplest possible example of a stochastic differential equation
that has generic features of neuronal dynamics.
\end{abstract}
\maketitle

\section{Introduction}

Neuronal networks are interesting physical systems in various respects:
they operate outside thermodynamic equilibrium \citep{Dahmen16_031024},
a consequence of directed synaptic connections that prohibit detailed
balance \citep{Sompolinsky88_2}; they show relaxational dynamics
and hence do not conserve but rather constantly dissipate energy;
and they show collective behavior that self-organizes as a result
of exposure to structured, correlated inputs and the interaction among
their constituents. But their analysis is complicated by three fundamental
properties: Neuronal activity is stochastic, the input-output transfer
function of single neurons is nonlinear, and networks show massive
recurrence \citep{Braitenberg91} that gives rise to strong interaction
effects. They hence bear similarity with systems that are investigated
in the field of (quantum) many particle systems. Here, as well, (quantum)
fluctuations need to be taken into account and the challenge is to
understand collective phenomena that arise from the nonlinear interaction
of their constituents. Not surprisingly, similar methods can in principle
be used to study these two \textit{a priori} distinct system classes
\citep{ZinnJustin96,Hohenberg77,Taeuber14,Chow15,Hertz16_033001}.

But so far, the techniques employed within theoretical neuroscience
just begin to harvest this potential. Here, we adapt methods from
statistical field theory and functional renormalization group techniques
to the study of neuronal dynamics.

A reader may wonder in which cases functional methods that we review
and extend here are needed to study neuronal network dynamics. In
fact, in many cases, simpler techniques could be sufficient: as long
one is sure that the dominant behavior of a network is not affected
by fluctuations, a mean-field approximation is enough \citep{Vreeswijk96,Brunel00_183}.
For example, if one wants to know the stationary firing rates of neurons
in typical network settings (but see \citep{Ocker17_1} for an exception).
Or if one seeks to understand first-order phase transitions, where
the mean activity suddenly changes: The transition from a quiescent
to a highly active state in a bistable neuronal network is a prime
example \citep{Brunel00_183}; the activation of attractors embedded
into the connectivity of a Hopfield network is a second \citep{Hopfield82}.

An inherent danger of the mean-field approximation, though, is that
by construction it is ``too self-consistent''; this means that there
is no way of determining its limit of validity as long as it is not
embedded into a wider framework that assesses the importance of fluctuations.
The loop expansion, briefly reviewed here, is this wider framework
(see also \citep{Amit84} section 6.4 for a discussion of this point
in the context of Landau theory).

Even if fluctuations are important, it may be sufficient to use linear
response theory around the mean-field approximation. This approach
has indeed been pursued quite successfully \citep{Brunel00_183,Lindner05_505,Shea-Brown08,Boustani_09,Pernice11_e1002059,Ostojic11_e1001056,Trousdale12_e1002408,Tetzlaff12_e1002596,Helias13_023002},
for example, to explain why correlations are weak in inhibition-dominated
networks \citep{Renart10_587,Tetzlaff12_e1002596}.

So why bother about methods that go beyond this widely established,
simple, and successful methodology?

If it was clear that the considerable variability observed in the
brain was meaningless noise, one should not invest any work in going
beyond linear response theory. But there is surely no neuroscientist
who would sign this statement. On the contrary, experiments have clearly
shown that fluctuations of neuronal activity are indeed linked in
a meaningful way to the animal's behavior (see, e.g., \citep{Riehle97_1950,Cohen09_1079,Kilavik09_12653}).
Treating fluctuations in linear response theory means that they follow
linear equations: Fluctuating signals that are simultaneously injected
at various places of the brain then travel through it without mutual
interference. The function the brain performs at this level of approximation
is thus that of a linear filter; clearly a too limited view to explain
most of its functions.

The diagrammatic techniques considered here were developed over decades
within the physics community to capture the interaction of fluctuations
in nonlinear systems and they are widely used in many fields \citep{Duclu17}.
In the neuronal context, they thus enable the investigation of potential
functional roles of the variability observed in neuronal systems.

The finding of signatures of criticality in neuronal activity is an
example of a network state that is dominated by fluctuations, a scenario
where mean-field and linear response theory clearly break down. Parallel
recordings in cell cultures \citep{Beggs04} and \textit{in vivo}
\citep{Tkacik14_e1003408} show power law distributions in the numbers
of co-active neurons, suggestive of scale free dynamics, which typically
indicates a continuous phase transition. Critical states also have
consequences for information processing: reservoir computing \citep{Jaeger01_echo,Maass02_2531}
with random networks close to criticality indeed shows the highest
computational performance \citep{Legenstein07_323}, maximizing the
wealth of transformations.

The renormalization group was one of the major achievements of theoretical
physics of the past century to understand collective phenomena close
to continuous phase transitions \citep{Wilson75_773}. So far, however,
these methods have rarely been employed to neuronal networks. The
need for a formalism of dynamical critical phenomena of neuronal networks
has explicitly been articulated by \citet[p. 288]{Mora2011}: ``Except
in a few cases, the mathematical language that we use to describe
criticality in statistical systems is quite different from the language
that we use in dynamical systems. Efforts to understand, for example,
current data on networks of neurons will force us to address the relations
between statistical and dynamical criticality more clearly.''

The current work presents such a formalism. The functional renormalization
group \citep{Wegner73_401,WETTERICH93_90,Berges02_223} allows the
study of critical fluctuations in networks that operate outside thermodynamic
equilibrium: dynamical critical phenomena \citep{Taeuber14}. This
method has witnessed successes in condensed matter physics in problems
ranging from classical and quantum critical phenomena over the explorations
of the ground states of interacting many-body systems to the improved
determination of effective model parameters from \textit{ab initio}
theories. It systematically improves the physical description beyond
mean-field theory by including fluctuations and by removing ambiguities.

To showcase the importance of continuous phase transitions we here
demonstrate that networks with connectivity close to the point of
balance between excitation and inhibition exhibit critical fluctuations.
We reduce the network model to a spatially homogeneous version of
``model A'' of the seminal characterization of dynamical critical
phenomena by \citet{Hohenberg77}.

Techniques similar to the ones reviewed and extended in the current
article have already been used in computational neuroscience. One
technical motivation comes from the need to study disordered systems;
for example, networks with randomly drawn connections. The formulation
with help of a generating functional is useful to investigate the
impact of connectivity structure on dynamics. For example, \citet{Marti18_062314}
study the slowdown of fluctuations in networks with an overrepresentation
of symmetric connections, as they are observed in cortex \citep{Song05_0507}.
Their analysis rests on the Martin-Siggia-Rose-de Dominicis-Janssen
(MSRDJ) formalism \citep{Martin73,janssen1976_377,dedominicis1976_247}
presented here (see also \citep{Chow15,Hertz16_033001} for concise
reviews). \citet{Crisanti18_062120} compute the transition to chaos
and the stability of mean-field solutions in deterministic random
networks. \citet{Schuecker18_041029} find a novel dynamical state,
between the breakdown of linear stability and the onset of chaos that
has optimal sequential memory. \citet{Dahmen19_13051} describe a
state of critical dynamics in neuronal networks that is hidden within
the high-dimensional space of all neurons.

Another motivation comes from the study of correlated activity between
pairs of cells. The loopwise expansion as a systematic extension of
the commonly performed mean-field approximation has been applied to
understand fluctuations in recurrent networks of neurons with discrete
activation states \citep{Buice07_031118,Buice10_377}, to obtain fluctuation
corrections to the mean activity and higher order correlations, structure-dynamics
relationships in stochastic spiking networks \citep{Ocker17_1}, and
to study spiking networks of quadratic integrate-and-fire models \citep{Buice13_1}.
The recent work by \citet{Brinkman18_e1006490} addresses the pertinent
question whether and how hidden units bias the estimation of connectivity
from correlations. These works use different variants of the systematic
fluctuation expansion discussed here on minimal examples.

In the context of critical phenomena, the study by Di Santo et al.
\citep{diSanto18_1356} is worth mentioning. The authors propose a
field theory of neuronal activity on a two-dimensional sheet, reminiscent
of a Ginzburg-Landau theory \citep[model A]{Hohenberg77}. So far,
this model has been investigated by help of the mean-field approximation
and a first-order transition has been found. An investigation of critical
phenomena of a continuous phase transition in this model would require
renormalization group methods that we discuss here. Formally, their
model resembles the Kardar-Parisi-Zhang equation \citep{Kardar84},
one of the most prominent models of dynamical criticality which requires
the functional renormalization group \citep{Canet10,Canet11} or mode
coupling theory \citep{Frey96_4424}.

The aim of the current paper is thus to review a sequence of formulations
for the statistics in neuronal-inspired stochastic dynamics, coherently
presented in the order of increasing self-consistency: mean-field
approximation, loop expansion, and functional renormalization group.
To our knowledge, advanced methods such as the functional renormalization
group, have so far not been applied to neuronal systems. We feel that
a work explaining this method in the context of neuronal dynamics
with help of minimal examples and in relation to the simpler approaches,
the mean-field approximation and the loop expansion, may be helpful
for researchers who are interested in neuronal variability.

The presentation together with more widely used approaches also allows
us to address some of the more subtle points that we could not find
documented in the literature. In particular, we discuss issues of
convexity of the cumulant-generating functional, the need for the
Legendre-Fenchel transform in systems with dynamic symmetry breaking,
the different definitions of the effective potential in equilibrium
and nonequilibrium systems and its use to investigate phase transitions.
We also point out particularities of the MSRDJ formalism compared
to the field theories more commonly covered in text books.

\section{Results}

\subsection{Outline of the paper}

The outline of the paper is as follows: in \prettyref{subsec:Stochastic-rate-equations}
we define the class of stochastic differential equations that have
been used in the literature to describe neuronal dynamics and we define
the simplest nontrivial example that will serve us throughout the
paper to illustrate the essence of the respective methods. We briefly
review the simplest approximation, the mean-field solution of a stochastic
differential equation. The following sections then develop self-consistent
schemes of increasing complexity.

These more elaborate schemes rely on a representation of the stochastic
system in terms of a cumulant-generating functional and its Legendre
transform, the effective action. We therefore introduce in \prettyref{subsec:Generating_functional_OM}
the Onsager-Machlup (OM) field theory, which has a rigorous basis
for stochastic differential equations with Gaussian noise. \prettyref{subsec:Stochastic-dynamics-as-variational}
explains the central role of Legendre transforms for the construction
of self-consistency equations.

We proceed to the MSRDJ-field theory in \prettyref{subsec:Martin-Siggia-Rose-de-Dominicis-Janssen},
which has computational advantages compared to OM. We show in \prettyref{subsec:Effective-equation-of-motion}
that the resulting self-consistency equation takes the form of an
effective, deterministic, integro-differential equation for the mean
value of the stochastic process. The concept of vertex functions is
introduced as a set of time nonlocal coupling kernels that appear
in this equation. Our first main result is the relation of the effective
actions in the OM and the MSRDJ formulation; thus we extend the definition
of the OM effective action for non-Gaussian noise.

These formal prerequisites allow us in \prettyref{subsec:Loop-expansion}
to briefly review the loop expansion as a systematic extension of
the mean-field approximation; it enables practical and systematic
calculations of vertex functions and of probabilistic quantities,
which, by construction, are self-consistent on the level of the first
moments. In \prettyref{subsec:time_dep} we present the effective
self-consistency equation for the mean value of the exemplary stochastic
process in one loop approximation; it shows how fluctuations influence
the relaxation back to baseline after a small perturbation and provides
an intuition for the meaning of vertex functions. A stochastic linear
convolution equation is presented that has the same second-order statistics
as the full nonlinear system; it explains the meaning of self-energy
corrections for stochastic dynamics.

The most advanced method of self-consistency covered here is the functional
renormalization group, shown in \prettyref{subsec:Functional-Renormalization-Group}.
It extends the self-consistency up to arbitrary orders of the vertex
functions. In particular, in \prettyref{subsec:BMW}, we present a
new interpretation of the BMW approximation \citep{Blaizot06_571}
within the vertex expansion. We show that all terms of the bare action
flowing in this altered BMW-fRG scheme lead to an improvement over
the one-loop result.

Finally, \prettyref{subsec:Analyzing-bifurcations} visits the problem
of self-consistency from the perspective of bifurcations. On the example
of a neuronal population dynamics close to the loss of balance, we
show that a bifurcation point in the deterministic model corresponds
to a continuous phase transition in the stochastic dynamics, illustrating
the use of the OM effective action for nonequilibrium dynamics.

In \prettyref{sec:Discussion} we discuss the presented concepts in
comparison to other approaches and provide an outlook towards applications
within theoretical neuroscience.

\subsection{Stochastic rate equations inspired by neuronal dynamics\label{subsec:Stochastic-rate-equations}}

The study of rate equations has a long history in theoretical neuroscience
\citep{Wilson1972,Amari72_643}. Initially these equations described
the time evolution of the average number of active neurons in a given
time interval: a set of deterministic, coupled differential equations.
To cater for fluctuations of neuronal activity \citep{Softky93},
stochastic models are needed. Markov models, for example, describe
the stochastic evolution of the active number of neurons \citep{Ginzburg94,Vreeswijk96,Buice07_031118}.
Such Markov jump processes may be approximated by stochastic differential
equations using a Kramers-Moyal expansion \citep{Weber17_046601}.

But also the dynamics of deterministic spiking network models \citep{Brunel99}
has been shown heuristically to be approximated by effective stochastic
differential equations \citep{Brunel03a,Ostojic09_10234,Ledoux11_1,Helias13_023002,Grytskyy13_131}.
Typically the resulting equations describe population averages, and
thus comprise only a single or a few components. Stochastic spiking
network models, moreover, can be treated within a variant of the the
field theoretical formalism \citep{Andreanov2006_030101,Lefvre2007_07024}
used in the current work \citep{Ocker17_109,Brinkman18_e1006490}.

Lastly, stochastic differential equations may be regarded as a direct
generalization of their deterministic counterparts in their own right.
For example, the classical model by Sompolinsky, Crisanti and Sommers
\citep{Sompolinsky88_259} has been extended to stochastic dynamics
\citep{Schuecker18_041029}.

A typical network dynamics is of the form
\begin{align}
dx_{i}(t)+x_{i}(t)\,dt & =\sum_{j}J_{ij}\phi(x_{j}(t))\,dt+dW_{i}(t),\label{eq:N_dim}
\end{align}
with a stochastic increment $dW\left(t\right)$ as a centered Gaussian
with first and second moments given by
\[
\left\langle dW_{i}\left(t\right)\right\rangle =0,\ \left\langle dW_{i}(t)dW_{j}(s)\right\rangle =D\,\delta_{i,j}\,\delta_{t,s}dt.
\]
Here $\phi$ is the gain function which is thought to transfer the
internal state variable $x$ (e.g. the membrane potential) into the
output (the firing rate). The term $x\,dt$ shall describe the leaky
dynamics of neurons.

The aim of this article is to survey methods to construct self-consistent
solutions to such stochastic nonlinear differential equations. The
$N$ components in \eqref{eq:N_dim} do not qualitatively increase
the difficulty \textendash{} rather the interplay of fluctuations
and the nonlinearity is the cause of complications. To illustrate
the concepts in the simplest but nontrivial setting, in the remainder
of this article we therefore mostly concentrate on the scalar equation

\begin{equation}
dx(t)=f\left(x\left(t\right)\right)\,dt+dW(t).\label{eq:SDE_general}
\end{equation}
This is, moreover, identical to \eqref{eq:N_dim} for the case of
a fully connected network $J_{ij}=N^{-1}J$ and perfectly correlated
stochastic increments $dW_{i}$ across neurons, if one defines $f\left(x\right)=-x+J\,\phi\left(x\right)$.
For the function $\phi$ we assume an expansive nonlinearity of convex
down shape. This is a typical result of neurons firing in the fluctuation-driven
regime with low firing rates \citep[eq. (2.6) and Fig. 2]{Bressloff12}.
For specific calculations in minimal examples, we consider a quadratic
form for the gain function that is thus the simplest approximation
of this qualitative shape. Similar approximations have recently been
used to study critical avalanche dynamics in neuronal networks \citep[eq. (1)]{diSanto18_1356}.

Of course, besides the interpretation as the firing rate or activity
of a neuron or of a population of neurons, here $x$ could as well
denote a magnetization, the concentration of some chemical substance,
the number of animals that reproduce and die, or the value of a stock
\citep{Gardiner85,VanKampen07,Kleinert09}. 

\subsubsection*{Mean-field approximation}

This article investigates a sequence of approximation techniques to
compute the statistics of the system self-consistently. We present
methods in order of increasing complexity and accuracy, starting with
the simplest possible self-consistent approximation: the neglect of
fluctuations altogether, which leads to a self-consistency equation
for the mean $\langle x\rangle$. Throughout the text we use $\langle x^{n}\rangle$
to denote the $n$-th moment of $x$ and $\langle\langle x^{n}\rangle\rangle$
to denote the $n$-th cumulant \citep{Gardiner85}.

This simplest self-consistent approximation to the stochastic differential
equation \eqref{eq:SDE_general} consists in neglecting the noise
and instead solving the ordinary differential equation
\begin{equation}
\frac{d}{dt}x\left(t\right)=f\left(x\left(t\right)\right).\label{eq:deterministic_diffeq}
\end{equation}
Finding the stationary solution to this differential equation amounts
to a fixed point problem, that is $f(x_{0})=0$. Small fluctuations
around that solution can, to first approximation, be accounted for
by linearizing \prettyref{eq:SDE_general} around $x_{0}$. By writing
$x\left(t\right)=x_{0}+\delta x\left(t\right)$, we obtain
\begin{align}
\frac{d}{dt}\,\delta x\left(t\right)= & f^{\prime}\left(x_{0}\right)\delta x\left(t\right)+\xi\left(t\right)+\mathcal{O}\left(\delta^{2}x\left(t\right)\right),\label{eq:linear_response_fluct}
\end{align}
where $\xi$ is a centered Gaussian white noise with variance $D$
(formally the derivative of $dW$). Denoting the Fourier transform
of $x$ as $X$, we describe the first and second-order statistics
of these small fluctuations as
\begin{align}
0=\left\langle \delta X\left(\omega\right)\right\rangle = & \frac{\left\langle \xi\left(\omega\right)\right\rangle }{i\omega-f^{\prime}\left(x_{0}\right)},\nonumber \\
\left\langle \delta X\left(\omega\right)\delta X\left(\omega^{\prime}\right)\right\rangle = & \frac{2\pi D\,\delta\left(\omega+\omega^{\prime}\right)}{\left(i\omega-f^{\prime}\left(x_{0}\right)\right)\left(i\omega^{\prime}-f^{\prime}\left(x_{0}\right)\right)}.\label{eq:variance_linear_response}
\end{align}
This approach is, however, restricted to small noise amplitudes, and
cannot be straight-forwardly generalized. From a conceptual point
of view, this approximation is furthermore not self-consistent, because
we solve the deterministic equation \prettyref{eq:deterministic_diffeq}
to determine the first moment $x_{0}$ and then study fluctuations
around it to determine the second-order statistics as \prettyref{eq:variance_linear_response}.
Such fluctuations would, in turn, affect the mean value, due to the
nonlinear parts of $f$. Continuing the Taylor expansion of $f$ in
\prettyref{eq:linear_response_fluct} to second order, this \textit{ad
hoc} approach would then yield a correction to the mean given by 
\begin{align}
\langle x\rangle & \simeq x_{0}+\frac{1}{2}\frac{f^{\prime\prime}\left(x_{0}\right)}{f^{\prime}\left(x_{0}\right)}\frac{D}{2f^{\prime}\left(x_{0}\right)},\label{eq:naive_correction_mean}
\end{align}
because the variance of the process, by \eqref{eq:variance_linear_response},
is $\langle\delta x^{2}\rangle=\frac{D}{-2f^{\prime}(x_{0})}$. Thus
we get an approximation for the mean that is inconsistent with the
value $x_{0}$ that we assumed to perform the approximation in the
first place. The common thrust of the methods surveyed in the remainder
of this article is to strive for self-consistency of the statistics
and to systematically compute such fluctuation corrections that are
self-consistent also on the level of higher moments.

\subsection{Generating functionals for stochastic differential equations\label{subsec:Generating_functional_OM}}

\subsubsection*{Onsager-Machlup path-integral}

To study the system more systematically, we introduce the path-integral
formalism, starting with its Onsager-Machlup (OM) formulation \citep{Onsager53,Stratonovich60}.
We assign a probability $p\left[x\right]\D x$ to every path $x\left(t\right)$,
where we define the integral measure ${\cal D}x$ as
\[
\int{\cal D}x\,\ldots\coloneqq\lim_{M\rightarrow\infty}\int dx_{t_{0}}\ldots\int dx_{t_{M-1}}\ldots,
\]
where $t_{0},\ldots,t_{M-1}$ is a uniform discretization of the time
axis into segments of length $\Delta t$ that scales inversely with
$M$. We here stick to the Itô convention, which means that we evaluate
the integrand at the beginning $t_{i}$ of every subinterval $\left[t_{i},t_{i+1}\right)$.
For additive noise, as it appears in \eqref{eq:SDE_general}, all
choices for a discretization converge to the same limit \citep[chap. 4.3.6.]{Gardiner85}.
Furthermore, we define $p\left[x\right]=\frac{1}{{\cal Z}}\exp\left(S_{\mathrm{OM}}\left[x\right]\right)$
via
\begin{align}
S_{\mathrm{OM}}[x] & =-\frac{1}{2}\lim_{M\rightarrow\infty}\sum_{i=0}^{M-1}\Big(\frac{x_{i+1}-x_{i}}{\Delta t}-f\left(x_{i}\right)\Big)D^{-1}\,\Big(\frac{x_{i+1}-x_{i}}{\Delta t}-f\left(x_{i}\right)\Big)\Delta t\nonumber \\
 & =-\frac{1}{2}\int dt\,\big(\partial_{t}x-f\left(x\left(t\right)\right)\big)D^{-1}\,\big(\partial_{t}x-f\left(x\left(t\right)\right)\big)\label{eq:Def_OM-action}
\end{align}
\citep{Onsager53,Stratonovich60,Graham77_281,Hunt81_976,Weber17_046601},\footnote{Note that the notation as an integral is meant symbolically: For concrete
calculations of the path integral, one always has to use the discrete
version with a finite sum, perform the integrations and draw the limit
afterwards.} and 
\[
\mathcal{Z}^{-1}\coloneqq\int{\cal D}x\,\exp\left(S_{\mathrm{OM}}\left[x\right]\right)
\]
is chosen such that the probability $p\left[x\right]$ is properly
normalized. The probability of the occurrence of deviations from the
solution fulfilling $\partial_{t}x=f\left(x\right)$ are suppressed
exponentially. Allowing for arbitrary time-dependent solutions $x(t)$,
for example by fixing the initial point $x(0)=x_{0}$ and the final
point $x(T)=x_{T}$, $p[x]$ determines the probability for any path
between these points; applied to the dynamics of the membrane potential
of a neuron, it can be used to determine the probability to exceed
the firing threshold. The rate of escape is, to leading exponential
order, given by $p[x^{\ast}]$, where $x^{\ast}$ minimizes $S_{\mathrm{OM}}$.
A thorough discussion of this kind of setting is beyond the scope
of this work; good introductions can be found e.g. in \citep[section 10.5]{Altland01}
or \citep[section 5]{Bressloff09_1488}.

The moments of the ensemble of paths
\begin{equation}
\left\langle x(t_{1})\cdots x(t_{n})\right\rangle \coloneqq\int\mathcal{D}x\,p\left[x\right]\,x(t_{1})\cdots x(t_{n})\label{eq:correlation}
\end{equation}
can be expressed as functional derivatives with respect to $j\left(t\right)$
of the moment-generating functional{\footnotesize{}}\footnote{Note the sign convention of the action which is defined without a
minus sign in front. We will stick to this convention throughout this
paper.}
\begin{equation}
Z\left[j\right]\coloneqq\int\mathcal{D}x\,p\left[x\right]\,\exp\left(\int dt\,j\left(t\right)x\left(t\right)\right),\label{eq:def_Z}
\end{equation}
{\footnotesize{} }evaluated at $j\left(t\right)=0$. The cumulant
generating function (or Helmholtz free energy) 
\begin{align}
W & =\ln Z\label{eq:def_W_OM}
\end{align}
encodes the statistics in terms of cumulants, the derivatives of $W$.
This is more efficient than encoding with moments, because higher
order cumulants do not contain information already contained in lower
orders.

\subsection{Stochastic dynamics as a variational problem\label{subsec:Stochastic-dynamics-as-variational}}

With the expressions for the actions $S_{\mathrm{OM}}$ one can calculate
moments and cumulants of activity as derivatives of the respective
functionals $Z$ and $W$. For a self-consistent determination
of the mean activity, it is, however, beneficial to consider the variational
problem of some functional $\Gamma_{\mathrm{OM}}[x^{\ast}]$ that
assumes stationary points at the true mean value $\xmean\left(t\right)\equiv\langle x\left(t\right)\rangle$.
To calculate it, we then have to solve the so-called equation of state
$\frac{\delta}{\delta x^{\ast}}\Gamma_{\mathrm{OM}}=0$ self-consistently,
where $\frac{\delta}{\delta x^{\ast}}$ denotes a functional derivative.

Indeed such a functional is readily defined via the Legendre-Fenchel
transform
\begin{align}
\Gamma_{\mathrm{OM}}[x^{\ast}] & :=\sup_{j}\,j^{\T}x^{\ast}-W[j],\label{eq:Gamma_OM}
\end{align}
where $x^{\T}y=\int_{-\infty}^{\infty}x(t)\,y(t)\,dt$ denotes the
inner product with respect to time. The so-defined $\Gamma_{\mathrm{OM}}$
is the effective action (or Gibbs free energy) \citep{ZinnJustin96}\footnote{Generalizing this approach, we could also allow for a nonvanishing
source $j$, then minimizing $\Phi[x^{\ast};j]\coloneqq\Gamma_{\mathrm{OM}}[x^{\ast}]-j^{\T}x^{\ast}.$}. It is central to the study of phase transitions, which reduces to
finding the stationary points or regions of $\Gamma_{\mathrm{OM}}$
(see \citep{Vasiliev98}, i.p. Chapter 6). The variational formulation
naturally solves the problem that derivatives of $W$ become multi-valued
at first-order phase transitions; when $W$ has a cusp and thus the
system has multiple states with different values for the mean $\langle x\left(t\right)\rangle$
at the same set of parameters. We study an example where spontaneous
symmetry breaking causes such a cusp in \prettyref{subsec:rel_Gamma_OM_MSRDJ}.

Since $W[j]$ is convex down in $j$, the Legendre-Fenchel transform
in \eqref{eq:Gamma_OM} is well-defined. Note that the Legendre-Fenchel
transform is a generalization of the Legendre transform for cases
where $W$ has a nondifferentiable point $j_{c}$ (see also \prettyref{sec:Convexity-and-spontaneous-symm-breaking}
for a proof of convexity of $W$). In such a case the mean of the
field $\langle x\rangle$ takes different values if $j$ approaches
$j_{c}$ from the left or from the right; such systems show an abrupt
change of the solution as a function of some control parameter, such
as $j$; an example is multi-stability in an attractor network. At
the corresponding points $\langle x\rangle$, $\Gamma_{\mathrm{OM}}$
has a flat segment, but is continuously differentiable everywhere
\footnote{At least in physically reasonable settings: A discontinuity in the
derivative in $\Gamma$ means that $W$, in turn, would have a flat
segment. In such systems, changing the source field would not affect
the mean; also fluctuations would vanish completely.} and is thus analytically simpler than the nonanalytical $W$ (for
a more detailed discussion on convexity, spontaneous symmetry breaking,
and the necessity of the Legendre-Fenchel transform, see \prettyref{sec:Convexity-and-spontaneous-symm-breaking}).
Another favorable property of the effective action is that symmetries
of $S_{\mathrm{OM}}$ are also symmetries of $\Gamma_{\mathrm{OM}}$,
giving rise to Ward-Takahashi identities \citep{Amit84} and the study
of Goldstone fluctuations in symmetry broken states of systems that
admit a continuous symmetry.

The simplest approximation to the effective action is the tree-level
approximation. In correspondence to \eqref{eq:deterministic_diffeq}
we replace the integral over all configurations $x$ in the definition
of $Z$ in \eqref{eq:def_Z} by its supremum, which yields $W[j]\simeq\ln\,\sup_{x}\exp(S_{\mathrm{OM}}[x]+j^{\T}x)-\ln{\cal Z}\left[0\right]$.
The monotonicity of $\exp$ and the involution property of the Legendre
transform \eqref{eq:Gamma_OM} then yield
\begin{align}
\Gamma_{\mathrm{OM}}[x^{\ast}] & \simeq-S_{\mathrm{OM}}[x^{\ast}]+\mathrm{const.}\label{eq:Gamma_OM_tree}
\end{align}
The name ``tree-level'' comes from the fact that if expanded in
$x^{\ast}$, only diagrams of tree shape contribute (see e.g. \citet[p. 128]{ZinnJustin96}
or \citet[Appendix: Equivalence of loopwise expansion and infinite resummation]{Helias19_10416}).
In the evaluation of the integral, one therefore neglects all fluctuations.
Practically computing corrections in the OM formalism is complicated
by the action involving a second-order differential operator. In the
following section we review a formalism that circumvents this difficulty.

\subsection{Martin-Siggia-Rose-de Dominicis-Janssen path-integral\label{subsec:Martin-Siggia-Rose-de-Dominicis-Janssen}}

It has been realized by \citet{Martin73} that, computing response
functions simultaneously together with correlation functions \eqref{eq:correlation},
simplifies practical computations. This is achieved by introducing
a second field, the response field $\tx$ by expressing the OM action
with Gaussian noise \eqref{eq:Def_OM-action} as 
\begin{align}
S_{\mathrm{OM}}[x] & =\underset{\tx}{\mathrm{extremize}}\,S_{\mathrm{MSRDJ}}[x,\tilde{x}].\label{eq:relation_actions}
\end{align}
$S_{\mathrm{MSRDJ}}$ is the Martin-Siggia-Rose-de Dominicis-Janssen
(MSRDJ) action \citep{Martin73,dedominicis1976_247,janssen1976_377,DeDominicis78_4913},
defined on the $2M$-dimensional phase space as
\begin{align}
S_{\mathrm{MSRDJ}}[x,\tilde{x}]:= & \tilde{x}^{\T}\left(\partial_{t}x-f\left(x\right)\right)+\tilde{x}^{\T}\frac{D}{2}\tilde{x}.\label{eq:S_MSRDJ}
\end{align}
Alternatively, one may obtain this result by performing a Hubbard-Stratonovitch
transform, that is by using the identity $e^{-\frac{x^{2}}{2}}=\frac{1}{i\sqrt{2\pi}}\int_{-i\infty}^{i\infty}e^{\frac{\tilde{x}^{2}}{2}+\tilde{x}x}d\tilde{x}$
with the response field as an additional auxiliary variable $\tilde{x}$.
The MSRDJ formulation has the advantage that only a single first-order
differential operator in time appears.

As for the OM form, we define the cumulant generating functional in
the MSRDJ formalism as

\begin{equation}
W\left[j,\tj\right]=\ln\int\D x\:\int\D\tx\,\exp\left(S_{\mathrm{MSRDJ}}\left[x,\tx\right]+j^{\T}x+\tj^{\T}\tx\right).\label{eq:Cumu_gen_fct_MSRDJ}
\end{equation}
Compared to its OM-form, $W\left[j,\tj\right]$ in addition incorporates
the response properties of the system as differentiating once with
respect to $\tj$ and $j$ each, respectively, yields the response
function, the deviation of the mean of the process caused by a $\delta$-shaped
inhomogeneity. This follows from comparing \eqref{eq:S_MSRDJ} with
\eqref{eq:Cumu_gen_fct_MSRDJ} to see that $\tj$ can as well be regarded
as an inhomogeneity in the stochastic differential equation \eqref{eq:SDE_general}
of the form
\begin{equation}
dx(t)=\Big[f\left(x\left(t\right)\right)-\tj(t)\Big]\,dt+dW(t).\label{eq:SDE_general_with_response_source}
\end{equation}
This form also allows the extension of the MSRDJ formalism to non-Gaussian
noise: If the stochastic increments $W$ have a cumulant-generating
functional $W_{\xi}(j)$, the last term $\tilde{x}^{\T}\frac{D}{2}\tilde{x}$
in the action \eqref{eq:S_MSRDJ} becomes $W_{\xi}(-\tx)$ \citep{Chow15}.
The form \eqref{eq:SDE_general_with_response_source} also shows that
$W[j,\tj]$ is real for real-valued $j$ and $\tj$; this is because
\eqref{eq:Cumu_gen_fct_MSRDJ}, once $\tx$ is integrated, is identical
to the OM form \eqref{eq:def_W_OM}, $W[j]=\langle e^{j^{\T}x}\rangle\in\mathbb{R}$,
for $j\in\mathbb{R}$, where $x$ solves the SDE \eqref{eq:SDE_general_with_response_source}
and thus $x\in\mathbb{R}.$

 The effective action in the MSRDJ formalism $\Gamma\left[x,\tx\right]$
is defined in analogy to \eqref{eq:Gamma_OM} as

\begin{align}
\Gamma\left[x^{\ast},\tx^{\ast}\right] & =\tj^{\T}\tx^{\ast}+\sup_{j}\,j^{\T}x^{\ast}-W\left[j,\tj\right],\label{eq:Gamma_MSRDJ}
\end{align}
where $\tj$ is chosen such that the right hand side is stationary.
Since $W$ is convex down in $j$, taking the Legendre-Fenchel transform
with regard to $j$ is involutive; this even holds for $W[j,\tj]$
that are nondifferentiable in $j$. We show in \prettyref{sub:Convexity-MSRDJ}
that the transform from $j$ to $x^{\ast}$ renders the resulting
functional differentiable in $\tj$, given the system is in thermodynamic
equilibrium or given that linear response functions of cumulants
of arbitrary order exist.

Like $Z$ and $W$, $\Gamma$ contains the full information of the
system, including effects from noise-driven fluctuations. The definition
of $\Gamma$ as the Legendre transform of $W$ implies the identities
\begin{alignat}{1}
\Gamma_{x}^{\left(1\right)}\left[x^{\ast},\tx^{\ast}\right] & :=\frac{\delta}{\delta x^{\ast}}\Gamma\left[x^{\ast},\tx^{\ast}\right]=j,\label{eq:Def_eq_of_state}\\
\Gamma_{\tx}^{\left(1\right)}\left[x^{\ast},\tx^{\ast}\right] & :=\frac{\delta}{\delta\tx^{\ast}}\Gamma\left[x^{\ast},\tx^{\ast}\right]=\tj,\nonumber 
\end{alignat}
which are implicit equations for $x^{\ast}$ and $\tx^{\ast}$, the
equations of state. For the physically relevant value $j=0$ of the
source field, normalization in systems with conserved probability
implies that the first equation \prettyref{eq:Def_eq_of_state} always
admits a solution $\tx^{\ast}\equiv0$ (\prettyref{subsec:Appendix_Normalization-Deker-Haake}).
We further show in \prettyref{sub:Convexity-MSRDJ} that the second
equation is then equivalent to the requirement that the fluctuations
around the true mean value average to zero within the OM formalism;
the additional Legendre transform from $\tj$ to $\tx$, which is
not always well-defined \citep{Andersen00_1979}, can therefore be
regarded a formal step only that does not require convexity of $W[j,\tj]$
in $\tj$.

Approximating $\Gamma\simeq-S$ up to the tree level, reduces \eqref{eq:Def_eq_of_state}
to the naive mean-field approximation \eqref{eq:deterministic_diffeq}
showing their tight relation. The resulting path maximizes the probability.
Because it ignores fluctuations, we also refer to it as the saddle
point approximation, or the mean-field approximation.

The use of the MSRDJ formalism simplifies the calculations of the
effective action with respect to the OM formalism. As a consequence
of the response fields being only auxiliary variables, their expectation
values vanish, i.e. $\left\langle \tx^{n}\right\rangle =0\ \forall n$
for solutions with stationary statistics (for a proof see \citet{Coolen00_arxiv_II}
or \prettyref{subsec:Appendix_Normalization-Deker-Haake} here).

To see why the Legendre transform is closely linked with the construction
of self-consistent solutions for the mean values of the fields $x^{\ast}$
and $\tx^{\ast}$, it is instructive to rewrite \prettyref{eq:Gamma_MSRDJ}
analogously with $y=(x,\tx)$ and $k=(j,\tj)$ as

\begin{align}
\Gamma[y^{\ast}] & =-\ln\,\int_{y}\,\exp(S[y]+k^{\T}(y-y^{\ast}))\label{eq:Def_Gamma_alternative}\\
\text{with }\frac{\delta\Gamma}{\delta k} & \stackrel{!}{=}0.\nonumber 
\end{align}

\footnote{In the following the symbol ``$\stackrel{!}{=}$'' denotes ``is
supposed to equal''; that is, the argument of the function is to
be determined such that equality holds.}The latter condition enforces that $\langle y-y^{\ast}\rangle\stackrel{!}{=}0$,
so we integrate over ensembles of configurations that obey this constraint;
in other words, the mean values for both fields $x$ and $\tx$ take
the values given by the argument of $\Gamma$. The right hand side
of \prettyref{eq:Def_Gamma_alternative} will hence depend via $y^{\ast}$
on the self-consistently determined value. In \prettyref{subsec:Analyzing-bifurcations},
we will show that this step is crucial to study systems at bifurcations.

\subsection{Effective equation of motion, Vertex functions\label{subsec:Effective-equation-of-motion}}

To see how the equations of state \prettyref{eq:Def_eq_of_state}
lead to self-consistency equations, we expand $\Gamma$ around a reference
point $(\bar{x},\bar{\tx})$
\begin{align}
\Gamma\left[x^{\ast},\tx^{\ast}\right] & =\sum_{n=0}^{\infty}\sum_{m=0}^{\infty}\frac{1}{n!m!}\frac{\delta^{n+m}\Gamma}{\left(\delta x^{\ast}\right)^{n}\left(\delta\tilde{x}^{\ast}\right)^{m}}\left[\bar{x},\bar{\tilde{x}}\right]\,\delta x^{n}\delta\tx^{m},\label{eq:vertex_expansion_gamma}
\end{align}
where we introduced the derivatives $\frac{\delta^{n+m}\Gamma}{\left(\delta x^{\ast}\right)^{n}\left(\delta\tilde{x}^{\ast}\right)^{m}}$,
the vertex functions, as covectors and the deflections $\delta x(t):=x^{\ast}(t)-\bar{x}$
and $\delta\tilde{x}(t):=\tilde{x}^{\ast}(t)-\bar{\tilde{x}}$ together
with the notation
\begin{align}
 & \frac{\delta^{n+m}\Gamma}{\left(\delta x^{\ast}\right)^{n}\left(\delta\tilde{x}^{\ast}\right)^{m}}\left[\bar{x},\bar{\tx}\right]\,\delta x^{n}\delta\tx^{m}\nonumber \\
\coloneqq & \Pi_{i=1}^{n}\int dt_{i}\,\Pi_{j=1}^{m}\int ds_{j}\left.\frac{\delta^{n+m}\Gamma\left[x^{\ast},\tx^{\ast}\right]}{\delta x^{\ast}\left(t_{1}\right)..\delta x^{\ast}\left(t_{n}\right)\delta\tx^{\ast}\left(s_{1}\right)..\delta\tx^{\ast}\left(s_{m}\right)}\right|_{x^{\ast}=\bar{x},\,\tx^{\ast}=\bar{\tx}}\,\delta x\left(t_{1}\right)..\delta x\left(t_{n}\right)\,\delta\tx\left(s_{1}\right)..\delta\tx\left(s_{m}\right)\label{eq:convolution}\\
\eqqcolon & \Gamma_{\underbrace{x\ldots x}_{n\mathrm{-times}}\underbrace{\tx\ldots\tx}_{m\mathrm{-times}}}^{\left(n+m\right)}\,\delta x^{n}\delta\tx^{m}.\nonumber 
\end{align}
We determine the true mean values $\bar{x}$ and $\bar{\tx}$ of the
two fields by solving the two implicit equations $\Gamma^{\left(1\right)}\left[x^{\ast},\tx^{\ast}\right]\overset{!}{=}0$
(compare \eqref{eq:Def_eq_of_state}). All further Taylor coefficients
(or Volterra kernels) in \eqref{eq:vertex_expansion_gamma} have physical
meanings. $\Gamma_{\tx\tx}^{\left(2\right)}$ includes all corrections
to the Gaussian component of the noise and the mixed second-order
derivatives are the inverse of the response functions. Consequently,
$\Gamma^{\left(2\right)}$ contains the corrections to the second
cumulant since it is the inverse of $W^{\left(2\right)}$. The Taylor
coefficients of order $n$ describe the interdependence of measurements
at $n$ points in time. We make this more explicit by considering
$\Gamma_{\tx xx}^{\left(3\right)}$ in the second equation of state
$\Gamma_{\tx}^{\left(1\right)}\left[x^{\ast},\tx^{\ast}\right]=\tilde{j}(t)$
(see second line in \prettyref{eq:Def_eq_of_state}). We use the decomposition
of the effective action into the action and the fluctuation correction
$\Gamma=-S+\Gammafl$ and expand $\Gamma_{\tx}^{\left(1\right)}$
in a Volterra series as shown for $\Gamma$ in \prettyref{eq:vertex_expansion_gamma}.
Then, the second equation of state takes the form
\begin{align}
\tilde{j}(t)= & -\left(\frac{\partial}{\partial t}-f^{\prime}(\bar{x})\right)\delta x\left(t\right)-D\delta\tx\left(t\right)+\frac{1}{2}f^{\prime\prime}\left(\bar{x}\right)\delta x\left(t\right)\delta x\left(t\right)+\ldots\nonumber \\
 & +\int ds\,\Giitofl\left(t,s\right)\,\delta x\left(s\right)\nonumber \\
 & +\frac{1}{2}\int ds\int du\,\Giiitoofl\left(t,s,u\right)\,\delta x\left(s\right)\delta x\left(u\right)+\ldots.\label{eq:eq_of_state_deriv_tx}
\end{align}
The first equation of state \eqref{eq:Def_eq_of_state} $\Gamma_{x}^{\left(1\right)}=j\equiv0$
admits the solution $\tilde{x}^{\ast}=0$ if probability is conserved
(see, e.g., \prettyref{subsec:Appendix_Normalization-Deker-Haake}).
Thus, $\Gammafl$ accounts for the corrections due to the noise. Additionally,
we neglect all higher order terms as well as the remaining components
of $\Gamma^{\left(3\right)}$, which are subleading, as we discuss
in \prettyref{subsec:Functional-Renormalization-Group} after \prettyref{eq:flow_eq_x_star}.
Looking at the fluctuation corrections in \prettyref{eq:eq_of_state_deriv_tx},
we notice that in general the noisy system exhibits interactions that
are nonlocal in time (cf. \prettyref{eq:convolution}) even if the
deterministic system does not contain such terms. As a consequence,
it is not possible to define a potential for which $\partial_{t}x\left(t\right)=-\partial_{x}V\left(x\right)$
even if we set $\tx=0$, in contrast to the tree-level approximation.
The first occurrence of an effective equation of motion as \eqref{eq:eq_of_state_deriv_tx}
in the context of neuronal networks has been presented in \citep[eqs. (42) and (43)]{Buice07_051919}
using the Doi-Peliti formalism \citep{Doi76_1465,Peliti85_1469} applied
to Markovian systems with discrete state spaces.

 We call the derivatives appearing in \prettyref{eq:vertex_expansion_gamma}
``full vertices'' or ``full vertex functions'' as opposed to those
of the action $S$, which we refer to as ``bare vertices''. The
vertex functions do not only serve as means to calculate cumulants,
as we show next, but can also be interpreted directly. For example,
those with only one derivative with respect to $\tx$ (and at least
one with respect to $x$) can be seen as temporal kernels in an effective
differential equation for the mean \prettyref{eq:eq_of_state_deriv_tx}.
Vertices with more derivatives with respect to $\tx$ represent effective
noise terms. Henceforth, we will therefore focus our attention on
vertices to obtain effective descriptions of nonlinear stochastic
systems.

\subsection{Extracting statistical dependencies from vertex functions}

The computation of the effective action $\Gamma$ or its Taylor coefficients,
the vertex functions $\Gamma^{(n)}$, ultimately serves the goal to
compute observables, the statistics of $x$.

The second cumulant, the covariance $W^{(2)},$ obeys the relation
\begin{align}
\Gamma^{\left(2\right)}\left[x^{\ast},\tx^{\ast}\right] & =\begin{pmatrix}\frac{\delta^{2}\Gamma\left[x^{\ast},\tx^{\ast}\right]}{\delta x^{\ast2}} & \frac{\delta^{2}\Gamma\left[x^{\ast},\tx^{\ast}\right]}{\delta x^{\ast}\,\delta\tx^{\ast}}\\
\frac{\delta^{2}\Gamma\left[x^{\ast},\tx^{\ast}\right]}{\delta\tx^{\ast}\,\delta x^{\ast}} & \frac{\delta^{2}\Gamma\left[x^{\ast},\tx^{\ast}\right]}{\delta\tx^{\ast2}}
\end{pmatrix}=\begin{pmatrix}\frac{\delta^{2}W\left[j,\tj\right]}{\delta j^{2}} & \frac{\delta^{2}W\left[j,\tj\right]}{\delta j\,\delta\tj}\\
\frac{\delta^{2}W\left[j,\tj\right]}{\delta\tj\,\delta j} & \frac{\delta^{2}W\left[j,\tj\right]}{\delta\tj^{2}}
\end{pmatrix}^{-1}=\left[W^{\left(2\right)}\right]^{-1},\label{eq:Gamma2_inv_W2}
\end{align}
which follows by differentiating \prettyref{eq:Def_eq_of_state}.
Differentiating the latter relation $n-1$-times with respect to $j$,
using $\frac{\partial}{\partial j}=\frac{\partial x^{\ast}}{\partial j}\frac{\partial}{\partial x^{\ast}}=\left(\Gamma^{\left(2\right)}\right)^{-1}\frac{\partial}{\partial x^{\ast}}$,
and repeated application of \prettyref{eq:Gamma2_inv_W2} yields expressions
for the $n$-th cumulant, expressed in terms of derivatives of $\Gamma$
(see also \citep[p. 115ff]{NegeleOrland98}).

The resulting expressions have the form of tree graphs, with vertex
functions forming the nodes and edges formed by the full propagators
$\big(\Gamma^{(2)}\big)^{-1}$ (see, e.g., \citet[Section 6.3]{ZinnJustin96}
or \citet[Section XIII]{Helias19_10416}). The third-order cumulants,
for example, follow as 
\begin{align*}
W_{abc}^{(3)} & =-\sum_{a^{\prime}b^{\prime}c^{\prime}}\Gamma_{a^{\prime}b^{\prime}c^{\prime}}^{(3)}\big(\Gamma^{(2)}\big)_{a^{\prime}a}^{-1}\,\big(\Gamma^{(2)}\big)_{b^{\prime}b}^{-1}\,\big(\Gamma^{(2)}\big)_{c^{\prime}c}^{-1}\\
\\
 & =-\quad\Diagram{\vertexlabel^{a}\\
fd & p & f\vertexlabel_{c}\\
\vertexlabel_{b}fu
}
\quad.
\end{align*}

Depending on the choice of the sources $a,b,c$, we either get the
third-order cumulant of the variable $x$ (for $a=j(s)$, $b=j(t)$,
$c=j(u)$) or the second-order response kernel of the mean to a perturbation
of the system (for $a=j(s)$, $b=\tj(t)$, $c=\tj(u)$), or the change
of the autocorrelation due to a perturbation at linear order (for
$a=j(s)$, $b=j(t)$, $c=\tj(u)$). The combination with $a,b,c$
each equal to a $\tj$ vanishes identically in stationary states,
because the moments of $\tx$ all vanish (see \prettyref{sec:General-properties-of_MSRDJ-Gamma}).

Even though we here consider the dynamics of a single neuron, the
formalism transparently extends to compute higher order statistics
of neuronal activity also across different neurons; the only difference
being that the source fields and original field will obtain an additional
index that identifies the respective neuron. Computing such correlations
is a topic of considerable interest in neuroscience \citep{Schneidman06_1007}.

\subsection{Relation between the effective actions $\Gamma_{\mathrm{OM}}$ and
$\Gamma_{\mathrm{MSRDJ}}$\label{subsec:rel_Gamma_OM_MSRDJ}}

A notable advantage of the MSRDJ formalism is that it can be easily
generalized to arbitrary noise statistics. By integrating out the
response field $\tx$ we can define the corresponding OM action by
\[
\SOM\left[x\right]=\ln\,\int\D\tx\,\exp\left(\SMSRDJ\left[x,\tx\right]\right).
\]
We show in \prettyref{subsec:App_MSRDJ_OM_Gamma} that if the effective
actions exist in both formalisms, they are related by\textcolor{black}{
\begin{equation}
\Gamma_{\mathrm{OM}}[x^{\ast}]=\underset{\tx^{\ast}}{\mathrm{extremize}}\,\Gamma_{\mathrm{MSRDJ}}[x^{\ast},\tx^{\ast}].\label{eq:Rel_eff_act_MSRDJ_OM}
\end{equation}
}Choosing $\tx^{\ast}$ in \prettyref{eq:Rel_eff_act_MSRDJ_OM}
so that it extremizes $\Gamma_{\mathrm{MSRDJ}}$, conserves the full
information on the fluctuations. The more convenient MSRDJ formalism
can therefore be used to perform the actual computation and only subsequently
one obtains the physically and probabilistically interpetable OM form.

The definition \prettyref{eq:Rel_eff_act_MSRDJ_OM} hence makes the
physically interpretable OM effective action available even if an
OM action $S_{\mathrm{OM}}$ cannot be formulated in the non-Gaussian
case. Therefore, relating approximations of the respective effective
actions can be nontrivial: Cooper et al. derived that $\Gamma_{\mathrm{MSRDJ}}$
and $\Gamma_{\mathrm{OM}}$ of the KPZ model yield the same effective
equations if one performs a saddle-point approximation in the auxiliary
fields of the Hubbard-Stratonovitch transform of the non-Gaussian
parts of the respective actions \citep{Cooper16}. We here provide
a general relation between the two effective actions, that is valid
in full generality beyond specific approximations. It may therefore
be used to check whether a pair of approximations, each formulated
for one of the two effective actions, is equivalent. The finding by
Cooper et al. \citep{Cooper16} is one such pair of equivalent approximations.

As a minimal non-Gaussian example, we study the influence of a nonvanishing
third-order cumulant of the noise defined by its cumulant generating
function
\begin{equation}
W_{\xi}\left(y\right)=\frac{D}{2}y^{2}+\frac{\alpha}{3!}y^{3}+\mathcal{O}\left(y^{4}\right)\label{eq:Non_Gaussian_noise_cum_gen_fct}
\end{equation}
on the generalized OM-action, where we assume that $\frac{\alpha}{D^{2}}\ll1$
and that we can neglect all higher order terms $\mathcal{O}\left(y^{4}\right)$.
A straight-forward perturbation calculation in $\alpha$, shown in
\prettyref{subsec:App_S_OM_and_S_MSRDJ}, demonstrates that 
\begin{equation}
\SOM\left[x\right]=\frac{1}{2}\ln\left(\frac{2\pi}{D}\right)\underbrace{-\frac{\left(\dot{x}-f\left(x\right)\right)^{2}}{2D}-\frac{\alpha}{3!}\frac{\left(\dot{x}-f\left(x\right)\right)^{3}}{D^{3}}}_{=\underset{\tx}{\mathrm{extremize}}\ \SMSRDJ\left[x,\tx\right]+\mathcal{O}\left(\alpha^{2}\right)}+\frac{\alpha}{2D^{2}}\left(\dot{x}-f\left(x\right)\right)+\mathcal{O}\left(\alpha^{2}\right).\label{eq:Non_Gaussian_OM_action}
\end{equation}
So, while eq. \prettyref{eq:Rel_eff_act_MSRDJ_OM} holds for arbitrary
statistics of the noise, the analogous relation for the respective
actions \prettyref{eq:relation_actions} does not in this case. Therefore,
we encounter the interesting case that the effective action in the
OM formalism might be easier to determine than the corresponding action.
An example where the noise is non-Gaussian the is stochastic dynamics
of pulse-coupled (spiking) network models. Here typically the statistics
of the noise is close to the Poisson process \citep{Ocker17_1,Brinkman18_e1006490}.

A simple special case arises if the noise, including all fluctuation
corrections, remains Gaussian. The field $\tx$ then still appears
quadratically in the effective action. Extremizing with respect to
$\tx$ in \eqref{eq:Rel_eff_act_MSRDJ_OM} is then identical to performing
the integral over $\tx$. A corollary is that under these conditions,
the OM effective action has the same form as its tree-level approximation
\eqref{eq:Def_OM-action}, only with vertices $S^{(n)}$ replaced
by effective vertices $-\Gamma^{(n)}$. For example, the noise matrix
$D=S_{\tx\tx}^{(2)}$ is replaced by $-\Gamma_{\tx\tx}^{(2)}$, an
approximation that is valid if corrections of order $\mathcal{O}\left(\tx^{3}\right)$
are small.

\subsection{Loop expansion\label{subsec:Loop-expansion}}

The effective action characterizes the state of a stochastic system.
Fluctuations provide corrections to the effective action that are
commonly defined as $\Gamma=-S+\Gammafl$. A standard approach in
statistical physics and quantum field theory to obtain these corrections
is the loop expansion. For an introduction, consult for example the
books by Kleinert \citep[chap. 3.23.]{Kleinert09} or Zinn-Justin
\citep[chap. 7.7.ff]{ZinnJustin96}. This technique has first been
applied in the context of neuronal networks by Buice and Chow \citep{Buice07_031118};
see e.g. \citep{Hertz16_033001,Helias19_10416} for recent reviews.

We here briefly outline the loop expansion on the concrete example
for four reasons. First, all terms in the action are at least of linear
order in the response field $\tx$; its equation of state therefore
always admits a trivial solution $\txmean\equiv0$ and the $\tx$$\tx$-propagator
vanishes. These are features specific to the MSRDJ formalism that
deserve some comments. Second, to illustrate that the Feynman diagrams
(sometimes referred to as Mayer graphs \citep{Mayer77,Vasiliev98})
in the one-loop approximation are essentially the same as those that
appear in the less standard functional renormalization group method.
Third, the loop expansion gives us the leading order fluctuation corrections
beyond tree level, $\Gamma=-S$. It will allow us in subsequent sections
to derive an effective deterministic equation for the mean and a linear
convolution equation for the variance of the process. One-loop corrections
also show which new vertices are generated along the renormalization
group flow. And fourth, the loop expansion provides a systematic way
to derive self-consistency equations for the mean of a process, an
idea that is conceptually continued in the functional renormalization
group approach to arbitrary orders of the statistics.

In contrast to mean-field theory, the loopwise expansion can be improved
systematically because it is an expansion in a parameter that measures
the fluctuation strength, often related to the system size. Here,
we consider a one-dimensional system, but it also depends on a small
parameter that organizes the loop expansion. In \prettyref{subsec:Loop_exp_small_parameter}
we demonstrate that adding a loop to a given diagram introduces an
additional factor which equals the product of the strength of the
nonlinearity $\beta$ squared, see \prettyref{eq:def_f_l_beta}, and
the variance of the noise $D$. Thus, in our case the loop expansion
amounts to an expansion in terms of powers of $\beta^{2}D$.

Because the solvable part of our theory is Gaussian, we express
cumulants of higher order by cumulants of order two in $x$ and $\tx$.
We call

\begin{fmffile}{struct_man_propagator}
\fmfset{thin}{0.75pt}
\fmfset{decor_size}{2mm}
\fmfcmd{style_def wiggly_arrow expr p = cdraw (wiggly p); shrink (0.8); cfill (arrow p); endshrink; enddef;}
\fmfcmd{style_def majorana expr p = cdraw p; cfill (harrow (reverse p, .5)); cfill (harrow (p, .5)) enddef;
		style_def alt_majorana expr p = cdraw p; cfill (tarrow (reverse p, .55)); cfill (tarrow (p, .55)) enddef;}
\begin{align*}
	\Delta \left(t - s \right) &=
	\begin{pmatrix}
		\langle x\left(t\right)x\left(s\right) \rangle & \langle x\left(t\right)\tilde{x}\left(s\right) \rangle \\
		\langle \tilde{x}\left(t\right)x\left(s\right) \rangle & \langle \tilde{x}\left(t\right)\tilde{x}\left(s\right) \rangle 
	\end{pmatrix} =
	\begin{pmatrix}
		\hspace{0.5cm}	
		\parbox{20mm}{
		\begin{fmfgraph*}(40,20)
			\fmfbottomn{v}{2}
			\fmffreeze
			\fmfshift{(0.0,0.1w)}{v1}
			\fmfshift{(0.0,0.1w)}{v2}
			\fmfv{l=$x\left(t\right)$, l.a=90, l.d=0.04w}{v1}
			\fmfv{l=$x\left(s\right)$, l.a=90, l.d=0.04w}{v2}
			\fmf{alt_majorana}{v1,v2}
		\end{fmfgraph*}
		} &
		\parbox{20mm}{
		\begin{fmfgraph*}(40,20)
			\fmfbottomn{v}{2}
			\fmffreeze
			\fmfshift{(0.0,0.1w)}{v1}
			\fmfshift{(0.0,0.1w)}{v2}
			\fmfv{l=$x\left(t\right)$, l.a=90, l.d=0.02w}{v1}
			\fmfv{l=$\tilde{x}\left(s\right)$, l.a=90, l.d=0.02w}{v2}
			\fmf{plain_arrow}{v2,v1}
		\end{fmfgraph*}
		} \\
		\hspace{0.5cm}	
		\parbox{20mm}{
		\begin{fmfgraph*}(40,20)
			\fmfbottomn{v}{2}
			\fmffreeze
			\fmfshift{(0.0,0.1w)}{v1}
			\fmfshift{(0.0,0.1w)}{v2}
			\fmfv{l=$\tilde{x}\left(t\right)$, l.a=90, l.d=0.02w}{v1}
			\fmfv{l=$x\left(s\right)$, l.a=90, l.d=0.02w}{v2}
			\fmf{plain_arrow}{v1,v2}
		\end{fmfgraph*}
		} &
		0
	\end{pmatrix}
\end{align*}
\end{fmffile}the propagators of the theory, where we have chosen a representation
in time, but we will often switch to frequency space (and back). Further
ingredients are the bare interaction vertices which, in general, are
given by the nonquadratic components of the action, in our case the
Taylor coefficients of the term $\tx\,f\left(x\right)$. We provide
the translation between diagrammatic expressions and their algebraic
counterparts in \prettyref{tab:feyn_diag_elements}.
\begin{table}
\begin{tabular}{>{\centering}m{4cm}>{\centering}m{5cm}>{\centering}m{5cm}>{\centering}m{3cm}}
\toprule 
\addlinespace
graphical representation & algebraic term (loop expansion) & algebraic term (fRG) & meaning\tabularnewline\addlinespace
\midrule
\addlinespace
\addlinespace
\includegraphics[scale=0.8]{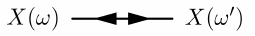} & $\Delta_{XX}^{0}\left(\omega,\omega^{\prime}\right)=\frac{2\pi D}{\omega^{2}+m^{2}}\delta\left(\omega+\omega^{\prime}\right)$ & $\Delta_{XX,\lambda}\left(\omega,\omega^{\prime}\right)=$$\left[\Gamma_{\lambda}^{\left(2\right)}\right]_{XX}^{-1}\left(\omega,\omega^{\prime}\right)$ & $xx$-component of the bare / full propagator\tabularnewline\addlinespace
\addlinespace
\addlinespace
\includegraphics[scale=0.8]{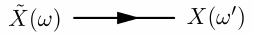} & $\Delta_{\tX X}^{0}\left(\omega,\omega^{\prime}\right)=\frac{2\pi}{i\omega+m}\delta\left(\omega+\omega^{\prime}\right)$ & $\Delta_{\tX X,\lambda}\left(\omega,\omega^{\prime}\right)=\left[\Gamma_{\lambda}^{\left(2\right)}\right]_{\tX X}^{-1}\left(\omega,\omega^{\prime}\right)$ & $\tx x$-component of the bare / full propagator\tabularnewline\addlinespace
\addlinespace
\addlinespace
\includegraphics[scale=0.8]{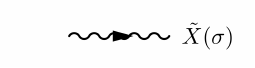} & - & - & external (amputated) leg\tabularnewline\addlinespace
\addlinespace
\addlinespace
\includegraphics[scale=0.8]{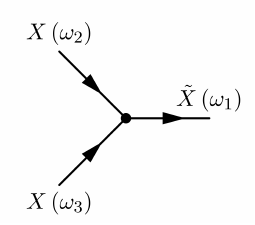} & $\frac{1}{1!2!}S_{\tX XX}^{\left(3\right)}\left(\omega_{1},\omega_{2},\omega_{3}\right)=-\frac{\beta}{\left(2\pi\right)^{2}}\delta\left(\omega_{1}+\omega_{2}+\omega_{3}\right)$ & $\frac{1}{1!2!}\Gamma_{\tX XX,\lambda}^{\left(3\right)}\left(\omega_{1},\omega_{2},\omega_{3}\right)$ & bare / full three-point interaction vertex\tabularnewline\addlinespace
\addlinespace
\addlinespace
\includegraphics[scale=0.8]{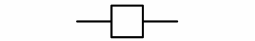} & - & $\frac{\partial\Rlbd}{\partial\lambda}$ & derivative of the regulator term\tabularnewline\addlinespace
\bottomrule
\addlinespace
\end{tabular}

\caption{\label{tab:feyn_diag_elements}Translation between graphical elements
of Feynman diagrams and corresponding algebraic terms. Amputated legs
do not introduce an additional factor in the algebraic expression
but indicate the value of the field $X$ (ingoing leg) or $\tilde{X}$
(outgoing leg) at which we evaluate the expression. A diagram contributing
to an $n$-point interaction $\frac{\delta^{n}}{\delta\protect\tX\left(\sigma_{1}\right)\cdots\delta X\left(\sigma_{n}\right)}\protect\Gammafl$
has $n$ external amputated legs where the number of ingoing and outgoing
legs corresponds to the number of functional derivatives with respect
to $X$ and $\protect\tX$. The generalization of the interaction
vertex to higher order interactions is straight forward (see \ref{sec:Effective-potential-in-bistable}
for an example). The prefactor is that of a usual Taylor expansion.}
\end{table}

Henceforth, we consider the corrections to the mean value, the variance
and one of the three-point vertices in the neuroscientific case, where
$f\left(x\right)=-x+J\,g\left(x\right)$. For small activities we
can expand the gain function and keep only its linear and quadratic
terms. This is in line with the observation that activation functions
are typically convex in the vicinity of the working point \citep{Roxin11_16217}.
We define 
\[
g\left(x\right)=x+\alpha x^{2}.
\]
Therefore, the only bare vertex is $S_{\tx xx}^{\left(3\right)}$.
We choose this quadratic nonlinearity also for pedagogical reasons,
as it constitutes the simplest nontrivial example which is suitable
to demonstrate the methods. For practical calculations it is convenient
to switch the parametrization to 
\begin{equation}
f\left(x\right)=-lx+\beta x^{2},\label{eq:def_f_l_beta}
\end{equation}
where $l=1-J>0$ and $\beta=\alpha J>0$. Then in frequency domain
the bare propagator reads
\begin{align}
\Delta^{0}(\omega,\omega^{\prime}) & =\left(-S^{\left(2\right)}\right)^{-1}=\left(\begin{pmatrix}0 & -i\omega+m\\
i\omega+m & -D
\end{pmatrix}\,\frac{\delta(\omega+\omega^{\prime})}{2\pi}\right)^{-1}\nonumber \\
 & =\begin{pmatrix}\frac{D}{\omega^{2}+m^{2}} & \frac{1}{-i\omega+m}\\
\frac{1}{i\omega+m} & 0
\end{pmatrix}2\pi\delta(\omega+\omega^{\prime}),\label{eq:propagator}
\end{align}
where $m=-l+2\beta x^{\ast}$ plays the role of a mass-like term in
the theory. For the definition of the Fourier transform, as we use
it throughout this paper, see \prettyref{subsec:App_def_fourier}.
Similarly, for the interaction vertex we obtain\begin{fmffile}{struct_man_int_vertex}
\fmfset{thin}{0.75pt}
\fmfset{decor_size}{2mm}
\fmfcmd{style_def wiggly_arrow expr p = cdraw (wiggly p); shrink (0.8); cfill (arrow p); endshrink; enddef;}
\fmfcmd{style_def majorana expr p = cdraw p; cfill (harrow (reverse p, .5)); cfill (harrow (p, .5)) enddef;
		style_def alt_majorana expr p = cdraw p; cfill (tarrow (reverse p, .55)); cfill (tarrow (p, .55)) enddef;}
\begin{align*}
	\parbox{20mm}{
	\begin{fmfgraph*}(40,40)
		\fmfleft{l1,l2}
		\fmfright{r}
		\fmftop{c}
		\fmffreeze
		\fmfshift{(0.0,-0.5w)}{c}
		\fmfshift{(0.0,0.1w)}{l1}
		\fmfshift{(0.0,-0.1w)}{l2}
		\fmfdot{c}
		\fmfv{l=$X\left(\omega_3\right)$, l.a=-90, l.d=0.04w}{l1}
		\fmfv{l=$X\left(\omega_2\right)$, l.a=90, l.d=0.04w}{l2}
		\fmfv{l=$\tilde{X}\left(\omega_1\right)$, l.a=90, l.d=0.04w}{r}
		\fmf{plain_arrow}{l1,c}
		\fmf{plain_arrow}{l2,c}
		\fmf{plain_arrow}{c,r}
	\end{fmfgraph*}
	}
&= \frac{1}{2}S^{\left(3\right)}_{\tilde{X}XX} = -\frac{\beta}{\left(2\pi\right)^2}\delta\left(\omega_1+\omega_2+\omega_3\right).
\end{align*}
\end{fmffile}The frequencies are conserved at each vertex and each propagator.
Since in the action $S\left[x,\tx\right]$ the function $f\left(x\right)$
is multiplied with one $\tx$, see \eqref{eq:S_MSRDJ}, we conclude
that the $n$-th Taylor coefficient of $f\left(x\right)$ leads to
an interaction vertex with $n$ incoming $x$-legs and one outgoing
$\tx$-leg.

\subsubsection*{One-loop correction to the mean value}

The first correction to the mean value is given by the contribution
known as the tadpole diagram that consists of one interaction vertex
whose two incoming legs are connected by an undirected propagator
($\Delta_{XX}$): \begin{fmffile}{struct_man_1l_tadpole}
\fmfset{thin}{0.75pt}
\fmfset{decor_size}{2mm}
\fmfcmd{style_def wiggly_arrow expr p = cdraw (wiggly p); shrink (0.8); cfill (arrow p); endshrink; enddef;}
\fmfcmd{style_def majorana expr p = cdraw p; cfill (harrow (reverse p, .5)); cfill (harrow (p, .5)) enddef;
		style_def alt_majorana expr p = cdraw p; cfill (tarrow (reverse p, .55)); cfill (tarrow (p, .55)) enddef;}
\begin{align}
	\Gamma_{\tilde{X},\mathrm{fl.}}^{\left(1\right)}\left(\sigma\right) &= \left(-1\right)
	\parbox{20mm}{
	\begin{fmfgraph*}(40,40)
		\fmfleft{l}
		\fmftop{c}
		\fmffreeze
		\fmfshift{(0.0,-0.5w)}{c}
		\fmfdot{c}
		\fmf{wiggly_arrow}{c,l}
		\fmf{alt_majorana, tension=0.55}{c,c}
	\end{fmfgraph*}
	}
	\label{struct_man_1l_tadpole} \\
&= \frac{2\pi\beta}{\left(2\pi\right)^2}\int d\omega\,\frac{D}{\omega^2 + m^2} \delta\left(\sigma\right) = \frac{\beta D}{2|m|}\delta\left(\sigma\right).\nonumber
\end{align}
\end{fmffile}Henceforth, the $\sigma_{i}$ denote the external frequencies of the
derivatives of the effective action, which are represented by wiggly
lines. In \prettyref{subsec:App_loop_expansion_vertices_parameter}
we provide a brief summary of how to evaluate the Feynman diagrams
that we use in this paper. Intuitively, the above diagram represents
the na\"{i}ve estimation for the correction to the mean \eqref{eq:naive_correction_mean}
that is obtained by taking the expectation value of the quadratic
nonlinearity (vertex with two incoming legs) over the Gaussian fluctuations
around the stationary mean (propagator connecting these two legs).
A conceptual difference is, though, that the mean value $\xmean$
here affects the point about which we perform the linear response
approximation, thus it appears on the right-hand side through the
value of the mass term $m\left(\xmean\right)$ in the propagators:
The approximation is hence self-consistent in the mean, defining $\xmean$
as the solution of the equation of state \eqref{eq:Def_eq_of_state}
as
\begin{align}
0=\tj & =-S_{\tx}^{(1)}+\Gamma_{\mathrm{fl},\tx}^{(1)}\nonumber \\
 & =f\left(\xmean\right)+\frac{\beta D}{2\,|m\left(\xmean\right)|}\label{eq:eq_state_j_tilde}\\
 & =-l\xmean+\beta\xmean^{2}+\frac{\beta D}{2|-l+2\beta\xmean|}.
\end{align}
But here, too, we take into account the nonlinearity only by considering
its effect of shifting the mean value and therefore the linear order
($-l\rightarrow m\left(\xmean\right)$), as in our na\"{i}ve approximation
in linear response in \prettyref{eq:linear_response_fluct}, and then
calculate the expectation value of the nonlinearity using this approximation.
This contribution is indeed the only one to consider at this loop
order as we demonstrate in \prettyref{subsec:App_loop_expansion_vertices_parameter}.
The result for the corrected mean value as a function of the strength
of the nonlinearity is shown in \prettyref{fig:Mean-variance-propagators}A.
Due to conservation of frequencies at the vertex, there are only corrections
for the zero frequency mode of the mean value.

\subsubsection*{One-loop correction to the variance and higher order statistics}

\begin{figure}
\begin{centering}
\includegraphics[width=1\textwidth]{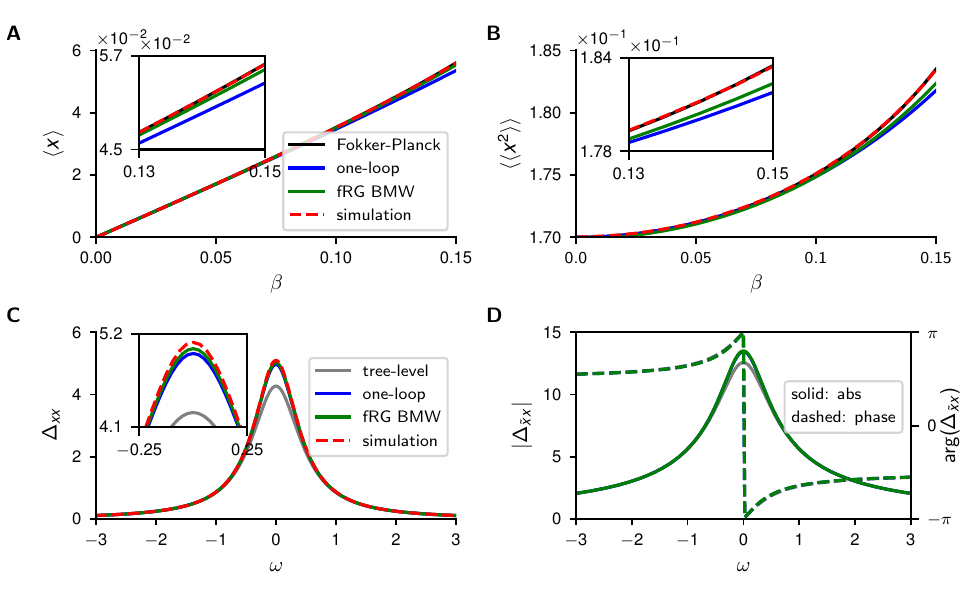}
\par\end{centering}
\caption{Mean (\textbf{A}) and variance (\textbf{B}) as functions of the strength
$\beta$ of the nonlinearity for different methods. Parameters: $l=0.5$,
$D=0.17$. \textbf{C} Power spectrum of the system from simulations
compared to different approximations. \textbf{D} Absolute value (solid
lines) and phase (dashed lines) of the response function. Its zero-mode
corresponds to the integrated response of a neural network to a delta-shaped
perturbation, for other frequencies the response is weighted according
to the respective mode. The results of the one-loop and the fRG BMW
approximation coincide at this resolution. For comparison between
simulations and theory results of the response of the system was subject
to small (but not infinitesimally small) perturbations; see \prettyref{subsec:time_dep},
\prettyref{fig:relaxation}. Parameters for (C) and (D): $l=0.5$,
$D=0.17$, and $\beta=0.15$.\label{fig:Mean-variance-propagators}}
\end{figure}

To compute the one-loop corrections to the variance, we determine
the corrections to the propagator by using the relation $\Delta=\left(\Gamma^{\left(2\right)}\right)^{-1}$\prettyref{eq:Gamma2_inv_W2}.
The first-order correction to $\Gamma^{\left(2\right)}$ is given
by the sum of all one-loop diagrams with two external legs. The diagrams
and the corresponding expressions can be found in \prettyref{subsec:one_loop_propagator}
and finally yield
\[
\Delta_{xx}\left(t\right)=-\left[\frac{D\left(m_{1}^{2}-\left(2m\right)^{2}+A\right)}{2\left(m_{1}^{2}-m_{2}^{2}\right)m_{1}}e^{m_{1}\lvert t\rvert}+\frac{D\left(\left(2m\right)^{2}-m_{2}^{2}-A\right)}{2\left(m_{1}^{2}-m_{2}^{2}\right)m_{2}}e^{m_{2}\lvert t\rvert}\right],
\]
for the correlation function, where
\begin{align*}
A & =2\beta^{2}D/m\\
m & =-l+2\beta\xmean\\
m_{1/2} & =3m/2\pm\sqrt{m^{2}/4-A}.
\end{align*}
The value at $t=0$ equals the variance $\llangle x\rrangle$ which
is plotted in \prettyref{fig:Mean-variance-propagators}B. In this
figure we compare the mean value and the variance computed using various
methods, where we use the solution by the Fokker-Planck equation \citep{Risken96}
as the ground truth, which indeed agrees very well with the simulated
results. In particular, we observe that the one-loop approximation
is markedly better than the tree-level approximation, which predicts
$\left\langle x\right\rangle =0$ and $\llangle x^{2}\rrangle=0.17$
for all $\beta$. Panel C and D show the power spectrum ($\Delta_{xx}\left(\sigma\right)$)
and the response function ($\Delta_{\tx x}\left(\sigma\right)$) of
the system. Like the correction to the tree-point vertex (compare
\prettyref{subsec:one_loop_three_point_vertex}), the corrections
to the propagator are largest for $\sigma_{i}=0$ and decay algebraically
with increasing frequency. In particular the fluctuation corrections
to $\Delta_{\tx x}\left(\sigma\right)$ lead to an elevation of the
power at low frequencies. This qualitative effect depends only on
coarse features of the system, such as the convexity of the nonlinearity.
One may therefore expect qualitatively similar features in networks
which operate in regimes in which the neuronal transfer function is
expansive, as typical for the low rate regimes of the cortex \citep{Roxin11_16217}.

In principle we could try to find a more accurate approximation of
the effective action by going to higher loop orders. However, this
already becomes unwieldy at the next order because the number of diagrams
quickly increases and the integration of the loop momenta becomes
numerically expensive. Instead, we will use renormalization group
techniques to obtain self-consistency at arbitrary levels of the statistics.

Even though the one-loop calculations are not confined to a certain
parameter range, we may not disregard the fact that strictly speaking
there is no stationary solution for our particular choice for $g\left(x\right)$,
because $x$ escapes towards infinity almost surely for $t\rightarrow\infty$.
However, for finite times and $l/\beta\gg\sqrt{D}$, the second unstable
fixed point $x_{1}=l/\beta$ is far away from the stable fixed point
$x_{0}=0$ measured in units of the fluctuations. Therefore, the escape
probability is negligible and we confine our analysis to this case
by effectively setting the escape probability to zero; in particular
for the otherwise exact Fokker-Planck solution to which we compare
the results of the loop expansion. As we show in \prettyref{subsec:App_parameter-reduction},
the position $x_{1}$ of the second unstable fixed point and the strength
of the nonlinearity are controlled by the same effective parameter.
In the presented example one therefore cannot increase the magnitude
of fluctuation corrections without also increasing the escape rate.
\prettyref{fig:Critical-point-bistable} shows a different system
that has notable fluctuation corrections. A rigorous analysis of the
complete setting including escape is a problem on its own requiring
the introduction of a probability for a path conditioned on the requirement
that it has not escaped to infinity. For a leak term equal to zero,
this analysis has recently been performed \citep{Ornigotti18_032127}.
Another possibility is to consider the time-dependent problem, as
is done, for example, in the context of laser physics \citep{Filip16_065401}.

The stochastic differential equation considered here is similar to
a typical differential equation describing the evolution of the membrane
potential of a neuron fed by fluctuating input from the network. Examples
are the quadratic integrate-and-fire neuron \citep{Latham00a} or
the exponential integrate-and-fire neuron \citep{Brette-2005_3637}.
The escape of the dynamical variable across the second, unstable fixed
point here denotes the firing of the cell. The biophysical mechanism
of the repolarization subsequently resets the membrane potential to
a low value after the escape. The firing rate of a model can in principle
be computed by determining this probability of escape. One-loop corrections
to the escape probability, however, require the computation of Gaussian
fluctuations around the most likely escape path; the path itself is
here a function of time. Thus it is more complicated than determining
the stationary statistics considered here (see, e.g., \citep{Elgart04_041106}
for more details on rare events in meta-stable systems). 

We will demonstrate in the next chapter by comparing to simulations
that neglecting the possibility of escaping is justified in the presence
of a leak term for sufficiently low noise and nonlinearity.

\subsection{Time dependence of statistics\label{subsec:time_dep}}

Applications often also require the study of the time-dependent response
of a system. In the context of neuronal networks, for example, we
would like to quantify the response of the system to an applied stimulus.
It is \textit{a priori} not clear what the effect of noise is for
a response that is driven by a transient stimulus. The simplest approximation
\prettyref{eq:linear_response_fluct} that neglects the effect of
the noise also provides us with the lowest order approximation of
such a response. A special ``input'' is the noise-mediated influence
of the past of the mean value of $x$ on itself. The following example
illustrates how the effective equation of motion \prettyref{eq:eq_of_state_deriv_tx}
with vertex functions computed in one-loop approximation explains
the nontrivial interplay of noise and nonlinearities.

\paragraph{}

\paragraph*{Relaxation of a small departure from the mean\protect \\
}

As an example, we consider the response of the system to a stimulus,
represented by the deflection of the system from its mean value at
time $t_{0}=0$ by setting $\tilde{j}(t)=-\delta x(0)\,\delta(t)$
in \eqref{eq:eq_of_state_deriv_tx} and examine the equation of motion
that describes its relaxation back to the baseline. We derive the
equation of motion by solving the second equation of state, for example
in the form of \eqref{eq:eq_of_state_deriv_tx}, for $\partial_{t}\delta x\left(t\right)$.
We ensure that we consider only nonescaping trajectories by setting
$\tx\equiv0$ (see \prettyref{subsec:Appendix_Normalization-Deker-Haake}).
By inserting the tree-level approximation $\Gamma_{\mathrm{fl.}}=0$
into \eqref{eq:eq_of_state_deriv_tx}, we obtain 
\[
\partial_{t}\delta x\left(t\right)=-l\delta x\left(t\right)+\beta\delta x\left(t\right)^{2},
\]
which we can integrate analytically
\begin{align*}
\delta x\left(t\right) & =\frac{c}{e^{lt}+\frac{c\beta}{l}\left(1-e^{lt}\right)}=ce^{-\left(l-c\beta\right)t+\mathcal{O}\left(\left(lt\right)^{2}\right)}+\mathcal{O}\left(\left(\frac{\beta c}{l}\right)^{2}\right),\,\mathrm{where}\,c=\delta x\left(0\right).
\end{align*}
We notice that the second term in the denominator is due to the nonlinearity
and leads to a slower relaxation of the system back to its mean value
compared to the time constant $l^{-1}$ of the linear part of the
dynamics. In the previous section \prettyref{subsec:Loop-expansion}
we computed the one-loop corrections to $\Gammafl$ which we can insert
into \eqref{eq:eq_of_state_deriv_tx} to obtain a one-loop approximation
of the equation of motion
\begin{align}
\partial_{t}\delta x\left(t\right)= & m\,\delta x\left(t\right)+\beta\,\delta x\left(t\right)^{2}\nonumber \\
 & -\frac{2\beta^{2}D}{m}\int_{t_{0}}^{t}dt^{\prime}\,\underbrace{e^{2m\left(t-t^{\prime}\right)}}_{\propto\Giitofl\left(t,t^{\prime}\right)}\,\delta x\left(t^{\prime}\right)\nonumber \\
 & -\frac{8\beta^{3}D}{m}\int_{t_{0}}^{t}dt^{\prime}\int_{t_{0}}^{t}dt^{\prime\prime}\,\underbrace{H\left(t^{\prime}-t^{\prime\prime}\right)e^{2m\left(t-t^{\prime\prime}\right)}}_{\propto\Giiitoofl\left(t,t^{\prime},t^{\prime\prime}\right)}\delta x\left(t^{\prime}\right)\delta x\left(t^{\prime\prime}\right),\label{eq:IDE_noisy_relaxation_one_loop}
\end{align}
where $H\left(t\right)$ denotes the Heaviside step function. Let
us inspect the single terms in \eqref{eq:IDE_noisy_relaxation_one_loop}
in more detail: The first line is the tree-level contribution. $\Gamma_{\mathrm{fl.}}^{\left(2\right)}$
mediates a linear self-feedback for the departure $\delta x$ of the
process from its stationary value. One of the interpretations of $\Gamma_{\tx xx}^{\left(3\right)}$,
writing it as $\Gamma_{\tx xx}^{\left(3\right)}\left(t,t^{\prime},t^{\prime\prime}\right)=\frac{\delta}{\delta x\left(t^{\prime\prime}\right)}\Gamma_{\tx x}^{\left(2\right)}\left(t,t^{\prime}\right)$,
is that it quantifies the change of the linear response kernel of
the self-energy at times $t,t^{\prime}$ due to a change of the activity
at time $t^{\prime\prime}$. This shows that only the interplay between
an interaction and noise, as apparent by the prefectors composed of
both $\beta$ and $D$, creates a self-interaction of the mean that
is nonlocal in time. This phenomenon is also generally observed if
certain degrees of freedom are implicitly taken into account to describe
the quantity of interest (see \citep{Zwanzig01}, e.g., e.g. chap.
1.6).

\begin{figure}
\begin{centering}
\includegraphics[width=1\textwidth]{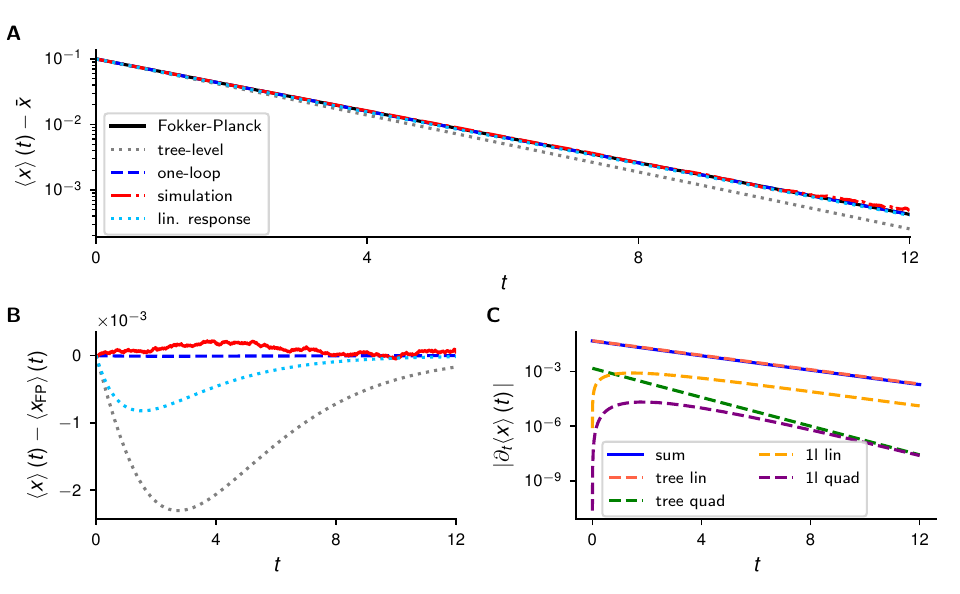}
\par\end{centering}
\caption{Relaxation after deflection described by effective equation of motion.
The one-loop result is given by \prettyref{eq:IDE_noisy_relaxation_one_loop}
and the Fokker-Planck result by the coupled differential equations
\prettyref{eq:eom_fp_mean} and \prettyref{eq:eom_fp_var}. For the
simulations, we let the activity of the system decay to its baseline
and then stimulate it by a small perturbation $\delta x_{0}$ applied
to the current value of $x$. Subsequently, the activity relaxes back
to its stationary state. \textbf{A} Mean activity averaged over $2.9\cdot10^{9}$
trials of the relaxation process. \textbf{B} Difference between the
various approximations and the simulation to the Fokker-Planck result.
The linear response (dotted light blue) contains solely the terms
in \prettyref{eq:IDE_noisy_relaxation_one_loop} that are linear in
$\delta x$. \textbf{C} Different contributions to the right hand
side of \eqref{eq:IDE_noisy_relaxation_one_loop}. Sum shows the full
right hand side (blue); linear part in $\delta x$ of the tree-level
term (dashed salmon pink); quadratic part in $\delta x$ of the tree-level
terms (dashed green); linear term of one-loop correction (dashed yellow);
quadratic term of one-loop corrections (dashed violet). \label{fig:relaxation}}
\end{figure}

An alternative way to arrive at \prettyref{eq:IDE_noisy_relaxation_one_loop}
is to derive ODEs for the first two moments from the Fokker-Planck
equation \citep{Risken96} and to use a Gaussian closure. The loop
expansion then amounts to a Taylor expansion of the Fokker-Planck
solution in $\delta x$ and assuming that $\llangle x^{3}\rrangle\ll\{\llangle x^{2}\rrangle\llangle x\rrangle,\,\llangle x\rrangle^{3}\}$
(known as Gaussian closure), as we show in detail in \prettyref{subsec:Appendix_time-dep_FP}.
In \prettyref{fig:relaxation} we compare the full Fokker-Planck solution
with Gaussian closure to the one-loop result. Indeed, the semilogarithmic
plot of the relaxation as a function of time shows an elevated time
constant due to fluctuations compared to the tree-level approximation.
Analyzing the origin of the elevated time constant, \prettyref{fig:relaxation}C
compares the different contributions to the right-hand side of \eqref{eq:IDE_noisy_relaxation_one_loop}.
The linear part of the tree-level approximation yields the largest
contribution. For sufficiently long times, we find that the linear
part of the one-loop correction comes next with opposite sign compared
to the term stemming from the bare interaction: This shows that only
the cooperation of the nonlinearity with the fluctuations in the system
causes this correction and is more important than the nonlinearity
itself, the quadratic tree-level contribution. Furthermore, the three-point
interaction gets enhanced by the fluctuation corrections, so that
in total, the relaxation is slower than in the deterministic system.

In the context of neuronal dynamics this example shows how the response
of a system to a stimulus is shaped by the presence of nonlinearities
and noise. An increase of the timescale of the response may be employed
by such systems to implement increased memory for past stimuli. In
heterogeneous networks, where the linear part of the dynamics is given
by a matrix that couples different neurons as $dx(t)=-x(t)+J\,x(t)\,dt+\ldots$,
the effective leak term $m=J-1$ correspondingly becomes a matrix.
The $i$-th eigendirection of the matrix then evolves with timescale
$(\lambda_{i}-1)^{-1}$. For random connectivity, the eigenvalues
are typically circularly distributed in the complex plane \citep{Sommers88};
thus a quasi-continuum of timescales appears already in tree-level
approximation \citep[Fig 6]{Dahmen16_031024}. The one loop corrections
to the self-energy generate additional timescales, as shown in eq.
\eqref{eq:IDE_noisy_relaxation_one_loop} and also \prettyref{eq:prop_1l_time_txx}
and \prettyref{eq:prop_1l_time_xx} for the one-dimensional case.
The emergence of multiple timescales has been discussed previously
in the literature. Some works proposed multiple adaptation mechanisms
as the origin \citep{LaCamera06_3448}. Others have shown that time
scales of responses to stimuli may change systematically within the
hierarchy of a complex neuronal network, from fast time-scales in
early areas, to long ones in higher levels of the hierarchy \citep{Chaudhuri2015_419}.
Loop corrections obtain an important meaning when considering the
influence of nonrecorded neurons on the correlation structure of the
observed ones \citep{Brinkman18_e1006490}: A one-loop correction
to the self-energy in this context contains the reverberation of activity
within the network, thus including the indirect effect due to the
presence of nonrecorded neurons. Such reverberations generate additional
timescales in the mutual coupling kernel between individual neurons,
mediated via the nonrecorded intermediate cells.

\paragraph*{Power spectrum}

So far we have discussed an effective equation of motion for the mean
value of the process that also allowed us to obtain an interpretation
for the various vertex functions. We can ask a corresponding question
for the second-order statistics, namely, whether there is a linear
system that possesses the same second-order statistics as the full
nonlinear system. Such a reduction may be useful to obtain insights
into the structure of network fluctuations and also to reduce the
complexity of stochastic nonlinear systems to simpler, linear ones.

Indeed, we can use the expression for the Hessian \prettyref{eq:Gamma2_inv_W2}
of the effective action to obtain the action of a linear system that,
up to second order, reproduces the stationary statistics of the full
system. To this end, we define 
\begin{equation}
S_{\mathrm{lin}}:=-\left(\delta\tilde{x}^{\T}\Gamma_{\tx x}^{\left(2\right)}\delta x+\frac{1}{2}\delta\tx^{\T}\Gamma_{\tx\tx}^{\left(2\right)}\delta\tx\right).\label{eq:Gamma2_lin}
\end{equation}
The corresponding equation of motion reads
\begin{align}
\frac{d}{dt}\delta x\left(t\right) & =-l\delta x\left(t\right)+\int dt^{\prime}\Gamma_{\tx x,\mathrm{fl.}}^{\left(2\right)}\left(t,t^{\prime}\right)\delta x\left(t^{\prime}\right)+\xi\left(t\right),\label{eq:eom_linear_sys}\\
\mathrm{where}\,\langle\xi\left(t\right)\rangle & =0\:\nonumber \\
\mathrm{and}\:\langle\langle\xi\left(t\right)\xi\left(t^{\prime}\right)\rangle\rangle & =D\delta\left(t-t^{\prime}\right)+\Gamma_{\tx\tx,\mathrm{fl.}}^{\left(2\right)}\left(t,t^{\prime}\right).
\end{align}
By construction this stochastic integro-differential equation \eqref{eq:eom_linear_sys}
reproduces the stationary variance $\Delta_{xx}(t,s)=W_{xx}^{(2)}(t,s)=\langle\delta x(t)\delta x(s)\rangle$
as well as the linear response $\Delta_{x\tx}(t,s)=W_{x\tx}^{(2)}(t,s)=\langle\delta x(t)\delta\tx(s)\rangle$
of the full system, because the solution of the Gaussian system \prettyref{eq:Gamma2_lin}
implicitly inverts the kernel $\Gamma^{(2)}$, which, by \prettyref{eq:Gamma2_inv_W2},
yields the covariance matrix $W^{(2)}$. We could also take into account
the effect of transient values of $x\left(t\right)$ on the variance
to obtain a corresponding reduction that is valid in the nonstationary
case. For this, however, we would need to know $\Gamma_{\tx x,\mathrm{fl.}}^{\left(2\right)}$
evaluated at arbitrary $x\left(t\right)$. In this case, it is therefore
more convenient to use the second Legendre transform that treats the
variance as given, just like $x^{\ast}$ in the case of the first
Legendre transform, as shown by Bravi and co-workers \citep{Bravi16}.

The construction of a linear system leads to a new perspective of
the effect of the nonlinearity: Up to the second cumulant, the nonlinear
system is equivalent to a linear one with a specific causal memory
kernel and a corresponding nonwhite Gaussian noise term, caused by
the self-energy correction $\Gamma_{\tx\tx}^{(2)}$.

\textbf{}

\subsection{Functional Renormalization Group\label{subsec:Functional-Renormalization-Group}}

The loopwise expansion, by virtue of approximating the effective action,
yields self-consistent equations for the mean. But we saw above that
also the second-order statistics and the higher order vertex functions
experience fluctuation corrections. One would therefore like to have
a scheme that is self-consistent with regard to these higher order
vertex functions, too.

One possible approach that has lead to reasonable results, is to correct
the mean, the propagator and the interaction vertex by the one-loop
results and therein replace the bare quantities by the corrected ones
to gain an even better approximation. This procedure is repeated until
the result eventually converges. This approach corresponds to taking
into account only specific diagrams with infinitely many loops and
is called self-consistent one-loop approximation. It typically corrects
the mean value and the self-energy while keeping the interaction vertices
at their bare values; it is then known as the ``Hartree-Fock approximation''
\citep{Vasiliev98,Hertz16_033001}. But of course this scheme can
be extended to arbitrary order of the vertex functions. A formal way
to derive such approximations systematically is by multiple Legendre
transforms, an idea going back to the seminal work by \citet{DeDominicis64_14}:
One re-expresses interaction potentials in terms of connected correlation
functions. Parquet equations are, for example, obtained by the fourth
Legendre transform of an even theory \citep[see e.g. ][for a review, especially chap. 6.2.10]{Vasiliev98}.

We here want to follow a different scheme that is inherently self-consistent
to arbitrary desired orders, the functional renormalization group
(fRG). The fRG scheme naturally takes into account fluctuation corrections
by renormalizing the mean value, the propagators, and all interaction
vertices simultaneously.

Technically, the functional renormalization group (fRG) \citep{WETTERICH93_90}
is an alternative way to calculate $\Gamma$. It is one of the exact
renormalization group (eRG) schemes \citep{Morris94_2411,Bagnuls01_91,Berges02_223},
in essence going back to the seminal work by \citet{Wegner73_401}.
It does not rely on an expansion in a small parameter, in contrast
to the loopwise expansion, but it is nevertheless represented by diagrams
with a one-loop structure. The technique induces an infinite hierarchy
of coupled differential equations for $\Gamma^{(n)}$, so that in
practice, we have to apply approximations, typically by truncating
the hierarchy. Yet, this technique is, as all exact renormalization
group schemes, exact only on the level of the full functional $\Gamma\left[x,\tx\right]$
and not for a particular truncation in terms of a subset of $\Gamma^{(n)}$.

The essential technical trick of the fRG is to simplify the theory
by adding an initially large quadratic term $-\frac{1}{2}y^{\T}R_{\lambda}y$
in the fields $y$, parametrized by the regulator $\Rlbd$, to the
action. It is a differentiable function of a so called flow parameter
$\lambda$ and can be chosen arbitrarily up to the following properties:
\begin{align}
\lim_{\lambda\rightarrow\Lambda}|R_{\lambda}| & =\infty\ \mathrm{and}\ \lim_{\lambda\ssearrow0}\Rlbd=0.\label{eq:Defining_properties_regulator}
\end{align}
The first property ensures that the theory for $\lambda=\Lambda$
has no fluctuations and its vertices correspond to the ones of the
bare action, while for $\lambda=0$, the original system is recovered.
For systems that exhibit symmetries it is often necessary that the
regulator is consistent with these symmetries of the effective action,
so that they are conserved during the flow. To interpolate between
the two limits of the noninteracting and the full system, a functional
differential equation for the effective action is derived by differentiating
with respect to $\lambda$. This is the Wetterich equation \citep{WETTERICH93_90},
whose derivation for our setting, in particular for the presence of
the response field, we will sketch in the following adhering to the
conventions of \citet{Berges02_223}. We will derive it for the effective
action evaluated at stationary $X^{\ast}$ and $\tX^{\ast}$ so that
the resulting equation boils down to an ODE.

Since the regulator $\Rlbd$ is intended to suppress fluctuations,
it is sufficient for our case to add it to the off-diagonal terms
of the free part of the action, defining
\begin{align}
S_{\lambda}\left[X,\tilde{X}\right]= & S_{0}\left[X,\tilde{X}\right]+\Delta S_{\lambda}\left[X,\tilde{X}\right]+S_{\mathrm{int}}\left[X,\tilde{X}\right],\quad\mathrm{where}\nonumber \\
\Delta S_{\lambda}\left[X,\tilde{X}\right]= & -\frac{1}{2}\int\frac{d\omega}{2\pi}\,\begin{pmatrix}X\left(-\omega\right)\\
\tilde{X}\left(-\omega\right)
\end{pmatrix}\begin{pmatrix}0 & \frac{1}{2}R_{\lambda}\\
\frac{1}{2}R_{\lambda} & 0
\end{pmatrix}\begin{pmatrix}X\left(\omega\right)\\
\tilde{X}\left(\omega\right)
\end{pmatrix}\label{eq:action_regulator}\\
S_{\mathrm{int}}\left[X,\tilde{X}\right]= & -\beta\int\frac{d\omega}{2\pi}\,\int\frac{d\omega^{\prime}}{2\pi}\,\tilde{X}\left(\omega\right)X\left(\omega^{\prime}\right)X\left(-\omega-\omega^{\prime}\right).\nonumber 
\end{align}
By \prettyref{eq:propagator}, the regulator modifies the leak term
$m$, thus controlling the variance of the fluctuations. A general
discussion on the choice of frequency-dependent regulators can be
found in \citep{Duclut17_012107}: The $XX$-diagonal element must
always be zero to maintain normalization (see also \prettyref{subsec:Appendix_Normalization-Deker-Haake}).
A regulator on the $\tX\tX$ element corresponds to a modification
of the second cumulant of the driving noise. For systems in equilibrium,
the fluctuation-dissipation theorem constrains the choice of the regulator
further.

For the choice in \prettyref{eq:action_regulator}, the bare propagator
reads
\[
\Delta_{\lambda}^{0}\left(\omega,\omega^{\prime}\right)=\begin{pmatrix}\frac{D}{\omega^{2}+\left(m+\frac{1}{2}\Rlbd\right)^{2}} & \frac{1}{-i\omega+m+\frac{1}{2}\Rlbd}\\
\frac{1}{i\omega+m+\frac{1}{2}\Rlbd} & 0
\end{pmatrix}2\pi\delta\left(\omega+\omega^{\prime}\right).
\]
We notice that $\Rlbd$ has to be negative to avoid a vanishing leak
term (since $m<0)$ and thus a fluctuation singularity at $\omega=0$
along the RG trajectory. We define $\tilde{\Gamma}_{\lambda}\left[X^{\ast},\tilde{X}^{\ast}\right]$
as the Legendre transform of the cumulant-generating functional, given
by
\begin{eqnarray*}
W_{\lambda}\left[J,\tilde{J}\right] & = & \ln\int\D X\,\int\D\tilde{X}\,\exp\left(S_{\lambda}\left[X,\tilde{X}\right]+J^{\T}X+\tilde{J}^{\T}\tilde{X}\right),
\end{eqnarray*}
where we used the abbreviation $J^{\T}X=\int d\omega\,J(\omega)X(\omega)$.
Defining $\Gamma_{\lambda}$, we remove the ``direct'' effect of
the regulator
\begin{equation}
\Gamma_{\lambda}\left[X^{\ast},\tilde{X}^{\ast}\right]:=\tilde{\Gamma}_{\lambda}\left[X^{\ast},\tilde{X}^{\ast}\right]+\Delta S_{\lambda}\left[X^{\ast},\tilde{X}^{\ast}\right],\label{eq:def_eff_action_fRG}
\end{equation}
so that $\lim_{\lambda\rightarrow0}\Gamma_{\lambda}=\Gamma$ and
$\lim_{\lambda\rightarrow\Lambda}\Gamma_{\lambda}=-S$. The latter
limit follows, because at $\lambda=\Lambda$ the regulator suppresses
all fluctuations, so that the mean-field approximation $\tilde{\Gamma}_{\Lambda}\simeq-S_{\Lambda}=-\big(S+\Delta S_{\Lambda}\big)$
becomes exact. Consequently, by definition \prettyref{eq:def_eff_action_fRG},
one then obtains the stated limit \citep[see also equations 2.9 to 2.12 in][]{Berges02_223}.
To derive the flow equation for the effective action of the theory
defined in \prettyref{eq:action_regulator}, we take the partial derivative
of $W_{\lambda}$ with respect to the flow parameter and deduce from
this the respective derivatives of $\tilde{\Gamma}_{\lambda}$ and
$\Gamma_{\lambda}$ (for details consult \prettyref{subsec:Derivation-of-fRG-flow}),
which results in the Wetterich equation

\begin{fmffile}{wetterich_eq}
\fmfset{thin}{0.75pt}
\fmfset{decor_size}{4mm}
\begin{align}
	\frac{\partial \Gamma_\lambda\left[ X^\ast, \tilde{X}^\ast \right]}{\partial \lambda} =\, 
	& \frac{1}{2} \mathrm{Tr} \left\{ \Delta_{\tilde{X}X,\lambda} \frac{\partial R_\lambda}{\partial \lambda} \right\}
	& \nonumber\\
	\label{Wetterich_equation}
	=\, & \frac{1}{2} \mathrm{Tr} \left\{ \left[ \Gamma^{\left( 2 \right)}_\lambda \left[X^\ast, \tilde{X}^\ast \right] + 
		\begin{pmatrix}0 & \frac{1}{2}R_{\lambda}\\ \frac{1}{2}R_{\lambda} & 0 \end{pmatrix} \right]^{-1}_{\tilde{X}X} \frac{\partial R_\lambda}{\partial \lambda} \right\} = \frac{1}{2} \quad
	 \parbox{10mm}{
		\begin{fmfgraph*}(30,40)
			\fmfsurroundn{v}{3}
			\fmfv{d.s=square, d.filled=empty}{v1}
			\fmf{plain, right=0.58}{v1,v2}
			\fmf{plain, right=0.58}{v3,v1}
			\fmf{plain_arrow, right=0.58}{v2,v3}
		\end{fmfgraph*}		
	}\mkern30mu,
\end{align}
\end{fmffile}where in this section, lines denote full propagators
\begin{equation}
\Dlbd\coloneqq\left(\tilde{\Gamma}_{\lambda}^{\left(2\right)}\right)^{-1}=\left(\Gamma_{\lambda}^{\left(2\right)}+\begin{pmatrix}0 & \frac{1}{2}R_{\lambda}\\
\frac{1}{2}R_{\lambda} & 0
\end{pmatrix}\right)^{-1}\label{eq:propagator_fRG}
\end{equation}
and open squares represent $\derivRlbd$. The translation between
graphical representations and algebraic expressions is also shown
in \prettyref{tab:feyn_diag_elements}. For the final result ($\lambda=0$),
the choice of the concrete form of $\Rlbd$ is arbitrary as long as
it fulfills \prettyref{eq:Defining_properties_regulator} and does
not lead to acausal terms in the action. The interpretation of $\Gamma_{\lambda}$
along the trajectory of the flow equations, however, depends on the
regulator.

The simplest choice is the uniform regulator, for example $\Rlbd=-\lambda$.
In this case all frequencies get damped equally. Its equivalence to
an additional leak term {[}compare \prettyref{eq:S_MSRDJ} and \eqref{eq:propagator_fRG}{]}
bears a second interpretation: We may as well interpret each point
along the solution of the flow equation as one system with a different
value for the leak term. In this context, a vanishing value and hence
a pole in the propagator at vanishing frequency becomes meaningful
again: it corresponds to a critical point where fluctuations dominate
the system behavior.

The simplest measure to extract from $\Gamma_{\lambda}$ is the mean
value $\xmean_{\lambda}$ defined by the equation of state $\tilde{\Gamma}_{\tx,\lambda}^{\left(1\right)}\left[x_{\lambda}^{\ast},\tilde{x}_{\lambda}^{\ast}\right]=0\,$
\footnote{The condition $\Gamma_{\tx,\lambda}^{\left(1\right)}\left[x_{\lambda}^{\ast},\tilde{x}_{\lambda}^{\ast}\right]=0$
would of course lead to the same result because we are eventually
interested in $\lambda=0$, where both quantities agree, but using
$\tilde{\Gamma}$ leads to the occurrence of the propagator including
the regulator term in \prettyref{eq:flow_eq_x_star_1_without_tilde},
which is more convenient.}. Differentiating this equation with respect to $\lambda$ leads to

\begin{align}
0 & =\frac{d}{d\lambda}\tilde{\Gamma}_{\lambda}^{\left(1\right)}\left(\omega\right)=\frac{\partial\tilde{\Gamma}_{\lambda}^{\left(1\right)}\left(\omega\right)}{\partial\lambda}+\int d\omega\,\tilde{\Gamma}_{\lambda}^{\left(2\right)}\left(\omega,\omega^{\prime}\right)\frac{\partial}{\partial\lambda}\begin{pmatrix}\Xmeanlbd\left(\omega^{\prime}\right)\\
\Xtmeanlbd\left(\omega^{\prime}\right)
\end{pmatrix}\nonumber \\
\Leftrightarrow\frac{\partial}{\partial\lambda}\begin{pmatrix}\Xmeanlbd\left(\sigma\right)\\
\Xtmeanlbd\left(\sigma\right)
\end{pmatrix} & =-\Delta_{\lambda}\left(\sigma\right)\left[\frac{\partial\Gamma_{\lambda}^{\left(1\right)}\left(-\sigma\right)}{\partial\lambda}+\frac{1}{4\pi}\frac{\partial R_{\lambda}}{\partial\lambda}\begin{pmatrix}\Xmeanlbd\left(\sigma\right)\\
\Xtmeanlbd\left(\sigma\right)
\end{pmatrix}\right].\label{eq:flow_eq_x_star_1_without_tilde}
\end{align}
To obtain \prettyref{eq:flow_eq_x_star_1_without_tilde}, we have
multiplied the line above by the propagator $\Delta_{\lambda}$ and
used $\Delta_{\lambda}\tilde{\Gamma}_{\lambda}^{\left(2\right)}=1$.
We observe that we now need a flow equation for $\tilde{\Gamma}^{\left(1\right)}$,
which we obtain by differentiating (\ref{Wetterich_equation}) with
respect to $\left(x,\tilde{x}\right)$ leading to\begin{fmffile}{Gammai}
\fmfset{thin}{0.75pt}
\fmfset{decor_size}{4mm}
\fmfcmd{style_def wiggly_arrow expr p = cdraw (wiggly p); shrink (0.9); cfill (arrow p); endshrink; enddef;}
\fmfcmd{style_def majorana expr p = cdraw p; cfill (harrow (reverse p, .5)); cfill (harrow (p, .5)) enddef;
		style_def alt_majorana expr p = cdraw p; cfill (tarrow (reverse p, .55)); cfill (tarrow (p, .55)) enddef;}
\begin{align}
	\frac{\partial \Gamma^{\left( 1 \right)}_{X,\lambda}\left( \sigma \right)}{\partial \lambda}
	& =\,-\frac{1}{2}\, \parbox{10mm}{
		\begin{fmfgraph*}(60,60)
			\fmfleft{i}
			\fmfright{r}
			\fmfdot{v}
			\fmf{plain_arrow, right=1.0, tension=0.4}{r,v,r}	
			\fmf{wiggly_arrow, tension=1.0, label=$\sigma$}{i,v}
			\fmfv{d.s=square, d.filled=empty}{r}
		\end{fmfgraph*}		
	} \nonumber \\
	\label{flow_eq_Gamma_x}
	& = \,-\frac{1}{2} \int\! \frac{d\omega}{2\pi}\, \Gamma^{\left( 3 \right)}_{\tilde{X}XX,\lambda}
		\left(\sigma,-\omega,\omega \right) \Delta_{\tilde{X}X,\lambda}\left(\omega \right) \frac{\partial R_\lambda}{\partial\lambda} \Delta_{\tilde{X}X,\lambda}\left(\omega \right)
	\\
	\frac{\partial \Gamma^{\left( 1 \right)}_{\tilde{X},\lambda}\left( \sigma \right)}{\partial \lambda}
	& =\,-\frac{1}{2}\, \parbox{30mm}{
		\begin{fmfgraph*}(60,60)
			\fmfleft{i}
			\fmfright{r}
			\fmfdot{v}
			\fmf{plain_arrow, right=1.0, tension=0.4}{r,v}
			\fmf{alt_majorana, right=1.0, tension=0.4}{v,r}	
			\fmf{wiggly_arrow, tension=1.0, label=$\sigma$, l.side=left}{v,i}
			\fmfv{d.s=square, d.filled=empty}{r}
		\end{fmfgraph*}		
	} \nonumber \\
	\label{flow_eq_Gamma_xt}
	& = \,-\frac{1}{2} \int\! \frac{d\omega}{2\pi}\, \Gamma^{\left( 3 \right)}_{\tilde{X}XX,\lambda}
		\left(\sigma,-\omega,\omega \right) \Delta_{XX,\lambda}\left(\omega \right) \frac{\partial R_\lambda}{\partial\lambda} \Delta_{\tilde{X}X,\lambda}\left(\omega \right).
\end{align}
\end{fmffile}In the last step for both diagrams we defined $\Delta_{\lambda}\left(\omega,\omega^{\prime}\right)\eqqcolon\Delta_{\lambda}\left(\omega\right)\delta\left(\omega+\omega^{\prime}\right)$
and $R_{\lambda}\left(\omega,\omega^{\prime}\right)\eqqcolon R_{\lambda}\delta\left(\omega+\omega^{\prime}\right)/\left(2\pi\right)$.
Since the three point vertex conserves momentum $\Gamma_{\lambda}^{\left(3\right)}\left(\omega,\omega^{\prime},\omega^{\prime\prime}\right)\propto\delta\left(\omega+\omega^{\prime}+\omega^{\prime\prime}\right)$,
the external momentum is fixed at $\sigma=0$. Inserting these equations
into \prettyref{eq:flow_eq_x_star_1_without_tilde} yields the final
results, which are equivalent to \citep[eq. (21)]{Schuetz06} and
\citep[eq. (7.94), fig. 7.7]{Kopietz10}. Because of the closed response
loop in (\ref{flow_eq_Gamma_x}), the right-hand side of this equation
is always identically zero leading to $\Xtmeanlbd=0\,\forall\,\lambda$.
The same applies to all other diagrams with one loop and one external
leg different from (\ref{flow_eq_Gamma_x}) or (\ref{flow_eq_Gamma_xt}),
because they also contain either $\Delta_{\tx\tx}=0$, response loops,
or a vertex $\Gamma_{XXX,\lambda}^{\left(3\right)}$. The last is
always identically zero as shown in \prettyref{sec:General-properties-of_MSRDJ-Gamma}.
So, only (\ref{flow_eq_Gamma_xt}) contains information on the flow
of $\Xmeanlbd$, finally yielding
\begin{equation}
\frac{\partial}{\partial\lambda}\Xmeanlbd\left(\sigma\right)=-\Delta_{X\tilde{X},\lambda}\left(\sigma\right)\left[\frac{\partial\Gamma_{\tilde{X},\lambda}^{\left(1\right)}\left(-\sigma\right)}{\partial\lambda}+\frac{1}{4\pi}\frac{\partial R_{\lambda}}{\partial\lambda}\Xmeanlbd\left(\sigma\right)\right].\label{eq:flow_eq_x_star}
\end{equation}
The right-hand side of this equation depends on $\Dlbd$ and $\Giiitoolbd$
via (\ref{flow_eq_Gamma_xt}) which in turn are also defined by flow
equations containing vertices of the respective next two orders. This
induces an infinite hierarchy. A first approximation of the mean value
$\Xmeanlbd$ is gained by truncating the hierarchy after $\Gamma_{\lambda}^{\left(1\right)}$,
that is using the bare quantities for $\Dlbd$ and $\Giiitoolbd$
and integrating the flow \prettyref{eq:flow_eq_x_star}. We then also
get a corrected value for the variance by inserting $\Xmeanlbd$ into
$\Dbaroo$. The flow equations at this level of approximation can
be integrated exactly: They recover the one-loop approximation.

We can improve the accuracy by taking into account the flow of higher
derivatives of $\Gamma$. In this work we included the flow of the
self-energy and the interaction vertex $\Gamma_{\tX XX}^{\left(3\right)}$,
but neither the one of $\Gamma_{\tX\tX X}^{\left(3\right)}$ and $\Gamma_{\tX\tX\tX}^{\left(3\right)}$
nor that of all all higher order vertices. The one loop correction
of $\Gamma_{\tilde{X}\tilde{X}X}^{(3)}$ ($\Gamma_{\tilde{X}\tilde{X}\tilde{X}}^{(3)}$)
involves two (three) $xx$-propagators, so that, in systems with small
fluctuations, which scale with $D$, this diagram is less important
than the others. Compared to the one-loop correction of $\Giittfl$,
$\Gamma_{\tilde{X}\tilde{X}X}^{\left(3\right)}$ bears the same number
of $xx$-propagators, but one additional interaction, which scales
with the other small factor $\beta$. Therefore we neglect the corrections
to $\Gamma_{\tilde{X}\tilde{X}X}^{(3)}$ and $\Gamma_{\tilde{X}\tilde{X}\tilde{X}}^{(3)}$
but not to $\Giiitoolbd$. In conclusion, we renormalize exactly those
terms that also appear in the bare action. Under these constraints,
the nonvanishing and non-negligible diagrams for the self-energy are
given by\begin{fmffile}{Gammaii_struct}
\fmfset{thin}{0.75pt}
\fmfset{decor_size}{4mm}
\fmfcmd{style_def wiggly_arrow expr p = cdraw (wiggly p); shrink (0.9); cfill (arrow p); endshrink; enddef;}
\fmfcmd{style_def majorana expr p = cdraw p; cfill (harrow (reverse p, .5)); cfill (harrow (p, .5)) enddef;
		style_def alt_majorana expr p = cdraw p; cfill (tarrow (reverse p, .55)); cfill (tarrow (p, .55)) enddef;}
\begin{align*}
	\frac{\partial \Gamma^{\left( 2 \right)}_{\tilde{X}X,\lambda}\left( \sigma_1,\sigma_2 \right)}{\partial \lambda}
	& =\,\frac{1}{2}\, \parbox{5mm}{
		\begin{fmfgraph*}(110,40)
			\fmfleft{i}
			\fmfright{o}
			\fmfdot{v1}
			\fmfdot{v2}
			\fmftop{r}
			\fmfbottom{phr}
			\fmfv{d.s=square, d.filled=empty}{r}	
			\fmf{wiggly_arrow, tension=1.0, label=$\sigma_2$}{i,v1}
			\fmf{wiggly_arrow, tension=1.0, label=$\sigma_1$}{v2,o}
			\fmf{plain_arrow, right=1.0, tension=0.4}{v1,v2}
			\fmf{plain_arrow, right=0.4, tension=0.4}{r,v1}
			\fmf{alt_majorana, right=0.6, tension=0.4}{v2,r}
			\fmf{phantom, right=0.5, tension=0.4}{v1,phr,v2}
		\end{fmfgraph*}		
	}
	 & \mkern-48mu +  \frac{1}{2}\, \parbox{10mm}{
		\begin{fmfgraph*}(110,40)
			\fmfleft{i}
			\fmfright{o}
			\fmfdot{v1}
			\fmfdot{v2}
			\fmftop{phr}
			\fmfbottom{r}
			\fmfv{d.s=square, d.filled=empty}{r}	
			\fmf{wiggly_arrow, tension=1.0, label=$\sigma_2$}{i,v1}
			\fmf{wiggly_arrow, tension=1.0, label=$\sigma_1$}{v2,o}
			\fmf{alt_majorana, right=1.0, tension=0.4}{v2,v1}
			\fmf{plain_arrow, right=0.4, tension=0.4}{v1,r,v2}
			\fmf{phantom, right=0.5, tension=0.4}{v2,phr,v1}
		\end{fmfgraph*}		
	}\\ \vphantom{a} \\
	& +\,\frac{1}{2}\, \parbox{10mm}{
		\begin{fmfgraph*}(110,40)
			\fmfleft{i}
			\fmfright{o}
			\fmfdot{v1}
			\fmfdot{v2}
			\fmftop{phr}
			\fmfbottom{r}
			\fmfv{d.s=square, d.filled=empty}{r}	
			\fmf{wiggly_arrow, tension=1.0, label=$\sigma_2$}{i,v1}
			\fmf{wiggly_arrow, tension=1.0, label=$\sigma_1$}{v2,o}
			\fmf{plain_arrow, left=1.0, tension=0.4}{v1,v2}
			\fmf{plain_arrow, right=0.4, tension=0.4}{r,v2}
			\fmf{alt_majorana, right=0.4, tension=0.4}{v1,r}
			\fmf{phantom, right=0.5, tension=0.4}{v2,phr,v1}
		\end{fmfgraph*}		
	}
	\\ \vphantom{A} \\
	\frac{\partial \Gamma^{\left( 2 \right)}_{\tilde{X}\tilde{X},\lambda}\left( \sigma_1,\sigma_2 \right)}{\partial \lambda}
	& =\,\frac{1}{2}\, \parbox{20mm}{
	\begin{fmfgraph*}(110,40)
			\fmfleft{i}
			\fmfright{o}
			\fmfdot{v1}
			\fmfdot{v2}
			\fmftop{phr}
			\fmfbottom{r}
			\fmfv{d.s=square, d.filled=empty}{r}	
			\fmf{wiggly_arrow, tension=1.0, label=$\sigma_1$, l.side=left}{v1,i}
			\fmf{wiggly_arrow, tension=1.0, label=$\sigma_2$}{v2,o}
			\fmf{alt_majorana, right=1.0, tension=0.4}{v2,v1}
			\fmf{alt_majorana, right=0.6, tension=0.4}{v1,r}
			\fmf{plain_arrow, right=0.4, tension=0.4}{r,v2}
			\fmf{phantom, right=0.5, tension=0.4}{v1,phr,v2}
		\end{fmfgraph*}		
	} & +\, \sigma_1 \leftrightarrow \sigma_2 . 
\end{align*}
\end{fmffile}
\vspace{0.1cm}
The translation of these diagrams is shown in \prettyref{subsec:Flow_eq_Gii_Giii}
together with the respective diagrams for the interaction vertex.

In general, the diagrams have the same form as those that appear in
the fluctuation expansion, except for the presence of a single regulator
in one of the propagator lines. The combination of two propagators
``sandwiching'' a regulator $\Delta_{\lambda}\left(\omega,-\omega\right)\derivRlbd\left(\omega,-\omega\right)\Delta_{\lambda}\left(\omega,-\omega\right)$
is called ``single scale propagator'' because the regulator is often
chosen in a way such that its derivative is peaked around frequencies
with $\left|\omega\right|=\lambda$, thus contributing at a single
scale. Due to the one-loop structure of the diagrams and the conservation
of frequencies at the vertices, we have to perform one integral over
an internal frequency for every possible combination of fixed external
frequencies. Therefore, the numerical evaluation of $\Gamma^{\left(n\right)}$
becomes increasingly computationally expensive for higher orders because
at $n$-th order we have $n-1$ independent external frequencies.
For practical computations we thus have to truncate the hierarchy
after $n=3$ if we want to keep the full frequency dependence of $\Gamma^{\left(n\right)}$.
Even at this order, the integration takes many hours on a usual desktop
PC when we choose the external frequencies to range from $-25$ to
$25$ with a resolution of $0.1$. Therefore, it is legitimate to
ask if one can reduce the number of required frequency-integrals by
assuming a simplified frequency-dependence of higher order vertices.

\subsubsection{The BMW scheme\label{subsec:BMW}}

A scheme that assumes a simplified momentum dependence of higher order
vertices has been suggested by Blaizot, Méndez, and Wschebor (BMW)
\citep{Blaizot06_571,Blaizot06a_051116}. It has been successfully
applied for example to the Kardar-Parisi-Zhang model \citep{Canet10}.
The principal idea is to neglect the frequency dependence of the effective
action as much as possible. The most radical choice in this respect
would be to assume it to be constant, which is known as the local
potential approximation (LPA). BMW refined this scheme by including
the exact frequency-dependence of all vertices up to a certain order
$s$, which are functions of $s-1$ external frequencies, and to approximate
vertices of the next two orders by evaluating the additional derivatives
at their zero frequency components obtaining a partial differential
equation \citep{Benitez12_026707}. We will pursue a different route
by deriving approximate flow equations for $\Gamma^{\left(s+1\right)}$
and $\Gamma^{\left(s+2\right)}$ with the simplified frequency dependence.

More precisely, we consider a typical contribution of a vertex within
the Wetterich flow equation for the vertex $\Gamma^{(s)}$ containing
\begin{align}
 & \Gamma_{\lambda}^{(s+1)}(\sigma_{1},\ldots,\sigma_{s}+\omega,-\omega),\label{eq:vertex_s-1}\\
 & \Gamma_{\lambda}^{(s+2)}(\sigma_{1},\ldots,\sigma_{s},\omega,-\omega),\nonumber 
\end{align}
where $\omega$ is the loop frequency, that represents the frequency
at the regulator. For simplicity we do not specify different components
of the field. For the approximation we assume the vertices to depend
only weakly on $\omega$ and therefore set $\omega=0$ in \eqref{eq:vertex_s-1},
replacing the vertices by
\begin{align}
 & \Gamma_{\lambda}^{(s+1)}(\sigma_{1},\ldots\sigma_{s},0),\label{eq:approx_vertices-1}\\
 & \Gamma_{\lambda}^{(s+2)}(\sigma_{1},\ldots\sigma_{s},0,0).\nonumber 
\end{align}
The frequency dependence on $\omega$ in the propagators and the regulator
is, however, kept.

The second step of the BMW scheme allows the closure of the system:
Due to the vanishing momentum on one or two legs of the vertices \eqref{eq:approx_vertices-1},
the vertex functions with $s+1$ and $s+2$ legs can be expressed
as the derivative of $\Gamma_{\lambda}^{\left(s\right)}$ with respect
to a uniform (background) field $X_{0}^{\ast}\coloneqq X^{\ast}\left(\sigma=0\right)$:
\begin{align}
\Gamma_{\lambda}^{\left(s+1\right)}\left(\sigma_{1},\ldots,\sigma_{s},0\right) & =\frac{\delta}{\delta X_{0}^{\ast}}\Gamma_{\lambda}^{\left(s\right)}\left(\sigma_{1},\ldots,\sigma_{s}\right)\label{eq:eq_bmw_original-1}\\
\Gamma_{\lambda}^{\left(s+2\right)}\left(\sigma_{1},\ldots,\sigma_{s},0,0\right) & =\frac{\delta^{2}}{\delta X_{0}^{\ast}{}^{2}}\Gamma_{\lambda}^{\left(s\right)}\left(\sigma_{1},\ldots,\sigma_{s}\right).\label{eq:eq_bmw_original_2-1}
\end{align}
If we now use \eqref{eq:eq_bmw_original-1} and \eqref{eq:eq_bmw_original_2-1}
to replace $\Gamma_{\lambda}^{\left(s+1\right)}$ and $\Gamma_{\lambda}^{\left(s+2\right)}$
by the ordinary derivatives of $\Gamma_{\lambda}^{\left(s\right)}$
with respect to the zero-modes of $X$, $\tilde{X}$, we close the
set of flow equations and additionally we take into account the flow
of order $s+1$ and $s+2$, at least approximately. Since we reduced
the number of independent frequencies by one (or two, respectively),
the computation time decreases significantly.

But the resulting equation is a partial differential equation in $\lambda$
and $X_{0}^{\ast}$ \citep[eq. (19)]{Benitez12_026707}; we hence
have to evaluate the derivatives with respect to $X_{0}^{\ast}$ for
every step at which we compute $\partial_{\lambda}\Gamma^{\left(s\right)}$.
Below we develop an alternative scheme that circumvents this complication
and entirely stays within the realm of a vertex expansion.

\begin{figure}
\begin{centering}
\includegraphics[scale=0.4]{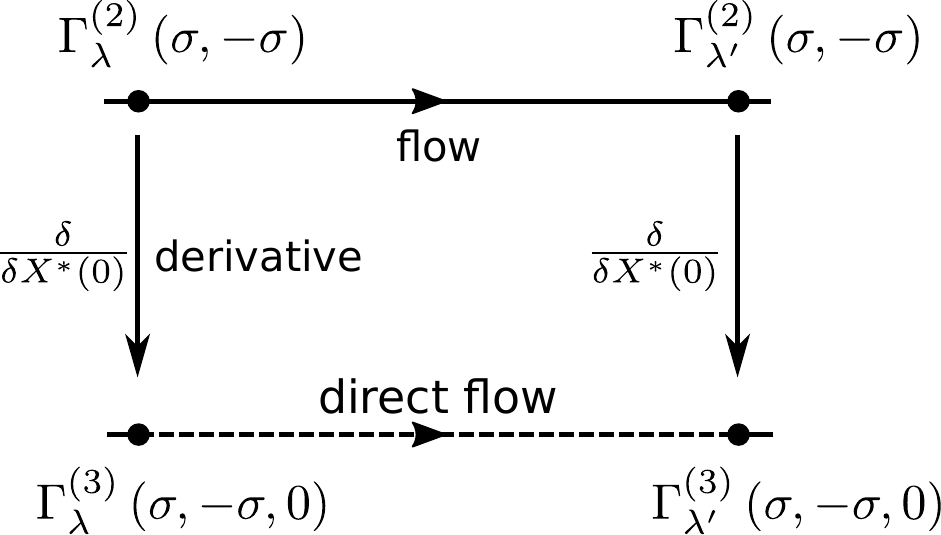}
\par\end{centering}
\caption{Implementation of the BMW scheme within the vertex expansion. In the
original formulation (top) we have to evaluate the derivatives $\delta/\delta X^{\ast}(0)$
numerically at each $\lambda$. In the new interpretation (bottom)
we derive an additional flow equation for $\Gamma^{\left(3\right)}\left(\sigma,-\sigma,0\right)$
(dashed line).\label{fig:bmw_scheme}}
\end{figure}

\subsubsection{Removing the PDE}

We aim to apply the BMW approximation at order $s=2$, thus keeping
the full frequency dependence of the self-energy. However, within
the vertex expansion scheme it remains unclear how to compute the
derivatives of $\Gamma^{\left(2\right)}$ numerically. This is because
at each given $\lambda$ we know the value of $\Gamma^{(2)}$ only
for the true mean value $\Xmeanlbd(0)$, but not in a vicinity around
it; we therefore cannot approximate the derivative by a ratio of finite
differences.

We can circumvent this problem by deriving an additional flow equation
for $\Gamma_{\lambda}^{\left(3\right)}\left(\sigma_{1},\sigma_{2},0\right)$
(illustrated in \prettyref{fig:bmw_scheme}; the more complex situation
with $x$ and $\tilde{x}$ fields will be addressed below) and in
principle also for $\Gamma_{\lambda}^{\left(4\right)}\left(\sigma_{1},\sigma_{2},0,0\right)$,
which we neglect in our model because the largest contribution of
$\Gamma^{\left(4\right)}$ is suppressed by a factor $\beta$ due
to an additional interaction vertex. We obtain this flow equation
by differentiating the one for $\Gamma_{\lambda}^{\left(2\right)}$
with respect to $X^{\ast}\left(0\right)$, i.e. $\frac{\delta}{\delta X^{\ast}\left(0\right)}\partial_{\lambda}\Gamma_{\lambda}^{\left(2\right)}=\partial_{\lambda}\Gamma_{\lambda}^{\left(3\right)}$
and then setting the frequencies of the original three-point vertices
at those legs to zero that are connected to the single scale propagator,
in line with the BMW scheme. But we need to keep the dependence on
$\omega$ of the vertices that emerge when we differentiate a propagator
by the background field, so that we treat the frequency dependence
of this additional vertex like that of the regulator in the original
diagram. Otherwise we would make an additional approximation on top
of BMW. Thus, drawing only the first argument of the propagators,
in diagrammatic language we obtain\begin{fmffile}{bmw_simple}
\fmfset{thin}{0.75pt}
\fmfset{decor_size}{4mm}
\fmfcmd{style_def wiggly_arrow expr p = cdraw (wiggly p); shrink (0.9); cfill (arrow p); endshrink; enddef;}
\fmfcmd{style_def majorana expr p = cdraw p; cfill (harrow (reverse p, .5)); cfill (harrow (p, .5)) enddef;
		style_def alt_majorana expr p = cdraw p; cfill (tarrow (reverse p, .55)); cfill (tarrow (p, .55)) enddef;}
\begin{align*}
	&\frac{\delta}{\delta X(0)} \frac{\partial \Gamma^{\left( 2 \right)}_{\lambda}\left( \sigma_1,-\sigma_1 \right)}{\partial \lambda}
	 =\,\frac{\delta}{\delta X(0)}\,\frac{1}{2}\, \parbox{10mm}{
		\begin{fmfgraph*}(110,40)
			\fmfleft{i}
			\fmfright{o}
			\fmfdot{v1}
			\fmfdot{v2}
			\fmftop{r}
			\fmfbottom{phr}
			\fmfv{d.s=square, d.filled=empty}{r}	
			\fmf{wiggly, tension=0.8, label=$\sigma_1$}{i,v1}
			\fmf{wiggly, tension=0.8, label=$-\sigma_1$, l.side=left}{o,v2}
			\fmf{plain, right=1.0, tension=0.4, label=$\sigma_{1}+\omega$}{v1,v2}
			\fmf{plain, right=0.4, tension=0.4, label=$\omega$}{v2,r,v1}
			\fmfv{l=$(1)$, l.a=0, l.d=0.02w}{v1}
			\fmfv{l=$(2)$, l.a=180, l.d=0.02w}{v2}
			\fmf{phantom, right=0.5, tension=0.4}{v1,phr,v2}
		\end{fmfgraph*}		
	}
	\\ \vphantom{A} \\ \vphantom{a}\\
	& \overset{\mathrm{BMW}}{=}\,-\frac{1}{2}\quad\parbox{35mm}{
		\begin{fmfgraph*}(90,80)
			\fmfcurved
			\fmfsurroundn{v}{8}
			\fmffreeze
			\fmfshift{(-0.1w,-0.1w)}{v2}
			\fmfshift{(0.1w,-0.1w)}{v4}
			\fmfshift{(0.0w,0.2w)}{v7}
			\fmfshift{(-0.23w,-0.05w)}{v1}
			\fmfv{d.s=square, d.filled=empty}{v1}
			\fmfdot{v2}
			\fmfdot{v4}
			\fmfdot{v7}
			\fmftopn{i}{2}
			\fmfbottom{o}
			\fmf{wiggly, tension=1.0}{i2,v2}
			\fmf{wiggly, tension=1.0}{i1,v4}
			\fmf{wiggly, tension=1.0}{v7,o}
			\fmflabel{$0$}{i2}
			\fmflabel{$\sigma_1$}{i1}
			\fmflabel{$-\sigma_1$}{o}
			\fmfv{l=$(4)$, l.a=90}{v7}
			\fmfv{l=$(5)$, l.a=-130, l.d=0.04w}{v2}
			\fmfv{l=$(3)$, l.a=-50, l.d=0.04w}{v4}
			\fmf{plain, right=0.5, label=$\omega$}{v2,v4}
			\fmf{plain, right=0.5, label=$\sigma_{1}+\omega$, l.d = 0.01}{v4,v7}
			\fmf{plain, right=0.3, label=$\omega$}{v7,v1}
			\fmf{plain, right=0.3, label=$\omega$}{v1,v2}
		\end{fmfgraph*}		
	}  -\frac{1}{4}\quad \parbox{35mm}{
		\begin{fmfgraph*}(90,80)
			\fmfcurved
			\fmfsurroundn{v}{8}
			\fmffreeze
			\fmfshift{(-0.1w,-0.1w)}{v2}
			\fmfshift{(0.1w,-0.1w)}{v4}
			\fmfshift{(0.0w,0.2w)}{v7}
			\fmfshift{(-0.23w,-0.05w)}{v1}
			\fmfshift{(0.0w,-0.05w)}{v3}
			\fmfv{d.s=square, d.filled=empty}{v3}
			\fmfdot{v2}
			\fmfdot{v4}
			\fmfdot{v7}
			\fmftopn{i}{2}
			\fmfbottom{o}
			\fmf{wiggly, tension=1.0}{i2,v2}
			\fmf{wiggly, tension=1.0}{i1,v4}
			\fmf{wiggly, tension=1.0}{v7,o}
			\fmflabel{$0$}{o}
			\fmflabel{$-\sigma_1$}{i2}
			\fmflabel{$\sigma_1$}{i1}
			\fmfv{l=$(8)$, l.a=90}{v7}
			\fmfv{l=$(7)$, l.a=-130, l.d=0.04w}{v2}
			\fmfv{l=$(6)$, l.a=-50, l.d=0.04w}{v4}
			\fmf{plain, right=0.3, label=$\omega$}{v2,v3,v4}
			\fmf{plain, right=0.5, label=$\sigma_{1}+\omega$, l.d = 0.01}{v4,v7}
			\fmf{plain, right=0.5, label=$\sigma_{1}+\omega$}{v7,v2}
		\end{fmfgraph*}		
	} -\frac{1}{2} \quad \parbox{35mm}{
		\begin{fmfgraph*}(110,40)
			\fmfcurved
			\fmfleftn{i}{2}
			\fmfright{o}
			\fmfdot{v1}
			\fmfdot{v2}
			\fmftop{r}
			\fmfbottom{phr}
			\fmfv{d.s=square, d.filled=empty}{r}	
			\fmf{wiggly, tension=0.8}{i1,v1}
			\fmf{wiggly, tension=0.8}{i2,v1}
			\fmf{wiggly, tension=0.8}{o,v2}
			\fmflabel{$\sigma_1$}{i1}
			\fmflabel{$0$}{i2}
			\fmflabel{$-\sigma_1$}{o}
			\fmf{plain, right=1.0, tension=0.4, label=$\sigma_{1}+\omega$}{v1,v2}
			\fmf{plain, right=0.4, tension=0.4, label=$\omega$}{v2,r,v1}
			\fmfv{l=$(9)$, l.a=0, l.d=0.02w}{v1}
			\fmfv{l=$(10)$, l.a=180, l.d=0.02w}{v2}
			\fmf{phantom, right=0.5, tension=0.4}{v1,phr,v2}
		\end{fmfgraph*}		
	}\\ \vphantom{A}\\
	&\quad + \sigma_1 \leftrightarrow -\sigma_1
\end{align*}
\end{fmffile}where the vertex functions are given by
\begin{align*}
\left(1\right):\; & \Gamma_{\lambda}^{\left(3\right)}\left(\sigma_{1},\omega,-\sigma_{1}-\omega\right) & \left(2\right):\; & \Gamma_{\lambda}^{\left(3\right)}\left(-\sigma_{1},-\omega,\sigma_{1}+\omega\right)\\
\left(3\right):\; & \Gamma_{\lambda}^{\left(3\right)}\left(\sigma_{1},0,-\sigma_{1}\right) & \left(4\right):\; & \Gamma_{\lambda}^{\left(3\right)}\left(-\sigma_{1},0,\sigma_{1}\right) & \left(5\right):\; & \Gamma_{\lambda}^{\left(3\right)}\left(\omega,-\omega,0\right)\\
\left(6\right):\; & \Gamma_{\lambda}^{\left(3\right)}\left(\sigma_{1},0,-\sigma_{1}\right) & \left(7\right):\; & \Gamma_{\lambda}^{\left(3\right)}\left(-\sigma_{1},0,\sigma_{1}\right) & \left(8\right):\; & \Gamma_{\lambda}^{\left(3\right)}\left(\sigma_{1}+\omega,-\sigma_{1}-\omega,0\right).\\
\left(9\right):\; & \Gamma_{\lambda}^{\left(4\right)}\left(\sigma_{1},0,-\sigma_{1},0\right) & \left(10\right):\; & \Gamma_{\lambda}^{\left(3\right)}\left(-\sigma_{1},0,\sigma_{1}\right)
\end{align*}
The crucial point to notice is that all vertices that appear in the
diagrams have only two nonzero frequencies. The set of differential
equations is therefore closed; the last diagram requires a four-point
vertex for which we could obtain a flow equation analogously. So we
have found an explicit flow equation for $\Gamma_{\lambda}^{\left(3\right)}\left(\sigma_{1},\sigma_{2}=-\sigma_{1},0\right)$,
indicated by the dashed line in \prettyref{fig:bmw_scheme}. As a
consequence we have to solve a coupled set of ODEs instead of a single
PDE.

If we do not want to neglect the flow of the four point vertex completely,
we can differentiate the diagrams once again, which leads to a flow
equation of $\Gamma_{\lambda}^{\left(4\right)}\left(\sigma_{1},\sigma_{2},0,0\right)$.
This flow equation then depends on $\Gamma_{\lambda}^{\left(3\right)}\left(\sigma_{1},\sigma_{2},0\right)$,
$\Gamma_{\lambda}^{\left(4\right)}\left(\sigma_{1},\sigma_{2},0,0\right)$
and $\Gamma_{\lambda}^{\left(5\right)}\left(\sigma_{1},\sigma_{2},0,0,0\right)$.
Due to the emergence of the latter, the set of equations can be closed
only if we truncate the series at some order (unlike the original
BMW scheme).

The scheme described above also generalizes to the case where we have
two fields components $x$, $\tx$ and the corresponding propagators
$\Delta_{\tx x}$, $\Delta_{xx}$. We then get two sets of 12 diagrams
each (compare \prettyref{subsec:Flow_eq_Gii_Giii}) the first of which
describing the flow of the two one-dimensional sections $\Giiitoofl\left(0,\sigma_{1},-\sigma_{1}\right)$
(type 1) and the second one that of $\Giiitoofl\left(\sigma_{1},-\sigma_{1},0\right)$
(type 2). Every diagram consists of three three-point interaction
vertices that are either of type 1 or of type 2. Thus, the common
flow of the two sections is computed consistently within this approximation.
\begin{figure}
\begin{centering}
\includegraphics[width=1\textwidth]{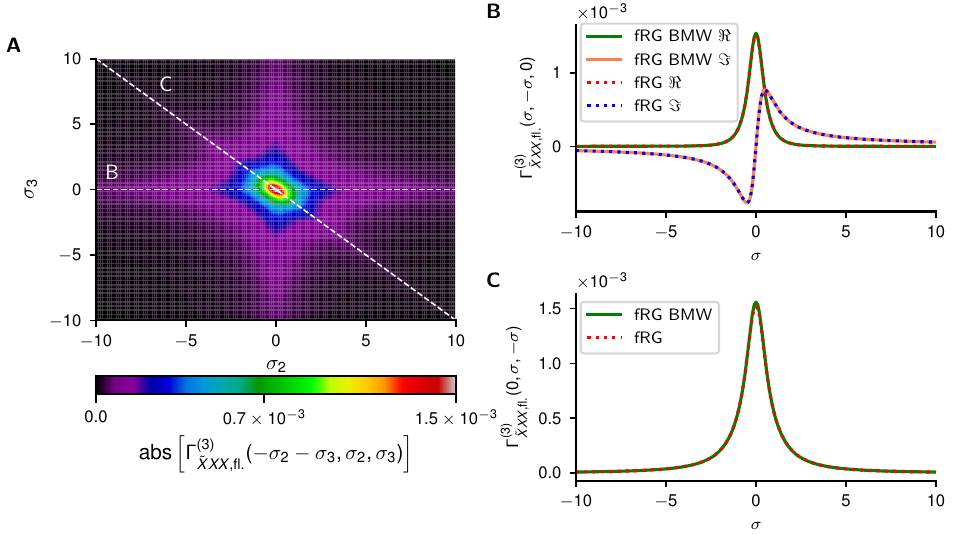}
\par\end{centering}
\caption{$\protect\Giiitoofl$ computed by fRG schemes. \textbf{A} Full frequency
dependence as a result of the calculation that takes into account
the flow of the mean value, $\Gamma_{\mathrm{fl.}}^{\left(2\right)}$,
and $\protect\Giiitoofl$. \textbf{B}, \textbf{C} $\Gamma_{\tilde{X}XX,\lambda,\mathrm{fl.}}^{\left(3\right)}\left(-\sigma_{2}-\sigma_{3},\sigma_{2},\sigma_{3}\right)$
along the sections $\sigma_{3}=0$ (\textbf{B}) and $\sigma_{2}=-\sigma_{3}$
(\textbf{C}), as indicated by the white dashed lines in (\textbf{A});
comparison to the respective types of vertices that appear in the
BMW approximation. $\Gamma_{\tilde{X}XX,\lambda,\mathrm{fl.}}^{\left(3\right)}\left(-\sigma_{2},\sigma_{2},0\right)$
in panel (B) quantifies the change of the linear response function
due to an altered constant mean activity (indicated by the derivative
with respect to the zero mode $\sigma_{3}=0$); in more neuroscientific
terms: it shows the dependence of the susceptibility (the neuron's
linear response strength to an input) on the baseline activity to
linear order. $\Gamma_{\tilde{X}XX,\lambda,\mathrm{fl.}}^{\left(3\right)}\left(0,\sigma_{2},-\sigma_{2}\right)$
in panel (C) is somewhat complementary: It is the lowest order term
describing the fluctuation-mediated effect of time-dependent deviations
on the constant part of the mean activity.  This term, for example,
quantifies the change of the constant baseline activity due to a small
sinusoidal stimulus with frequency $\sigma_{2}$; the linear order
of this response averages out over time, but the quadratic response
does not. Note that $\Gamma_{\tilde{X}XX,\lambda,\mathrm{fl.}}^{\left(3\right)}\left(0,\sigma_{2},-\sigma_{2}\right)\in\mathbb{R}$
because its Fourier transform $\Gamma_{\protect\tx xx,\lambda,\mathrm{fl.}}^{\left(3\right)}\left(0,t_{2},-t_{2}\right)$
is real by definition and symmetric because the last two arguments
are those of two $x$ at different time points, which are interchangeable.\label{fig:fRG_three_point_vertex}}
\end{figure}

\prettyref{fig:fRG_three_point_vertex} compares the three point vertices
obtained from the truncated flow equation to the result from the BMW
approximation. The latter of course only yields the three-point vertex
$\Gamma_{\tX XX}^{(3)}(\sigma_{1},\sigma_{2},\sigma_{3})$ along the
one-dimensional sections $-\sigma_{1}=\sigma_{2}+\sigma_{3}=0$ (type
1, panel C) and $\sigma_{3}=0$ (type 2, panel B). The agreement between
the two approximations is high. This result is to be expected, since
the fluctuation corrections \textit{per se} are small in the regime
considered, so that the bare vertices still constitute the largest
contributions to any fluctuation correction.

\subsection{Analyzing bifurcations by effective potentials\label{subsec:Analyzing-bifurcations}}

A fundamental question when considering neuronal dynamics is the stability
of the system and the global network state that emerges if the system
is left at rest. While in deterministic systems such a consideration
reduces to finding fixed points, typically of a set of differential
equations, in stochastic systems the situation is more complicated
due to the presence of fluctuations. To determine the fluctuation
corrections on the stationary statistics of a stochastic system and
to study the stability of the found solutions, it is convenient to
introduce what is known as the effective potential.

In this section we introduce two different approaches and apply them
to the example of a bistable system. Limiting the study to stationary
solutions $\xbarstar\coloneqq x^{\ast}\left(t\right)=\mathrm{const.}$,
such that $X\left(\omega\right)=2\pi\delta\left(\omega\right)\xbarstar$,
we can use the OM effective action \eqref{eq:Def_OM-action} to define
the effective potential 
\begin{equation}
U_{\mathrm{OM}}\left(\xbarstar\right)\coloneqq\frac{1}{T}\Gamma_{\mathrm{OM}}\left(\xbarstar\right),\label{eq:eff_po}
\end{equation}
with $T$ being the total time during which we observe the system.
The effective potential inherits the property that stationary points
correspond to the true mean value of the system from the effective
action.  The effective potential further plays the role of a rate
function \citep{Touchette09} that describes departures of the temporal
average from the ensemble average in the limit of long observation
times \citep[Section III]{Eyink96_3419}.

In tree level approximation \eqref{eq:Gamma_OM_tree} and with \eqref{eq:Def_OM-action}
we have

\begin{align}
U_{\mathrm{OM},0}\left(\xbarstar\right)=-\frac{1}{T}S_{\mathrm{OM}}\left(\xbarstar\right)= & \frac{1}{2TD}\int_{0}^{T}dt\,\left(-f\left(\xbarstar\right)\right)^{2}=\frac{1}{2D}\left(f\left(\xbarstar\right)\right)^{2}.\label{eq:effective_potential}
\end{align}
For $D>0$, its curvature at the minimum $\xbarstar=\bar{x}$ equals
the inverse of the zero-frequency fluctuations $\left\langle X\left(0\right)^{2}\right\rangle -\left\langle X\left(0\right)\right\rangle ^{2}$
around the mean, which we deduce by using $f\left(\bar{x}\right)=0$
from
\begin{align}
U_{\mathrm{OM},0}^{\prime\prime}(\bar{x}) & =\frac{\partial^{2}}{\partial\xbarstar^{2}}U_{0}\left(\xbarstar\right)\Big|_{\bar{x}}=\frac{\left(f^{\prime}\left(\bar{x}\right)\right)^{2}}{D},\label{eq:U_ddash}
\end{align}
compared to the covariance \eqref{eq:variance_linear_response} in
linear response.

This relation holds beyond this lowest order approximation: The curvature
of $U_{\mathrm{OM}}$ at a stationary point is the inverse of the
zero frequency mode of the correlation, which follows from \prettyref{eq:inv_Hessians_frequency}
at the end of \prettyref{subsec:App_MSRDJ_OM_Gamma}.

\subsubsection*{}

\subsubsection*{Computing the effective potential in the MSRDJ formalism}

For systems in thermodynamic equilibrium, the deterministic force
appearing in \prettyref{eq:SDE_general} can be written as $f\left(x\right)=-V^{\prime}(x)$.
As a consequence, the stationary distribution obeys 
\begin{align}
p(x) & \propto\exp\big(-\frac{2}{D}V(x)\big).\label{eq:p_Boltzmann}
\end{align}
For such systems, de Dominicis has defined an effective potential
\citep{DeDominicis78_4913}
\begin{align}
U_{\mathrm{DD}}[\bar{x}^{\ast}] & =V(\bar{x}^{\ast})+\ldots\label{eq:U_DD-1}
\end{align}
(see \prettyref{subsec:Effective-action-for-equilibrium} i.p. Eq.
\prettyref{eq:eff_pot_DeDominicis-1}); here ``$\ldots$'' are fluctuation
corrections. For a typical network dynamics in multiple dimensions,
the deterministic force can usually not be written in such a form.
For example, if one considers a coupling term $\sum_{j}J_{ij}x_{j}$
between neurons as in eq. \eqref{eq:N_dim}; only for a symmetric
matrix $J_{ij}=J_{ji}$ it is the derivative of the potential $V(x)=\frac{1}{2}\sum_{ij}x_{i}J_{ij}x_{j}$.
\begin{figure}
\includegraphics{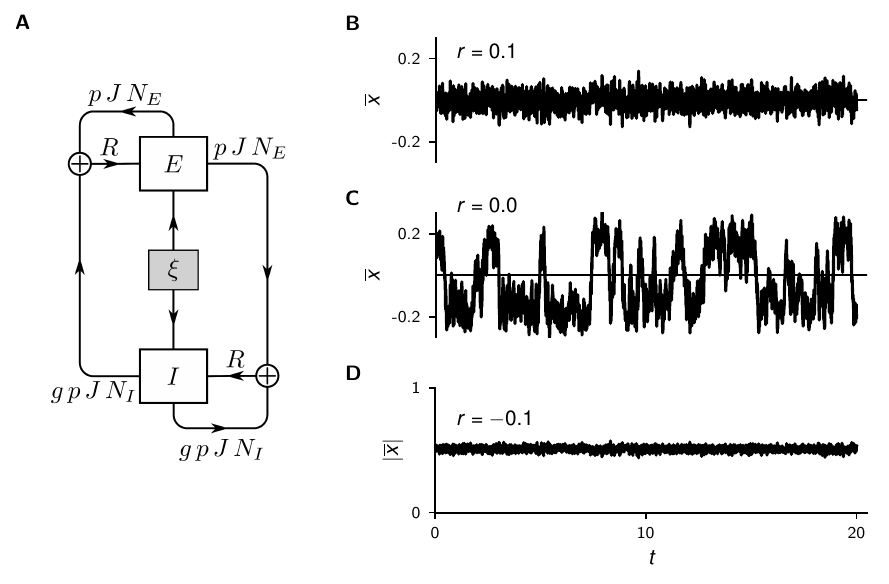}

\caption{\textbf{Diverging fluctuations in an excitatory-inhibitory network.
A} Sketch of the network consisting of $N=N_{E}+N_{I}$ neurons composed
of an excitatory (E) and an inhibitory (I) population. The connection
probability is given by $p$ and the synaptic weight by $J$ ($gJ$)
if the presynaptic neuron belongs to the excitatory (inhibitory) population.
Each neuron has a fixed outdegree of $p\,N$. Thus, the average recurrent
input weight to each neuron equals $R=pJ(N_{E}+gN_{I})$. All neurons
are driven by external white noise ($\xi$) with zero mean and standard
deviation $\sigma$. Panels \textbf{B} to \textbf{D} show the simulation
results of the population-averaged firing rate $\bar{x}(t)=N^{-1}\,\sum_{i}x_{i}(t)$
as a function of time for three different values of $r=1-R$. \textbf{B}
Network in the balanced state ($r>0)$ shows small fluctuations around
a vanishing mean value. \textbf{C} Close to the critical point ($r=0$)
the fluctuations increase considerably. \textbf{D} In the bistable
regime ($r<0$) the fluctuations decrease again and are no longer
centered around zero but around $\pm c$, where $c>0$. For the simulation
we used rate neurons with a nonlinearity of the form $\phi(x)=\tanh(x)$.
The other parameters are: $N_{E}=800$, $N_{I}=200$, $J=N^{-\frac{1}{2}}$,
$p=0.1$, $\sigma=0.1$. The simulations were performed with NEST
\citep{Jordan19_2605422}.\textcolor{brown}{\label{fig:Diverging-fluctuations-in}}}
\end{figure}
\begin{figure}
\includegraphics{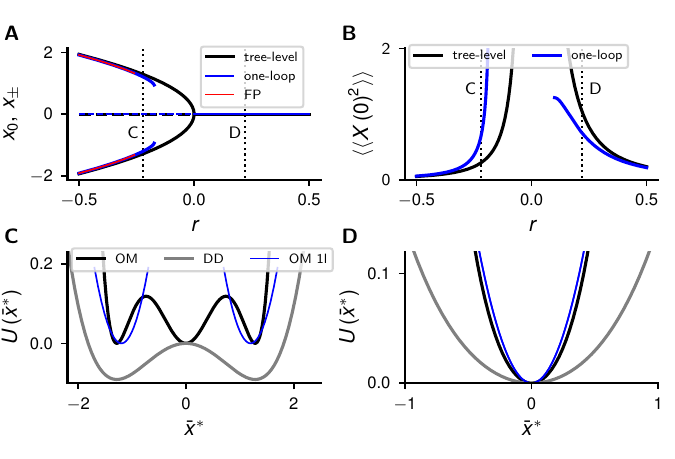}

\caption{\textbf{Critical point at the loss of balance.} Dynamical equation
\eqref{eq:Langevin_third_order-1} as a model of the population activity.
\textbf{A }Stationary points of the deterministic system (tree level,
black) and the stochastic system (one-loop, blue) as a function of
$r=1-R$. For $r<0$, the fixed point $x_{0}=0$ is unstable (dashed
horizontal lines). Exact solution from Fokker-Planck equation (FP)
in red. The dotted vertical lines denote the values of $r$ that are
used in panel C and D. \textbf{B }Zero-frequency variance $\left\langle X\left(0\right)^{2}\right\rangle -\left\langle X\left(0\right)\right\rangle ^{2}=U_{\mathrm{OM}}^{\prime\prime}(\bar{x}^{\ast})^{-1}$
of the stochastic system as a function of $r$ at the stable stationary
points determined from the curvature of the tree-level approximation
(black) and one loop approximation (blue) of $U_{\mathrm{OM}}$. The
one-loop result for $r>0$ is shown only for $r>r_{\mathrm{c}}$,
where $r_{\mathrm{c}}$ is the value of the leak term below which
the one-loop corrections are greater than the tree-level contributions
(Ginzburg criterion \citep[chap. 6.4]{Amit84}), i.e. $|\protect\Giitofl|>|r|$.
\textbf{C }Effective potentials $U_{\mathrm{OM}}$ \eqref{eq:U_OM}
and $U_{\mathrm{DD}}$ \eqref{eq:U_DD} in the broken symmetry phase
($r=-0.22$, left black dotted vertical line in A and B). Tree level
approximation ($U_{\mathrm{OM}}$: black, $U_{\mathrm{DD}}$: gray)
and one-loop approximation of $U_{\mathrm{OM}}$ \eqref{eq:U_OM_one_loop}
(blue); one-loop approximation is expanded up to second order in $\delta x=x^{\ast}-x_{\pm}^{1}$,
where $x_{\pm}^{1}\protect\neq0$ is the one-loop stationary point
of $U_{\mathrm{OM}}$ shown in A (blue curve). \textbf{D} Same as
C, but in the symmetric phase ($r=0.22$, right black dotted vertical
line in A and B); one-loop result expanded in $\text{\ensuremath{\delta x}}$
around $x_{0}=0$. For all panels $u=0.4$ and $D=0.05$.\label{fig:Critical-point-bistable}}
\end{figure}

To illustrate the usefulness of these two approaches, the de Dominics
equilibrium effective potential \eqref{eq:U_DD-1} and the OM form
\eqref{eq:eff_po}, we study the dynamics of an excitatory-inhibitory
network of rate neurons (\prettyref{fig:Diverging-fluctuations-in}A),
which is a generalization \citep{Kadmon15_041030} of the classical
model by Sompolinsky, Crisanti, and Sommers \citep{Sompolinsky88_259}.
The model is given by an $N$-dimensional stochastic differential
equation of the general form \eqref{eq:N_dim}. It describes the activity
of $N_{E}$ excitatory and $N_{I}$ inhibitory nonlinear neurons in
a sparse random network with a fixed number of outgoing connections
$p\cdot N$ for each neuron. Here $p$ denotes the connection probability
and $N$ the number of neurons. Nonzero connections $J_{ij}$ take
on the values $J$ or $gJ$ depending on whether neuron $j$ is excitatory
or inhibitory, with $g<0$. We here choose $\phi=\tanh$, as in the
original work \citep{Sompolinsky88_259}. Focusing on the population-averaged
activity $x(t)=\frac{1}{N}\sum_{i}x_{i}(t)$ only, \eqref{eq:N_dim}
can be rewritten as
\begin{align}
dx(t)+x(t)\,dt & =\frac{1}{N}\sum_{i}\sum_{j}J_{ij}\phi(x_{j}(t))\,dt+dW(t)\nonumber \\
 & =pJ\left(\sum_{j\in E}\phi(x_{j}(t))+g\sum_{j\in I}\phi(x_{j}(t))\right)\,dt+dW(t)\nonumber \\
 & =pJ\left(N_{E}+gN_{I}\right)\phi(x(t))\,dt+\mathcal{O}(\delta x_{i}(t))dt+dW(t),\label{eq:SCS}
\end{align}
where $dW(t)=\frac{1}{N}\sum_{i}W_{i}(t)$ is the noise component
projected in the population direction $(1,...,1)$, and $\delta x_{i}(t)=x_{i}(t)-x(t)$
is the deviation of each individual neuron activity from the population
average $x(t)$. The terms $\mathcal{O}(\delta x_{i}(t))$ arise from
a Taylor expansion of the nonlinearity $\phi$ around $x(t)$. We
notice that the population-averaged activity is driven by fluctuations
$\delta x_{i}(t)$ of individual neurons and external fluctuations
$dW_{i}(t)$. The former arise from the recurrent processing of uncorrelated
external inputs $dW_{i}(t)$ via random connections $J_{ij}$. Numerical
simulations show that the variance of the population activity strongly
depends on the ratio $R:=pJ\left(N_{E}+gN_{I}\right)$ between excitation
and inhibition in the network (\prettyref{fig:Diverging-fluctuations-in}B-D).
In particular, the size and timescale of fluctuations strongly increase
for $R\rightarrow1$ (\prettyref{fig:Diverging-fluctuations-in}C),
the point where excitation and inhibition in the network exactly balance
the neuronal leak term on the left side of \eqref{eq:SCS}. Moreover,
for $R>1$ we observe a non-zero mean activity (\prettyref{fig:Diverging-fluctuations-in}D),
even though the external input has zero mean and the nonlinearity
is point-symmetric. These observations point towards a phase transition
in the model.

A fully self-consistent treatment of the model including the colored-noise
fluctuations $\delta x_{i}$ of individual neurons is possible. To
this end the MSRDJ formalism needs to be combined with a disorder
average \citep{Sompolinsky88_259,Schuecker18_041029}. However, in
order to expose the phase transition elicited from the interplay of
general fluctuations on the right hand side of \eqref{eq:SCS} and
the ratio $R$ between excitation and inhibition, it is sufficient
to only consider white noise fluctuations. Then the population dynamics
reduces to
\begin{align}
dx(t)+x(t)\,dt & =R\,\phi(x(t))\,dt+dW(t)\nonumber \\
 & \approx R\,\big(x(t)-\frac{1}{3}\,x^{3}(t)\big)\,dt+dW(t),\label{eq:Langevin_third_order-1}
\end{align}
where we expanded the nonlinearity $\phi(x)=\tanh(x)\approx x-\frac{1}{3}\,x^{3}$,
which can be done for small external noise amplitudes $D$ defined
as $\left\langle dW(t)dW(s)\right\rangle =D\,\delta_{t,s}dt$. \prettyref{eq:Langevin_third_order-1}
describes the stereotypical setting of a bistable system \citep[see]["Model A"]{Hohenberg77}
for which both the de Dominics equilibrium effective potential \eqref{eq:U_DD-1}
and the OM form \eqref{eq:eff_po} are applicable to study phase transitions
(\prettyref{fig:Critical-point-bistable}). The deterministic part
can be written as the gradient of the potential $V(x)=\frac{r}{2}x^{2}+\frac{u}{12}\,x^{4}$;
the system is hence identical to fluctuations around the equilibrium
state of a Ginzburg-Landau model \citep[model A]{Hohenberg77}. In
this analogy, the parameter $r=1-R$ plays the role of the reduced
temperature \textendash{} when it vanishes, the system is at a critical
point \textemdash{} and $u=R$ is the strength of the interaction.
The fix points in the noiseless case ($D=0$) are
\begin{equation}
x_{0}:=0,\ x_{\pm}:=\pm\sqrt{-\frac{3r}{u}},\text{ for }r<0;u>0,\label{eq:fixed_points}
\end{equation}
The trivial fix point $x_{0}=0$ is stable as long as $r>0$. As $r$
becomes negative, the two stable fixed points $x_{\pm}$ come into
existence. They move out of zero in a continuous manner; the hallmark
of a continuous phase transition, shown in \prettyref{fig:Critical-point-bistable}A.
So the system becomes bistable if the level of recurrent positive
feedback is high enough; namely at $R>1$; the deterministic system
shows a pitchfork bifurcation. This is what happens in a network if
excitation becomes dominant, so that the balanced state \citep{Vreeswijk96}
is destabilized and a pair of stable states, one at high and another
at low activity appear \citep{Brunel00_183}. Furthermore, one notices
that the timescale of fluctuations diverges if $R\to1$, $r\to0$,
because the leak term in \eqref{eq:Langevin_third_order-1} vanishes.

This analysis shows that the network at the point of balance between
excitatory and inhibitory coupling can be mapped to the prototypical
model of a continuous dynamics phase transition, the time-dependent
$x^{4}$-theory \citep[see]["Model A"]{Hohenberg77}. A difference
between the original model A and the network studied here is that
the random connectivity leads to an effective, spatially homogeneous
all-to-all coupling. As is known from the general theory of critical
phenomena, the divergence of fluctuations often leads to considerable
deviations from the above mean-field analysis; for example, the mismatch
of critical exponents. In the following, we therefore illustrate how
to assess fluctuation corrections on the example of the simplified
dynamics \prettyref{eq:Langevin_third_order-1}.

In the stochastic system, the effective potential can be used to investigate
this continuous phase transition. The corresponding effective potential
in tree-level approximation \eqref{eq:effective_potential} takes
the form
\begin{align}
U_{\mathrm{OM},0}\left(\bar{x}^{\ast}\right) & =\frac{1}{2D}\big(V^{\prime}(\bar{x}^{\ast})\big)^{2}\label{eq:U_OM}\\
 & =\frac{1}{2D}(\bar{x}^{\ast})^{2}\,\big(r+\frac{u}{3}\,(\bar{x}^{\ast})^{2}\big)^{2},\nonumber 
\end{align}
shown in \prettyref{fig:Critical-point-bistable}B. This effective
potential differs from the one constructed by de Dominicis. The latter
with \eqref{eq:DeDominicis_eff_action-1} yields 
\begin{align}
U_{\mathrm{DD},0}(\bar{x}^{\ast}) & =V(\bar{x}^{\ast})\label{eq:U_DD}\\
 & =\frac{1}{2}(\bar{x}^{\ast})^{2}\,\big(r+\frac{u}{6}\,(\bar{x}^{\ast})^{2}\big),\nonumber 
\end{align}
shown in \prettyref{fig:Critical-point-bistable}C and D for $r<0$
and $r>0$, respectively. However, the fixed points \eqref{eq:fixed_points}
are the stationary points of $V$ and thus of $U_{\mathrm{DD}}$.
These are also stationary points of $U_{\mathrm{OM}}$. In addition,
$U_{\mathrm{OM}}$ has the stationary solutions that are roots of
$0=V^{\prime\prime}(\bar{x}^{\ast})=r+u\,(\bar{x}^{\ast})^{2}$, namely
$\bar{x}_{\pm}^{\ast\prime}=\sqrt{-\frac{r}{u}}$ (as minima). These
are the inflection points of $V$, which denote the points at which
the static theory \eqref{eq:p_Boltzmann} has a diverging propagator
\citep[see also][discussion at the end of section 6.4]{Amit84}.

The vicinity of the joint stationary points of $U_{\mathrm{DD}}$
and $U_{\mathrm{OM}}$ can also be seen in the light of equal-time
versus frequency-zero fluctuations. The curvature of $U_{\mathrm{DD}}$
and hence, to leading order, of $V$ \textemdash{} by \eqref{eq:p_Boltzmann}
\textemdash{} is the inverse of the equal-time covariance $\big(\frac{2}{D}U_{\mathrm{DD}}^{\prime\prime}(\bar{x}^{\ast})\big)^{-1}=\langle x(t)x(t)\rangle-\langle x(t)\rangle\langle x(t)\rangle$.
The curvature of $U_{\mathrm{OM}}$ at a stationary point yields \textemdash{}
by \eqref{eq:U_ddash} and \eqref{eq:variance_linear_response} \textemdash{}
the covariance of the zero frequency fluctuations $\big(U_{\mathrm{OM}}^{\prime\prime}(\bar{x}^{\ast})\big)^{-1}=\left\langle X\left(0\right)^{2}\right\rangle -\left\langle X\left(0\right)\right\rangle ^{2}$.
These two relations show that points of vanishing curvature in both
cases \eqref{eq:U_OM} and \eqref{eq:U_DD} signify the divergence
of fluctuations; hence a critical point. In the current example, both
effective potentials show that such a fluctuation infinity at the
fixed point $\bar{x}^{\ast}=0$ appears if $r=0$; at the point where
the network dynamics changes from inhibition dominance ($r>0$) to
excitation dominance ($r<0$), shown in \prettyref{fig:Critical-point-bistable}B.

Computing the effective potential in one-loop approximation (see
\prettyref{sec:Effective-potential-in-bistable}, equation \prettyref{eq:U_OM_one_loop}),
the divergence of the fluctuations appears at smaller $r<0$, whereas
for positive $r$ fluctuations are reduced compared to the tree-level
approximation. In this example we see a considerable correction caused
by the fluctuations. Thus, the point where balance is lost, the transition
temperature $r_{\mathrm{c}}$, is shifted towards smaller $r$, similar
to the Ginzburg-Landau model where fluctuation corrections reduce
the critical value $r_{\mathrm{c}}$. There, the one-loop corrections
to the variance, shown in \prettyref{fig:Critical-point-bistable}B,
are considerable.

The one-loop corrections to the effective potential diverge as $r\to0,$
as expected at a continuous phase transition. The solution to the
equation of state disappears already far above $r=0$, as shown in
\prettyref{fig:Critical-point-bistable}A. This shows that the behavior
of the system close to $r\simeq0$ is indeed strongly fluctuation-driven
and qualitatively different from the simple bifurcation in its deterministic
counterpart, the tree-level approximation. This simple example illustrates
that deterministic and stochastic models of neuronal activity may
show qualitatively quite different behavior in particular at such
critical points. The details of the calculations for this model are
presented in \prettyref{sec:Effective-potential-in-bistable}. 

For $r<0$ the system thus possesses two degenerate solutions. If
the external drive $\tilde{j}$ to the system is varied, we observe
a first-order phase transition: as $\tilde{j}$ crosses zero, the
mean jumps over from one local minimum of $U$ to the other. The true
$U_{\mathrm{OM}}$ inherits the convexity of $\Gamma_{\mathrm{OM}}$,
and should hence have a flat segment between the two local minima
in \prettyref{fig:Critical-point-bistable}C. The nonconvexity of
the approximations \prettyref{eq:U_OM} and \prettyref{eq:U_DD} is
an artifact of the simple approximation used here; in particular whether
there is a local minimum in $U_{\mathrm{DD}}$ or a local maximum
in $U_{\mathrm{OM}}$ as $\bar{x}^{\ast}=0$ is inconsequential; both
would be replaced by a straight line in the convex envelope of $U$;
the latter is obtained because $W$, computed as the Legendre transform
of the (nonconvex) approximation of $\Gamma_{\mathrm{OM}/\mathrm{DD}}$,
has different left and right-sided derivatives. Transforming back
one obtains a convex approximation of $\Gamma$ and hence $U$ (see
\prettyref{sec:Convexity-and-spontaneous-symm-breaking} for a detailed
discussion of convexity and differentiability of $W$).

Note also that on a global scale, the identification of $U_{\mathrm{OM}}$
with an energy landscape is not possible, because it assumes stationarity
and is therefore only valid near a stable fixed point. As a consequence,
the maxima of $U_{\mathrm{OM}}$ do not indicate borders of the basin
of attraction of this stable fixpoint, contrary to what would be expected
for an energy. Moreover, unstable fixed points of the system will
show up as minima of $U_{\mathrm{OM}}$, as is obvious from the tree
level approximation \eqref{eq:Gamma_OM_tree} that is positive semidefinite
and vanishes whenever $f(x^{\ast})=0$.

\paragraph*{}

\section{Discussion\label{sec:Discussion}}

This article surveys methods to obtain self-consistent approximations
by functional and diagrammatic techniques for stochastic differential
equations as they appear in models of neuronal networks. Besides a
systematic introduction, going from simple to more complex methods,
we present three main new findings.

First, we expose the fundamental relation between the Onsager-Machlup
(OM) effective action, which has a direct physical and probabilistic
interpretation, and the Martin - Siggia - Rose - de Dominicis - Janssen
(MSRDJ) effective action, which is computationally favorable. The
general exposition of this fundamental link, to our knowledge, has
been missing in the literature; it has earlier surfaced in specific
problems in certain approximations \citep{Cooper16}. In particular,
the derivation of the OM effective action from the corresponding MSRDJ
effective action naturally extends the definition of the former beyond
Gaussian noise. The OM effective action in addition allows the analysis
of bifurcations in stochastic systems. These can be studied conveniently
by help of the corresponding effective potential, which exposes whether
the stochastic system makes a first-order phase transition or a continuous
phase transition and which allows the assessment of fluctuations at
the transition. We show for the neuroscientifically important example
of the balanced state \citep{Vreeswijk96}, that the loss of balance,
which in the deterministic system causes a pitchfork bifurcation,
in the stochastic system becomes a continuous phase transition dominated
by fluctuations. We also expose the relation to the de Dominicis effective
potential \citep{DeDominicis78_4913} in equilibrium systems.

Second we derive two effective equations that are equivalent to the
stochastic nonlinear system. The first is a deterministic integro-differential
equation that captures the time evolution of the mean of the process.
A related equation has previously been derived within the Doi-Peliti
formalism of Markovian dynamics \citep{Buice07_031118}. The second
is a stochastic, but linear integro-differential equation that has
identical second-order statistics as the full system. These effective
equations serve us here to provide an intuitive interpretation of
the meaning of various vertex functions and to show how to relate
stochastic nonlinear models to effective deterministic or stochastic
linear systems.

Third, we develop a truncation scheme for the hierarchy of flow equations
that arises in the functional renormalization group, which is based
on the BMW scheme \citep{Blaizot06_571}. We here transfer this method
from the derivative expansion to the vertex expansion, and demonstrate
that this scheme yields a closed set of flow equations for the vertex
functions which accurately captures the statistics of the system.
The presented scheme is generic and may therefore be employed beyond
the application to neuronal dynamics.

The link between the OM and the MSRDJ formalism also allows us to
comment on a set of more subtle points. We carefully consider the
convexity of the cumulant-generating functional $W$ and discuss physically
relevant cases in which $W$ becomes nondifferentiable as a result
of degeneracy, for example by spontaneous symmetry breaking as it
appears in attractor networks or in networks that show bi-stability:
the existence of a convex set of solutions to the equation of state.
The relation to the OM effective action enables us to address the
question whether the effective action in the MSRDJ formalism is well-defined.
To our knowledge, this is still an open question \citep[see also][]{Duclu17}.
The work by \citet{Andersen00_1979} presents a mathematically rigorous
version of the MSR operator formalism \citep{Martin73} and concludes
that there are cases where the Legendre transform cannot be applied.
The problem in defining a Legendre transform for both sources $j$
and $\tilde{j}$ at once is their mutual dependence. This necessitates
\citet{Andersen00_1979} to consider an ensemble of paths with an
initial period of trivial dynamics (see i.p. his section V and his
Appendix D). In the path-integral formulation that we follow here,
albeit not mathematically rigorous in a strict sense, we are able
to address the problem from another view point. We separate the Legendre
transform into two steps. The first, which can rigorously be done
thanks to the convexity of $W$ in $j$, and a second, which is in
fact only needed formally: we show that the solutions of the equation
of state obtained from the MSRDJ formalism fulfill the requirement
$\langle x\rangle=x^{\ast}$, as requested by the well-defined Onsager
Machlup effective action. What hence remains open is to show that
all solutions of the OM equation of state also solve the the MSRDJ
equation of state.

The model systems studied in the current paper are intentionally left
simple to illustrate the techniques in a minimal setting. In the following
we therefore provide a slightly wider outlook for potential applications
that are of relevance to the study of neuronal networks.

The initial part of this paper reformulates the problem of finding
self-consistency equations by help of the effective action. We here
apply the standard approach known from quantum field theory and statistical
physics \citep{Vasiliev98}. This technique yields self-consistent
equations for the mean of the process that incorporate fluctuation
corrections. Applications to neuronal networks include the study of
bifurcations in the network dynamics, as we demonstrate here. Pitchfork
bifurcations, for example, are responsible for the occurrence of
multi-stability, the basis of classical attractor networks \citep{Hopfield82}.
The effective action allows us to transfer the concept of a bifurcation
in a deterministic differential equation \citep{Strogatz94} to a
stochastic system: We need to investigate the bifurcations of the
stationary points of the effective action, instead of studying the
differential equation itself. A pitchfork bifurcation, the transition
from a regime with a unique solution to one with multiple fixed points,
corresponds in the stochastic system to a critical point; the effective
action changes from having a single minimum to exhibiting a flat segment
that, beyond the bifurcation point, admits a continuum of stationary
states. Traversing the bifurcation point, the curvature vanishes
and hence fluctuations diverge. We here showed that the OM effective
action clearly exposes this fundamental property in the example of
a network at the point where feedback changes from dominance of inhibition
to dominance of excitation. Beyond the transition point, the system
may be brought to jump from one end of the plateau to the other -
showing a first-order phase transition as an external parameter is
varied; the network is bistable.

The study of transitions to oscillatory states, as they are ubiquitously
observed in neuronal systems \citep{Buzsaki04_1926}, would require
the computation of the effective action as a functional of a field
with Fourier components at nonzero frequencies. If the stationary
point of the functional is assumed at a constant field configuration
at one side of the bifurcation and by an oscillatory state on the
other side, we have the stochastic analog of a Hopf bifurcation. These
bifurcations play a central role for the generation of oscillations
\citep{Brunel03a} and for the appearance of spatio-temporal waves
which are observed in cortical networks \citep{Takahashi15,Denker18_1,Senk18_arxiv_06046v1}.
The formalism exposed here makes the influence of noise on such bifurcations
accessible. For example, the recently found phase transition at the
onset of an oscillatory state \citep{diSanto18_1356} could be analyzed
within this framework. Bifurcations in neuronal networks in the presence
of symmetries can be studied by means of the equivariant branching
lemma \citep{Barreiro17}. So far these techniques neglect the influence
of the noise altogether (i.e. employ the tree-level approximation)
and therefore their applicability is limited to network states with
weak noise. The formulation of bifurcations in terms of the effective
action would allow an extension of this method to study how symmetries
constrain bifurcations between fluctuation-dominated states.

A closely related point is the transition between multiple stable
states, such as up- and down states \citep{Steriade93,Destexhe2007_334}
or the different states of an attractor network \citep{Hopfield82}.
The average noise-driven paths of transitions between such meta-stable
states allows the assessment of the statistics of transitions between
multiple states, for example to quantify the vulnerability to noise
of information encoded in the activation of an attractor. Technically,
one here seeks escape solutions to the equation of state which have
nonvanishing values for the response field $\tx$ \citep{Bressloff09_1488,Bressloff17_033206}\citep[see ][i.p. Chapter 10 for a review]{Altland01}.
 In the setting of given initial value and free endpoint (relaxation),
however, we can use that $\tx=0$ and we can limit ourselves to small
deviations $\delta x$ of the physical variable. This enables the
computation of the effective equation of motion for the mean value
as a Taylor expansion of the equation of state. The theoretical prediction
agrees with the simulation reasonably well. A related approach was
here used to provide an approximation for the effective potential:
if the approximation of the MSRDJ effective action is quadratic in
the response field, extremizing $\tx$ is equivalent to integrating
out the response field to obtain the OM effective action. This technique
has an advantage over the computation of $\Gamma[x,\tx]$ for arbitrary
values of $\tx\neq0$, because in the latter case closed response
loops do not vanish, neither do the propagators $\Delta_{\tx\tx}$,
thus proliferating the number of diagrams to compute.

The loop expansion is shortly reviewed here, because it provides qualitative
insights into the leading order of the fluctuation corrections. In
the context of neuronal networks, the seminal work by \citet{Buice07_051919}
has introduced this technique to the study of neuronal networks. Since
the structure of the one-loop diagrams is identical to those that
appear in the functional renormalization group, one may use this method
to check which additional vertices are produced along the RG flow,
thus providing information for a good ansatz for the effective action.
We here show that the loop expansion for the considered example is
an expansion $\propto\beta^{2}D$, where $D$ is the amplitude of
the noise and $\beta$ the prefactor of the nonlinearity. In our example,
fluctuations are small so that the one-loop result is already quite
accurate. The sign of the fluctuation corrections together with the
form of the effective equations for the mean and for the second-order
fluctuations, moreover expose qualitative mechanisms that arise from
the fluctuation corrections: we show that a convex nonlinearity always
causes a positive shift of the mean of the process and that the additional
linear memory kernel that arises from the self-energy has a sign that
diminishes the leak term, thus causing a slower relaxation of the
system; the interplay of noise and nonlinearity thus prolongs the
memory of a stimulus within the system. For small deflections from
the steady state, this indirect contribution, moreover, typically
dominates over the effect of the nonlinearity \textit{per se}. Nevertheless,
many studies of neuronal networks neglect this feedback, and keep
the nonlinear terms at their mean-field level. This approach has been
shown to yield good results \citep[eq. (6)]{Brinkman18_e1006490},
\citep[eq. (3.8)]{Bravi17_045010} if fluctuations are not too strong.
This is in line with our results provided that the noise level is
low, because the linear memory kernel scales with $D\beta^{2}$, as
can be seen in \prettyref{eq:IDE_noisy_relaxation_one_loop}. Therefore
for $D\beta^{2}\ll1$, while $\beta=\mathcal{O}\left(1\right)$, it
might indeed be sufficient to consider the deterministic effect of
the nonlinearity, but not the interaction with the noise. However,
for $D\beta^{2}=\mathcal{O}\left(1\right)$ while $\beta\ll1$, the
nonlinear effects by themselves are negligible, but their interaction
with the noise induces a significant memory term. Setting $\beta$
exactly to zero obviously makes both effects vanish, which for very
noisy environments might lead to the erroneous conclusion that the
nonlinearity without the noise is the reason for the deviation from
the linear case. For large times especially, the linear noise-mediated
component due to its ``memory'' wins over the deterministic nonlinear
part. Correspondingly we show that the power spectrum at low frequencies
is enhanced. Convex nonlinearities of the gain functions of neurons,
that are required for these qualitative features to hold, are typical
in regimes in which neurons are driven by fluctuations \citep{Amit-1997_373}.

Networks with disordered connectivity, where connections are drawn
randomly, are commonly treated in mean-field approximation \citep{Sompolinsky88_259,Amit-1997_373,Vreeswijk98}.
The loop expansion is a principled way to go beyond this lowest order
approximation. The approximation is typically performed with help
of auxiliary fields, the physical meaning of which is the time-lagged
autocorrelation function of the input to a neuron \citep[see e.g.][Appendix A, eq. (A6)]{Schuecker18_041029}.
Constructing the effective action in these fields allows the systematic
computation of fluctuation corrections to the mean-field solution
\citep[see e.g.][Appendix A, eq. (A8)]{Schuecker18_041029}.

The functional renormalization group (fRG) approach is presented here
in the context of neuronal dynamics as a method that overcomes the
limitations of the loop expansion with regard to self-consistency
that is restricted to the mean. Instead, all vertex functions are
potentially renormalized by fluctuation corrections, thus in principle
allowing a fully self-consistent treatment also including corrections
to the propagators and the interaction vertices. In the regime of
weak fluctuations, we show that the results are slightly superior
to the one-loop approximation, but here do not yield qualitatively
new results. Computing higher order loop approximations is, moreover,
inherently difficult, whereas all integrals in the fRG approach always
have one-loop structure. The fRG approach, however, leads to improved
results over the one-loop approximation even though the frequency-resolution
of higher order vertices is limited by the adapted BMW approximation
introduced here. Given in addition that similar ansätze have been
successfully applied to spatially extended systems like the KPZ model
\citep{Canet10}, we believe that the insights gained by this work
will prove useful in studying neuronal networks embedded in space.
Here, also more sophisticated approaches might become useful, for
example, the decomposition of vertices in so-called channels, characterized
by the way they can be separated into two pieces by cutting two lines
(particle-particle, particle-hole, and crossed particle-hole in solid
state-physics terms). For a momentum-independent bare interaction,
the contribution of each channel has a characteristic momentum structure,
which can be used to drastically reduce the numerical effort both
in fRG \citep{Husemann09_195125,Wentzell16_arxiv} and parquet calculations
\citep{Eckhardt18_075143}.

From a conceptual point of view, the functional renormalization group
is interesting, because the flow generates new interaction vertices
that are not contained in the original model. Thus, this method shows
how the description of a neuronal network changes as fluctuations
are integrated out: It exposes which effective interactions are generated.
A specific feature of a flat regulator in frequency domain is its
direct physical interpretation as a leak term of the neuronal dynamics.
It controls the relaxation rate. Each point along the renormalization
group flow therefore corresponds to a physical system with a different
neuronal timescale. This insight may be used to study the approach
towards a critical point at which the leak term vanishes and the timescale
of fluctuations diverges. A more general discussion of frequency-dependent
regulators can be found in \citet{Duclut17_012107}.

This view is complementary to the typical application of an RG analysis,
where mostly a momentum-dependent regulator is used so that the short-ranged
degrees of freedom are subsequently integrated out. A rescaling of
the momenta then yields identical momentum ranges before and after
this marginalization, so that fixed points may occur \citep[see e.g.][i. p. "The Scaling Form of the RG Equation of the Dimensionless Potential"]{Delamotte12}.
In this view, each point on the RG trajectory represents the same
system, just described at a different level of coarse graining.

Despite its simplicity, the here-considered model exposes two fundamental
properties: First, the fluctuation corrections to the self-energy
$\Giitofl$ shift the point of transition with regard to mean-field
theory. The latter predicts criticality at the point of vanishing
leak term $m=0$. The self-energy corrections reduce the leak term,
thus promoting critical fluctuations. The critical point is therefore
reached already at a nonzero negative value $m_{c}<0$ of the leak
term. Qualitatively, the behavior of the self-energy corrections is
therefore opposite to the best known text book model of criticality,
the $\varphi^{4}$ theory, where the transition is delayed to a negative
mass term \citep[e.g. ][eq. 6.26]{Amit84}. In addition, a second
mechanism causes a shift of the transition point, which is absent
in an even theory as the $\varphi^{4}$ model: Fluctuation corrections
to the mean value increase the mean value $\bar{x}$. Thus, the effective
leak term $m(\bar{x})$ is weakened, further promoting the approach
to the critical point.

These generic observations only depend on the assumption of an expansive
nonlinear neuronal gain function, so that we expect qualitatively
similar results for example in a (fully or densely connected) network.
The shift of the transition point obviously depends on the amplitude
of the noise. It is known that fluctuations vary in neuronal networks
in response to stimuli \citep{Churchland10,LitvinKumar12_e1002667}.
Thus, neuronal systems may dynamically change their distance to the
critical point within short periods of time.

A particularly interesting feature of the approach to the critical
point are trajectories that depart from the stationary mean towards
the location of the second, unstable, fixed point. Their dynamics
slows down not only due to the reduced leak term by the two mechanisms
described above, but also due to passing the vicinity of the second,
unstable fixed point. In neuronal systems this mechanism may be useful
to generate transient behavior on slow timescales, beyond the slow
down of fluctuations close to the stable fixed point. One may speculate
if such mechanisms play a role in long transient behavior observed
in delayed response tasks \citep{Obayashi03_233}.

Recently a Ginzburg-Landau type theory of neuronal activity has been
formulated by Di Santo et al. \citep{diSanto18_1356}. A bit earlier,
\citet{Henningson17} succeeded in fitting a simpler, linear model
- a leaky heat equation with additional Gaussian noise - to the recordings
of subthreshold fluctuations in acute hippocampal brain slices from
rat. Such models, expressed as partial stochastic differential equations,
naturally fall into the realm of the statistical field theoretical
methods discussed here. In particular the study of second-order phase
transitions has come into reach now \textendash{} either by employing
the established formalism of statistical field theory based on Wilson's
renormalization group for nonequilibrium stochastic dynamics \citep[see][for an authoritative review]{Hohenberg77}
or by the functional renormalization group methods presented here.
Whether or not nontrivial fixed points in neuronal networks are accessible
by former methods that rely on the closeness of the fixed point to
a Gaussian one is so far unclear. The closely related Kardar-Parisi-Zhang
model \citep{Kardar84}, for example, exhibits fixed points in the
strong coupling regime, which is therefore only accessible by nonperturbative
methods, for example the fRG approach presented here \citep{Wiese1998}.

The currently employed theory of second-order phase transitions in
neuronal networks follows two main themes. The first employs dynamic
models like branching processes. It is determined by the branching
parameter, the average number of downstream descendants produced by
the current activity \citep{Priesemann14_80,Wilting18_2325}. It relates
second-order phase transitions to the transition originally studied
in the sandpile model \citep{Bak87}. In a similar spirit, \citet{Buice07_051919}
have introduced a network of neurons, which can be either active,
quiescent, or refractory and are described by a master equation. This
model also shows a dynamic phase transition, so this is part of the
first theme. In the second theme, experimentally measured activity
is compared to equilibrium ensembles, such as pairwise maximum entropy
models \citep{Martignon95,Schneidman06_1007}; the discrepancy between
the nonequilibrium dynamics of neuronal networks and this latter approach
based on equilibrium thermodynamics has been identified as a pressing
problem \citep{Mora2011}. Following a field theoretical approach
would in particular allow the study of critical exponents in models
where neuronal activity unfolds in a spatially extended field, representing
a coarse-grained view on the activity of mesoscopic numbers of neurons
at each space point. Thus it would enable experimental predictions,
for example with regard to the spatial structure of correlated activity.
For the Manna sandpile model \citep{Manna91_L363} such a continuous
theory has been formulated and it has been found to belong to the
universality class of directed percolation with a conserved quantity
(C-DP) \citep{Vespignani98_5676,Dickmann98_5095,LeDoussal15_110601,Wiese16_042117};
the conservation of the number of sand grains here gives rise to the
conserved quantity. Belonging to the C-DP universality class, the
Manna sandpile model in particular features an absorbing state. Also
the three-state neural dynamics by Buice and Cowan can be shown to
belong to the C-DP universality class \citep{Buice07_051919}. Therefore,
this model has an absorbing state, too. In neuronal networks, though,
we typically see ongoing activity; the absence of an absorbing state
would therefore give rise to a different structure of the effective
field theory, possibly also affecting the universality class. An alternative
approach is therefore to start at the biophysics of neuronal networks
on the microscopic level and to derive the structure of an effective
long-range field theory. Investigating such models, the functional
renormalization group has proven one of the few tools that make nonperturbative
RG fixed points accessible \citep{Canet11}.

So far field theoretical methods have been applied to neuronal networks
with Markovian dynamics on discrete state spaces \citep{Buice07_051919,Buice10_377},
employing the Doi-Peliti \citep{Doi76_1465} formalism or the alternative
approach by Biroli et al. \citep{Andreanov2006_030101,Lefvre2007_07024},
which is closer to the MSRDJ formulation used here. Also neuronal
dynamics described by stochastic differential equations \citep{Harish15_e1004266,Chow15,Hertz16_033001,Schuecker18_041029,Marti18_062314,Crisanti18_arxiv}
and stochastically spiking models (nonlinear Hawkes processes \citep{Hawkes71_438,Hawkes71_83})
have recently been formulated by field theoretical methods \citep{Ocker17_1}.
For quadratic integrate-and-fire models in the mean driven regime,
a mapping to a coupled set of phase oscillators, moreover, allows
the application of the MSRDJ formalism \citep{Buice13_1,Qiu18_arxiv}.
The renormalization group methods that we presented here can directly
be applied to these systems.

Networks of leaky integrate-and-fire models \citep{Stein67a} in the
fluctuation driven, asynchronous irregular state \citep{Amit-1997_373,Brunel00_183}
are, however, inherently complicated to treat by field theoretical
methods; the reason is that an action for such models is cumbersome
to define due to the hard threshold and reset of the membrane potential.
This model and its biophysically more realistic extensions \citep{Brette-2005_3637},
however, form a kind of gold standard. It would therefore be a major
step to treat networks of such models by systematic approaches as
they are offered by field theory. So far methods for this central
model are constrained to \textit{ad-hoc} mean-field approximations,
typically resting on the annealed approximation of the connectivity
\citep{Amit97,Brunel00_183}; in particular this mean-field based
approach prohibits a systematic study of critical phenomena.

In summary, the current work imports methods from established fields
of physics into the field of theoretical neuroscience that we think
have a high potential to solve some of the technical difficulties
that arise in the study of neuronal networks in the presence of fluctuations,
nonlinearities, phase transitions, and critical phenomena.
\begin{acknowledgments}
We thank PierGianLuca Porta Mana for fruitful discussions. We also
express our gratitude towards the three anonymous reviewers, who provided
highly valuable feedback on the contents and the presentation; in
particular, for raising the question of convexity and well-definedness
of the MSRDJ effective action. Their comments and advice substantially
improved our manuscript. We are grateful to our colleagues in the
NEST developer community for continuous collaboration. All network
simulations carried out with NEST (\href{http://www.nest-simulator.org}{http://www.nest-simulator.org}).

This work was partly supported by the Exploratory Research Space seed
funds MSCALE and G:(DE-82)ZUK2-SF-CLS002 (partly financed by Hans
Herrmann Voss Stiftung) of the RWTH university; the Jülich-Aachen
Research Alliance Center for Simulation and Data Science (JARA-CSD)
School for Simulation and Data Science (SSD); the Helmholtz Association:
Young Investigator's grant VH-NG-1028; European Union\textquoteright s
Horizon 2020 Framework Programme for Research and Innovation under
the Specific Grant Agreement No. 785907 (Human Brain Project SGA2);
Jülich Aachen Research Alliance (JARA); the German Federal Ministry
for Education and Research (BMBF Grant No. 01IS19077A).
\end{acknowledgments}

\section{Appendix}

\subsection{Normalization and escape\label{subsec:Appendix_Normalization-Deker-Haake}}

It is often stated that in stationary settings all moments of the
response field vanish \citep[p. 38]{Coolen00_arxiv_II}\citep{Sompolinsky82_6860}.
This statement follows from the normalization of the probability functional
$p[x|\tilde{j}]$ for all paths, $\int\D x\,p[x|\tj]=1$. The latter
is given by
\begin{align}
p[x|\tj] & =\int\D\tx\,\exp(S[x,\tx]+\tj^{\T}\tx).\label{eq:p_x}
\end{align}
The normalization can therefore also be written as $Z[0,\tilde{j}]\equiv1$,
which, upon $n$-fold differentiation by $\tj$, yields 
\begin{align}
\langle\tx(t_{1})\cdots\tx(t_{n})\rangle & \equiv0\quad\forall t_{1},\ldots,t_{n},n.\label{eq:DekerHaake}
\end{align}
Note, however, that this holds only for paths for which either the
endpoint or the starting point is given. Fixing both, as is done for
escape problems \citep[see e.g. ][Chapter 10]{Altland01}, effectively
leads to an additional term in the action and therefore to nonzero
moments of powers of the response field: specifying the end point
of the path, we implicitly restrict the ensemble of paths in the integral
appearing in $Z$. In the path integral formulation, we include the
initial point $x_{0}$ and the final point $x_{T}$ as
\begin{align*}
\Z[j,\tilde{j}] & =\int\D x\int\D\tx\,\exp(S[x,\tx]-\tx^{\T}\delta(\circ)\,x_{0}+j^{\T}x+\tj^{\T}\tx)\,\delta(x(T)-x_{T}).
\end{align*}
The presence of the initial condition $x_{0}$ obviously does not
affect the normalization - it can as well be absorbed into a shift
of $\tilde{j}\to\tilde{j}-\delta(\circ)x_{0}$. Fixing the final condition
$x_{T}$, however, leads to the relation
\begin{align*}
\Z[0,\tilde{j}] & =p(x(T)=x_{T}|x_{0},\tj),
\end{align*}
which is the conditional probability to reach the final point $x_{T}$
from the initial point $x_{0}$ given the inhomogeneity $\tilde{j}$.
This probability is not necessarily independent of $\tilde{j}$, so
that in general nonzero moments \eqref{eq:DekerHaake} appear.

\subsection{General properties of the effective action in the MSRDJ-formalism\label{sec:General-properties-of_MSRDJ-Gamma}}

Multiple derivatives of $\Gamma$ with respect to $x$ \textbf{only}
are always zero, as we will demonstrate in this section. We have seen
in \prettyref{subsec:Appendix_Normalization-Deker-Haake} that if
we do not specify an endpoint for $x$, the expectation value of all
powers of $\tx$ vanish. As a consequence, we see from \prettyref{eq:Gamma2_inv_W2}
that $\Gamma_{xx}^{\left(2\right)}\left[x^{\ast},0\right]=0$. Extending
our analysis to higher order derivatives of $\Gamma$ not involving
$\tx$, we observe that
\begin{align}
\Gamma_{xxx}^{\left(3\right)} & =-\sum_{y\in\left\{ x,\tx\right\} }\sum_{y_{1},y_{2},y_{3}}\Gamma_{x_{1}y_{1}}^{\left(2\right)}\Gamma_{x_{2}y_{2}}^{\left(2\right)}\Gamma_{x_{3}y_{3}}^{\left(2\right)}W_{y_{1},y_{2},y_{3}}\nonumber \\
 & \overset{\Gamma_{xx}^{\left(2\right)}=0}{=}-\sum_{\tx_{1},\tx_{2},\tx_{3}}\Gamma_{x_{1}\tx_{1}}^{\left(2\right)}\Gamma_{x_{2}\tx_{2}}^{\left(2\right)}\Gamma_{x_{3}\tx_{3}}^{\left(2\right)}\underbrace{W_{\tx_{1},\tx_{2},\tx_{3}}}_{=0}=0.\label{eq:Vanishing_Gamma_n_x}
\end{align}
Similarly, we can argue for all higher order vertices: They can be
decomposed into ``tree diagrams'' with derivatives of $W$ as nodes
and $\Gamma^{\left(2\right)}$ as connecting elements. A tree diagram
is defined by having the property to not include loops and especially,
that means that two nodes are connected by at most one element $\Gamma^{\left(2\right)}$.
Because we have only $x$'s at the external legs and $\Gamma_{xx}^{\left(2\right)}=0$,
we have all $W^{\left(n\right)}$ connected to external legs to be
with respect to $\tj$. Therefore, the only possibility to ``justify''
$j$-derivatives are $\Gamma_{x\tx}^{\left(2\right)}$-components
acting as internal connecting elements providing one $j$-derivative
each. Since the graphs are tree-like, we have exactly one $W^{\left(n\right)}$-node
more than connecting elements. However, we need at least one $j$-derivative
at every node to prevent that it vanishes. We deduce that the contribution
of at least one of the nodes is zero. This demonstrates that 
\[
\Gamma_{\underbrace{x,\dots,x}_{n\,\mathrm{times}}}^{\left(n\right)}=0\ \forall n.
\]

\subsection{Convexity and spontaneous symmetry breaking\label{sec:Convexity-and-spontaneous-symm-breaking}}

Cumulant-generating functions are convex. This can be seen from the
Hoelder inequality that holds for two nonnegative sequences $g_{k},h_{k}\ge0$
with $\alpha+\beta=1$ and $0\le\alpha,\beta\le1$
\begin{align}
\sum_{k}(g_{k})^{\alpha}(h_{k})^{\beta} & \le(\sum_{k}g_{k})^{\alpha}(\sum_{k}h_{k})^{\beta}\label{eq:Hoelder}
\end{align}
and from the fact that probabilities are positive, so that one can
always define an ``action'' as the log probability $p(x)=:e^{S(x)}$
(we here omit the normalization by the partition function for brevity
that would read $p(x)=e^{S(x)-\ln\mathcal{Z}}$). We here follow a
modified version of the argument in \citep{Goldenfeld92}; a similar
proof can be found in \citep{Eyink96_3419}. Applied to the moment-generating
function $Z(j)=\langle e^{j^{\T}x}\rangle$ one gets with a generalization
of Hoelder's inequality for infinite-dimensional spaces
\begin{align}
Z(\alpha j_{1}+\beta j_{2}) & =\langle e^{(\alpha j_{1}+\beta j_{2})\,x}\rangle=\int_{x}e^{(\alpha j_{1}+\beta j_{2})\,x+S(x)}\label{eq:proof_convexity_W}\\
 & \stackrel{\alpha+\beta=1}{=}\int_{x}\,e^{\alpha(S(x)+j_{1}^{\T}x)}\,e^{\beta(S(x)+j_{2}^{\T}x)}\nonumber \\
 & =\int_{x}\,\big(e^{S(x)+j_{1}^{\T}x}\big)^{\alpha}\,\big(e^{S(x)+j_{2}^{\T}x}\big)^{\beta}\nonumber \\
 & \stackrel{\text{Hoelder}}{\le}\big(\int_{x}e^{S(x)+j_{1}^{\T}x}\big)^{\alpha}\,\big(\int_{y}e^{S(y)+j_{2}^{\T}y}\big)^{\beta}\nonumber \\
 & =Z(j_{1})^{\alpha}\,Z(j_{2})^{\beta}.\nonumber 
\end{align}
So consequently the cumulant-generating function $W=\ln Z$
\begin{align}
W(\alpha j_{1}+\beta j_{2}) & \le\alpha\,W(j_{1})+\beta\,W(j_{2})\label{eq:convexity_W}
\end{align}
has a graph that is always below its chord; it is convex down.

In the case that $W(j)$ is differentiable, this means that its Hessian
is positive definite (a corresponding short proof can be found in
\citep[p. 166]{ZinnJustin96}). The definition of the effective action
by the Legendre-Fenchel transform instead of the ordinary Legendre
transform is only required if $W$ is nonanalytic; if it has a cusp
at a certain value $j^{\ast}$. Such a cusp corresponds to the situation
of spontaneous symmetry breaking: the mean value that is conjugate
to $j$ is different if $j^{\ast}$ is approached from above or below:
\begin{align*}
\langle x^{+}\rangle & :=\lim_{j\searrow j^{\ast}}W^{(1)}(j)\\
 & \neq\lim_{j\nearrow j^{\ast}}\,W^{(1)}(j)\\
 & =:\langle x^{-}\rangle.
\end{align*}
The authoritative book by Vasiliev contains a more detailed discussion
of the role of Legendre transforms in the study of phase transitions
\citep[i.p. section 6]{Vasiliev98}.

The Legendre transform $\mathcal{L}$ of any function $f(j)$ is convex
down. This is because for
\begin{align*}
g(x) & :=\sup_{j}\,j^{\T}x-f(j)
\end{align*}
we find with $\alpha+\beta=1$ that
\begin{align}
g(\alpha x_{a}+\beta x_{b}) & =\sup_{j}\,j^{\T}(\alpha x_{a}+\beta x_{b})-(\alpha+\beta)\,f(j)\label{eq:convexity_g}\\
 & \le\sup_{j_{a}}\,\alpha\big(j_{a}^{\T}x_{a}-f(j_{a})\big)+\sup_{j_{b}}\,\beta\big(j_{b}^{\T}x_{b}-f(j_{b})\big)\nonumber \\
 & =\alpha\,g(x_{a})+\beta\,g(x_{b}),\nonumber 
\end{align}
which shows that $g$ is convex down. Convexity of $f$ is not required
here. This general result holds in particular for the effective action
defined as
\begin{align}
\Gamma\left(x^{\ast}\right) & :=\sup_{j}j^{\T}x^{\ast}-W\left(j\right),\label{eq:def_Gamma_W}
\end{align}
which therefore is convex down, too.

Furthermore, the Legendre transform is an involution on convex functions;
applied twice we come back to the original function. However, in the
case that the original function was not convex, the result would be
the convex envelope of the original function \citep[see e.g.][i.p. Fig 9 and surrounding text]{Touchette09}).
So far, this issue cannot arise if we were able to compute $\Gamma$
or $W$ exactly; both functions are convex and therefore are the Legendre
transforms of one another. Moreover, $\Gamma(x^{\ast})$ for a typical
physical system is in addition differentiable everywhere. For if it
had a cusp, this would mean that $W(j)$ has a flat segment; the value
of the source $j$ would not affect the mean $\langle x\rangle$ for
values within this segment and all fluctuations would vanish, an untypical
behavior (thinking of $j$ being the external magnetic field and $\langle x\rangle$
the magnetization): so even if $W$ is nonanalytic in some point $j_{c}$,
its Legendre transform is analytic.

An issue arises when approximations are made. We will illustrate the
point with help of the simplest tree-level approximation, the loopwise
approximation to lowest order. The approximation of the effective
action then is
\begin{align*}
\Gamma_{0}\left(x^{\ast}\right) & =-S\left(x^{\ast}\right),
\end{align*}
which is not necessarily convex; let us think of the action $S(x)=-\frac{1}{2}x^{2}+\frac{1}{4}x^{4}$,
for example, which has two minima at $x_{\pm}^{\ast}=\pm1$ and an
intermediate local maximum at $x=0$, clearly a nonconvex function,
which is meant to approximate the convex function $\Gamma$. Legendre-transforming
this approximation to $W_{0}(j)=\mathcal{L}\{\Gamma_{0}\}(j)=\mathcal{L}\{-S\}(j)$
yields a function in $j$ with a cusp at $j^{\ast}=0$, because the
supremum operation in $W(j)=\sup_{x}j\,x+S(x)$ for $j<0$ finds the
supremum at $x\le-1$, and for $j>0$ finds the supremum at $x\ge1$.
As $j$ moves through zero from below, the point of the supremum thus
jumps from $x=-1$ to $x=+1$. Since the position of the supremum
is the left-sided (for $j<0$) or the right-sided (for $j>0$) slope
of $W(j)$, the latter function has a cusp at zero
\begin{align}
\langle x^{\pm}\rangle\equiv W^{(1)}(0\pm) & =\pm1.\label{eq:cusp_W}
\end{align}
We may perform the Legendre transform explicitly by finding the supremum
as the solution to $\partial_{x}(S(x)+j\,x)=0$ for $x\in(-\infty,-1]$
for $j<0$ and within $x\in[1,\infty)$ for $j>0$. Transforming back
to $\Gamma_{0}^{\ast}:=\mathcal{L}\{W_{0}\}$, we obtain the convex
envelope of the original $\Gamma_{0}$: The two minima at $x_{\pm}^{\ast}=\pm1$
are joined by a straight line. This follows directly from the cusp
of $W$ in \eqref{eq:cusp_W}, because computing $\Gamma_{0}^{\ast}(x^{\ast})=\sup_{j}j\,x-W(j)$
for $x^{\ast}\in(-1,1)$ always assumes the supremum for $j$ at $j=0$;
so the resulting function $\Gamma^{\ast}$ is flat on this segment,
the value of which is $W(0)$. These relations are easiest appreciated
graphically. A detailed explanation including graphical illustrations
can for example be found in \citep[i.p. Fig 9]{Touchette09}.

The fRG, in particular in its implementation with the derivative expansion,
has the interesting property that the resulting partial differential
equation for the effective action (or effective potential, the effective
action at vanishing Fourier mode $k$) becomes convex as the flow
parameter evolves \citep[see e.g.][i.p. Fig 2.15 and discussing text]{Delamotte12}.
This is in contrast to simpler approximation schemes, such as the
loop expansion, as we have illustrated above on its lowest order approximation.
Still, as more loop corrections are incorporated, the plateau becomes
successively flatter.

\subsection{Convexity of the MSRDJ cumulant-generating functional\label{sub:Convexity-MSRDJ}}

In the form \prettyref{eq:def_W_OM}, $W[j]$ is a cumulant
generating functional of the field $x$ and thus, by \prettyref{eq:convexity_W},
a convex functional in $j$. We can therefore perform the Legendre
transform with regard to $j$ to obtain the effective action

\begin{align}
\Gamma_{1}[x^{\ast},\tj] & :=\sup_{j}j^{\T}x^{\ast}-W_{\mathrm{MSRDJ}}[j,\tj]\label{eq:half_legendre}
\end{align}
as in \prettyref{eq:def_Gamma_W} only that we left $\tj$ as a parameter
indicating some additional input. By the convexity of $W$ in $j$
it is assured that the mapping between $x^{\ast}$ and $j$ is one-to-one
and the Legendre transform with regard to $j$ is involutive.

\subsubsection*{Necessity of the Legendre-Fenchel transform}

Let us first show that the cumulant-generating functional $W[j,\tj]$
indeed may have nonanalytical behavior that requires the use of the
Legendre-Fenchel transform rather than the ordinary Legendre transform;
this happens, for example, in a bistable system. For concreteness,
let us assume the stochastic differential equation of the form 
\begin{align}
dx(t) & =-V^{\prime}(x)\,dt+dW(t)\label{eq:diffeq_bistable}
\end{align}
with $V(x)=-\frac{1}{2}x^{2}+\frac{1}{4}x^{4}$ and initial condition
$x(0)=0$ on the local maximum of $V$. For $\tj=0$ depending on
the realization of the noise $dW$, the system will move close to
either of the two minima $x_{\pm}=\pm1$ of the potential $V$; for
sufficiently small noise, the system will stay close to the spontaneously
chosen minimum for prolonged times.

We first consider the analytical properties of $W$ in $j$ at $\tj=0$.
The presence of the source term $\int_{0}^{T}j(t)x(t)\,dt$ for $j=\epsilon$
assigns a different probability to the paths $x(t)=x_{\pm}$, namely
\begin{align}
p[x(t) & =x_{+}=1]/p[x(t)=x_{-}=-1]=e^{2\epsilon T}.\label{eq:ratio_prob}
\end{align}
So in the $T\to\infty$ limit, a nonzero source $j$ suppresses either
of these symmetric solutions in the integration measure. Here time
$T$ plays a similar role as system size for spontaneous symmetry
breaking in static thermodynamics \citep[see e.g.][i.p. section 2.9]{Goldenfeld92}.
Therefore,
\begin{align*}
\lim_{T\to\infty}\,\frac{1}{T}\int_{0}^{T}\langle x(t)\rangle_{\pm}\,dt=\frac{1}{T} & \lim_{\epsilon\searrow0}\,\frac{1}{\pm\epsilon}\big(W[j(t)=\pm\epsilon,0]-W[0,0]\big)\\
\simeq & \pm1
\end{align*}
yields a different left and right-sided derivative and hence mean
value. So indeed one needs a Legendre-Fenchel transform to define
$\Gamma_{1}$ as in \eqref{eq:half_legendre}.

\subsubsection*{Differentiability of $\Gamma_{1}$ in $\protect\tj$}

We now consider the dependence of $\Gamma_{1}$ on the source $\tj\neq0$.
Such a nonzero source term corresponds to an inhomogeneity on the
right-hand side of \prettyref{eq:diffeq_bistable}
\begin{align}
dx(t)= & -V^{\prime}(x)\,dt+dW(t)-\tj\label{eq:inhom_diffeq_equi}\\
= & -(V(x)+\tj x)^{\prime}\,dt+dW(t).\nonumber 
\end{align}
In a system in thermodynamic equilibrium, we may write the stationary
probability distribution of \prettyref{eq:diffeq_bistable} as
\begin{align}
p(x(t)) & \propto\exp\big(-\frac{2}{D}\,V(x(t))-\frac{2}{D}\tj\,x(t)\big).\label{eq:equilibrium}
\end{align}
The additional inhomogeneity $-\tj$ in \eqref{eq:inhom_diffeq_equi}
can therefore be regarded as a modified source term $\left(j(t)-\frac{2}{D}\tj\right)x(t)$.
We therefore have
\begin{align*}
W[j,\tj] & =\ln\,\int\D x\,\exp\Big(\ldots+\int\big(j(t)-\frac{2}{D}\tj\big)\,x(t)\,dt\Big).
\end{align*}
The Legendre transform with regard to $j$ eliminates the nondifferentiability
in $\tj$. This is because the left and right-sided derivatives are
identical, as both limits 
\begin{align*}
\frac{\partial\Gamma_{1}[x^{\ast};\pm\epsilon]}{\partial\epsilon}= & \lim_{\epsilon\searrow0}\,\frac{1}{\pm\epsilon}\Big(\sup_{j}j^{\T}x^{\ast}-W[j,\pm\epsilon]-\Big(\sup_{k}k^{\T}x^{\ast}-W[k,0]\Big)\Big)\\
= & \lim_{\epsilon\searrow0}\,\frac{1}{\pm\epsilon}\Big(\sup_{j}j^{\T}x^{\ast}-W[\underbrace{j\mp\frac{2}{D}\epsilon}_{\hat{j}},0]-\Big(\sup_{k}k^{\T}x^{\ast}-W[k,0]\Big)\Big)\\
= & \lim_{\epsilon\searrow0}\,\frac{1}{\pm\epsilon}\Big(\sup_{\hat{j}}(\hat{j}\pm\frac{2}{D}\epsilon)^{\T}x^{\ast}-W[\hat{j},0]-\Big(\sup_{k}k^{\T}x^{\ast}-W[k,0]\Big)\Big)\\
= & \lim_{\epsilon\searrow0}\,\frac{1}{\pm\epsilon}\Big(\pm\epsilon{}^{\T}\frac{2}{D}x^{\ast}\Big)=\frac{2}{D}x^{\ast}
\end{align*}
exist for nonzero $D$ and are identical. We here used that a one-dimensional
dynamics can always be considered as following an equilibrium distribution
of the form \eqref{eq:equilibrium}. For systems for which the right-hand
side is not given by the gradient of a potential (in contrast to \prettyref{eq:inhom_diffeq_equi}),
it is less clear that the Legendre transform with respect to $j$
is differentiable with respect to $\tj$. However, it is plausible
that the effective action depends smoothly on the input $-\tj$.

In the case of a nonequilibrium system, we assume that we have a
cumulant-generating functional with potentially a nonanalytical cusp
at $j_{c}$; since left and right-sided derivatives are equal almost
everywhere {[}the set of points where $W$ is nonanalytic has measure
zero\foreignlanguage{english}{ \citep[see e.g.][section 25]{Rockafellar70}}{]},
it is sufficient to assume a single such point as $j=j_{c}$ where
$W^{(1)}[j_{c}+,\tj]\neq W^{(1)}[j_{c}-,\tj]$. To study the derivative
in $\tj$ we assume that the presence of a nonzero $\tj=\pm\epsilon$
has an infinitesimal effect on potentially all cumulants; that is
to say, we assume that we can expand the cumulant-generating functional
of the perturbed system 
\begin{align*}
W[j,\pm\epsilon] & =W[j,0]\pm\epsilon\,\left\{ \begin{array}{cc}
\sum_{n=0}^{\infty}\frac{G_{n}^{+}j^{n}}{n!}, & j>j_{c}\\
\sum_{n=0}^{\infty}\frac{G_{n}^{-}j^{n}}{n!}, & j<j_{c}
\end{array}\right\} +\mathcal{O}\left(\epsilon^{2}\right)
\end{align*}
This assumption is equivalent to stating that we assume all linear
response Green's functions $G_{n}$ for cumulants of arbitrary order
$n$ to exist. For a given physical system this assumption has to
be checked; however, it is quite reasonable to assume to hold for
typical systems. The zeroth order terms $G_{0}$ must be chosen such
that $W[0,\pm\epsilon]=0$ (due to normalization) and that $W[j,\pm\epsilon]$
is continuous at $j_{c}$; for otherwise $W$ would be nonconvex,
thus in contradiction to being a cumulant-generating functional. Here
the notation $\epsilon G_{n}^{\pm}j^{n}$ is to read $\int ds\,\epsilon(s)\,\prod_{i=1}^{n}\{\int dt_{i}\,j(t_{i})\}\,G^{\pm}(s,t_{1},\ldots,t_{n})$.

Considering the left- and right-sided derivative by $\tj$

\begin{align}
\frac{\partial\Gamma_{1}[x^{\ast};\pm\epsilon]}{\partial(\pm\epsilon)}= & \lim_{\epsilon\searrow0}\,\frac{1}{\pm\epsilon}\Big(\sup_{j}\Big(j^{\T}x^{\ast}-W[j,0]\pm\sum_{n=0}^{\infty}\frac{\epsilon G_{n}^{\pm}j^{n}}{n!}\Big)-\sup_{k}\Big(k^{\T}x^{\ast}-W[k,0]\Big)+\text{\ensuremath{\mathcal{O}(\epsilon^{2})}}\Big),\label{eq:deriv_Gam1_tilj}
\end{align}
we need to distinguish three cases: 1.) If $x^{\ast}$ is such that
the supremum in \eqref{eq:half_legendre} is assumed at a point $j\neq j_{c}$,
$W$ is differentiable in $j$. So $j^{\ast}(x^{\ast})$ is a local
maximum of $j^{\T}x^{\ast}-W[j]$, hence the first variation by $j$
vanishes, so that we get, w.l.o.g. assuming $j>j_{c}$,
\begin{align}
 & \frac{d}{d(\pm\epsilon)}\,\big(j^{\T}x^{\ast}-W[j,0]\pm\sum_{n=1}^{\infty}\frac{\epsilon\,G_{n}^{+}j^{n}}{n!}\big)\big|_{\epsilon=0}\nonumber \\
= & \underbrace{\frac{\partial j^{\T}}{\partial(\pm\epsilon)}\,x^{\ast}-\underbrace{\frac{\partial W^{\T}}{\partial j}}_{x^{\ast}}\,\frac{\partial j}{\partial(\pm\epsilon)}}_{\equiv0\text{ vanishing variation }}+\sum_{n=1}^{\infty}\frac{G_{n}^{+}j^{n}}{n!}=\sum_{n=1}^{\infty}\frac{G_{n}^{+}}{n!}\,j^{n}.\label{eq:deriv_diffable}
\end{align}
So the left and right sided derivatives by $\epsilon$ are identical;
$\Gamma_{1}[x^{\ast};\tj]$ is differentiable in $\tj$ for such $x^{\ast}$.

2.) If the supremum in \eqref{eq:half_legendre} is assumed at a point
$j=j_{c}$, we have (in the unperturbed system with $\epsilon=0$)
\begin{align*}
W^{(1)}[j_{c}-,0] & \le x^{\ast}\le W^{(1)}[j_{c}+,0].
\end{align*}
In the case that the unequal signs hold strictly as ``$<$'' the
$x^{\ast}$ form an open set; to it is clear that one can find an
$\epsilon$ small enough so that it also holds that

\begin{align}
W^{(1)}[j_{c}-,0] & \pm\sum_{n=1}^{\infty}\frac{\epsilon G_{n}^{-}j_{c}^{n-1}}{n-1!}<x^{\ast}<W^{(1)}[j_{c}+,0]\pm\sum_{n=1}^{\infty}\frac{\epsilon G_{n}^{+}j_{c}^{n-1}}{n-1!}.\label{eq:inner}
\end{align}
Hence for all such $\epsilon$ one has the supremum at $j=j_{c}$.
Therefore the first term in the derivative \eqref{eq:deriv_Gam1_tilj}
evaluates to
\begin{align*}
 & j_{c}^{\T}x^{\ast}-W[j_{c},0]\pm\sum_{n=0}^{\infty}\frac{\epsilon G_{n}^{-}j_{c}^{n}}{n!}=j_{c}^{\T}x^{\ast}-W[j_{c},0]\pm\sum_{n=0}^{\infty}\frac{\epsilon G_{n}^{+}j_{c}^{n}}{n!},
\end{align*}
where equality in the latter condition holds due to continuity of
the convex functions $W[j,\pm\epsilon]$ in $j$, as stated above.
The second term correspondingly assumes the supremum at $k=j_{c}$,
so that the result of the limit in \eqref{eq:deriv_Gam1_tilj} is
\begin{align}
\frac{\partial\Gamma_{1}[x^{\ast};\pm\epsilon]}{\partial(\pm\epsilon)}=\sum_{n=0}^{\infty}\frac{G_{n}^{-}j_{c}^{n}}{n!} & =\sum_{n=0}^{\infty}\frac{G_{n}^{+}j_{c}^{n}}{n!},\label{eq:deriv_inner}
\end{align}
independent of whether we take the left or the right sided derivative.

3.) The last case to be checked is if $x^{\ast}=W^{(1)}[j_{c}-,0]$
or $x^{\ast}=W^{(1)}[j_{c}+,0]$. It is sufficient to consider one
case, say, ``$+$''. If $\epsilon$ is such that $x^{\ast}$ moves
into the inner region so that \eqref{eq:inner} holds, the derivation
under point 2.) shows that the derivative evaluates to \eqref{eq:deriv_inner}.
In the other case $x^{\ast}>W^{(1)}[j_{c}+,0]\pm\sum_{n=1}^{\infty}\frac{\epsilon G_{n}^{+}j_{c}^{n-1}}{n-1!}$
and hence we have a local maximum $j(x^{\ast})$, as in case 1.);
the derivative is hence \eqref{eq:deriv_diffable} with $j=j_{c}$
in the limit; so the left- and right-sided derivatives both yield
the same result.

In summary, under the assumption of the existence of linear response
Green's functions for all cumulants follows that $\Gamma_{1}[x^{\ast};\tj]$
is everywhere differentiable in $\tj$; a Legendre transform with
regard to $\tj$ is thus sufficient. The Legendre-Fenchel transform
from $j$ to $x^{\ast}$ is, however, required in systems that show
spontaneous symmetry breaking.

\subsubsection*{Consistency of Legendre transform in $\protect\tj$}

It is left to be checked that the additional transform from $\tj$
to $\tx^{\ast}$ 
\begin{align*}
\Gamma[x^{\ast},\tx^{\ast}] & =\Gamma_{1}[x^{\ast};\tj]-\tj^{\T}\tx^{\ast}
\end{align*}
is also such that a one-to-one relationship exists. Here a proof of
the convexity of $\Gamma_{1}[x^{\ast};\tj]$ seems not to be possible;
to the contrary, it can be shown rigorously that for certain systems
that the Legendre transform with regard to both, $j$ and $\tj$,
is not well defined \citep{Andersen00_1979}. For the special case
of an Ornstein-Uhlenbeck process it is simple to check that $\Gamma_{1}$
is convex down in $\tj$. In general this is, however, not true. We
therefore here instead demonstrate a weaker condition: We show that
the equation of state that follows from a formal definition of the
MSRDJ effective action is indeed identical to that of the OM effective
action. The latter, as stated above, can be defined rigorously. This
is done by deriving the nontrivial part of the equation of state of
the MSRDJ formalism \eqref{eq:eq_of_state_deriv_tx} directly from
the well-defined effective action $\Gamma_{1}$.

We start by rewriting the definition of the Legendre-Fenchel transform
as
\begin{align*}
\Gamma_{1}[x^{\ast};\tj] & =-\sup_{j}\,\ln\int_{x}\exp\big(S_{\mathrm{OM}}[x;\tj]+j(x-x^{\ast})\big).
\end{align*}
The supremum assumed at the physical value $j=0$ implies that $x^{\ast}=\langle x\rangle$
equals the mean of the process. This condition, with $\delta x=x-x^{\ast}$,
is equivalently given by

\begin{align}
0 & \stackrel{!}{=}\langle\delta x\rangle\equiv\int_{\delta x}\delta x\,\exp\big(S_{\mathrm{OM}}[x^{\ast}+\delta x;\tj]\big),\label{eq:sup_condition}
\end{align}
where we assumed the Onsager-Machlup form of the action, but kept
the dependence on the source $\tj$ (as in \eqref{eq:SDE_general_with_response_source})
\begin{align*}
S_{\mathrm{OM}}[x;\tj] & =-\frac{1}{2D}\big(\partial_{t}x-f(x)+\tj\big)^{\T}\big(\partial_{t}x-f(x)+\tj\big).
\end{align*}
We have seen in \prettyref{eq:relation_actions} that we may express
this action as the Legendre transform with respect to the auxiliary
field $\tx$; this is so because $S_{\mathrm{OM}}$ is convex in $\partial_{t}x-f(x)+\tj$
and every convex function can be written as a Legendre transform of
a suitably chosen function (namely its Legendre transform)
\begin{align}
S_{\mathrm{OM}}[x;\tj] & =\underset{\tx}{\mathrm{extremize}}\,\tx^{\T}\,\big(\partial_{t}x-f(x)+\tj\big)+\frac{D}{2}\,\tx^{\T}\tx.\label{eq:S_OM_with_tx}
\end{align}
The extremum is hence attained at $\tx=-D^{-1}\big(\partial_{t}x-f(x)+\tj\big)$.
This allows us to rewrite the Onsager-Machlup action, expanded in
$\delta x=x-x^{\ast}$, as
\begin{align*}
S_{\mathrm{OM}}[x^{\ast}+\delta x] & =\frac{1}{2}\,\underbrace{\frac{-1}{D}\big(\partial_{t}x-f(x)+\tj\big)^{\T}}_{\tx^{\T}}\,\big(\partial_{t}x^{\ast}-f(x^{\ast})+\tj+\partial_{t}\delta x-f^{(1)}(x^{\ast})\,\delta x+\sum_{n=2}^{\infty}\frac{f^{(n)}(x^{\ast})}{n!}\,\delta x^{n}\big)\\
 & =\underset{\tx}{\mathrm{extremize}}\,\tx^{\T}\,\big(\partial_{t}\delta x-f^{(1)}(x^{\ast})\,\delta x\big)+\frac{D}{2}\tx^{\T}\tx\\
 & +\tx^{\T}\,\big(\partial_{t}x^{\ast}-f(x^{\ast})+\tj\big)\\
 & +\tx^{\T}\,\sum_{n=2}^{\infty}\frac{f^{(n)}(x^{\ast})}{n!}\,\delta x^{n}.
\end{align*}
The second line here contains all terms bi-linear in $\delta x$ and
$\tx$; so it defines the propagator. The third line can be regarded
as a shift of the mean of the noise; a term that is linear in $\tx$.
The last line contains the non-Gaussian terms that produce corrections.
If we neglected these corrections, the remaining terms would correspond
to an Ornstein-Uhlenbeck process $\delta x$ that is driven by a noise
$\xi$ with mean $\langle\xi\rangle=\partial_{t}x^{\ast}-f(x^{\ast})+\tj$
and variance $D$. Because the noise is chosen to be Gaussian, we
can rewrite the extremum condition \eqref{eq:S_OM_with_tx} as an
integral over $\tx$:
\begin{align}
0\stackrel{!}{=}\langle\delta x\rangle & \equiv\int_{\delta x,\tx}\,\delta x\,\exp\Big(\tx^{\T}\,\big(\partial_{t}-f^{(1)}(x^{\ast})\big)\,\delta x+\frac{D}{2}\tx^{\T}\tx\label{eq:condition_deltax_zero}\\
 & \phantom{\equiv\int_{x,\tx}\delta x\,\exp\Big(}+\tx^{\T}\,\big(\partial_{t}x^{\ast}-f(x^{\ast})+\tj\big)+\tx^{\T}\,\sum_{n=2}^{\infty}\frac{f^{(n)}}{n!}\,\delta x^{n}\Big).\nonumber 
\end{align}
In the following we will show that this equation is fulfilled if the
term $\partial_{t}x^{\ast}-f(x^{\ast})+\tj$ is represented by the
negative of all one-line irreducible diagrams with one uncontracted
$\tx$-leg; we denote the sum of these diagrams by $\Xi$. If we assume
(as we normally do) that the representation by diagrams is convergent,
this can be seen as the defining property of $\Gamma_{\mathrm{MSRDJ,}\tx,\mathrm{fl.}}^{\left(1\right)}$
that we formally obtain from a Legendre transform with respect to
both, $j$ and $\tj$. We can then conclude that $\partial_{t}x^{\ast}-f(x^{\ast})+\tj=-\Xi=\Gamma_{\mathrm{MSRDJ,}\tx,\mathrm{fl.}}^{\left(1\right)}$.
We will first demonstrate how to obtain this result from a formal
Legendre transform also with regard to $\tj$. Afterwards, we will
demonstrate that indeed the identification of $\partial_{t}x^{\ast}-f(x^{\ast})+\tj=-\Xi$
solves \prettyref{eq:condition_deltax_zero}.

The equations of state derived from the formally performed Legendre
transform of $W[j,\tj]$ with regard to $j$ and $\tj$ to $\Gamma_{\mathrm{MSRDJ}}[x^{\ast},\tx^{\ast}]$
are
\begin{align}
\tj(t)=\frac{\delta\Gamma_{\mathrm{MSRDJ}}}{\delta\tx^{\ast}(t)}\rightarrow\quad\partial_{t}x^{\ast}-f(x^{\ast})+D\,\tx^{\ast}+\tj & =\Gamma_{\mathrm{MSRDJ,}\tx,\mathrm{fl.}}^{\left(1\right)}[x^{\ast},\tx^{\ast}],\label{eq:pair_eq_state}\\
j(t)=\frac{\delta\Gamma_{\mathrm{MSRDJ}}}{\delta x^{\ast}(t)}\rightarrow-\partial_{t}\tx^{\ast}-f^{\prime}(x^{\ast})\,\tx^{\ast}+j & =\Gamma_{\mathrm{MSRDJ,}x,\mathrm{fl.}}^{\left(1\right)}[x^{\ast},\tx^{\ast}].\nonumber 
\end{align}
The second equation for the physically relevant value $j\equiv0$
admits the solution $\tx^{\ast}\equiv0$. This is so because the left-hand
side is linear in $\tx^{\ast}$ and the right-hand side vanishes for
$\tx^{\ast}=0$, because one cannot produce any nonvanishing diagrams
with one amputated $x$-leg. One can therefore insert $\tx^{\ast}\equiv0$
into the first equation of state to obtain a single nontrivial equation
\begin{align}
\partial_{t}x^{\ast}-f(x^{\ast})+\tj & =\Gamma_{\mathrm{MSRDJ,}\tx,\mathrm{fl.}}^{\left(1\right)}[x^{\ast},0],\label{eq:eq_state_consistency}
\end{align}
which shows the first part of our assertion by the usual proof that
$\Gammafl$ is composed of one-line irreducible diagrams alone (see
e.g. \citet{Kleinert09} sec 3.23.6 or \citet{Helias19_10416} section
XIV).

We hence are left to show the diagrammatic statement, namely that
\begin{align}
0= & \int_{\delta x,\tx}\,\delta x\,\exp\Big(\tx^{\T}\,\big(\partial_{t}-f^{(1)}(x^{\ast})\big)\,\delta x+\frac{D}{2}\tx^{\T}\tx\label{eq:condition_deltax_zero-2}\\
 & \phantom{\equiv\int_{x,\tx}\delta x\,\exp\Big(}-\tx^{\T}\,\Xi\left(x^{\ast}\right)+\tx^{\T}\,\sum_{n=2}^{\infty}\frac{f^{(n)}}{n!}\,\delta x^{n}\Big).\nonumber 
\end{align}
The Gaussian part in the first line defines the usual propagators
$\Delta_{xx}=\langle xx\rangle$ and $\Delta_{\tx x}=\langle\tx x\rangle$,
whereas $\Delta_{\tx\tx}\equiv0$. A nonvanishing contribution requires
the explicitly present term $\delta x$ in the integrand to be contracted.
It cannot be contracted by the propagator $\Delta_{xx}$, because
the other leg of the propagator would need to connect to a $\delta x$
from an interaction vertex of the form $\tx\,\delta x^{n}$; contracting
the remaining $\tx$ would hence require at least one closed response
loop formed by $\Delta_{\tx x}$, so the contribution vanishes. For
the same reason there cannot appear any tadpole subdiagrams that are
attached by a $\delta x$. The only possibility is hence to contract
the explicitly present $\delta x$ with an $\tx$ of an interaction
vertex by the propagator $\Delta_{\tx x}$. Thus the produced diagrammatic
corrections to the mean are of the form of tadpole diagrams
\begin{align*}
 & \Delta_{x\tx}\text{graph with single uncontracted \ensuremath{\tx}}.
\end{align*}
We defined $-\Xi$ to contain all one-line irreducible diagrams with
one amputated $\tx$-leg and negative sign. The presence of the term
$-\tx\,\Xi\left(x^{\ast}\right)$ hence cancels all irreducible tadpole
diagrams with a single amputated $\tx$-leg. Likewise, reducible contributions
cannot appear, because any reducible diagram would contain at least
one tadpole sub-diagram; but these subdiagrams are canceled by the
presence of $-\tx\Xi\left(x^{\ast}\right)$ as well. As a result,
we conclude that no diagrams remain and hence \prettyref{eq:condition_deltax_zero-2}
holds. This proves that the solution to the equations of state \eqref{eq:eq_state_consistency},
obtained from the formal joint Legendre transform with regard to both,
$j$ and $\tj$, solves condition \prettyref{eq:condition_deltax_zero}
(even if it is not clear that this is the only solution).

In summary, the specific feature of the MSRDJ formalism that the expectation
value of $\tx$ vanishes cures the fact that the Legendre transform
with respect to $\tj$ is not necessarily well-defined, because it
allows the reduction from the pair of equations of state \eqref{eq:pair_eq_state}
to a single one \eqref{eq:eq_state_consistency}. The latter can be
derived in the well-defined OM formalism, as shown above. The Legendre
transform from $\Gamma_{1}[x^{\ast};\tj]$ to $\Gamma[x^{\ast},\tx^{\ast}]$
can therefore be considered a formal step, merely used to simplify
the diagrammatic derivation of the equation of state.

\subsection{The effective action in the MSRDJ and the Onsager-Machlup-formalism\label{subsec:App_MSRDJ_OM_Gamma}}

Considering $\text{\ensuremath{\Gamma}}$ as a potential whose extremal
points are the solutions of a differential equation and that $\tx=0$
is the extremizing solution, we might conclude from $\Gamma_{x\dots x}^{\left(n\right)}=0$
(shown in \prettyref{eq:Vanishing_Gamma_n_x}) that $\Gamma$ is constant
(or at least nonanalytic in $\left(\xmean,0\right)$). However, $\Gamma$
is clearly nonconstant. It turns out that setting $\tx=0$ is correct
for the stationary point $\xmean$, but does not give us the true
shape of the ``energy landscape'' for a different $x$ in a neighborhood
of $\xmean$. This is so because it is forbidden to set $\tx$ to
a constant value prior to the calculation of the statistics of $x$;
instead we must integrate it out, since it is just a Hubbard-Stratonovich
auxiliary field used to formulate a constraint. Only for Gaussian
noise, this leads to the OM action \prettyref{eq:Def_OM-action}.
For arbitrary noise distributions, we can define the cumulant-generating
function without the need to first define the $\mathrm{OM}$ action
by writing
\begin{equation}
W_{\mathrm{OM}}\left[j\right]=\ln\int\D x\:\int\D\tx\,\exp\left(S\left[x,\tx\right]+j^{T}x\right)\label{eq:Cumulant_gen_fct_OM}
\end{equation}
This makes sense because we introduced $\tx$ as an auxiliary variable
to represent the delta-distribution and $\tj$ to measure the response
function. If we are not interested in the latter, but just in the
statistics of $x$, we can drop $\tj$ in \prettyref{eq:Cumu_gen_fct_MSRDJ}
to obtain \prettyref{eq:Cumulant_gen_fct_OM}. The OM-type effective
action then takes the form
\begin{align}
\Gamma_{\mathrm{OM}}\left[x^{\ast}\right] & =\underset{j}{\sup}\,j^{\T}x^{\ast}-W_{\mathrm{OM}}\left[j\right]\label{eq:app_def_gamma_OM}\\
 & =\left[j^{\T}x^{\ast}-W_{\mathrm{OM}}\left[j\right]\right]_{j\text{ such that }x^{\ast}=\frac{\partial}{\partial j}W_{\mathrm{OM}}\left[j\right]}.\nonumber 
\end{align}
Let us expose the connection to the MSRDJ-formalism more clearly by
performing the Legendre transform with respect to $j$ and $\tj$
gradually instead of simultaneously:
\begin{align*}
\Gamma_{1}\left[x^{\ast},\tj\right] & \coloneqq\sup_{j}\,j^{\T}x^{\ast}-W\left[j,\tj\right]\\
\Gamma_{2}\left[x^{\ast},\tx^{\ast}\right] & \coloneqq\left.\tj\tx^{\ast}+\Gamma_{1}\left[x^{\ast},\tj\right]\right|_{\tj\mathrm{\ such\ that\ }\tx^{\ast}=-\partial_{\tj}\Gamma_{1}\left[x^{\ast},\tj\right]}.
\end{align*}
In order to define $\Gamma_{2}$, we had to assume that the relation
$\tj\rightarrow\partial_{\tj}\Gamma_{1}\left[x^{\ast},\tj\right]$
is invertible. This is given if $\Gamma_{1}$ is convex in $\tj$.
So this is a sufficient condition, but not a necessary one \footnote{Consider for example the function $f:\,\mathbb{R}^{2}\rightarrow\mathbb{R}:\ \left(x,y\right)^{\T}\rightarrow\frac{1}{2}\left(x^{2}-y^{2}\right)$,
which is not convex, but Legendre-transformable (namely on itself).}. We easily see that $\Gamma_{\mathrm{OM}}\left[x^{\ast}\right]=\Gamma_{1}\left[x^{\ast},0\right]$.
For the identification of $\Gamma_{2}$, observe that for the elimination
of $j$ obtaining $\Gamma_{1}$, we choose $j\left[x^{\ast},\tj\right]$
such that
\[
x^{\ast}=\frac{\partial W}{\partial j}\left[j\left[x^{\ast},\tj\right],\tj\right].
\]
Note that by this notation, we have also lifted possible ambiguities
due to a multivalued derivative of $W$ with respect to $j$, because
we have fixed $W^{\left(1\right)}$ to $x^{\ast}$ (see also \prettyref{sec:Convexity-and-spontaneous-symm-breaking}).
In the second step, yielding $\Gamma_{2}$, we determine $\tj$ in
the following way:
\begin{align*}
\tx^{\ast} & =-\frac{d}{d\tj}\Gamma_{1}\left[x^{\ast},\tj\right]=-\frac{d}{d\tj}\left\{ j\left[x^{\ast},\tj\right]x^{\ast}-W\left[j\left[x^{\ast},\tj\right],\tj\right]\right\} \\
 & =-\frac{\partial j}{\partial\tj}x^{\ast}+\underbrace{\frac{\partial W}{\partial j}\left[j\left[x^{\ast},\tj\right],\tj\right]}_{=x^{\ast}}\,\frac{\partial j}{\partial\tj}+\frac{\partial W}{\partial\tj}\left[j\left[x^{\ast},\tj\right],\tj\right]=\frac{\partial W}{\partial\tj}\left[j\left[x^{\ast},\tj\right],\tj\right],
\end{align*}
which leads to the identification $\Gamma_{\mathrm{MSRDJ}}\left[x^{\ast},\tilde{x}^{\ast}\right]=\Gamma_{2}\left[x^{\ast},\tilde{x}^{\ast}\right]$.
Therefore, we obtain $\Gamma_{\mathrm{OM}}$ by performing the Legendre
transform of $\Gamma_{\mathrm{MSRDJ}}$ with respect to $\tx^{\ast}$
and setting $\tj=0$ afterwards, which is equivalent to finding the
$\tx^{\ast}$ extremizing $\Gamma_{\mathrm{MSRDJ}}\left[x^{\ast},\tilde{x}^{\ast}\right]$
for given $x^{\ast}$, or, in other words,
\[
\Gamma_{\mathrm{OM}}[x^{\ast}]=\underset{\tilde{x}}{\mathrm{extremize}}\,\Gamma_{\mathrm{MSRDJ}}[x^{\ast},\tx]\eqqcolon\underset{\tilde{x}}{\mathrm{extremize}}\,\Gamma[x^{\ast},\tx].
\]
The Legendre transform \prettyref{eq:app_def_gamma_OM} implies 
\begin{equation}
\frac{\delta^{2}}{\delta x\left(t\right)\delta x\left(t^{\prime}\right)}\Gamma_{\mathrm{OM}}\left[x\right]=\left[\frac{\delta^{2}}{\delta j\left(t\right)\delta j\left(t^{\prime}\right)}W\left[j,\tj\right]\right]^{-1}.\label{eq:app_second_derivative_relation}
\end{equation}
So, expressed in words, the second derivative of $\Gamma_{\mathrm{OM}}$
at $x^{\ast}$ equals the inverse of $\left\langle \delta x\delta x\right\rangle $.
Note that all $\left[\cdot\right]^{-1}$ are meant as the inverse
of operators acting on functions. For our model, in frequency domain
these inversions are simple (matrix) inversions, whereas in time domain
this amounts to finding the Green's function of the operator or, in
other words, solving a differential equation. Therefore, we can relate
the integrated covariances, given by the zero mode of the covariances
in Fourier space by Fourier-transforming \prettyref{eq:app_second_derivative_relation}
and inverting it:
\begin{equation}
\frac{\delta^{2}}{\delta J\left(\omega\right)\delta J\left(\omega^{\prime}\right)}W\left[J,\tilde{J}\right]=\left[\frac{\delta^{2}}{\delta X\left(\omega\right)\delta X\left(\omega^{\prime}\right)}\Gamma_{\mathrm{OM}}\left[X\right]\right]^{-1}.\label{eq:inv_Hessians_frequency}
\end{equation}

\subsection{Relation between $\protect\SOM$ and $\protect\SMSRDJ$ in case of
non-Gaussian noise\label{subsec:App_S_OM_and_S_MSRDJ}}

In this section, we demonstrate that approximations of $\Gamma_{\mathrm{OM}}$
and $\Gamma_{\mathrm{MSRDJ}}$ are not as simply related as their
exact counterparts. For non-Gaussian noise, even the comparison of
the tree-level approximations of the effective actions in both formalisms
yield the counterintuitive results
\begin{equation}
-\SOM\left[x\right]\neq\underset{\tilde{x}}{\mathrm{extremize}}\,-\SMSRDJ\left[x,\tx\right]\text{ in general}.\label{eq:Non_relation_SOM_SMSRDJ}
\end{equation}
To see this as an example, consider the SDE:
\[
\frac{d}{dt}x=f\left(x\right)+\xi,
\]
with the cumulant-generating function of the noise $\xi$ including
a nonvanishing third-order cumulant of strength $\alpha$ as defined
in \prettyref{eq:Non_Gaussian_noise_cum_gen_fct}. Then, the MSRDJ-action
is given by
\[
\SMSRDJ\left[x,\tx\right]=\tx\left(\dot{x}-f\left(x\right)\right)+W_{\xi}\left(\tx\right).
\]
We want to calculate
\[
\SOM[x]\coloneqq\ln\left(\int d\tx\exp\left(\SMSRDJ\left[x,\tx\right]\right)\right)
\]
 to linear order in $\alpha$. We expand $\SMSRDJ$ around the saddle
point $\tx_{0}\left[x\right]$, defined by
\[
\frac{\partial\SMSRDJ}{\partial\tx}\left[x,\tx_{0}\left[x\right]\right]=0,
\]
which leads to
\[
\tx_{0}\left[x\right]=\frac{f\left(x\right)-\dot{x}}{D}-\frac{\alpha}{2}\frac{\left(f\left(x\right)-\dot{x}\right)^{2}}{D^{2}}+\mathcal{O}\left(\alpha^{2}\right).
\]
By expanding $\tx_{0}\left[x\right]$ in $\alpha$ and inserting it
into $S$ and $\frac{\partial^{2}S}{\partial\tx^{2}}$, we observe
that
\begin{align}
\SMSRDJ\left[x,\tx_{0}\left[x\right]\right] & =-\frac{\left(\dot{x}-f\left(x\right)\right)^{2}}{2D}+\frac{\alpha}{3!}\frac{\left(f\left(x\right)-\dot{x}\right)^{3}}{D^{3}}+\mathcal{O}\left(\alpha^{2}\right)\label{eq:SMSRDJ_with_xtilde_saddle_point}\\
\frac{\partial^{2}\SMSRDJ}{\partial\tx^{2}}\left[x,\tx_{0}\left[x\right]\right] & =D\left(1-\frac{\alpha}{D^{2}}\left(\dot{x}-f\left(x\right)\right)\right)+\mathcal{O}\left(\alpha^{2}\right).\nonumber 
\end{align}
Computing the contribution from the fluctuations around the stationary
point with respect to $\tx$ yields
\begin{align*}
 & \int d\tx\exp\left(\SMSRDJ\left[x,\tx\right]\right)\\
= & \int d\tx\exp\left(\SMSRDJ\left[x,\tx_{0}\right]+\frac{\partial^{2}\SMSRDJ}{\partial\tx^{2}}\left[x,\tx_{0}\right]\frac{\left(\tx-\tx_{0}\left[x\right]\right)^{2}}{2}+\underbrace{\frac{\partial^{3}\SMSRDJ}{\partial\tx^{3}}\left[x,\tx_{0}\right]}_{=\alpha}\frac{\left(\tx-\tx_{0}\left[x\right]\right)^{3}}{3!}\right)\\
\overset{y\coloneqq\left(\tx-\tx_{0}\left[x\right]\right)}{=} & \exp\left(\SMSRDJ\left[x,\tx_{0}\left[x\right]\right]\right)\int dy\exp\left(\frac{1}{2}\frac{\partial^{2}\SMSRDJ}{\partial\tx^{2}}\left[x,\tx_{0}\right]y^{2}\right)\left(1+\frac{1}{3!}\alpha y^{3}\right)+\mathcal{O}\left(\alpha^{2}\right)\\
= & \exp\left(\SMSRDJ\left[x,\tx_{0}\left[x\right]\right]\right)\sqrt{\frac{2\pi}{\frac{\partial^{2}\SMSRDJ}{\partial\tx^{2}}\left[x,\tx_{0}\right]}}+\mathcal{O}\left(\alpha^{2}\right)\\
= & \exp\left(\SMSRDJ\left[x,\tx_{0}\left[x\right]\right]\right)\sqrt{\frac{2\pi}{D\left(1-\frac{\alpha}{D^{2}}\left(\dot{x}-f\left(x\right)\right)\right)}}+\mathcal{O}\left(\alpha^{2}\right)\\
= & \exp\left(-\frac{\left(\dot{x}-f\left(x\right)\right)^{2}}{2D}+\frac{\alpha}{3!}\frac{\left(f\left(x\right)-\dot{x}\right)^{3}}{D^{3}}\right)\sqrt{\frac{2\pi}{D}}\left(1+\frac{\alpha}{2D^{2}}\left(\dot{x}-f\left(x\right)\right)\right)+\mathcal{O}\left(\alpha^{2}\right).
\end{align*}
So, in total, we have
\[
\SOM\left[x\right]=\frac{1}{2}\ln\left(\frac{2\pi}{D}\right)\underbrace{-\frac{\left(\dot{x}-f\left(x\right)\right)^{2}}{2D}-\frac{\alpha}{3!}\frac{\left(\dot{x}-f\left(x\right)\right)^{3}}{D^{3}}}_{=\sup_{\tx}\SMSRDJ\left[x,\tx\right]+\mathcal{O}\left(\alpha^{2}\right)}+\frac{\alpha}{2D^{2}}\left(\dot{x}-f\left(x\right)\right)+\mathcal{O}\left(\alpha^{2}\right).
\]

This is the announced result \prettyref{eq:Non_Gaussian_OM_action}.
We see that the fluctuations around the saddle-point value of the
action lead to additional terms contributing to $\SOM$, not only
constant, but also $x$-dependent ones. We can reformulate this to
\prettyref{eq:Non_relation_SOM_SMSRDJ}, so the relation we obtained
for the respective effective actions does not hold for the actions,
i.e. the tree-level approximations of the effective actions.

\subsection{The loop expansion for vertices up to order three, the propagator
and its small parameter\label{subsec:App_loop_expansion_vertices_parameter}}

In this section, we explain in more detail how to translate the Feynman
diagrams into algebraic expressions and which diagrams contribute
to the first order of the loop expansion. This is of course textbook
knowledge \citep{ZinnJustin96,Kleinert09,Helias19_10416}. However,
we find it useful to recapitulate the calculation in our notation
preparing the introduction of the functional renormalization group
and furthermore, it gives us the opportunity to show what is the small
parameter in our model. This is unclear \textit{a priori} because
its action is not multiplied by a small constant, as usually assumed
in the context of a loop expansion.

\subsubsection{General structure of diagrams contributing to $\protect\Gammafl$}

To obtain all diagrams that contribute to the $l$-loop correction
to $\frac{\delta^{n}}{\delta X{}^{n}}\frac{\delta^{m}}{\delta\tX^{m}}\Gammafl$
we draw all possible connected, one-particle irreducible (1PI) diagrams
with $l$ loops and $n$ ingoing and $m$ outgoing external legs.
One-particle irreducible diagrams are those that cannot be separated
into two unconnected pieces by cutting a single propagator. This is
an advantage in terms of practical computations of the Legendre transform
$\Gammafl$ over the cumulant-generating function $W$ where one has
to account for all connected diagrams, leading to a larger number
of terms to evaluate \citep[A self-contained proof as an induction over the number of loops can be found in ][]{Helias19_10416}.

In our case the number of diagrams reduces further, because in the
Itô discretization of the stochastic differential equation, loops
that are constructed out of only directed propagators ($\Delta_{\tX X}$)
pointing all in the same direction evaluate to zero \citep{Hertz16_033001}.
Therefore, all diagrams that contain such a response loop vanish.
This is the reason why the diagram shown in eq. (\ref{struct_man_1l_tadpole})
is indeed the only one to be considered for the one-loop contribution
to the one-point vertex. Usually there are several ways of connecting
propagators and interaction vertices that lead to the same diagram.
We account for this fact by multiplying each diagram with a prefactor
that equals the number of equivalent diagrammatic representations.
We will demonstrate how to determine the multiplicity and how to translate
diagrams into algebraic expressions by means of the example of the
one-loop contributions to the propagator.

\paragraph{Example: one-loop fluctuation corrections to $\protect\Gammafl^{\left(2\right)}$}

The first fluctuation corrections to $\Gammafl^{\left(2\right)}$
consist of those diagrams that have two interaction vertices and two
amputated legs. The latter property defines what is known as ``self-energy
corrections'',\begin{fmffile}{struct_man_1l_propagator}
\fmfset{thin}{0.75pt}
\fmfset{decor_size}{4mm}
\fmfcmd{style_def wiggly_arrow expr p = cdraw (wiggly p); shrink (0.8); cfill (arrow p); endshrink; enddef;}
\fmfcmd{style_def majorana expr p = cdraw p; cfill (harrow (reverse p, .5)); cfill (harrow (p, .5)) enddef;
		style_def alt_majorana expr p = cdraw p; cfill (tarrow (reverse p, .55)); cfill (tarrow (p, .55)) enddef;}
\begin{align*}
	\Gamma_{\tilde{X}X,\mathrm{fl.}}^{\left(2\right)}\left(\sigma_1, \sigma_2\right)
		& =\,(-8)\,\parbox{30mm}{
		\begin{fmfgraph*}(80,80)
			\fmfleft{o}
			\fmfright{i}
			\fmfcurved
			\fmfsurroundn{v}{2}
			\fmffreeze
			\fmfshift{(0.22w,0.w)}{v2}
			\fmfshift{(-0.22w,0.w)}{v1}
			\fmfdotn{v}{2}
			\fmf{wiggly_arrow, tension=1.0, label=$\sigma_1$}{v2,o}
			\fmf{wiggly_arrow, tension=1.0, label=$\sigma_2$}{i,v1}
			\fmf{alt_majorana, left=0.9, tension=0.4}{v2,v1}
			\fmf{plain_arrow, left=0.9, tension=0.4}{v1,v2}
		\end{fmfgraph*}
	}
	= \left[\Gamma_{X\tilde{X},\mathrm{fl.}}^{\left(2\right)}\left(\sigma_1, \sigma_2\right)\right]^\ast \\
	\frac{1}{2!}\Gamma_{\tilde{X}\tilde{X},\mathrm{fl.}}^{\left(2\right)}\left(\sigma_1, \sigma_2\right)
		&=\,(-2)\,\parbox{30mm}{
	\begin{fmfgraph*}(80,80)
			\fmfleft{o}
			\fmfright{i}
			\fmfcurved
			\fmfsurroundn{v}{2}
			\fmffreeze
			\fmfshift{(0.22w,0.w)}{v2}
			\fmfshift{(-0.22w,0.w)}{v1}
			\fmfdotn{v}{2}
			\fmf{wiggly_arrow, tension=1.0, label=$\sigma_1$}{v2,o}
			\fmf{wiggly_arrow, tension=1.0, label=$\sigma_2$, label.side=left}{v1,i}
			\fmf{alt_majorana, left=0.9, tension=0.4}{v2,v1}
			\fmf{alt_majorana, left=0.9, tension=0.4}{v1,v2}
		\end{fmfgraph*}
	}.
\end{align*}
\end{fmffile}In principle, we could also draw a diagram of this shape with two
ingoing external legs which, however, would contain a response loop
and therefore vanishes. Thus, there is no contribution to $\Gamma_{XX,\mathrm{fl.}}^{\left(2\right)}$.
The prefactors are determined as follows. For the first diagram we
have two interaction vertices to choose from to get the external outgoing
leg. Moreover, we have to select one of the two ingoing legs from
the remaining vertex to be the external ingoing leg. Finally, there
are two possibilities to connect the internal legs of the two vertices:
either as shown above or cross-connected. In total this gives a multiplicity
of $2^{3}=8$. The minus sign is due to the sign in our definition
of the effective action and hence is present in all diagrams. The
prefactor of the second diagram stems from the two possibilities to
connect the internal legs. Using this counting scheme, we have to
include the factors of the Taylor expansion on the lefthand side ($\frac{1}{2!}$
for two times the derivative with respect to $\tX$). 

Now we use \prettyref{tab:feyn_diag_elements} to obtain the algebraic
expression for the diagrams where we have to integrate over all internal
frequencies. Taking into account all frequency conservations at the
propagators and the interaction vertices we get for the first diagram
\begin{align*}
\Giitofl= & -8\int d\omega\,\frac{\beta^{2}}{\left(2\pi\right)^{4}}\frac{2\pi D}{\omega^{2}+m^{2}}\frac{2\pi}{-i\left(\omega+\sigma_{2}\right)+m}\,\delta\left(\sigma_{1}+\sigma_{2}\right)\\
= & 4\frac{\beta^{2}D}{2\pi m}\frac{1}{i\sigma_{1}+2m}\,\delta\left(\sigma_{1}+\sigma_{2}\right),
\end{align*}
where we used the residue theorem to solve the integral. With the
second diagram we proceed in the same way and obtain
\begin{align*}
\Giittfl= & 2\frac{D^{2}}{2\pi m}\frac{1}{\sigma_{1}^{2}+\left(2m\right)^{2}}\,\delta\left(\sigma_{1}+\sigma_{2}\right).
\end{align*}
Distributing the result for the mixed derivative evenly between the
two off-diagonal entries we can write the one-loop corrections to
the second derivative of the effective action as
\[
\Gamma_{\mathrm{\mathrm{fl.}}}^{\left(2\right)}\left(\sigma_{1},\sigma_{2}\right)=\begin{pmatrix}0 & \frac{1}{-i\sigma_{1}+2m}\\
\frac{1}{i\sigma_{1}+2m} & \frac{D}{\sigma_{1}^{2}+\left(2m\right)^{2}}
\end{pmatrix}\,\frac{2\beta^{2}D}{2\pi m}\delta(\sigma_{1}+\sigma_{2}).
\]

\subsubsection{The propagator in one-loop approximation\label{subsec:one_loop_propagator}}

Making use of the property of the Legendre transform $\Gamma^{\left(2\right)}\Delta=1$
{[}see \eqref{eq:Gamma2_inv_W2}{]}, we obtain the one-loop approximation
of the propagator and hence, the variance and the response functions
by solving the former identity for $\Delta$. In frequency domain
this yields
\begin{align}
1 & =\begin{pmatrix}\Gamma_{X\tilde{X}}^{(2)}\left(\omega\right)\Delta_{\tilde{X}X}\left(-\omega\right) & \Gamma_{X\tilde{X}}^{(2)}\left(\omega\right)\Delta_{\tilde{X}\tilde{X}}\left(-\omega\right)\\
\Gamma_{\tilde{X}X}^{(2)}\left(\omega\right)\Delta_{XX}\left(-\omega\right)-\Gamma_{\tilde{X}\tilde{X}}^{(2)}\left(\omega\right)\Delta_{\tilde{X}X}\left(-\omega\right)\qquad & \Gamma_{\tilde{X}X}^{(2)}\left(\omega\right)\Delta_{X\tilde{X}}\left(-\omega\right)-\Gamma_{\tilde{X}\tilde{X}}^{(2)}\left(\omega\right)\Delta_{\tilde{X}\tilde{X}}\left(-\omega\right)
\end{pmatrix}.\label{eq:gamma2_times_delta-1}
\end{align}
For $\Gamma^{\left(2\right)}$ we use the one-loop result that we
derived in the previous section
\[
\Gamma^{\left(2\right)}\left(\omega\right)=-S^{\left(2\right)}+\Gammafl^{\left(2\right)}=\begin{pmatrix}0 & \left(-i\omega+m+\frac{A}{-i\omega+2m}\right)\\
\left(i\omega+m+\frac{A}{i\omega+2m}\right) & D\left(1-\frac{A}{\omega^{2}+\left(2m\right)^{2}}\right)
\end{pmatrix}.
\]
The variables $m$ and $A$ are expressed in terms of the model parameters
as $m=-l+2\beta x^{\ast}$ and $A=2\beta^{2}D/m$. Equation \prettyref{eq:gamma2_times_delta-1}
is solved by
\begin{alignat*}{1}
\Delta_{\tilde{X}\tilde{X}}\left(\omega\right)= & \,0,\\
\Delta_{\tilde{X}X}\left(\omega\right)=\left[\Delta_{X\tilde{X}}\left(\omega\right)\right]^{*}= & \frac{i\omega+2m}{\left(i\omega+m\right)\left(i\omega+2m\right)+A},\\
\Delta_{XX}\left(\omega\right)= & \Delta_{x\tilde{x}}\left(\omega\right)D\Delta_{\tilde{x}x}\left(\omega\right)-\Delta_{x\tilde{x}}\left(\omega\right)\Gamma_{\tX X,\mathrm{fl.}}^{\left(2\right)}\left(\omega\right)DA\Gamma_{X\tX,\mathrm{fl.}}^{\left(2\right)}\left(\omega\right)\Delta_{\tilde{x}x}\left(\omega\right),
\end{alignat*}
where, we introduced the notation $\Delta\left(\omega,\omega^{\prime}\right)=\Delta\left(\omega\right)2\pi\delta\left(\omega+\omega^{\prime}\right).$
In time domain these equations read
\begin{align}
\Delta_{\tx x}\left(t,t^{\prime}\right)=\Delta_{x\tx}\left(t^{\prime},t\right)= & -H\left(t^{\prime}-t\right)\left[\frac{m_{1}-2m}{m_{1}-m_{2}}e^{m_{1}\lvert t-t^{\prime}\rvert}+\frac{2m-m_{2}}{m_{1}-m_{2}}e^{m_{2}\lvert t-t^{\prime}\rvert}\right]\label{eq:prop_1l_time_txx}\\
\Delta_{xx}\left(t^{\prime},t\right)= & -\left[\frac{D\left(m_{1}^{2}-\left(2m\right)^{2}+A\right)}{2\left(m_{1}^{2}-m_{2}^{2}\right)m_{1}}e^{m_{1}\lvert t-t^{\prime}\rvert}+\frac{D\left(\left(2m\right)^{2}-m_{2}^{2}-A\right)}{2\left(m_{1}^{2}-m_{2}^{2}\right)m_{2}}e^{m_{2}\lvert t-t^{\prime}\rvert}\right],\label{eq:prop_1l_time_xx}
\end{align}
where $m_{1/2}=3m/2\pm\sqrt{m^{2}/4-A}$ for which the relation $m_{2}<2m<m<m_{1}<0$
holds in case of a choice of parameters for which the classical fixed
point $x_{0}$ is stable.

\subsubsection{The three-point vertex in one-loop approximation\label{subsec:one_loop_three_point_vertex}}

The corrections to the three-point interaction vertex read as\begin{fmffile}{struct_man_1l_int_vertex}
\fmfset{thin}{0.75pt}
\fmfset{decor_size}{4mm}
\fmfcmd{style_def wiggly_arrow expr p = cdraw (wiggly p); shrink (0.8); cfill (arrow p); endshrink; enddef;}
\fmfcmd{style_def majorana expr p = cdraw p; cfill (harrow (reverse p, .5)); cfill (harrow (p, .5)) enddef;
		style_def alt_majorana expr p = cdraw p; cfill (tarrow (reverse p, .55)); cfill (tarrow (p, .55)) enddef;}
\begin{align*}
	\Gamma_{\tilde{X}XX,\mathrm{fl.}}^{\left(3\right)}\left(\sigma_{1},\sigma_{2},\sigma_{3}\right)
		& =\,-8\left[\vphantom{\Bigg|}\right.\, \parbox{20mm}{
		\begin{fmfgraph*}(60,60)
			\fmftopn{i}{2}
			\fmfbottom{o}
			\fmfcurved
			\fmfsurroundn{v}{3}
			\fmffreeze
			\fmfshift{(-0.2w,0.2w)}{v1}
			\fmfshift{(-0.0w,-0.25w)}{v2}
			\fmfshift{(0.265w,0.2w)}{v3}
			\fmfdotn{v}{3}
			\fmf{wiggly_arrow, tension=1.0, label=$\sigma_1$}{v3,o}
			\fmf{wiggly_arrow, tension=1.0, label=$\sigma_2$}{i1,v2}
			\fmf{wiggly_arrow, tension=1.0, label=$\sigma_3$}{i2,v1}
			\fmf{plain_arrow, left=0.5, tension=0.4}{v2,v1,v3}
			\fmf{alt_majorana, left=0.5, tension=0.4}{v3,v2}
		\end{fmfgraph*}
	}
	 \hspace{-0.0cm} + \quad\sigma_2 \leftrightarrow \sigma_3\quad
	 + \parbox{6mm}{
		\begin{fmfgraph*}(60,60)
			\fmftopn{i}{2}
			\fmfbottom{o}
			\fmfcurved
			\fmfsurroundn{v}{3}
			\fmffreeze
			\fmfshift{(-0.2w,0.2w)}{v1}
			\fmfshift{(-0.0w,-0.25w)}{v2}
			\fmfshift{(0.265w,0.2w)}{v3}
			\fmfdotn{v}{3}
			\fmf{wiggly_arrow, tension=1.0, label=$\sigma_1$}{v3,o}
			\fmf{wiggly_arrow, tension=1.0, label=$\sigma_2$}{i1,v2}
			\fmf{wiggly_arrow, tension=1.0, label=$\sigma_3$}{i2,v1}
			\fmf{plain_arrow, left=0.5, tension=0.4}{v1,v3}
			\fmf{plain_arrow, right=0.5, tension=0.4}{v2,v3}
			\fmf{alt_majorana, left=0.5, tension=0.4}{v2,v1}
		\end{fmfgraph*}
	}
	\hspace{1.6cm} \left. \vphantom{\Bigg|} \right] \\
&= -\frac{8\beta^{3}D}{\left(2\pi\right)^{2}m} \frac{\left(4m+i\sigma_{1}\right)}{\left(2m+i\sigma_{1}\right)\left(2m-i\sigma_{2}\right)\left(2m-i\sigma_{3}\right)} \delta\left(\sigma_{1}+\sigma_{2}+\sigma_{3}\right).
\end{align*}
\end{fmffile}Often, it is useful to go into the time domain, which yields
\begin{equation}
\Gamma_{\tilde{x}xx,\mathrm{fl.}}^{\left(3\right)}\left(t_{1},t_{2},t_{3}\right)=\begin{cases}
-8\beta^{3}\frac{D}{m}\,e^{2m\left(t_{1}-t_{3}\right)} & ,\,t_{1}>t_{2}>t_{3}\\
-8\beta^{3}\frac{D}{m}\,e^{2m\left(t_{1}-t_{2}\right)} & ,\,t_{1}>t_{3}>t_{2}\\
0 & ,\,\mathrm{else}
\end{cases}.\label{eq:three_point_time-1}
\end{equation}
For the BMW approximation we will need the two quantities $\Gamma_{\tilde{X}XX,\mathrm{fl.}}^{\left(3\right)}\left(\sigma_{1},-\sigma_{1},0\right)$
and $\Gamma_{\tilde{X}XX,\mathrm{fl.}}^{\left(3\right)}\left(0,\sigma_{2},-\sigma_{2}\right)$
which in one-loop approximation read
\begin{align*}
\Gamma_{\tilde{X}XX,\mathrm{fl.}}^{\left(3\right)}\left(\sigma_{1},-\sigma_{1},0\right)= & -\frac{4\beta^{3}D}{\left(2\pi\right)^{2}}\,\frac{4m+i\sigma_{1}}{m^{2}\left(2m+i\sigma_{1}\right)^{2}},\\
\Gamma_{\tilde{X}XX,\mathrm{fl.}}^{\left(3\right)}\left(0,\sigma_{2},-\sigma_{2}\right)= & -\frac{16\beta^{3}D}{\left(2\pi\right)^{2}m}\,\frac{1}{\sigma_{2}^{2}+4m^{2}}\,,
\end{align*}
which in time domain yields
\begin{alignat*}{1}
\Gamma_{\tilde{x}xx,\mathrm{fl.},1}^{\left(3\right)}\left(t_{1},t_{2}\right)= & -\frac{8\beta^{3}D}{2\pi m}\,H\left(t_{1}-t_{2}\right)\,\left(t_{1}-t_{2}-\frac{1}{2m}\right)e^{2m\left(t_{1}-t_{2}\right)},\\
\Gamma_{\tilde{x}xx,\mathrm{fl.},2}^{\left(3\right)}\left(t_{1},t_{2}\right)= & \frac{8\beta^{3}D}{2\pi m}\frac{1}{2m}e^{2m|t_{1}-t_{2}|}.
\end{alignat*}

A comparison between the fluctuation corrections $\Gammafl$ and the
corresponding terms in the action $S$ reveals that $\Giiotfl$ counteracts
the tree-level contribution $m$, whereas the $\tx\tx$- and the $\tx xx$-contributions
are enhanced. Therefore, the linear term $m$ of the differential
equation gets weakened such that for large noise it could even effectively
vanish. This would correspond to a second-order phase transition signaled
by the divergence of fluctuations \citep[given by the Ginzburg criterion, see e.g.][i.p. section 6.4]{Amit84}.
However, we are unable to explore this regime because the destabilization
of the trivial fix point is always accompanied by a breakdown of the
loop expansion, as we will demonstrate in the following.

\subsubsection{Reduction from three to one parameter\label{subsec:App_parameter-reduction}}

Obviously, we increase the escape probability by decreasing the leak
term, which amounts to approaching the critical point. These two effects
cannot be decoupled by appropriately redefining the noise level $D$.
We see this by rescaling the time as $s=l\,t$ and accordingly the
fields $\tilde{y}\left(s\right)=\sqrt{D/l}\,\tx\left(t\right)$ and
$y\left(s\right)=\sqrt{l/D}\,x\left(t\right)$ which leads to
\[
S\left[\tilde{y},y\right]=\int ds\,\left[\tilde{y}^{\T}\left(\partial_{s}+1\right)y-\beta^{\prime}\tilde{y}^{\T}y^{2}+\tilde{y}^{\T}\frac{1}{2}\tilde{y}\right],\hspace{1em}\textrm{where}\,\beta^{\prime}=\sqrt{\frac{D}{l^{3}}}.
\]
The unstable fixed point is then given by $x_{0}=1/\beta^{\prime}$
in the noise-less case. Therefore, $\beta^{\prime}$ is the only free
parameter of the model and the strength of the nonlinearity determines
also the distance between the stable and the unstable fixed point.
So it is impossible to find a set of parameters for which on the one
hand the system operates far from the unstable fixed point so that
the expansion around the stable fixed point is accurate and for which
on the other hand and concurrently the effect of the nonlinearity
is stronger than the linear part.

\subsubsection{The small parameter of the loop expansion\label{subsec:Loop_exp_small_parameter}}

\begin{figure}
\includegraphics[width=1\textwidth]{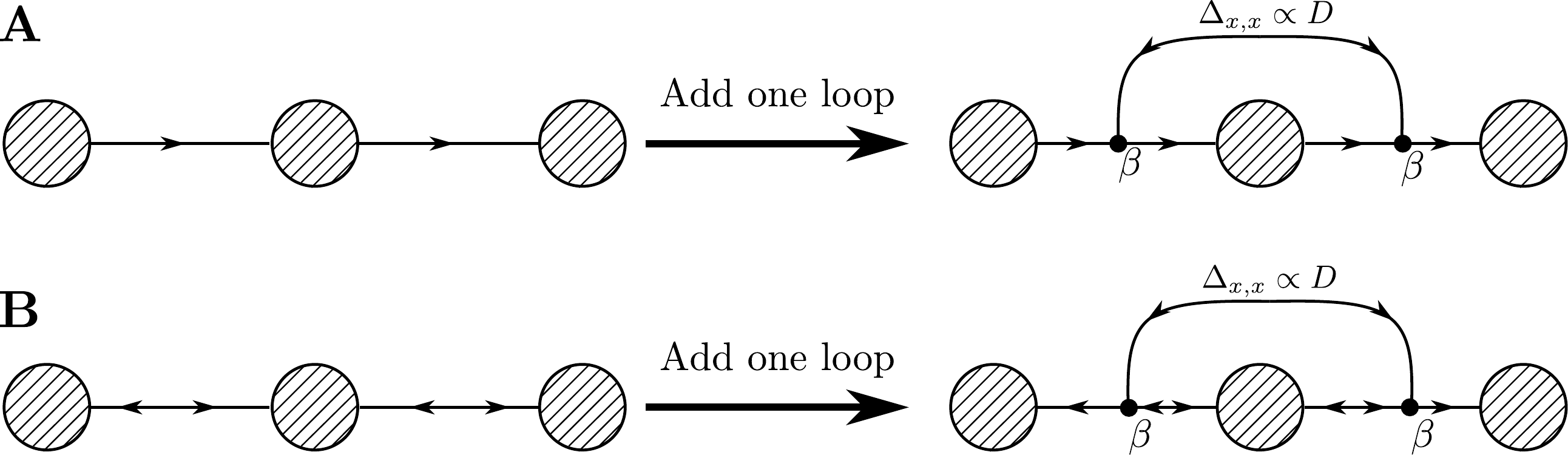}

\caption{\label{fig:Illustrative_loop_parameter}Adding an additional loop
to an arbitrary diagram: In (\textbf{A}), we attach the concerning
line to two $\Delta_{x\protect\tx}$-lines, in (\textbf{B}) to two
$\Delta_{xx}$-lines. The mixed case is analogous and therefore omitted.
Shaded circles denote arbitrary parts of a diagram.}
\end{figure}
In this section we will argue that the expansion of the effective
action in diagrams with an increasing number of loops is effectively
an expansion in terms of powers of $\beta^{2}D$. Therefore, diagrams
with higher number of loops contribute less important corrections
as long as the product of strength of the nonlinearity and the noise
variance is small.

We note that every additional loop requires exactly one $\Delta_{xx}$-propagator
and two interaction vertices. We see this as follows: Adding a loop
means attaching a propagator-line to two points in the original diagram.
This requires points at which exactly three lines meet; if we had
interactions higher than three, there could also be more. The interaction
bears the strength $\beta$ which leads to the part $\beta^{2}$ in
the loop expansion parameter. It is not clear \textit{a priori}, however,
that the line connecting these two interactions cannot be a $\Delta_{\tx x}$-
line. But \prettyref{fig:Illustrative_loop_parameter} demonstrates
that the form of the interaction with two ingoing lines and one outgoing
line always forces us to plug in a $\Delta_{xx}$-propagator, which
introduces the factor $D$. So, the loop expansion in our case is
an expansion in the parameter $\beta^{2}D$. This consideration compares
only different loop orders and not the first loop order to the tree
level, and therefore, this is not in contradiction to \prettyref{eq:eq_state_j_tilde}.

\subsection{Equation of motion for $\delta x$ from Fokker-Plank equation\label{subsec:Appendix_time-dep_FP}}

We start by multiplying the Fokker-Plank equation \citep{Risken96}
for a time-dependent density
\[
\tau\partial_{t}\,\rho\left(x,t\right)=-\partial_{x}\left(f(x)-\frac{D}{2}\partial_{x}\right)\,\rho\left(x,t\right)
\]
 by $x$ and integrating over $x$:
\begin{eqnarray*}
\partial_{t}\left\langle x\right\rangle \left(t\right) & = & -\int dx\,x\left(\partial_{x}\left(f\left(x\right)\rho\left(x,t\right)\right)-\frac{D}{2}\partial_{x}^{2}\rho\left(x,t\right)\right)\\
 & = & \int dx\,f\left(x\right)\rho\left(x,t\right)=-l\left\langle x\left(t\right)\right\rangle +\beta\left\langle x\left(t\right)^{2}\right\rangle ,
\end{eqnarray*}
where from the first to the second line we used partial integration.
From the second term only a derivative under an integral remains,
which vanishes because we assume that $\rho$ vanishes at the borders
of the integral - a property that we will use repeatedly in the following.
In the last equality, furthermore, we inserted $f\left(x\right)=-lx+\beta x^{2}$.
Now, we need an ODE for the second moment, which we obtain analogously:
\begin{eqnarray*}
\partial_{t}\left\langle x^{2}\right\rangle \left(t\right) & = & -\int dx\,x^{2}\left(\partial_{x}\left(f\left(x\right)\rho\left(x,t\right)\right)-\frac{D}{2}\partial_{x}^{2}\rho\left(x,t\right)\right)\\
 & = & \int dx\,\left(2xf\left(x\right)\rho\left(x,t\right)-Dx\partial_{x}\rho\left(x,t\right)\right)\\
 & = & \int dx\,\left(2xf\left(x\right)\rho\left(x,t\right)+D\rho\left(x,t\right)\right)\\
 & = & -2l\left\langle x^{2}\right\rangle \left(t\right)+2\beta\left\langle x^{3}\right\rangle \left(t\right)+D.
\end{eqnarray*}
Truncating the hierarchy of moments by a Gaussian closure, that is
approximating
\begin{eqnarray*}
\left\langle x^{3}\right\rangle  & = & \llangle x^{3}\rrangle+3\llangle x^{2}\rrangle\llangle x\rrangle+\llangle x\rrangle^{3}\\
 & = & \llangle x^{3}\rrangle+3\left\langle x^{2}\right\rangle \left\langle x\right\rangle -2\left\langle x\right\rangle ^{3}\approx3\left\langle x^{2}\right\rangle \left\langle x\right\rangle -2\left\langle x\right\rangle ^{3}
\end{eqnarray*}
leads to
\[
\partial\left\langle x^{2}\right\rangle \left(t\right)\approx-2l\left\langle x^{2}\right\rangle \left(t\right)+6\beta\left\langle x^{2}\right\rangle \left(t\right)\left\langle x\right\rangle \left(t\right)-4\beta\left(\left\langle x\right\rangle \left(t\right)\right)^{3}+D.
\]
Using $\llangle x^{2}\rrangle=\left\langle x^{2}\right\rangle -\left\langle x\right\rangle ^{2}$,
the equations for the first two cumulants read
\begin{align}
\partial_{t}\llangle x\rrangle\left(t\right) & =-l\llangle x\rrangle\left(t\right)+\beta\left(\llangle x^{2}\rrangle\left(t\right)+\left(\llangle x\rrangle\left(t\right)\right)^{2}\right),\label{eq:eom_fp_mean}\\
\partial_{t}\llangle x^{2}\rrangle\left(t\right) & =-2l\llangle x^{2}\rrangle\left(t\right)+4\beta\llangle x^{2}\rrangle\left(t\right)\llangle x\rrangle\left(t\right)+D.\label{eq:eom_fp_var}
\end{align}

We can interpret the one-loop equation of motion as an approximation
of the solution of the Fokker-Planck equations of motion \eqref{eq:eom_fp_mean}
and \eqref{eq:eom_fp_var}. To see this, we formally solve \eqref{eq:eom_fp_var}
and insert the result into \eqref{eq:eom_fp_mean}. It is convenient
to express the time dependence of $x$ and $\Delta$ as a deviation
from the stationary solution, defining 
\begin{align}
\left\langle x\right\rangle \left(t\right) & =\bar{x}+\delta x\left(t\right),\label{eq:Decomposition_x_x_star_delta_x}\\
\llangle x^{2}\rrangle\left(t\right) & =\bar{\Delta}+\delta\Delta\left(t\right),\label{eq:Decomposition_Delta_Delta_star_delta_Delta}\\
\bar{m} & =-l+2\beta\bar{x},\nonumber \\
m\left(t\right) & =-l+2\beta\left\langle x\right\rangle \left(t\right)=\bar{m}+2\beta\delta x\left(t\right).\nonumber 
\end{align}
Using these definitions, we obtain for the stationary case
\begin{alignat*}{2}
0 & =-l\bar{x}+\beta\bar{x}^{2}+\beta\bar{\Delta}\hspace{1em}\hspace{1em}\mathrm{and}\hspace{1em}\hspace{1em} & \bar{\Delta} & =-\frac{D}{2\bar{m}},
\end{alignat*}
which follows from \eqref{eq:eom_fp_mean} and \eqref{eq:eom_fp_var},
respectively. Expressing these equations in the new variables \prettyref{eq:Decomposition_x_x_star_delta_x}
and \prettyref{eq:Decomposition_Delta_Delta_star_delta_Delta} yields
\begin{align}
\partial_{t}\delta x\left(t\right)=\partial_{t}\llangle x\rrangle\left(t\right) & =-\left(l-2\beta\bar{x}\right)\delta x\left(t\right)+\beta\delta x\left(t\right)^{2}\text{\ensuremath{\underbrace{-l\bar{x}+\beta\bar{x}^{2}+\beta\bar{\Delta}}_{=0}}}+\beta\delta\Delta\left(t\right)\nonumber \\
 & =\bar{m}\delta x\left(t\right)+\beta\,\delta x\left(t\right)^{2}+\beta\,\delta\Delta\left(t\right)\label{eq:eom_fp_delta_x}
\end{align}
and
\begin{align*}
\partial_{t}\delta\Delta\left(t\right) & =\underbrace{2\bar{m}\bar{\Delta}+D}_{=0}+2\left(\bar{m}+2\beta\,\delta x\left(t\right)\right)\delta\Delta\left(t\right)+4\beta\bar{\Delta}\,\delta x\left(t\right)\\
 & =2m\left(t\right)\,\delta\Delta\left(t\right)+4\beta\bar{\Delta}\,\delta x\left(t\right).
\end{align*}
The latter equation turns out to be more convenient for our purpose
compared to \eqref{eq:eom_fp_var}. Solving it by variation of constants,
we obtain 
\begin{align}
\delta\Delta\left(t\right) & =4\beta\bar{\Delta}\int_{t_{0}}^{t}dt^{\prime\prime}\,\delta x\left(t^{\prime\prime}\right)\,e^{\int_{t^{\prime\prime}}^{t}dt^{\prime}\,2m\left(t^{\prime}\right)}\nonumber \\
 & =4\beta\bar{\Delta}\int_{t_{0}}^{t}dt^{\prime\prime}\,\delta x\left(t^{\prime\prime}\right)\,e^{2\bar{m}\left(t-t^{\prime\prime}\right)}\,e^{4\beta\int_{t^{\prime\prime}}^{t}dt^{\prime}\,\delta x\left(t^{\prime}\right)}\,,\label{eq:delta_Delta_explicit}
\end{align}
where we introduced the initial time $t_{0}$. We notice that $\llangle x^{2}\rrangle\left(t=t_{0}\right)=\Delta^{\ast},$
since $\delta\Delta\left(t=t_{0}\right)=0$. This corresponds to $x$
being distributed according to the stationary distribution with mean
value shifted by the initial deflection $\delta x\left(t_{0}\right)$.
If we assume that $\delta x\left(t\right)$ is small for all $t$,
we can expand the second exponential function in \prettyref{eq:delta_Delta_explicit}
and neglect terms of $\mathcal{O}\left(\delta x^{3}\right)$. Thus
\begin{align*}
\delta\Delta\left(t\right) & =4\beta\bar{\Delta}\int_{t_{0}}^{t}dt^{\prime\prime}\,\delta x\left(t^{\prime\prime}\right)\,e^{2\bar{m}\left(t-t^{\prime\prime}\right)}+16\bar{\Delta}\beta^{2}\int_{t_{0}}^{t}dt^{\prime\prime}\,\delta x\left(t^{\prime\prime}\right)\,e^{2\bar{m}\left(t-t^{\prime\prime}\right)}\int_{t^{\prime\prime}}^{t}dt^{\prime}\,\delta x\left(t^{\prime}\right)+\mathcal{O}\left(\delta x^{3}\right)\\
 & =4\beta\bar{\Delta}\int_{t_{0}}^{t}dt^{\prime}\,\delta x\left(t^{\prime}\right)\,e^{2\bar{m}\left(t-t^{\prime}\right)}+16\bar{\Delta}\beta^{2}\int_{t_{0}}^{t}dt^{\prime}\,\delta x\left(t^{\prime}\right)\int_{t_{0}}^{t^{\prime}}dt^{\prime\prime}\,\delta x\left(t^{\prime\prime}\right)\,e^{2\bar{m}\left(t-t^{\prime\prime}\right)}+\mathcal{O}\left(\delta x^{3}\right).
\end{align*}
If we insert this result into \eqref{eq:eom_fp_delta_x} we obtain
up to second order
\begin{align*}
\partial_{t}\delta x\left(t\right)= & \bar{m}\delta x\left(t\right)+\beta\,\delta x\left(t\right)^{2}\\
 & +4\beta^{2}\bar{\Delta}\int_{t_{0}}^{t}dt^{\prime}\,\delta x\left(t^{\prime}\right)\,e^{2\bar{m}\left(t-t^{\prime}\right)}\\
 & +16\bar{\Delta}\beta^{3}\int_{t_{0}}^{t}dt^{\prime}\,\delta x\left(t^{\prime}\right)\int_{t_{0}}^{t^{\prime}}dt^{\prime\prime}\,\delta x\left(t^{\prime\prime}\right)\,e^{2\bar{m}\left(t-t^{\prime\prime}\right)},
\end{align*}
which equals exactly the one-loop result \prettyref{eq:IDE_noisy_relaxation_one_loop}.

\subsection{Derivation of fRG-flow equations\label{subsec:Derivation-of-fRG-flow}}

For completeness, the derivation of the Wetterich equation \citep{WETTERICH93_90}
as presented in \citep{Berges02_223} is repeated in the following.
The difference of the current presentation is the additional presence
of the response field. We will use the property $\partial_{\lambda}\Gamma_{\lambda}=-\partial_{\lambda}W_{\lambda}$,
which holds generally for Legendre transforms of quantities depending
on a parameter, here $\lambda$ \foreignlanguage{english}{\citep[eq. (1.93)]{ZinnJustin96}}.
This yields, using the regulator of the form introduced in \prettyref{eq:action_regulator},

\begin{align}
 & \frac{\partial\tilde{\Gamma}_{\lambda}\left[X^{\ast},\tilde{X}^{\ast}\right]}{\partial\lambda}=-\frac{\partial W_{\lambda}\left[J,\tilde{J}\right]}{\partial\lambda}\nonumber \\
= & -\frac{1}{\mathcal{Z}\left[J,\tilde{J}\right]}\int\D X\,\int\D\tilde{X}\,\frac{\partial}{\partial\lambda}\Delta S_{\lambda}\left[X,\tilde{X}\right]\exp\left(S_{\lambda}\left[X,\tilde{X}\right]\right)\nonumber \\
 & \times\exp\left(J^{\T}X+\tilde{J}^{\T}\tilde{X}\right)\nonumber \\
= & \frac{1}{2}\int d\omega\,\int d\omega^{\prime}\,\langle X\left(\omega\right)\frac{\partial R_{\lambda}\left(\omega,\omega^{\prime}\right)}{\partial\lambda}\tilde{X}\left(\omega^{\prime}\right)\rangle\nonumber \\
= & \frac{1}{2}\int d\omega\,\int d\omega^{\prime}\,\left\{ \Delta_{\tilde{X}X,\lambda}\left(\omega^{\prime},\omega\right)\frac{\partial R_{\lambda}\left(\omega,\omega^{\prime}\right)}{\partial\lambda}+X^{\ast}\left(\omega\right)\frac{\partial R_{\lambda}\left(\omega,\omega^{\prime}\right)}{\partial\lambda}\tilde{X}^{\ast}\left(\omega^{\prime}\right)\right\} \nonumber \\
= & \frac{1}{2}\mathrm{Tr}\left\{ \Delta_{\tilde{X}X,\lambda}\frac{\partial R_{\lambda}}{\partial\lambda}\right\} +\frac{\partial}{\partial\lambda}\Delta S_{\lambda}\left[X^{\ast},\tilde{X}^{\ast}\right].\label{eq:Final_expression_Wetterich_eq_tilde}
\end{align}
In the third line we used $\langle\tilde{X}\left(\omega^{\prime}\right)X\left(\omega\right)\rangle=\langle\langle\tilde{X}\left(\omega\right)X\left(\omega^{\prime}\right)\rangle\rangle+\langle\tilde{X}\left(\omega^{\prime}\right)\rangle\langle X\left(\omega\right)\rangle=\Delta_{\tilde{X}X,\lambda}\left(\omega^{\prime},\omega\right)+X^{\ast}\left(\omega\right)\tilde{X}^{\ast}\left(\omega^{\prime}\right)$.
From the relation \prettyref{eq:def_eff_action_fRG} between $\tilde{\Gamma}_{\lambda}$
and $\Gamma_{\lambda}$ we arrive directly at the final form of the
Wetterich equation as presented in eq. (\ref{Wetterich_equation}).

\subsection{Flow equations for the self-energy and the interaction vertex\label{subsec:Flow_eq_Gii_Giii}}

The nonvanishing diagrams for the self-energy translate to

\begin{alignat*}{2}
\frac{\partial\Gamma_{\tilde{X}X,\lambda}^{\left(2\right)}\left(\sigma_{1},-\sigma_{1}\right)}{\partial\lambda} & = &  & \frac{1}{2}\int\frac{d\omega}{\left(2\pi\right)}\,\Gamma_{\tilde{X}XX,\lambda}^{\left(3\right)}\left(\sigma_{1},-\sigma_{1}-\omega,\omega\right)\Delta_{XX,\lambda}\left(\omega\right)\frac{\partial R_{\lambda}}{\partial\lambda}\Delta_{\tilde{X}X,\lambda}\left(\omega\right)\\
 &  &  & \times\Gamma_{XX\tilde{X},\lambda}^{\left(3\right)}\left(-\sigma_{1},-\omega,\sigma_{1}+\omega\right)\Delta_{\tilde{X}X,\lambda}\left(\sigma_{1}+\omega\right)\\
 & + &  & \frac{1}{2}\int\frac{d\omega}{\left(2\pi\right)}\,\Gamma_{\tilde{X}XX,\lambda}^{\left(3\right)}\left(\sigma_{1},-\sigma_{1}-\omega,\omega\right)\Delta_{XX,\lambda}\left(\omega\right)\Gamma_{XX\tilde{X},\lambda}^{\left(3\right)}\left(-\sigma_{1},-\omega,\sigma_{1}+\omega\right)\\
 &  &  & \times\Delta_{\tilde{X}X,\lambda}\left(\sigma_{1}+\omega\right)\frac{\partial R_{\lambda}}{\partial\lambda}\Delta_{\tilde{X}X,\lambda}\left(\sigma_{1}+\omega\right)\\
 & + &  & \frac{1}{2}\int\frac{d\omega}{\left(2\pi\right)}\,\Gamma_{\tilde{X}XX,\lambda}^{\left(3\right)}\left(\sigma_{1},-\sigma_{1}-\omega,\omega\right)\Delta_{X\tilde{X},\lambda}\left(\omega\right)\Gamma_{X\tilde{X}X,\lambda}^{\left(3\right)}\left(-\sigma_{1},-\omega,\sigma_{1}+\omega\right)\\
 &  &  & \times\Delta_{XX,\lambda}\left(\sigma_{1}+\omega\right)\frac{\partial R_{\lambda}}{\partial\lambda}\Delta_{\tilde{X}X,\lambda}\left(\sigma_{1}+\omega\right)\\
\frac{\partial\Gamma_{\tilde{X}\tilde{X},\lambda}^{\left(2\right)}\left(\sigma_{1},-\sigma_{1}\right)}{\partial\lambda} & = &  & \frac{1}{2}\int\frac{d\omega}{\left(2\pi\right)}\,\Gamma_{\tilde{X}XX,\lambda}^{\left(3\right)}\left(\sigma_{1},-\sigma_{1}-\omega,\omega\right)\Delta_{XX,\lambda}\left(\omega\right)\frac{\partial R_{\lambda}}{\partial\lambda}\Delta_{\tilde{X}X,\lambda}\left(\omega\right)\\
 &  &  & \times\Gamma_{\tilde{X}XX,\lambda}^{\left(3\right)}\left(-\sigma_{1},-\omega,\sigma_{1}+\omega\right)\Delta_{XX,\lambda}\left(\sigma_{1}+\omega\right)\\
 & + &  & \sigma_{1}\rightarrow-\sigma_{1}.
\end{alignat*}
The diagrams for the flow of the interaction vertex are given by\begin{fmffile}{Gammaiii_struct}
\fmfset{thin}{0.75pt}
\fmfset{decor_size}{4mm}
\fmfcmd{style_def wiggly_arrow expr p = cdraw (wiggly p); shrink (0.7); cfill (arrow p); endshrink; enddef;}
\fmfcmd{style_def majorana expr p = cdraw p; cfill (harrow (reverse p, .5)); cfill (harrow (p, .5)) enddef;
		style_def alt_majorana expr p = cdraw p; cfill (tarrow (reverse p, .55)); cfill (tarrow (p, .55)) enddef;}
\begin{align*}
	&\frac{\partial \Gamma^{\left( 3 \right)}_{\tilde{X}XX,\lambda}\left( \sigma_1,\sigma_2,\sigma_3 \right)}{\partial \lambda}\\
	& =\,-\frac{1}{2}\left[\vphantom{\Bigg|}\right.\, \parbox{10mm}{
		\begin{fmfgraph*}(90,80)
			\fmfcurved
			\fmfsurroundn{v}{8}
			\fmffreeze
			\fmfshift{(-0.1w,-0.1w)}{v2}
			\fmfshift{(0.1w,-0.1w)}{v4}
			\fmfshift{(0.0w,0.2w)}{v7}
			\fmfshift{(-0.23w,-0.05w)}{v1}
			\fmfv{d.s=square, d.filled=empty}{v1}
			\fmfdot{v2}
			\fmfdot{v4}
			\fmfdot{v7}
			\fmftopn{i}{2}
			\fmfbottom{o}
			\fmf{wiggly_arrow, tension=1.0, label=$\sigma_3$}{i2,v2}
			\fmf{wiggly_arrow, tension=1.0, label=$\sigma_2$}{i1,v4}
			\fmf{wiggly_arrow, tension=1.0, label=$\sigma_1$}{v7,o}
			\fmf{plain_arrow, right=0.5}{v2,v4,v7}
			\fmf{alt_majorana, right=0.4}{v7,v1}
			\fmf{plain_arrow, right=0.3}{v1,v2}
		\end{fmfgraph*}		
	}
	& +  \parbox{10mm}{
		\begin{fmfgraph*}(90,80)
			\fmfcurved
			\fmfsurroundn{v}{8}
			\fmffreeze
			\fmfshift{(-0.1w,-0.1w)}{v2}
			\fmfshift{(0.1w,-0.1w)}{v4}
			\fmfshift{(0.0w,0.2w)}{v7}
			\fmfshift{(-0.23w,-0.05w)}{v1}
			\fmfshift{(0.0w,-0.05w)}{v3}
			\fmfv{d.s=square, d.filled=empty}{v3}
			\fmfdot{v2}
			\fmfdot{v4}
			\fmfdot{v7}
			\fmftopn{i}{2}
			\fmfbottom{o}
			\fmf{wiggly_arrow, tension=1.0, label=$\sigma_3$}{i2,v2}
			\fmf{wiggly_arrow, tension=1.0, label=$\sigma_2$}{i1,v4}
			\fmf{wiggly_arrow, tension=1.0, label=$\sigma_1$}{v7,o}
			\fmf{plain_arrow, right=0.3}{v2,v3,v4}
			\fmf{plain_arrow, right=0.5}{v4,v7}
			\fmf{alt_majorana, right=0.5}{v7,v2}
		\end{fmfgraph*}		
	}&
	\\ \vphantom{A} \\
	& \hspace{1cm} + \parbox{10mm}{
		\begin{fmfgraph*}(90,80)
			\fmfcurved
			\fmfsurroundn{v}{8}
			\fmffreeze
			\fmfshift{(-0.1w,-0.1w)}{v2}
			\fmfshift{(0.1w,-0.1w)}{v4}
			\fmfshift{(0.0w,0.2w)}{v7}
			\fmfshift{(-0.23w,-0.05w)}{v1}
			\fmfshift{(-0.05w,-0.06w)}{v3}
			\fmfv{d.s=square, d.filled=empty}{v3}
			\fmfdot{v2}
			\fmfdot{v4}
			\fmfdot{v7}
			\fmftopn{i}{2}
			\fmfbottom{o}
			\fmf{wiggly_arrow, tension=1.0, label=$\sigma_3$}{i2,v2}
			\fmf{wiggly_arrow, tension=1.0, label=$\sigma_2$}{i1,v4}
			\fmf{wiggly_arrow, tension=1.0, label=$\sigma_1$}{v7,o}
			\fmf{alt_majorana, right=0.3}{v2,v3}
			\fmf{plain_arrow, right=0.3}{v3,v4}
			\fmf{plain_arrow, right=0.5}{v4,v7}
			\fmf{plain_arrow, left=0.5}{v2,v7}
		\end{fmfgraph*}		
	}
	& +  \parbox{10mm}{
		\begin{fmfgraph*}(90,80)
			\fmfcurved
			\fmfsurroundn{v}{8}
			\fmffreeze
			\fmfshift{(-0.1w,-0.1w)}{v2}
			\fmfshift{(0.1w,-0.1w)}{v4}
			\fmfshift{(0.0w,0.2w)}{v7}
			\fmfshift{(-0.23w,-0.05w)}{v1}
			\fmfshift{(0.0w,-0.05w)}{v3}
			\fmfshift{(0.22w,-0.06w)}{v5}
			\fmfv{d.s=square, d.filled=empty}{v5}
			\fmfdot{v2}
			\fmfdot{v4}
			\fmfdot{v7}
			\fmftopn{i}{2}
			\fmfbottom{o}
			\fmf{wiggly_arrow, tension=1.0, label=$\sigma_3$}{i2,v2}
			\fmf{wiggly_arrow, tension=1.0, label=$\sigma_2$}{i1,v4}
			\fmf{wiggly_arrow, tension=1.0, label=$\sigma_1$}{v7,o}
			\fmf{plain_arrow, right=0.5}{v2,v4}
			\fmf{plain_arrow, right=0.3}{v4,v5,v7}
			\fmf{alt_majorana, right=0.5}{v7,v2}
		\end{fmfgraph*}		
	}&
	\\ \vphantom{A} \\
	& \hspace{1cm} + \parbox{10mm}{
		\begin{fmfgraph*}(90,80)
			\fmfcurved
			\fmfsurroundn{v}{8}
			\fmffreeze
			\fmfshift{(-0.1w,-0.1w)}{v2}
			\fmfshift{(0.1w,-0.1w)}{v4}
			\fmfshift{(0.0w,0.2w)}{v7}
			\fmfshift{(-0.23w,-0.05w)}{v1}
			\fmfshift{(0.0w,-0.05w)}{v3}
			\fmfshift{(0.22w,-0.07w)}{v5}
			\fmfv{d.s=square, d.filled=empty}{v5}
			\fmfdot{v2}
			\fmfdot{v4}
			\fmfdot{v7}
			\fmftopn{i}{2}
			\fmfbottom{o}
			\fmf{wiggly_arrow, tension=1.0, label=$\sigma_3$}{i2,v2}
			\fmf{wiggly_arrow, tension=1.0, label=$\sigma_2$}{i1,v4}
			\fmf{wiggly_arrow, tension=1.0, label=$\sigma_1$}{v7,o}
			\fmf{alt_majorana, right=0.5}{v2,v4}
			\fmf{plain_arrow, right=0.3}{v4,v5,v7}
			\fmf{plain_arrow, left=0.5}{v2,v7}
		\end{fmfgraph*}		
	}
	& +  \parbox{10mm}{
		\begin{fmfgraph*}(90,80)
			\fmfcurved
			\fmfsurroundn{v}{8}
			\fmffreeze
			\fmfshift{(-0.1w,-0.1w)}{v2}
			\fmfshift{(0.12w,-0.02w)}{v4}
			\fmfshift{(0.0w,0.2w)}{v7}
			\fmfshift{(-0.23w,-0.05w)}{v1}
			\fmfshift{(0.0w,-0.05w)}{v3}
			\fmfshift{(0.22w,-0.15w)}{v5}
			\fmfv{d.s=square, d.filled=empty}{v5}
			\fmfdot{v2}
			\fmfdot{v4}
			\fmfdot{v7}
			\fmftopn{i}{2}
			\fmfbottom{o}
			\fmf{wiggly_arrow, tension=1.0, label=$\sigma_3$}{i2,v2}
			\fmf{wiggly_arrow, tension=1.0, label=$\sigma_2$}{i1,v4}
			\fmf{wiggly_arrow, tension=1.0, label=$\sigma_1$}{v7,o}
			\fmf{plain_arrow, left=0.5}{v4,v2,v7}
			\fmf{plain_arrow, right=0.3}{v5,v7}
			\fmf{alt_majorana, right=0.5}{v4,v5}
		\end{fmfgraph*}		
	}&\hspace{2.5cm}\left.\vphantom{\Bigg|}\right]\\
& +\sigma_2 \leftrightarrow \sigma_3.
\end{align*}
\end{fmffile}

\subsection{Effective potential from MSRDJ formalism: equilibrium systems\label{subsec:Effective-action-for-equilibrium}}

For systems in thermodynamic equilibrium, where the statistical weight
for each configuration $x(t)$ is of Boltzmann form (setting the inverse
temperature $\beta=\frac{2}{D}$)
\begin{align}
p(x) & \propto\exp\big(-\frac{2}{D}V(x)\big),\label{eq:p_Boltzmann-1}
\end{align}
a construction of such an effective action has been given in the seminal
work of de Dominicis \citep[Appendix]{DeDominicis78_4913}. In this
particular case, the Langevin equation 
\begin{align}
dx(t) & =-V^{\prime}(x)\,dt+dW(t)\label{eq:Langevin_equilibrium-1}
\end{align}
has an equilibrium distribution of the form \eqref{eq:p_Boltzmann-1},
given that the variance $D$ of the noise is $\langle dW(s)dW(t)\rangle=D\delta_{ts}dt$;
a condition that follows from demanding vanishing probability flux
in the associated Fokker-Planck equation \citep[see e.g.][]{Goldenfeld92};
it is one expression of the fluctuation-dissipation theorem that holds
in equilibrium systems.

Moreover, a linear term $\frac{D}{2}\,h\,x(t)$ in addition to the
potential $V$ leads to a source term $h^{\T}\tx$ in the MSRDJ action
$S[x,\tx]=\tx^{\T}(\partial_{t}x+V^{\prime}(x))+\frac{D}{2}\tx^{\T}\tx+h^{\T}\tx$
which corresponds to \eqref{eq:Langevin_equilibrium-1}. The equation
of state \eqref{eq:Def_eq_of_state} for $\tx$ admits a solution
$\tx\equiv0$, for which the equation of state for $x$ takes the
form
\begin{align}
h[x^{\ast}] & =\frac{\delta\Gamma}{\delta\tx^{\ast}(t)}=-\frac{\delta S}{\delta\tx^{\ast}(t)}+\ldots\label{eq:equ_of_state_equilibrium-1}\\
 & =\partial_{t}x^{\ast}+V^{\prime}(x^{\ast})+\ldots,\nonumber 
\end{align}
where $\ldots$ denote all fluctuation corrections. The construction
of the effective action by de Dominics proceeds by functionally integrating
the equation of state \citep[eq. A4]{DeDominicis78_4913}
\begin{align}
\Gamma_{\mathrm{DD}}[x^{\ast}] & \coloneqq\int_{0}^{x^{\ast}}\delta x\,h[x]\label{eq:DeDominicis_eff_action-1}\\
 & =\Gamma_{\mathrm{DD}}[0]+\frac{1}{2}x^{\ast\T}\partial_{t}x^{\ast}+V(x^{\ast})+\ldots.\nonumber 
\end{align}
The last step requires that the equation of state \eqref{eq:equ_of_state_equilibrium-1}
be the derivative of a functional; otherwise the functional integration
would not yield a unique result, independent of the integration path.
This is where the equilibrium properties, the existence of a Boltzmann
weight \eqref{eq:p_Boltzmann-1}, enter. The latter implies further
that the problem can be treated with statics alone. For a constant
solution $x^{\ast}(t)=\bar{x}^{\ast}$, the effective potential is
thus
\begin{align}
U_{\mathrm{DD}}[\bar{x}^{\ast}]:=\Gamma[\bar{x}^{\ast}] & =V(\bar{x}^{\ast})+\ldots\label{eq:eff_pot_DeDominicis-1}
\end{align}

\subsection{Effective potential in a bistable network model\label{sec:Effective-potential-in-bistable}}

We here compute the one-loop corrections to the OM effective potential
for the bistable system \eqref{eq:Langevin_third_order-1}. The MSRDJ
action is
\[
S[x,\widetilde{x}]=\tx^{\T}\left[\left(\partial_{t}+r\right)x+\frac{u}{3}\,x^{3}\right]+\frac{D}{2}\tx^{\T}\tx.
\]
To lowest order in the fluctuations we have $\Gamma_{0}\left[x^{\ast},\tx^{\ast}\right]=-S\left[x^{\ast},\tx^{\ast}\right]$
for the MSRDJ effective action and the OM effective action \eqref{eq:Rel_eff_act_MSRDJ_OM}
is $S_{\mathrm{OM}}[x^{\ast}]=\frac{1}{2D}\left[\left(\partial_{t}+r\right)x+\frac{u}{3}x^{3}\right]^{2}$.
Next, we assume a stationary solution $\bar{x}^{\ast}$ and compute
all corrections around vanishing response fields $\tx=0$.

\paragraph*{Corrections to the mean}

The only one-loop contribution to $\Gitfl$ is given by the tadpole
diagram\begin{fmffile}{struct_man_1l_tadpole_cubic}
\fmfset{thin}{0.75pt}
\fmfset{decor_size}{2mm}
\fmfcmd{style_def wiggly_arrow expr p = cdraw (wiggly p); shrink (0.8); cfill (arrow p); endshrink; enddef;}
\fmfcmd{style_def majorana expr p = cdraw p; cfill (harrow (reverse p, .5)); cfill (harrow (p, .5)) enddef;
		style_def alt_majorana expr p = cdraw p; cfill (tarrow (reverse p, .55)); cfill (tarrow (p, .55)) enddef;}
\begin{align*}
	\Gamma_{\tilde{x},\mathrm{fl.}}^{\left(1\right)} &= \left(-1\right)
	\parbox{20mm}{
	\begin{fmfgraph*}(40,40)
		\fmfleft{l}
		\fmftop{c}
		\fmffreeze
		\fmfshift{(0.0,-0.5w)}{c}
		\fmfdot{c}
		\fmf{wiggly_arrow}{c,l}
		\fmf{alt_majorana, tension=0.55}{c,c}
	\end{fmfgraph*}
	}\\
&= -\frac{uDx^\ast}{2|m|},
\end{align*}
\end{fmffile}where $m$ is a function of the mean value and given by $m\left(x\right)=-r-ux^{2}$.
Including this into the equation of state, its physically relevant
solutions read
\begin{align*}
x_{0} & =0\\
x_{\pm} & =\pm\sqrt{-2\frac{r}{u}+\sqrt{\left(\frac{r}{u}\right)^{2}-\frac{3}{2}\frac{D}{u}}}.
\end{align*}

\paragraph*{Self-energy}

The one-loop corrections to $\Gamma^{\left(2\right)}$ are given by
the following diagrams\begin{fmffile}{struct_man_1l_propagator_cubic}
\fmfset{thin}{0.75pt}
\fmfset{decor_size}{4mm}
\fmfcmd{style_def wiggly_arrow expr p = cdraw (wiggly p); shrink (0.8); cfill (arrow p); endshrink; enddef;}
\fmfcmd{style_def majorana expr p = cdraw p; cfill (harrow (reverse p, .5)); cfill (harrow (p, .5)) enddef;
		style_def alt_majorana expr p = cdraw p; cfill (tarrow (reverse p, .55)); cfill (tarrow (p, .55)) enddef;}
\begin{align*}
	\Gamma_{\tilde{X}X,\mathrm{fl.}}^{\left(2\right)}\left(\sigma_1, \sigma_2\right)
		& =\,(-1)\,\parbox{30mm}{
		\begin{fmfgraph*}(80,80)
			\fmfleft{o}
			\fmfright{i}
			\fmfcurved
			\fmfsurroundn{v}{2}
			\fmffreeze
			\fmfshift{(0.22w,0.w)}{v2}
			\fmfshift{(-0.22w,0.w)}{v1}
			\fmfdotn{v}{2}
			\fmf{wiggly_arrow, tension=1.0, label=$\sigma_1$}{v2,o}
			\fmf{wiggly_arrow, tension=1.0, label=$\sigma_2$}{i,v1}
			\fmf{alt_majorana, left=0.9, tension=0.4}{v2,v1}
			\fmf{plain_arrow, left=0.9, tension=0.4}{v1,v2}
		\end{fmfgraph*}
	}
	+ (-1) \,\parbox{30mm}{
		\begin{fmfgraph*}(80,80)
			\fmfstraight
			\fmfbottomn{v}{3}
			\fmffreeze
			\fmfshift{(0.0w,0.3w)}{v1}
			\fmfshift{(0.0w,0.3w)}{v2}
			\fmfshift{(0.0w,0.3w)}{v3}
			\fmfdot{v2}
			\fmf{wiggly_arrow, tension=1.0, label=$\sigma_1$}{v2,v1}
			\fmf{wiggly_arrow, tension=1.0, label=$\sigma_2$}{v3,v2}
			\fmf{alt_majorana, tension=0.65}{v2,v2}
		\end{fmfgraph*}
	}\\
	\Gamma_{\tilde{X}\tilde{X},\mathrm{fl.}}^{\left(2\right)}\left(\sigma_1, \sigma_2\right)
		&=\,-\frac{1}{2}\,\parbox{30mm}{
	\begin{fmfgraph*}(80,80)
			\fmfleft{o}
			\fmfright{i}
			\fmfcurved
			\fmfsurroundn{v}{2}
			\fmffreeze
			\fmfshift{(0.22w,0.w)}{v2}
			\fmfshift{(-0.22w,0.w)}{v1}
			\fmfdotn{v}{2}
			\fmf{wiggly_arrow, tension=1.0, label=$\sigma_1$}{v2,o}
			\fmf{wiggly_arrow, tension=1.0, label=$\sigma_2$, label.side=left}{v1,i}
			\fmf{alt_majorana, left=0.9, tension=0.4}{v2,v1}
			\fmf{alt_majorana, left=0.9, tension=0.4}{v1,v2}
		\end{fmfgraph*}
	}.
\end{align*}
\end{fmffile}The zero frequency components of the self-energy read
\begin{align*}
\int d\sigma_{2}\,\Giitofl\left(0,\sigma_{2}\right)= & \frac{1}{2\pi}\left[\frac{\left(ux^{\ast}\right)^{2}D}{m^{2}}+\frac{uD}{2m}\right]\\
\int d\sigma_{2}\,\Giittfl\left(0,\sigma_{2}\right)= & \frac{1}{4\pi}\frac{\left(ux^{\ast}\right)^{2}D^{2}}{m^{3}}.
\end{align*}

\paragraph*{Construction of corrections to the effective potential}

We have now expanded the effective action around the reference point
$\tx^{\ast}=0$ and $\bar{x}^{\ast}$, which takes the form
\begin{align*}
\Gamma\left[x^{\ast},\tx^{\ast}\right] & =\tx^{\ast}\left[\overbrace{-rx^{\ast}-\frac{u}{3}x^{\ast3}+\Gitfl}^{\overset{\mathrm{eq.\,of\,state}}{=}0}+\left(-\partial_{t}-r-ux^{\ast2}+\Giitofl\right)\left(\xbarstar-x^{\ast}\right)\right]\\
 & +\frac{1}{2}\tx^{\ast}\big(-D+\Giittfl\big)\tx^{\ast}+\mathcal{O}\left(\left(\xbarstar-x^{\ast}\right)^{2}\right).
\end{align*}
We compute only terms up to linear order in $\xbarstar-x^{\ast}$
because this generates all terms up to quadratic order in the effective
potential. This is enough to calculate its curvature which equals
the integrated covariance at the stationary points. The response field
still appears quadratically in $\Gamma\left[x^{\ast},\tx^{\ast}\right]$,
so that we can extremize $\tx^{\ast}$ by writing the OM effective
action with corrected vertices as
\begin{align*}
\Gamma_{\mathrm{OM}}[x^{\ast}] & =\underset{\tx^{*}}{\mathrm{extremize}}\,\Gamma[x^{\ast},\tx^{\ast}]\\
 & =\left[\left(-\partial_{t}-r-ux^{\ast2}+\Giitofl\right)\delta x^{\ast}\right]\\
 & \times\frac{1}{2}\,\Big[D-\Giittfl\Big]^{-1}\times\\
 & \times\left[\left(-\partial_{t}-r-ux^{\ast2}+\Giitofl\right)\delta x^{\ast}\right],
\end{align*}
where $\delta x^{\ast}=\xbarstar-x^{\ast}$. Computing the effective
potential we get
\begin{align}
U\left(\xbarstar\right) & =\Gamma_{\mathrm{OM}}\left[\xbarstar\right]/T\nonumber \\
 & =\frac{1}{2}\,\frac{\left[\left(r+ux^{\ast2}-D\big(\frac{ux^{\ast}}{m}\big)^{2}-\frac{uD}{2m}\right)\delta x^{\ast}\right]^{2}}{D-\frac{\left(ux^{\ast}D\right)^{2}}{2m^{3}}},\label{eq:U_OM_one_loop}
\end{align}
where $m=m\left(x^{\ast}\right)=-r-ux^{\ast2}$.

\subsection{Definition of the Fourier transform\label{subsec:App_def_fourier}}

By choosing the Fourier transform of the fields and the sources as
the inverse of each other, we arrive at a representation of our formulas
that look similar in time and frequency domain. Therefore we define
\begin{align*}
x\left(t\right)= & \int\frac{d\omega}{2\pi}\,e^{i\omega t}X\left(\omega\right), &  &  & j\left(t\right)= & \int d\omega\,e^{-i\omega t}J\left(\omega\right),\\
X\left(\omega\right)= & \int dt\,e^{-i\omega t}x\left(t\right), &  &  & J\left(\omega\right)= & \int\frac{dt}{2\pi}\,e^{i\omega t}j\left(t\right).
\end{align*}
Thus, we obtain $x^{T}j=\int dt\,x\left(t\right)j\left(t\right)=\int d\omega\,X\left(\omega\right)J\left(\omega\right)$.
Moreover, we get with $y\coloneqq\left(x,\tx\right)^{\T}$ for the
matrix $A$ of the quadratic part of the action $S_{0}\left[x,\tilde{x}\right]=\int dt\,\int dt^{\prime}\,y\left(t\right)A\left(t,t^{\prime}\right)y(t^{\prime})$
\begin{alignat*}{3}
A\left(t,t^{\prime}\right)= & \int d\omega\,\int d\omega^{\prime}\,e^{-i\left(\omega t+\omega^{\prime}t^{\prime}\right)}A\left(\omega,\omega^{\prime}\right), & \quad & \quad & A^{-1}\left(t,t^{\prime}\right)= & \int\frac{d\omega}{2\pi}\,\int\frac{d\omega^{\prime}}{2\pi}\,e^{i\left(\omega t+\omega^{\prime}t^{\prime}\right)}A^{-1}\left(\omega,\omega^{\prime}\right),\\
A\left(\omega,\omega^{\prime}\right)= & \int\frac{dt}{2\pi}\,\int\frac{dt^{\prime}}{2\pi}\,e^{i\left(\omega t+\omega^{\prime}t^{\prime}\right)}A\left(t,t^{\prime}\right), & \quad & \quad & A^{-1}\left(\omega,\omega^{\prime}\right)= & \int dt\,\int dt^{\prime}\,e^{-i\left(\omega t+\omega^{\prime}t^{\prime}\right)}A^{-1}\left(t,t^{\prime}\right),
\end{alignat*}
due to the chain rule for functional derivatives. From this we can
derive the following useful identities
\begin{alignat*}{2}
\int dt\,\int dt^{\prime}\,y\left(t\right)A\left(t,t^{\prime}\right)y\left(t^{\prime}\right) & = & \int d\omega\,\int d\omega^{\prime}\,Y\left(\omega\right)A\left(\omega,\omega^{\prime}\right)Y\left(\omega^{\prime}\right),\\
\int dt\,\int dt^{\prime}\,\bar{j}\left(t\right)A^{-1}\left(t,t^{\prime}\right)\bar{j}\left(t^{\prime}\right) & = & \int d\omega\,\int d\omega^{\prime}\,\bar{J}\left(\omega\right)A^{-1}\left(\omega,\omega^{\prime}\right)\bar{J}\left(\omega^{\prime}\right),
\end{alignat*}
\begin{align*}
\int ds\,A\left(t,s\right)A^{-1}\left(s,t^{\prime}\right)= & \delta\left(t-t^{\prime}\right) & \Leftrightarrow &  & \int d\sigma\,A\left(\omega,\sigma\right)A^{-1}\left(\sigma,\omega^{\prime}\right)= & \delta\left(\omega-\omega^{\prime}\right).
\end{align*}
For the interaction part of the action we obtain
\[
\int dt\,\tilde{x}\left(t\right)x^{2}\left(t\right)=\int\frac{d\omega}{2\pi}\int\frac{d\omega^{\prime}}{2\pi}\,\tilde{X}\left(\omega\right)X\left(\omega^{\prime}\right)X\left(-\omega-\omega^{\prime}\right).
\]

\end{document}

%% file: macros.tex
\global\long\def\Ftr#1#2{\mathcal{F}\left[#1\right]\left(#2\right)}%
\global\long\def\iFtr#1#2{\mathcal{F}^{-1}\left[#1\right]\left(#2\right)}%
\global\long\def\D{\mathcal{D}}%
\global\long\def\x{\mathbf{x}}%
\global\long\def\tx{\tilde{x}}%
\global\long\def\hx{\text{\ensuremath{\hat{\mathbf{x}}}}}%
\global\long\def\htx{\hat{\tilde{\mathbf{x}}}}%
\global\long\def\xmean{\text{\ensuremath{\check{x}}}}%
\global\long\def\Xmean{\text{\ensuremath{\check{X}}}}%
\global\long\def\txmean{\text{\ensuremath{\check{\tilde{x}}}}}%
\global\long\def\tXmean{\text{\ensuremath{\check{\tilde{X}}}}}%
\global\long\def\xbarstar{\text{\ensuremath{\bar{x}^{\ast}}}}%
\global\long\def\w{\mathbf{w}}%
\global\long\def\l{\mathbf{l}}%
\global\long\def\tl{\tilde{\mathbf{l}}}%
\global\long\def\hl{\hat{\mathbf{l}}}%
\global\long\def\htl{\hat{\tilde{\mathbf{l}}}}%

\global\long\def\hX{\hat{\mathbf{X}}}%
\global\long\def\htX{\hat{\tilde{\mathbf{X}}}}%
\global\long\def\hL{\hat{\mathbf{L}}}%
\global\long\def\htL{\hat{\tilde{\mathbf{L}}}}%
\global\long\def\cX{\mathcal{X}}%
\global\long\def\cK{\mathcal{K}}%
\global\long\def\cF{\mathcal{F}}%
\global\long\def\T{\mathrm{T}}%
\global\long\def\tj{\tilde{j}}%
\global\long\def\Gammafl{\Gamma_{\mathrm{fl.}}}%
\global\long\def\tX{\tilde{X}}%
\global\long\def\diag{\mathrm{diag}}%
\global\long\def\sc{\mathrm{sc}}%
\global\long\def\tr{\mathrm{tr}}%
\global\long\def\Gammafllam{\Gamma_{\mathrm{fl.}\lambda}}%
\global\long\def\Rlbd{R_{\lambda}}%
\global\long\def\derivRlbd{\frac{\partial R_{\lambda}}{\partial\lambda}}%
\global\long\def\Xstarlbd{X_{\lambda}^{\ast}}%
\global\long\def\Xmeanlbd{\check{X}_{\lambda}}%
\global\long\def\Xtmeanlbd{\check{\tilde{X}}_{\lambda}}%
\global\long\def\Xstarquadlbd{X_{\lambda}^{\ast2}}%
\global\long\def\Xtstarlbd{\tilde{X}_{\lambda}^{\ast}}%
\global\long\def\Siioo{S_{xx}^{\left(2\right)}}%
\global\long\def\Siito{S_{\tilde{x}x}^{\left(2\right)}}%
\global\long\def\Siiot{S_{x\tilde{x}}^{\left(2\right)}}%
\global\long\def\Siitt{S_{\tilde{x}\tilde{x}}^{\left(2\right)}}%
\global\long\def\Siiitoo{S_{\tilde{x}xx}^{\left(3\right)}}%
\global\long\def\Gitfl{\Gamma_{\tilde{x},\mathrm{fl.}}^{\left(1\right)}}%
\global\long\def\Giioofl{\Gamma_{xx,\mathrm{fl.}}^{\left(2\right)}}%
\global\long\def\Giitofl{\Gamma_{\tilde{x}x,\mathrm{fl.}}^{\left(2\right)}}%
\global\long\def\Giiotfl{\Gamma_{x\tilde{x},\mathrm{fl.}}^{\left(2\right)}}%
\global\long\def\Giittfl{\Gamma_{\tilde{x}\tilde{x},\mathrm{fl.}}^{\left(2\right)}}%
\global\long\def\Giiitoofl{\Gamma_{\tilde{x}xx,\mathrm{fl.}}^{\left(3\right)}}%
\global\long\def\Giiiitooofl{\Gamma_{\tilde{x}xxx,\mathrm{fl.}}^{\left(4\right)}}%
\global\long\def\Dlbd{\Delta_{\lambda}}%
\global\long\def\Dto{\Delta_{\tilde{X}X}}%
\global\long\def\Dot{\Delta_{X\tilde{X}}}%
\global\long\def\Doo{\Delta_{XX}}%
\global\long\def\Gitlbd{\Gamma_{\tilde{X},\lambda}^{\left(1\right)}}%
\global\long\def\Giioolbd{\Gamma_{XX,\lambda}^{\left(2\right)}}%
\global\long\def\Giitolbd{\Gamma_{\tilde{X}X,\lambda}^{\left(2\right)}}%
\global\long\def\Giiotlbd{\Gamma_{X\tilde{X},\lambda}^{\left(2\right)}}%
\global\long\def\Giittlbd{\Gamma_{\tilde{X}\tilde{X},\lambda}^{\left(2\right)}}%
\global\long\def\Giiitoolbd{\Gamma_{\tilde{X}XX,\lambda}^{\left(3\right)}}%
\global\long\def\Dtolbd{\Delta_{\tilde{X}X,\lambda}}%
\global\long\def\Dotlbd{\Delta_{X\tilde{X},\lambda}}%
\global\long\def\Doolbd{\Delta_{XX,\lambda}}%
\global\long\def\Dbarto{\Delta_{\tilde{X}X}^{0}}%
\global\long\def\Dbarot{\Delta_{X\tilde{X}}^{0}}%
\global\long\def\Dbaroo{\Delta_{XX}^{0}}%
\global\long\def\So{S_{X}^{\left(1\right)}}%
\global\long\def\St{S_{\tilde{X}}^{\left(1\right)}}%
\global\long\def\Stoo{S_{\tilde{X}XX}^{\left(3\right)}}%
\global\long\def\tz{\tilde{z}}%
\global\long\def\xtx{{\cal X}}%
\global\long\def\jtj{{\cal J}}%
\global\long\def\SMSRDJ{S_{\mathrm{MSRDJ}}}%
\global\long\def\SOM{S_{\mathrm{OM}}}%
\global\long\def\gmaMSR{\Gamma_{\mathrm{MSRDJ}}}%
\global\long\def\gmaOM{\Gamma_{\mathrm{OM}}}%
\global\long\def\Z{\mathcal{Z}}%

%% file: structure_manuscript.bbl
\begin{thebibliography}{154} \expandafter\ifx\csname natexlab\endcsname\relax\def\natexlab#1{#1}\fi \expandafter\ifx\csname bibnamefont\endcsname\relax   \def\bibnamefont#1{#1}\fi \expandafter\ifx\csname bibfnamefont\endcsname\relax   \def\bibfnamefont#1{#1}\fi \expandafter\ifx\csname citenamefont\endcsname\relax   \def\citenamefont#1{#1}\fi \expandafter\ifx\csname url\endcsname\relax   \def\url#1{\texttt{#1}}\fi \expandafter\ifx\csname urlprefix\endcsname\relax\def\urlprefix{URL }\fi \providecommand{\bibinfo}[2]{#2} \providecommand{\eprint}[2][]{\url{#2}}
\bibitem[{\citenamefont{Dahmen et~al.}(2016)\citenamefont{Dahmen, Bos, and   Helias}}]{Dahmen16_031024} \bibinfo{author}{\bibfnamefont{D.}~\bibnamefont{Dahmen}},   \bibinfo{author}{\bibfnamefont{H.}~\bibnamefont{Bos}}, \bibnamefont{and}   \bibinfo{author}{\bibfnamefont{M.}~\bibnamefont{Helias}},   \bibinfo{journal}{Phys Rev X} \textbf{\bibinfo{volume}{6}},   \bibinfo{pages}{031024} (\bibinfo{year}{2016}).
\bibitem[{\citenamefont{Sompolinsky}(1988)}]{Sompolinsky88_2} \bibinfo{author}{\bibfnamefont{H.}~\bibnamefont{Sompolinsky}},   \bibinfo{journal}{Physics Today} \textbf{\bibinfo{volume}{41}},   \bibinfo{pages}{70} (\bibinfo{year}{1988}).
\bibitem[{\citenamefont{Braitenberg and Sch\"{u}z}(1991)}]{Braitenberg91} \bibinfo{author}{\bibfnamefont{V.}~\bibnamefont{Braitenberg}} \bibnamefont{and}   \bibinfo{author}{\bibfnamefont{A.}~\bibnamefont{Sch\"{u}z}},   \emph{\bibinfo{title}{Anatomy of the Cortex: Statistics and Geometry}}   (\bibinfo{publisher}{Springer-Verlag}, \bibinfo{address}{Berlin, Heidelberg,   New York}, \bibinfo{year}{1991}), ISBN \bibinfo{isbn}{3-540-53233-1}.
\bibitem[{\citenamefont{Zinn-Justin}(1996)}]{ZinnJustin96} \bibinfo{author}{\bibfnamefont{J.}~\bibnamefont{Zinn-Justin}},   \emph{\bibinfo{title}{Quantum field theory and critical phenomena}}   (\bibinfo{publisher}{Clarendon Press, Oxford}, \bibinfo{year}{1996}).
\bibitem[{\citenamefont{Hohenberg and Halperin}(1977)}]{Hohenberg77} \bibinfo{author}{\bibfnamefont{P.~C.} \bibnamefont{Hohenberg}}   \bibnamefont{and} \bibinfo{author}{\bibfnamefont{B.~I.}   \bibnamefont{Halperin}}, \bibinfo{journal}{Rev. Mod. Phys.}   \textbf{\bibinfo{volume}{49}}, \bibinfo{pages}{435} (\bibinfo{year}{1977}).
\bibitem[{\citenamefont{Taeuber}(2014)}]{Taeuber14} \bibinfo{author}{\bibfnamefont{U.~C.} \bibnamefont{Taeuber}},   \emph{\bibinfo{title}{Critical dynamics: a field theory approach to   equilibrium and non-equilibrium scaling behavior}}   (\bibinfo{publisher}{Cambridge University Press}, \bibinfo{year}{2014}), ISBN   \bibinfo{isbn}{9780521842235}.
\bibitem[{\citenamefont{Chow and Buice}(2015)}]{Chow15} \bibinfo{author}{\bibfnamefont{C.}~\bibnamefont{Chow}} \bibnamefont{and}   \bibinfo{author}{\bibfnamefont{M.}~\bibnamefont{Buice}}, \bibinfo{journal}{J   Math. Neurosci} \textbf{\bibinfo{volume}{5}} (\bibinfo{year}{2015}).
\bibitem[{\citenamefont{Hertz et~al.}(2017)\citenamefont{Hertz, Roudi, and   Sollich}}]{Hertz16_033001} \bibinfo{author}{\bibfnamefont{J.~A.} \bibnamefont{Hertz}},   \bibinfo{author}{\bibfnamefont{Y.}~\bibnamefont{Roudi}}, \bibnamefont{and}   \bibinfo{author}{\bibfnamefont{P.}~\bibnamefont{Sollich}},   \bibinfo{journal}{Journal of Physics A: Mathematical and Theoretical}   \textbf{\bibinfo{volume}{50}}, \bibinfo{pages}{033001}   (\bibinfo{year}{2017}).
\bibitem[{\citenamefont{van Vreeswijk and Sompolinsky}(1996)}]{Vreeswijk96} \bibinfo{author}{\bibfnamefont{C.}~\bibnamefont{van Vreeswijk}}   \bibnamefont{and}   \bibinfo{author}{\bibfnamefont{H.}~\bibnamefont{Sompolinsky}},   \bibinfo{journal}{Science} \textbf{\bibinfo{volume}{274}},   \bibinfo{pages}{1724} (\bibinfo{year}{1996}).
\bibitem[{\citenamefont{Brunel}(2000)}]{Brunel00_183} \bibinfo{author}{\bibfnamefont{N.}~\bibnamefont{Brunel}}, \bibinfo{journal}{J.   Comput. Neurosci.} \textbf{\bibinfo{volume}{8}}, \bibinfo{pages}{183}   (\bibinfo{year}{2000}).
\bibitem[{\citenamefont{Ocker et~al.}(2017{\natexlab{a}})\citenamefont{Ocker,   Josi{\'c}, Shea-Brown, and Buice}}]{Ocker17_1} \bibinfo{author}{\bibfnamefont{G.~K.} \bibnamefont{Ocker}},   \bibinfo{author}{\bibfnamefont{K.}~\bibnamefont{Josi{\'c}}},   \bibinfo{author}{\bibfnamefont{E.}~\bibnamefont{Shea-Brown}},   \bibnamefont{and} \bibinfo{author}{\bibfnamefont{M.~A.} \bibnamefont{Buice}},   \bibinfo{journal}{PLOS Comput. Biol.} \textbf{\bibinfo{volume}{13}},   \bibinfo{pages}{1} (\bibinfo{year}{2017}{\natexlab{a}}).
\bibitem[{\citenamefont{Hopfield}(1982)}]{Hopfield82} \bibinfo{author}{\bibfnamefont{J.~J.} \bibnamefont{Hopfield}},   \bibinfo{journal}{PNAS} \textbf{\bibinfo{volume}{79}}, \bibinfo{pages}{2554}   (\bibinfo{year}{1982}).
\bibitem[{\citenamefont{Amit}(1984)}]{Amit84} \bibinfo{author}{\bibfnamefont{D.~J.} \bibnamefont{Amit}},   \emph{\bibinfo{title}{Field theory, the renormalization group, and critical   phenomena}} (\bibinfo{publisher}{World Scientific}, \bibinfo{year}{1984}).
\bibitem[{\citenamefont{Lindner and Longtin}(2005)}]{Lindner05_505} \bibinfo{author}{\bibfnamefont{B.}~\bibnamefont{Lindner}} \bibnamefont{and}   \bibinfo{author}{\bibfnamefont{A.}~\bibnamefont{Longtin}},   \bibinfo{journal}{Journal of Theoretical Biology}   \textbf{\bibinfo{volume}{232}}, \bibinfo{pages}{505} (\bibinfo{year}{2005}).
\bibitem[{\citenamefont{Shea-Brown et~al.}(2008)\citenamefont{Shea-Brown,   Josic, de~la Rocha, and Doiron}}]{Shea-Brown08} \bibinfo{author}{\bibfnamefont{E.}~\bibnamefont{Shea-Brown}},   \bibinfo{author}{\bibfnamefont{K.}~\bibnamefont{Josic}},   \bibinfo{author}{\bibfnamefont{J.}~\bibnamefont{de~la Rocha}},   \bibnamefont{and} \bibinfo{author}{\bibfnamefont{B.}~\bibnamefont{Doiron}},   \bibinfo{journal}{Phys. Rev. Lett.} \textbf{\bibinfo{volume}{100}},   \bibinfo{pages}{108102} (\bibinfo{year}{2008}).
\bibitem[{\citenamefont{{El Boustani} and Destexhe}(2009)}]{Boustani_09} \bibinfo{author}{\bibfnamefont{S.}~\bibnamefont{{El Boustani}}}   \bibnamefont{and} \bibinfo{author}{\bibfnamefont{A.}~\bibnamefont{Destexhe}},   \bibinfo{journal}{Neural Comput.} \textbf{\bibinfo{volume}{21}},   \bibinfo{pages}{46} (\bibinfo{year}{2009}).
\bibitem[{\citenamefont{Pernice et~al.}(2011)\citenamefont{Pernice, Staude,   Cardanobile, and Rotter}}]{Pernice11_e1002059} \bibinfo{author}{\bibfnamefont{V.}~\bibnamefont{Pernice}},   \bibinfo{author}{\bibfnamefont{B.}~\bibnamefont{Staude}},   \bibinfo{author}{\bibfnamefont{S.}~\bibnamefont{Cardanobile}},   \bibnamefont{and} \bibinfo{author}{\bibfnamefont{S.}~\bibnamefont{Rotter}},   \bibinfo{journal}{PLOS Comput. Biol.} \textbf{\bibinfo{volume}{7}},   \bibinfo{pages}{e1002059} (\bibinfo{year}{2011}).
\bibitem[{\citenamefont{Ostojic and Brunel}(2011)}]{Ostojic11_e1001056} \bibinfo{author}{\bibfnamefont{S.}~\bibnamefont{Ostojic}} \bibnamefont{and}   \bibinfo{author}{\bibfnamefont{N.}~\bibnamefont{Brunel}},   \bibinfo{journal}{PLOS Comput. Biol.} \textbf{\bibinfo{volume}{7}},   \bibinfo{pages}{e1001056} (\bibinfo{year}{2011}).
\bibitem[{\citenamefont{Trousdale et~al.}(2012)\citenamefont{Trousdale, Hu,   Shea-Brown, and Josic}}]{Trousdale12_e1002408} \bibinfo{author}{\bibfnamefont{J.}~\bibnamefont{Trousdale}},   \bibinfo{author}{\bibfnamefont{Y.}~\bibnamefont{Hu}},   \bibinfo{author}{\bibfnamefont{E.}~\bibnamefont{Shea-Brown}},   \bibnamefont{and} \bibinfo{author}{\bibfnamefont{K.}~\bibnamefont{Josic}},   \bibinfo{journal}{PLOS Comput. Biol.} \textbf{\bibinfo{volume}{8}},   \bibinfo{pages}{e1002408} (\bibinfo{year}{2012}).
\bibitem[{\citenamefont{Tetzlaff et~al.}(2012)\citenamefont{Tetzlaff, Helias,   Einevoll, and Diesmann}}]{Tetzlaff12_e1002596} \bibinfo{author}{\bibfnamefont{T.}~\bibnamefont{Tetzlaff}},   \bibinfo{author}{\bibfnamefont{M.}~\bibnamefont{Helias}},   \bibinfo{author}{\bibfnamefont{G.~T.} \bibnamefont{Einevoll}},   \bibnamefont{and} \bibinfo{author}{\bibfnamefont{M.}~\bibnamefont{Diesmann}},   \bibinfo{journal}{PLOS Comput. Biol.} \textbf{\bibinfo{volume}{8}},   \bibinfo{pages}{e1002596} (\bibinfo{year}{2012}).
\bibitem[{\citenamefont{Helias et~al.}(2013)\citenamefont{Helias, Tetzlaff, and   Diesmann}}]{Helias13_023002} \bibinfo{author}{\bibfnamefont{M.}~\bibnamefont{Helias}},   \bibinfo{author}{\bibfnamefont{T.}~\bibnamefont{Tetzlaff}}, \bibnamefont{and}   \bibinfo{author}{\bibfnamefont{M.}~\bibnamefont{Diesmann}},   \bibinfo{journal}{New J. Phys.} \textbf{\bibinfo{volume}{15}},   \bibinfo{pages}{023002} (\bibinfo{year}{2013}).
\bibitem[{\citenamefont{Renart et~al.}(2010)\citenamefont{Renart, {De La   Rocha}, Bartho, Hollender, Parga, Reyes, and Harris}}]{Renart10_587} \bibinfo{author}{\bibfnamefont{A.}~\bibnamefont{Renart}},   \bibinfo{author}{\bibfnamefont{J.}~\bibnamefont{{De La Rocha}}},   \bibinfo{author}{\bibfnamefont{P.}~\bibnamefont{Bartho}},   \bibinfo{author}{\bibfnamefont{L.}~\bibnamefont{Hollender}},   \bibinfo{author}{\bibfnamefont{N.}~\bibnamefont{Parga}},   \bibinfo{author}{\bibfnamefont{A.}~\bibnamefont{Reyes}}, \bibnamefont{and}   \bibinfo{author}{\bibfnamefont{K.~D.} \bibnamefont{Harris}},   \bibinfo{journal}{Science} \textbf{\bibinfo{volume}{327}},   \bibinfo{pages}{587} (\bibinfo{year}{2010}).
\bibitem[{\citenamefont{Riehle et~al.}(1997)\citenamefont{Riehle, Gr{\"u}n,   Diesmann, and Aertsen}}]{Riehle97_1950} \bibinfo{author}{\bibfnamefont{A.}~\bibnamefont{Riehle}},   \bibinfo{author}{\bibfnamefont{S.}~\bibnamefont{Gr{\"u}n}},   \bibinfo{author}{\bibfnamefont{M.}~\bibnamefont{Diesmann}}, \bibnamefont{and}   \bibinfo{author}{\bibfnamefont{A.}~\bibnamefont{Aertsen}},   \bibinfo{journal}{Science} \textbf{\bibinfo{volume}{278}},   \bibinfo{pages}{1950} (\bibinfo{year}{1997}).
\bibitem[{\citenamefont{Cohen and Maunsell}(2009)}]{Cohen09_1079} \bibinfo{author}{\bibfnamefont{M.~R.} \bibnamefont{Cohen}} \bibnamefont{and}   \bibinfo{author}{\bibfnamefont{J.~H.~R.} \bibnamefont{Maunsell}},   \bibinfo{journal}{Nat. Neurosci.} \textbf{\bibinfo{volume}{12}},   \bibinfo{pages}{1594} (\bibinfo{year}{2009}).
\bibitem[{\citenamefont{Kilavik et~al.}(2009)\citenamefont{Kilavik, Roux,   Ponce-Alvarez, Confais, Gr{\"u}n, and Riehle}}]{Kilavik09_12653} \bibinfo{author}{\bibfnamefont{B.~E.} \bibnamefont{Kilavik}},   \bibinfo{author}{\bibfnamefont{S.}~\bibnamefont{Roux}},   \bibinfo{author}{\bibfnamefont{A.}~\bibnamefont{Ponce-Alvarez}},   \bibinfo{author}{\bibfnamefont{J.}~\bibnamefont{Confais}},   \bibinfo{author}{\bibfnamefont{S.}~\bibnamefont{Gr{\"u}n}}, \bibnamefont{and}   \bibinfo{author}{\bibfnamefont{A.}~\bibnamefont{Riehle}},   \bibinfo{journal}{J. Neurosci.} \textbf{\bibinfo{volume}{29}},   \bibinfo{pages}{12653} (\bibinfo{year}{2009}).
\bibitem[{\citenamefont{Duclut}(2017)}]{Duclu17} \bibinfo{author}{\bibfnamefont{C.}~\bibnamefont{Duclut}},   \bibinfo{type}{Theses}, \bibinfo{school}{{Universit{\'e} Pierre et Marie   Curie - Paris VI}} (\bibinfo{year}{2017}),   \urlprefix\url{https://tel.archives-ouvertes.fr/tel-01690438}.
\bibitem[{\citenamefont{Beggs and Plenz}(2004)}]{Beggs04} \bibinfo{author}{\bibfnamefont{J.~M.} \bibnamefont{Beggs}} \bibnamefont{and}   \bibinfo{author}{\bibfnamefont{D.}~\bibnamefont{Plenz}}, \bibinfo{journal}{J.   Neurosci.} \textbf{\bibinfo{volume}{24}}, \bibinfo{pages}{5216}   (\bibinfo{year}{2004}).
\bibitem[{\citenamefont{Tka{\v{c}}ik et~al.}(2014)\citenamefont{Tka{\v{c}}ik,   Marre, Amodei, Schneidman, Bialek, and Berry~II}}]{Tkacik14_e1003408} \bibinfo{author}{\bibfnamefont{G.}~\bibnamefont{Tka{\v{c}}ik}},   \bibinfo{author}{\bibfnamefont{O.}~\bibnamefont{Marre}},   \bibinfo{author}{\bibfnamefont{D.}~\bibnamefont{Amodei}},   \bibinfo{author}{\bibfnamefont{E.}~\bibnamefont{Schneidman}},   \bibinfo{author}{\bibfnamefont{W.}~\bibnamefont{Bialek}}, \bibnamefont{and}   \bibinfo{author}{\bibfnamefont{M.~J.} \bibnamefont{Berry~II}},   \bibinfo{journal}{PLOS Comput. Biol.} \textbf{\bibinfo{volume}{10}},   \bibinfo{pages}{e1003408} (\bibinfo{year}{2014}).
\bibitem[{\citenamefont{Jaeger}(2001)}]{Jaeger01_echo} \bibinfo{author}{\bibfnamefont{H.}~\bibnamefont{Jaeger}}, \bibinfo{type}{Tech.   Rep.} \bibinfo{number}{GMD Report 148}, \bibinfo{institution}{German National   Research Center for Information Technology}, \bibinfo{address}{St. Augustin,   Germany} (\bibinfo{year}{2001}).
\bibitem[{\citenamefont{Maass et~al.}(2002)\citenamefont{Maass,   Natschl\"{a}ger, and Markram}}]{Maass02_2531} \bibinfo{author}{\bibfnamefont{W.}~\bibnamefont{Maass}},   \bibinfo{author}{\bibfnamefont{T.}~\bibnamefont{Natschl\"{a}ger}},   \bibnamefont{and} \bibinfo{author}{\bibfnamefont{H.}~\bibnamefont{Markram}},   \bibinfo{journal}{Neural Comput.} \textbf{\bibinfo{volume}{14}},   \bibinfo{pages}{2531} (\bibinfo{year}{2002}).
\bibitem[{\citenamefont{Legenstein and Maass}(2007)}]{Legenstein07_323} \bibinfo{author}{\bibfnamefont{R.}~\bibnamefont{Legenstein}} \bibnamefont{and}   \bibinfo{author}{\bibfnamefont{W.}~\bibnamefont{Maass}},   \bibinfo{journal}{Neural Networks} \textbf{\bibinfo{volume}{20}},   \bibinfo{pages}{323} (\bibinfo{year}{2007}).
\bibitem[{\citenamefont{Wilson}(1975)}]{Wilson75_773} \bibinfo{author}{\bibfnamefont{K.~G.} \bibnamefont{Wilson}},   \bibinfo{journal}{Rev. Mod. Phys.} \textbf{\bibinfo{volume}{47}},   \bibinfo{pages}{773} (\bibinfo{year}{1975}).
\bibitem[{\citenamefont{Mora and Bialek}(2011)}]{Mora2011} \bibinfo{author}{\bibfnamefont{T.}~\bibnamefont{Mora}} \bibnamefont{and}   \bibinfo{author}{\bibfnamefont{W.}~\bibnamefont{Bialek}},   \bibinfo{journal}{Journal of Statistical Physics}   \textbf{\bibinfo{volume}{144}}, \bibinfo{pages}{268} (\bibinfo{year}{2011}),   ISSN \bibinfo{issn}{1572-9613}.
\bibitem[{\citenamefont{Wegner and Houghton}(1973)}]{Wegner73_401} \bibinfo{author}{\bibfnamefont{F.~J.} \bibnamefont{Wegner}} \bibnamefont{and}   \bibinfo{author}{\bibfnamefont{A.}~\bibnamefont{Houghton}},   \bibinfo{journal}{Phys. Rev. A} \textbf{\bibinfo{volume}{8}},   \bibinfo{pages}{401} (\bibinfo{year}{1973}).
\bibitem[{\citenamefont{Wetterich}(1993)}]{WETTERICH93_90} \bibinfo{author}{\bibfnamefont{C.}~\bibnamefont{Wetterich}},   \bibinfo{journal}{Physics Letters B} \textbf{\bibinfo{volume}{30}},   \bibinfo{pages}{90} (\bibinfo{year}{1993}), ISSN \bibinfo{issn}{0370-2693}.
\bibitem[{\citenamefont{Berges et~al.}(2002)\citenamefont{Berges, Tetradis, and   Wetterich}}]{Berges02_223} \bibinfo{author}{\bibfnamefont{J.}~\bibnamefont{Berges}},   \bibinfo{author}{\bibfnamefont{N.}~\bibnamefont{Tetradis}}, \bibnamefont{and}   \bibinfo{author}{\bibfnamefont{C.}~\bibnamefont{Wetterich}},   \bibinfo{journal}{Physics Reports} \textbf{\bibinfo{volume}{363}},   \bibinfo{pages}{223} (\bibinfo{year}{2002}), ISSN \bibinfo{issn}{0370-1573},   \bibinfo{note}{renormalization group theory in the new millennium.}
\bibitem[{\citenamefont{Mart\'{\i} et~al.}(2018)\citenamefont{Mart\'{\i},   Brunel, and Ostojic}}]{Marti18_062314} \bibinfo{author}{\bibfnamefont{D.}~\bibnamefont{Mart\'{\i}}},   \bibinfo{author}{\bibfnamefont{N.}~\bibnamefont{Brunel}}, \bibnamefont{and}   \bibinfo{author}{\bibfnamefont{S.}~\bibnamefont{Ostojic}},   \bibinfo{journal}{Phys. Rev. E} \textbf{\bibinfo{volume}{97}},   \bibinfo{pages}{062314} (\bibinfo{year}{2018}).
\bibitem[{\citenamefont{Song et~al.}(2005)\citenamefont{Song, Sj\"{o}str\"{o}m,   Reigl, Nelson, and Chklovskii}}]{Song05_0507} \bibinfo{author}{\bibfnamefont{S.}~\bibnamefont{Song}},   \bibinfo{author}{\bibfnamefont{P.}~\bibnamefont{Sj\"{o}str\"{o}m}},   \bibinfo{author}{\bibfnamefont{M.}~\bibnamefont{Reigl}},   \bibinfo{author}{\bibfnamefont{S.}~\bibnamefont{Nelson}}, \bibnamefont{and}   \bibinfo{author}{\bibfnamefont{D.}~\bibnamefont{Chklovskii}},   \bibinfo{journal}{PLoS Biol.} \textbf{\bibinfo{volume}{3}},   \bibinfo{pages}{e68} (\bibinfo{year}{2005}).
\bibitem[{\citenamefont{Martin et~al.}(1973)\citenamefont{Martin, Siggia, and   Rose}}]{Martin73} \bibinfo{author}{\bibfnamefont{P.}~\bibnamefont{Martin}},   \bibinfo{author}{\bibfnamefont{E.}~\bibnamefont{Siggia}}, \bibnamefont{and}   \bibinfo{author}{\bibfnamefont{H.}~\bibnamefont{Rose}},   \bibinfo{journal}{Phys. Rev. A} \textbf{\bibinfo{volume}{8}},   \bibinfo{pages}{423} (\bibinfo{year}{1973}).
\bibitem[{\citenamefont{Janssen}(1976)}]{janssen1976_377} \bibinfo{author}{\bibfnamefont{H.-K.} \bibnamefont{Janssen}},   \bibinfo{journal}{Zeitschrift f{\"u}r Physik B Condensed Matter}   \textbf{\bibinfo{volume}{23}}, \bibinfo{pages}{377} (\bibinfo{year}{1976}).
\bibitem[{\citenamefont{De~Dominicis}(1976)}]{dedominicis1976_247} \bibinfo{author}{\bibfnamefont{C.}~\bibnamefont{De~Dominicis}},   \bibinfo{journal}{J. Phys. Colloques} \textbf{\bibinfo{volume}{37}},   \bibinfo{pages}{C1} (\bibinfo{year}{1976}).
\bibitem[{\citenamefont{Crisanti and   Sompolinsky}(2018{\natexlab{a}})}]{Crisanti18_062120} \bibinfo{author}{\bibfnamefont{A.}~\bibnamefont{Crisanti}} \bibnamefont{and}   \bibinfo{author}{\bibfnamefont{H.}~\bibnamefont{Sompolinsky}},   \bibinfo{journal}{Phys Rev E} \textbf{\bibinfo{volume}{98}},   \bibinfo{pages}{062120} (\bibinfo{year}{2018}{\natexlab{a}}).
\bibitem[{\citenamefont{Schuecker et~al.}(2018)\citenamefont{Schuecker,   Goedeke, and Helias}}]{Schuecker18_041029} \bibinfo{author}{\bibfnamefont{J.}~\bibnamefont{Schuecker}},   \bibinfo{author}{\bibfnamefont{S.}~\bibnamefont{Goedeke}}, \bibnamefont{and}   \bibinfo{author}{\bibfnamefont{M.}~\bibnamefont{Helias}},   \bibinfo{journal}{Phys Rev X} \textbf{\bibinfo{volume}{8}},   \bibinfo{pages}{041029} (\bibinfo{year}{2018}).
\bibitem[{\citenamefont{Dahmen et~al.}(2019)\citenamefont{Dahmen, Gr\"{u}n,   Diesmann, and Helias}}]{Dahmen19_13051} \bibinfo{author}{\bibfnamefont{D.}~\bibnamefont{Dahmen}},   \bibinfo{author}{\bibfnamefont{S.}~\bibnamefont{Gr\"{u}n}},   \bibinfo{author}{\bibfnamefont{M.}~\bibnamefont{Diesmann}}, \bibnamefont{and}   \bibinfo{author}{\bibfnamefont{M.}~\bibnamefont{Helias}},   \bibinfo{journal}{Proc. Nat. Acad. Sci. USA} \textbf{\bibinfo{volume}{116}},   \bibinfo{pages}{13051} (\bibinfo{year}{2019}),   \bibinfo{note}{10.1073/pnas.1818972116}.
\bibitem[{\citenamefont{Buice and Chow}(2007)}]{Buice07_031118} \bibinfo{author}{\bibfnamefont{M.~A.} \bibnamefont{Buice}} \bibnamefont{and}   \bibinfo{author}{\bibfnamefont{C.~C.} \bibnamefont{Chow}},   \bibinfo{journal}{Phys. Rev. E} \textbf{\bibinfo{volume}{76}},   \bibinfo{pages}{031118} (\bibinfo{year}{2007}).
\bibitem[{\citenamefont{Buice et~al.}(2010)\citenamefont{Buice, Cowan, and   Chow}}]{Buice10_377} \bibinfo{author}{\bibfnamefont{M.~A.} \bibnamefont{Buice}},   \bibinfo{author}{\bibfnamefont{J.~D.} \bibnamefont{Cowan}}, \bibnamefont{and}   \bibinfo{author}{\bibfnamefont{C.~C.} \bibnamefont{Chow}},   \bibinfo{journal}{Neural Comput.} \textbf{\bibinfo{volume}{22}},   \bibinfo{pages}{377} (\bibinfo{year}{2010}), ISSN \bibinfo{issn}{0899-7667}.
\bibitem[{\citenamefont{Buice and Chow}(2013)}]{Buice13_1} \bibinfo{author}{\bibfnamefont{M.~A.} \bibnamefont{Buice}} \bibnamefont{and}   \bibinfo{author}{\bibfnamefont{C.~C.} \bibnamefont{Chow}},   \bibinfo{journal}{PLOS Comput. Biol.} \textbf{\bibinfo{volume}{9}},   \bibinfo{pages}{1} (\bibinfo{year}{2013}).
\bibitem[{\citenamefont{Brinkman et~al.}(2018)\citenamefont{Brinkman, Rieke,   Shea-Brown, and Buice}}]{Brinkman18_e1006490} \bibinfo{author}{\bibfnamefont{B.~A.~W.} \bibnamefont{Brinkman}},   \bibinfo{author}{\bibfnamefont{F.}~\bibnamefont{Rieke}},   \bibinfo{author}{\bibfnamefont{E.}~\bibnamefont{Shea-Brown}},   \bibnamefont{and} \bibinfo{author}{\bibfnamefont{M.~A.} \bibnamefont{Buice}},   \bibinfo{journal}{PLOS Comput. Biol.} \textbf{\bibinfo{volume}{14}},   \bibinfo{pages}{e1006490} (\bibinfo{year}{2018}).
\bibitem[{\citenamefont{DiSanto et~al.}(2018)\citenamefont{DiSanto, Villegas,   Burioni, and Munoz}}]{diSanto18_1356} \bibinfo{author}{\bibfnamefont{S.}~\bibnamefont{DiSanto}},   \bibinfo{author}{\bibfnamefont{P.}~\bibnamefont{Villegas}},   \bibinfo{author}{\bibfnamefont{R.}~\bibnamefont{Burioni}}, \bibnamefont{and}   \bibinfo{author}{\bibfnamefont{M.~A.} \bibnamefont{Munoz}},   \bibinfo{journal}{Proceedings of the National Academy of Sciences}   \textbf{\bibinfo{volume}{115}}, \bibinfo{pages}{E1356}   (\bibinfo{year}{2018}), ISSN \bibinfo{issn}{0027-8424},   \eprint{http://www.pnas.org/content/early/2018/01/26/1712989115.full.pdf}.
\bibitem[{\citenamefont{Kardar et~al.}(1986)\citenamefont{Kardar, Parisi, and   Zhang}}]{Kardar84} \bibinfo{author}{\bibfnamefont{M.}~\bibnamefont{Kardar}},   \bibinfo{author}{\bibfnamefont{G.}~\bibnamefont{Parisi}}, \bibnamefont{and}   \bibinfo{author}{\bibfnamefont{Y.-C.} \bibnamefont{Zhang}},   \bibinfo{journal}{Phys. Rev. Lett.} \textbf{\bibinfo{volume}{56}},   \bibinfo{pages}{889} (\bibinfo{year}{1986}).
\bibitem[{\citenamefont{Canet et~al.}(2010)\citenamefont{Canet, Chat\'e,   Delamotte, and Wschebor}}]{Canet10} \bibinfo{author}{\bibfnamefont{L.}~\bibnamefont{Canet}},   \bibinfo{author}{\bibfnamefont{H.}~\bibnamefont{Chat\'e}},   \bibinfo{author}{\bibfnamefont{B.}~\bibnamefont{Delamotte}},   \bibnamefont{and} \bibinfo{author}{\bibfnamefont{N.}~\bibnamefont{Wschebor}},   \bibinfo{journal}{Phys. Rev. Lett.} \textbf{\bibinfo{volume}{104}},   \bibinfo{pages}{150601} (\bibinfo{year}{2010}).
\bibitem[{\citenamefont{Canet et~al.}(2011)\citenamefont{Canet, Chat\'e, and   Delamotte}}]{Canet11} \bibinfo{author}{\bibfnamefont{L.}~\bibnamefont{Canet}},   \bibinfo{author}{\bibfnamefont{H.}~\bibnamefont{Chat\'e}}, \bibnamefont{and}   \bibinfo{author}{\bibfnamefont{B.}~\bibnamefont{Delamotte}},   \bibinfo{journal}{J Phys. A} \textbf{\bibinfo{volume}{44}},   \bibinfo{pages}{495001} (\bibinfo{year}{2011}).
\bibitem[{\citenamefont{Frey et~al.}(1996)\citenamefont{Frey, T\"auber, and   Hwa}}]{Frey96_4424} \bibinfo{author}{\bibfnamefont{E.}~\bibnamefont{Frey}},   \bibinfo{author}{\bibfnamefont{U.~C.} \bibnamefont{T\"auber}},   \bibnamefont{and} \bibinfo{author}{\bibfnamefont{T.}~\bibnamefont{Hwa}},   \bibinfo{journal}{Phys. Rev. E} \textbf{\bibinfo{volume}{53}},   \bibinfo{pages}{4424} (\bibinfo{year}{1996}).
\bibitem[{\citenamefont{Blaizot   et~al.}(2006{\natexlab{a}})\citenamefont{Blaizot, M\'{e}ndez-Galain, and   Wschebor}}]{Blaizot06_571} \bibinfo{author}{\bibfnamefont{J.-P.} \bibnamefont{Blaizot}},   \bibinfo{author}{\bibfnamefont{R.}~\bibnamefont{M\'{e}ndez-Galain}},   \bibnamefont{and} \bibinfo{author}{\bibfnamefont{N.}~\bibnamefont{Wschebor}},   \bibinfo{journal}{Physics Letters B} \textbf{\bibinfo{volume}{632}},   \bibinfo{pages}{571 } (\bibinfo{year}{2006}{\natexlab{a}}), ISSN   \bibinfo{issn}{0370-2693}.
\bibitem[{\citenamefont{Wilson and Cowan}(1972)}]{Wilson1972} \bibinfo{author}{\bibfnamefont{H.~R.} \bibnamefont{Wilson}} \bibnamefont{and}   \bibinfo{author}{\bibfnamefont{J.~D.} \bibnamefont{Cowan}},   \bibinfo{journal}{Biophysical Journal} \textbf{\bibinfo{volume}{12}},   \bibinfo{pages}{1 } (\bibinfo{year}{1972}), ISSN \bibinfo{issn}{0006-3495}.
\bibitem[{\citenamefont{Amari}(1972)}]{Amari72_643} \bibinfo{author}{\bibfnamefont{S.-I.} \bibnamefont{Amari}},   \bibinfo{journal}{Systems, Man and Cybernetics, IEEE Transactions on}   \textbf{\bibinfo{volume}{SMC-2}}, \bibinfo{pages}{643}   (\bibinfo{year}{1972}), ISSN \bibinfo{issn}{2168-2909}.
\bibitem[{\citenamefont{Softky and Koch}(1993)}]{Softky93} \bibinfo{author}{\bibfnamefont{W.~R.} \bibnamefont{Softky}} \bibnamefont{and}   \bibinfo{author}{\bibfnamefont{C.}~\bibnamefont{Koch}}, \bibinfo{journal}{J.   Neurosci.} \textbf{\bibinfo{volume}{13}}, \bibinfo{pages}{334}   (\bibinfo{year}{1993}).
\bibitem[{\citenamefont{Ginzburg and Sompolinsky}(1994)}]{Ginzburg94} \bibinfo{author}{\bibfnamefont{I.}~\bibnamefont{Ginzburg}} \bibnamefont{and}   \bibinfo{author}{\bibfnamefont{H.}~\bibnamefont{Sompolinsky}},   \bibinfo{journal}{Phys. Rev. E} \textbf{\bibinfo{volume}{50}},   \bibinfo{pages}{3171} (\bibinfo{year}{1994}).
\bibitem[{\citenamefont{Weber and Frey}(2017)}]{Weber17_046601} \bibinfo{author}{\bibfnamefont{M.~F.} \bibnamefont{Weber}} \bibnamefont{and}   \bibinfo{author}{\bibfnamefont{E.}~\bibnamefont{Frey}},   \bibinfo{journal}{Reports on Progress in Physics}   \textbf{\bibinfo{volume}{80}}, \bibinfo{pages}{046601}   (\bibinfo{year}{2017}).
\bibitem[{\citenamefont{Brunel and Hakim}(1999)}]{Brunel99} \bibinfo{author}{\bibfnamefont{N.}~\bibnamefont{Brunel}} \bibnamefont{and}   \bibinfo{author}{\bibfnamefont{V.}~\bibnamefont{Hakim}},   \bibinfo{journal}{Neural Comput.} \textbf{\bibinfo{volume}{11}},   \bibinfo{pages}{1621} (\bibinfo{year}{1999}).
\bibitem[{\citenamefont{Brunel and Wang}(2003)}]{Brunel03a} \bibinfo{author}{\bibfnamefont{N.}~\bibnamefont{Brunel}} \bibnamefont{and}   \bibinfo{author}{\bibfnamefont{X.-J.} \bibnamefont{Wang}},   \bibinfo{journal}{J. Neurophysiol.} \textbf{\bibinfo{volume}{90}},   \bibinfo{pages}{415} (\bibinfo{year}{2003}).
\bibitem[{\citenamefont{Ostojic et~al.}(2009)\citenamefont{Ostojic, Brunel, and   Hakim}}]{Ostojic09_10234} \bibinfo{author}{\bibfnamefont{S.}~\bibnamefont{Ostojic}},   \bibinfo{author}{\bibfnamefont{N.}~\bibnamefont{Brunel}}, \bibnamefont{and}   \bibinfo{author}{\bibfnamefont{V.}~\bibnamefont{Hakim}}, \bibinfo{journal}{J.   Neurosci.} \textbf{\bibinfo{volume}{29}}, \bibinfo{pages}{10234}   (\bibinfo{year}{2009}).
\bibitem[{\citenamefont{Ledoux and Brunel}(2011)}]{Ledoux11_1} \bibinfo{author}{\bibfnamefont{E.}~\bibnamefont{Ledoux}} \bibnamefont{and}   \bibinfo{author}{\bibfnamefont{N.}~\bibnamefont{Brunel}},   \bibinfo{journal}{Front. Comput. Neurosci.} \textbf{\bibinfo{volume}{5}},   \bibinfo{pages}{1} (\bibinfo{year}{2011}).
\bibitem[{\citenamefont{Grytskyy et~al.}(2013)\citenamefont{Grytskyy, Tetzlaff,   Diesmann, and Helias}}]{Grytskyy13_131} \bibinfo{author}{\bibfnamefont{D.}~\bibnamefont{Grytskyy}},   \bibinfo{author}{\bibfnamefont{T.}~\bibnamefont{Tetzlaff}},   \bibinfo{author}{\bibfnamefont{M.}~\bibnamefont{Diesmann}}, \bibnamefont{and}   \bibinfo{author}{\bibfnamefont{M.}~\bibnamefont{Helias}},   \bibinfo{journal}{Front. Comput. Neurosci.} \textbf{\bibinfo{volume}{7}},   \bibinfo{pages}{131} (\bibinfo{year}{2013}).
\bibitem[{\citenamefont{Andreanov et~al.}(2006)\citenamefont{Andreanov, Biroli,   Bouchaud, and Lefevre}}]{Andreanov2006_030101} \bibinfo{author}{\bibfnamefont{A.}~\bibnamefont{Andreanov}},   \bibinfo{author}{\bibfnamefont{G.}~\bibnamefont{Biroli}},   \bibinfo{author}{\bibfnamefont{J.-P.} \bibnamefont{Bouchaud}},   \bibnamefont{and} \bibinfo{author}{\bibfnamefont{A.}~\bibnamefont{Lefevre}},   \bibinfo{journal}{Phys. Rev. E} \textbf{\bibinfo{volume}{74}},   \bibinfo{pages}{030101(R)} (\bibinfo{year}{2006}).
\bibitem[{\citenamefont{Lefevre and Biroli}(2007)}]{Lefvre2007_07024} \bibinfo{author}{\bibfnamefont{A.}~\bibnamefont{Lefevre}} \bibnamefont{and}   \bibinfo{author}{\bibfnamefont{G.}~\bibnamefont{Biroli}},   \bibinfo{journal}{Journal of Statistical Mechanics: Theory and Experiment}   \textbf{\bibinfo{volume}{2007}}, \bibinfo{pages}{P07024}   (\bibinfo{year}{2007}).
\bibitem[{\citenamefont{Ocker et~al.}(2017{\natexlab{b}})\citenamefont{Ocker,   Hu, Buice, Doiron, Josi{\'c}, Rosenbaum, and Shea-Brown}}]{Ocker17_109} \bibinfo{author}{\bibfnamefont{G.~K.} \bibnamefont{Ocker}},   \bibinfo{author}{\bibfnamefont{Y.}~\bibnamefont{Hu}},   \bibinfo{author}{\bibfnamefont{M.~A.} \bibnamefont{Buice}},   \bibinfo{author}{\bibfnamefont{B.}~\bibnamefont{Doiron}},   \bibinfo{author}{\bibfnamefont{K.}~\bibnamefont{Josi{\'c}}},   \bibinfo{author}{\bibfnamefont{R.}~\bibnamefont{Rosenbaum}},   \bibnamefont{and}   \bibinfo{author}{\bibfnamefont{E.}~\bibnamefont{Shea-Brown}},   \bibinfo{journal}{Current Opinion in Neurobiology}   \textbf{\bibinfo{volume}{46}}, \bibinfo{pages}{109}   (\bibinfo{year}{2017}{\natexlab{b}}).
\bibitem[{\citenamefont{Sompolinsky et~al.}(1988)\citenamefont{Sompolinsky,   Crisanti, and Sommers}}]{Sompolinsky88_259} \bibinfo{author}{\bibfnamefont{H.}~\bibnamefont{Sompolinsky}},   \bibinfo{author}{\bibfnamefont{A.}~\bibnamefont{Crisanti}}, \bibnamefont{and}   \bibinfo{author}{\bibfnamefont{H.~J.} \bibnamefont{Sommers}},   \bibinfo{journal}{Phys. Rev. Lett.} \textbf{\bibinfo{volume}{61}},   \bibinfo{pages}{259} (\bibinfo{year}{1988}).
\bibitem[{\citenamefont{Bressloff}(2012)}]{Bressloff12} \bibinfo{author}{\bibfnamefont{P.~C.} \bibnamefont{Bressloff}},   \bibinfo{journal}{Journal of Physics A: Mathematical and Theoretical}   \textbf{\bibinfo{volume}{45}}, \bibinfo{pages}{033001}   (\bibinfo{year}{2012}).
\bibitem[{\citenamefont{Gardiner}(1985)}]{Gardiner85} \bibinfo{author}{\bibfnamefont{C.~W.} \bibnamefont{Gardiner}},   \emph{\bibinfo{title}{Handbook of Stochastic Methods for Physics, Chemistry   and the Natural Sciences}} (\bibinfo{publisher}{Springer-Verlag},   \bibinfo{address}{Berlin}, \bibinfo{year}{1985}), \bibinfo{edition}{2nd} ed.,   ISBN \bibinfo{isbn}{3-540-61634-9, 3-540-15607-0}.
\bibitem[{\citenamefont{Van~Kampen}(2007)}]{VanKampen07} \bibinfo{author}{\bibfnamefont{N.~G.} \bibnamefont{Van~Kampen}},   \emph{\bibinfo{title}{Stochastic Processes in Physics and Chemistry, Third   Edition (North-Holland Personal Library)}} (\bibinfo{publisher}{North   Holland}, \bibinfo{year}{2007}), \bibinfo{edition}{3rd} ed.
\bibitem[{\citenamefont{Kleinert}(2009)}]{Kleinert09} \bibinfo{author}{\bibfnamefont{H.}~\bibnamefont{Kleinert}},   \emph{\bibinfo{title}{Path Integrals in Quantum Mechanics, Statistics,   Polymer Physics, and Financial Markets}} (\bibinfo{publisher}{World   Scientific}, \bibinfo{year}{2009}), \bibinfo{edition}{5th} ed.
\bibitem[{\citenamefont{Onsager and Machlup}(1953)}]{Onsager53} \bibinfo{author}{\bibfnamefont{L.}~\bibnamefont{Onsager}} \bibnamefont{and}   \bibinfo{author}{\bibfnamefont{S.}~\bibnamefont{Machlup}},   \bibinfo{journal}{Phys. Rev.} \textbf{\bibinfo{volume}{91}},   \bibinfo{pages}{1505} (\bibinfo{year}{1953}).
\bibitem[{\citenamefont{Stratonovich}(1960)}]{Stratonovich60} \bibinfo{author}{\bibfnamefont{R.~L.} \bibnamefont{Stratonovich}}, in   \emph{\bibinfo{booktitle}{Proc. Sixth All-Union Conf. Theory Prob. and Math.   Statist, Vilnius}} (\bibinfo{year}{1960}), pp. \bibinfo{pages}{471--482},   \bibinfo{note}{translation to english by A. N. Rossolimo published in   "Selected Transl. in Math. Statist. and Probability, Vol. 10, 1971"}.
\bibitem[{\citenamefont{Graham}(1977)}]{Graham77_281} \bibinfo{author}{\bibfnamefont{R.}~\bibnamefont{Graham}},   \bibinfo{journal}{Zeitschrift f{\"u}r Physik B Condensed Matter}   \textbf{\bibinfo{volume}{26}}, \bibinfo{pages}{281} (\bibinfo{year}{1977}),   ISSN \bibinfo{issn}{1431-584X}.
\bibitem[{\citenamefont{Hunt and Ross}(1981)}]{Hunt81_976} \bibinfo{author}{\bibfnamefont{K.~L.~C.} \bibnamefont{Hunt}} \bibnamefont{and}   \bibinfo{author}{\bibfnamefont{J.}~\bibnamefont{Ross}}, \bibinfo{journal}{The   Journal of Chemical Physics} \textbf{\bibinfo{volume}{75}},   \bibinfo{pages}{976} (\bibinfo{year}{1981}),   \eprint{https://doi.org/10.1063/1.442098}.
\bibitem[{\citenamefont{Altland and Simons}(2010)}]{Altland01} \bibinfo{author}{\bibfnamefont{A.}~\bibnamefont{Altland}} \bibnamefont{and}   \bibinfo{author}{\bibfnamefont{B.}~\bibnamefont{Simons}},   \emph{\bibinfo{title}{Concepts of Theoretical Solid State Physics}}   (\bibinfo{publisher}{Cambridge University Press}, \bibinfo{year}{2010}).
\bibitem[{\citenamefont{Bressloff}(2009)}]{Bressloff09_1488} \bibinfo{author}{\bibfnamefont{P.~C.} \bibnamefont{Bressloff}},   \bibinfo{journal}{SIAM J. appl. math.} \textbf{\bibinfo{volume}{70}},   \bibinfo{pages}{1488} (\bibinfo{year}{2009}).
\bibitem[{\citenamefont{Vasiliev}(1998)}]{Vasiliev98} \bibinfo{author}{\bibfnamefont{A.}~\bibnamefont{Vasiliev}},   \emph{\bibinfo{title}{Functional Methods in Quantum Field Theory and   Statistical Physics}} (\bibinfo{publisher}{Gordon and Breach Science   Publishers}, \bibinfo{year}{1998}).
\bibitem[{\citenamefont{Helias and Dahmen}(2019)}]{Helias19_10416} \bibinfo{author}{\bibfnamefont{M.}~\bibnamefont{Helias}} \bibnamefont{and}   \bibinfo{author}{\bibfnamefont{D.}~\bibnamefont{Dahmen}},   \bibinfo{journal}{arXiv}  (\bibinfo{year}{2019}), \bibinfo{note}{1901.10416   [cond-mat.dis-nn]}.
\bibitem[{\citenamefont{De~Dominicis}(1978)}]{DeDominicis78_4913} \bibinfo{author}{\bibfnamefont{C.}~\bibnamefont{De~Dominicis}},   \bibinfo{journal}{Phys. Rev. B} \textbf{\bibinfo{volume}{18}},   \bibinfo{pages}{4913} (\bibinfo{year}{1978}).
\bibitem[{\citenamefont{Andersen}(2000)}]{Andersen00_1979} \bibinfo{author}{\bibfnamefont{H.~C.} \bibnamefont{Andersen}},   \bibinfo{journal}{Journal of Mathematical Physics}   \textbf{\bibinfo{volume}{41}}, \bibinfo{pages}{1979} (\bibinfo{year}{2000}).
\bibitem[{\citenamefont{Coolen}(2000)}]{Coolen00_arxiv_II} \bibinfo{author}{\bibfnamefont{A.~C.~C.} \bibnamefont{Coolen}},   \bibinfo{journal}{arXiv:cond-mat/0006011}  (\bibinfo{year}{2000}).
\bibitem[{\citenamefont{Buice and Cowan}(2007)}]{Buice07_051919} \bibinfo{author}{\bibfnamefont{M.~A.} \bibnamefont{Buice}} \bibnamefont{and}   \bibinfo{author}{\bibfnamefont{J.~D.} \bibnamefont{Cowan}},   \bibinfo{journal}{Phys. Rev. E} \textbf{\bibinfo{volume}{75}},   \bibinfo{pages}{051919} (\bibinfo{year}{2007}).
\bibitem[{\citenamefont{Doi}(1976)}]{Doi76_1465} \bibinfo{author}{\bibfnamefont{M.}~\bibnamefont{Doi}},   \bibinfo{journal}{Journal of Physics A: Mathematical and General}   \textbf{\bibinfo{volume}{9}}, \bibinfo{pages}{1465} (\bibinfo{year}{1976}).
\bibitem[{\citenamefont{Peliti}(1985)}]{Peliti85_1469} \bibinfo{author}{\bibfnamefont{L.}~\bibnamefont{Peliti}}, \bibinfo{journal}{J.   Phys. France} \textbf{\bibinfo{volume}{46}}, \bibinfo{pages}{1469}   (\bibinfo{year}{1985}).
\bibitem[{\citenamefont{Negele and Orland}(1998)}]{NegeleOrland98} \bibinfo{author}{\bibfnamefont{J.~W.} \bibnamefont{Negele}} \bibnamefont{and}   \bibinfo{author}{\bibfnamefont{H.}~\bibnamefont{Orland}},   \emph{\bibinfo{title}{Quantum Many-Particle Systems}}   (\bibinfo{publisher}{New York: Perseus Books}, \bibinfo{year}{1998}).
\bibitem[{\citenamefont{Schneidman et~al.}(2006)\citenamefont{Schneidman,   Berry, Segev, and Bialek}}]{Schneidman06_1007} \bibinfo{author}{\bibfnamefont{E.}~\bibnamefont{Schneidman}},   \bibinfo{author}{\bibfnamefont{M.~J.} \bibnamefont{Berry}},   \bibinfo{author}{\bibfnamefont{R.}~\bibnamefont{Segev}}, \bibnamefont{and}   \bibinfo{author}{\bibfnamefont{W.}~\bibnamefont{Bialek}},   \bibinfo{journal}{Nature} \textbf{\bibinfo{volume}{440}},   \bibinfo{pages}{1007} (\bibinfo{year}{2006}).
\bibitem[{\citenamefont{Cooper and Dawson}(2016)}]{Cooper16} \bibinfo{author}{\bibfnamefont{F.}~\bibnamefont{Cooper}} \bibnamefont{and}   \bibinfo{author}{\bibfnamefont{J.~F.} \bibnamefont{Dawson}},   \bibinfo{journal}{Annals of Physics} \textbf{\bibinfo{volume}{365}},   \bibinfo{pages}{118 } (\bibinfo{year}{2016}), ISSN \bibinfo{issn}{0003-4916}.
\bibitem[{\citenamefont{Mayer and Goeppert-Mayer}(1977)}]{Mayer77} \bibinfo{author}{\bibfnamefont{J.~E.} \bibnamefont{Mayer}} \bibnamefont{and}   \bibinfo{author}{\bibfnamefont{M.}~\bibnamefont{Goeppert-Mayer}},   \emph{\bibinfo{title}{Statistical mechanics}} (\bibinfo{publisher}{John Wiley   \& Sons, Inc.}, \bibinfo{year}{1977}), \bibinfo{edition}{2nd} ed.
\bibitem[{\citenamefont{Roxin et~al.}(2011)\citenamefont{Roxin, Brunel, Hansel,   Mongillo, and van Vreeswijk}}]{Roxin11_16217} \bibinfo{author}{\bibfnamefont{A.}~\bibnamefont{Roxin}},   \bibinfo{author}{\bibfnamefont{N.}~\bibnamefont{Brunel}},   \bibinfo{author}{\bibfnamefont{D.}~\bibnamefont{Hansel}},   \bibinfo{author}{\bibfnamefont{G.}~\bibnamefont{Mongillo}}, \bibnamefont{and}   \bibinfo{author}{\bibfnamefont{C.}~\bibnamefont{van Vreeswijk}},   \bibinfo{journal}{J. Neurosci.} \textbf{\bibinfo{volume}{31}},   \bibinfo{pages}{16217} (\bibinfo{year}{2011}).
\bibitem[{\citenamefont{Risken}(1996)}]{Risken96} \bibinfo{author}{\bibfnamefont{H.}~\bibnamefont{Risken}},   \emph{\bibinfo{title}{The Fokker-Planck Equation}}   (\bibinfo{publisher}{Springer Verlag Berlin Heidelberg},   \bibinfo{year}{1996}),   \urlprefix\url{https://doi.org/10.1007/978-3-642-61544-3_4}.
\bibitem[{\citenamefont{Ornigotti et~al.}(2018)\citenamefont{Ornigotti, Ryabov,   Holubec, and Filip}}]{Ornigotti18_032127} \bibinfo{author}{\bibfnamefont{L.}~\bibnamefont{Ornigotti}},   \bibinfo{author}{\bibfnamefont{A.}~\bibnamefont{Ryabov}},   \bibinfo{author}{\bibfnamefont{V.}~\bibnamefont{Holubec}}, \bibnamefont{and}   \bibinfo{author}{\bibfnamefont{R.}~\bibnamefont{Filip}},   \bibinfo{journal}{Phys. Rev. E} \textbf{\bibinfo{volume}{97}},   \bibinfo{pages}{032127} (\bibinfo{year}{2018}).
\bibitem[{\citenamefont{Filip and Zem\'{a}nek}(2016)}]{Filip16_065401} \bibinfo{author}{\bibfnamefont{R.}~\bibnamefont{Filip}} \bibnamefont{and}   \bibinfo{author}{\bibfnamefont{P.}~\bibnamefont{Zem\'{a}nek}},   \bibinfo{journal}{Journal of Optics} \textbf{\bibinfo{volume}{18}},   \bibinfo{pages}{065401} (\bibinfo{year}{2016}).
\bibitem[{\citenamefont{Latham et~al.}(2000)\citenamefont{Latham, Richmond,   Nelson, and Nirenberg}}]{Latham00a} \bibinfo{author}{\bibfnamefont{P.~E.} \bibnamefont{Latham}},   \bibinfo{author}{\bibfnamefont{B.~J.} \bibnamefont{Richmond}},   \bibinfo{author}{\bibfnamefont{P.~G.} \bibnamefont{Nelson}},   \bibnamefont{and}   \bibinfo{author}{\bibfnamefont{S.}~\bibnamefont{Nirenberg}},   \bibinfo{journal}{J. Neurophysiol.} \textbf{\bibinfo{volume}{83}},   \bibinfo{pages}{808} (\bibinfo{year}{2000}).
\bibitem[{\citenamefont{Brette and Gerstner}(2005)}]{Brette-2005_3637} \bibinfo{author}{\bibfnamefont{R.}~\bibnamefont{Brette}} \bibnamefont{and}   \bibinfo{author}{\bibfnamefont{W.}~\bibnamefont{Gerstner}},   \bibinfo{journal}{J. Neurophysiol.} \textbf{\bibinfo{volume}{94}},   \bibinfo{pages}{3637} (\bibinfo{year}{2005}).
\bibitem[{\citenamefont{Elgart and Kamenev}(2004)}]{Elgart04_041106} \bibinfo{author}{\bibfnamefont{V.}~\bibnamefont{Elgart}} \bibnamefont{and}   \bibinfo{author}{\bibfnamefont{A.}~\bibnamefont{Kamenev}},   \bibinfo{journal}{Phys. Rev. E} \textbf{\bibinfo{volume}{70}},   \bibinfo{pages}{041106} (\bibinfo{year}{2004}).
\bibitem[{\citenamefont{Zwanzig}(2001)}]{Zwanzig01} \bibinfo{author}{\bibfnamefont{R.}~\bibnamefont{Zwanzig}},   \emph{\bibinfo{title}{Nonequilibrium Statistical Mechanics}}   (\bibinfo{publisher}{Oxford University Press}, \bibinfo{year}{2001}), ISBN   \bibinfo{isbn}{0-19-514018-4}.
\bibitem[{\citenamefont{Sommers et~al.}(1988)\citenamefont{Sommers, Crisanti,   Sompolinsky, and Stein}}]{Sommers88} \bibinfo{author}{\bibfnamefont{H.}~\bibnamefont{Sommers}},   \bibinfo{author}{\bibfnamefont{A.}~\bibnamefont{Crisanti}},   \bibinfo{author}{\bibfnamefont{H.}~\bibnamefont{Sompolinsky}},   \bibnamefont{and} \bibinfo{author}{\bibfnamefont{Y.}~\bibnamefont{Stein}},   \bibinfo{journal}{Phys. Rev. Lett.} \textbf{\bibinfo{volume}{60}},   \bibinfo{pages}{1895} (\bibinfo{year}{1988}).
\bibitem[{\citenamefont{La~Camera et~al.}(2006)\citenamefont{La~Camera, Rauch,   Thurbon, Lüscher, Senn, and Fusi}}]{LaCamera06_3448} \bibinfo{author}{\bibfnamefont{G.}~\bibnamefont{La~Camera}},   \bibinfo{author}{\bibfnamefont{A.}~\bibnamefont{Rauch}},   \bibinfo{author}{\bibfnamefont{D.}~\bibnamefont{Thurbon}},   \bibinfo{author}{\bibfnamefont{H.-R.} \bibnamefont{Lüscher}},   \bibinfo{author}{\bibfnamefont{W.}~\bibnamefont{Senn}}, \bibnamefont{and}   \bibinfo{author}{\bibfnamefont{S.}~\bibnamefont{Fusi}},   \bibinfo{journal}{Journal of Neurophysiology} \textbf{\bibinfo{volume}{96}},   \bibinfo{pages}{3448} (\bibinfo{year}{2006}), \bibinfo{note}{pMID: 16807345},   \eprint{https://doi.org/10.1152/jn.00453.2006}.
\bibitem[{\citenamefont{Chaudhuri et~al.}(2015)\citenamefont{Chaudhuri,   Knoblauch, Gariel, Kennedy, and Wang}}]{Chaudhuri2015_419} \bibinfo{author}{\bibfnamefont{R.}~\bibnamefont{Chaudhuri}},   \bibinfo{author}{\bibfnamefont{K.}~\bibnamefont{Knoblauch}},   \bibinfo{author}{\bibfnamefont{M.-A.} \bibnamefont{Gariel}},   \bibinfo{author}{\bibfnamefont{H.}~\bibnamefont{Kennedy}}, \bibnamefont{and}   \bibinfo{author}{\bibfnamefont{X.-J.} \bibnamefont{Wang}},   \bibinfo{journal}{Neuron} \textbf{\bibinfo{volume}{88}}, \bibinfo{pages}{419}   (\bibinfo{year}{2015}), ISSN \bibinfo{issn}{0896-6273}.
\bibitem[{\citenamefont{Bravi et~al.}(2016)\citenamefont{Bravi, Sollich, and   Opper}}]{Bravi16} \bibinfo{author}{\bibfnamefont{B.}~\bibnamefont{Bravi}},   \bibinfo{author}{\bibfnamefont{P.}~\bibnamefont{Sollich}}, \bibnamefont{and}   \bibinfo{author}{\bibfnamefont{M.}~\bibnamefont{Opper}},   \bibinfo{journal}{Journal of Physics A: Mathematical and Theoretical}   \textbf{\bibinfo{volume}{49}}, \bibinfo{pages}{194003}   (\bibinfo{year}{2016}).
\bibitem[{\citenamefont{De~Dominicis and Martin}(1964)}]{DeDominicis64_14} \bibinfo{author}{\bibfnamefont{C.}~\bibnamefont{De~Dominicis}}   \bibnamefont{and} \bibinfo{author}{\bibfnamefont{P.~C.}   \bibnamefont{Martin}}, \bibinfo{journal}{Journal of Mathematical Physics}   \textbf{\bibinfo{volume}{5}}, \bibinfo{pages}{14} (\bibinfo{year}{1964}).
\bibitem[{\citenamefont{Morris}(1994)}]{Morris94_2411} \bibinfo{author}{\bibfnamefont{T.~R.} \bibnamefont{Morris}},   \bibinfo{journal}{International Journal of Modern Physics A}   \textbf{\bibinfo{volume}{09}}, \bibinfo{pages}{2411} (\bibinfo{year}{1994}),   \eprint{https://doi.org/10.1142/S0217751X94000972}.
\bibitem[{\citenamefont{Bagnuls and Bervillier}(2001)}]{Bagnuls01_91} \bibinfo{author}{\bibfnamefont{C.}~\bibnamefont{Bagnuls}} \bibnamefont{and}   \bibinfo{author}{\bibfnamefont{C.}~\bibnamefont{Bervillier}},   \bibinfo{journal}{Physics Reports} \textbf{\bibinfo{volume}{348}},   \bibinfo{pages}{91 } (\bibinfo{year}{2001}), ISSN \bibinfo{issn}{0370-1573},   \bibinfo{note}{renormalization group theory in the new millennium. II}.
\bibitem[{\citenamefont{Duclut and Delamotte}(2017)}]{Duclut17_012107} \bibinfo{author}{\bibfnamefont{C.}~\bibnamefont{Duclut}} \bibnamefont{and}   \bibinfo{author}{\bibfnamefont{B.}~\bibnamefont{Delamotte}},   \bibinfo{journal}{Phys. Rev. E} \textbf{\bibinfo{volume}{95}},   \bibinfo{pages}{012107} (\bibinfo{year}{2017}).
\bibitem[{\citenamefont{Sch\"{u}tz and Kopietz}(2006)}]{Schuetz06} \bibinfo{author}{\bibfnamefont{F.}~\bibnamefont{Sch\"{u}tz}} \bibnamefont{and}   \bibinfo{author}{\bibfnamefont{P.}~\bibnamefont{Kopietz}},   \bibinfo{journal}{Journal of Physics A: Mathematical and General}   \textbf{\bibinfo{volume}{39}}, \bibinfo{pages}{8205} (\bibinfo{year}{2006}).
\bibitem[{\citenamefont{Kopietz et~al.}(2010)\citenamefont{Kopietz, Bartosch,   and Sch\"utz}}]{Kopietz10} \bibinfo{author}{\bibfnamefont{P.}~\bibnamefont{Kopietz}},   \bibinfo{author}{\bibfnamefont{L.}~\bibnamefont{Bartosch}}, \bibnamefont{and}   \bibinfo{author}{\bibfnamefont{F.}~\bibnamefont{Sch\"utz}},   \emph{\bibinfo{title}{Introduction to the Functional Renormalization Group}}   (\bibinfo{publisher}{Springer}, \bibinfo{year}{2010}).
\bibitem[{\citenamefont{Blaizot   et~al.}(2006{\natexlab{b}})\citenamefont{Blaizot, M\'{e}ndez-Galain, and   Wschebor}}]{Blaizot06a_051116} \bibinfo{author}{\bibfnamefont{J.-P.} \bibnamefont{Blaizot}},   \bibinfo{author}{\bibfnamefont{R.}~\bibnamefont{M\'{e}ndez-Galain}},   \bibnamefont{and} \bibinfo{author}{\bibfnamefont{N.}~\bibnamefont{Wschebor}},   \bibinfo{journal}{Phys. Rev. E} \textbf{\bibinfo{volume}{74}},   \bibinfo{pages}{051116} (\bibinfo{year}{2006}{\natexlab{b}}).
\bibitem[{\citenamefont{Benitez et~al.}(2012)\citenamefont{Benitez, Blaizot,   Chat\'e, Delamotte, M\'endez-Galain, and Wschebor}}]{Benitez12_026707} \bibinfo{author}{\bibfnamefont{F.}~\bibnamefont{Benitez}},   \bibinfo{author}{\bibfnamefont{J.-P.} \bibnamefont{Blaizot}},   \bibinfo{author}{\bibfnamefont{H.}~\bibnamefont{Chat\'e}},   \bibinfo{author}{\bibfnamefont{B.}~\bibnamefont{Delamotte}},   \bibinfo{author}{\bibfnamefont{R.}~\bibnamefont{M\'endez-Galain}},   \bibnamefont{and} \bibinfo{author}{\bibfnamefont{N.}~\bibnamefont{Wschebor}},   \bibinfo{journal}{Phys. Rev. E} \textbf{\bibinfo{volume}{85}},   \bibinfo{pages}{026707} (\bibinfo{year}{2012}).
\bibitem[{\citenamefont{Touchette}(2009)}]{Touchette09} \bibinfo{author}{\bibfnamefont{H.}~\bibnamefont{Touchette}},   \bibinfo{journal}{Physics Reports} \textbf{\bibinfo{volume}{478}},   \bibinfo{pages}{1} (\bibinfo{year}{2009}).
\bibitem[{\citenamefont{Eyink}(1996)}]{Eyink96_3419} \bibinfo{author}{\bibfnamefont{G.~L.} \bibnamefont{Eyink}},   \bibinfo{journal}{Phys. Rev. E} \textbf{\bibinfo{volume}{54}},   \bibinfo{pages}{3419} (\bibinfo{year}{1996}).
\bibitem[{\citenamefont{Jordan et~al.}(2019)\citenamefont{Jordan, M{\o}rk,   Vennemo, Terhorst, Peyser, Ippen, Deepu, Eppler, van Meegen, Kunkel   et~al.}}]{Jordan19_2605422} \bibinfo{author}{\bibfnamefont{J.}~\bibnamefont{Jordan}},   \bibinfo{author}{\bibfnamefont{H.}~\bibnamefont{M{\o}rk}},   \bibinfo{author}{\bibfnamefont{S.~B.} \bibnamefont{Vennemo}},   \bibinfo{author}{\bibfnamefont{D.}~\bibnamefont{Terhorst}},   \bibinfo{author}{\bibfnamefont{A.}~\bibnamefont{Peyser}},   \bibinfo{author}{\bibfnamefont{T.}~\bibnamefont{Ippen}},   \bibinfo{author}{\bibfnamefont{R.}~\bibnamefont{Deepu}},   \bibinfo{author}{\bibfnamefont{J.~M.} \bibnamefont{Eppler}},   \bibinfo{author}{\bibfnamefont{A.}~\bibnamefont{van Meegen}},   \bibinfo{author}{\bibfnamefont{S.}~\bibnamefont{Kunkel}},   \bibnamefont{et~al.}, \emph{\bibinfo{title}{Nest 2.18.0}}   (\bibinfo{year}{2019}),   \urlprefix\url{https://doi.org/10.5281/zenodo.2605422}.
\bibitem[{\citenamefont{Kadmon and Sompolinsky}(2015)}]{Kadmon15_041030} \bibinfo{author}{\bibfnamefont{J.}~\bibnamefont{Kadmon}} \bibnamefont{and}   \bibinfo{author}{\bibfnamefont{H.}~\bibnamefont{Sompolinsky}},   \bibinfo{journal}{Phys. Rev. X} \textbf{\bibinfo{volume}{5}},   \bibinfo{pages}{041030} (\bibinfo{year}{2015}).
\bibitem[{\citenamefont{Strogatz}(1994)}]{Strogatz94} \bibinfo{author}{\bibfnamefont{S.~H.} \bibnamefont{Strogatz}},   \emph{\bibinfo{title}{Nonlinear Dynamics and Chaos: with Applications to   Physics, Biology, Chemistry, and Engineering}} (\bibinfo{publisher}{Perseus   Books}, \bibinfo{address}{Reading, Massachusetts}, \bibinfo{year}{1994}),   ISBN \bibinfo{isbn}{0-201-54344-3}.
\bibitem[{\citenamefont{Buzs\'{a}ki and {Draguhn}}(2004)}]{Buzsaki04_1926} \bibinfo{author}{\bibfnamefont{G.}~\bibnamefont{Buzs\'{a}ki}} \bibnamefont{and}   \bibinfo{author}{\bibfnamefont{A.}~\bibnamefont{{Draguhn}}},   \bibinfo{journal}{Science} \textbf{\bibinfo{volume}{304}},   \bibinfo{pages}{1926} (\bibinfo{year}{2004}).
\bibitem[{\citenamefont{Takahashi et~al.}(2015)\citenamefont{Takahashi, Kim,   Coleman, Brown, Suminski, Best, and Hatsopoulos}}]{Takahashi15} \bibinfo{author}{\bibfnamefont{K.}~\bibnamefont{Takahashi}},   \bibinfo{author}{\bibfnamefont{S.}~\bibnamefont{Kim}},   \bibinfo{author}{\bibfnamefont{T.~P.} \bibnamefont{Coleman}},   \bibinfo{author}{\bibfnamefont{K.~A.} \bibnamefont{Brown}},   \bibinfo{author}{\bibfnamefont{A.~J.} \bibnamefont{Suminski}},   \bibinfo{author}{\bibfnamefont{M.~D.} \bibnamefont{Best}}, \bibnamefont{and}   \bibinfo{author}{\bibfnamefont{N.~G.} \bibnamefont{Hatsopoulos}},   \bibinfo{journal}{Nature Communications} \textbf{\bibinfo{volume}{6}}   (\bibinfo{year}{2015}).
\bibitem[{\citenamefont{Denker et~al.}(2018)\citenamefont{Denker, Zehl,   Kilavik, Diesmann, Brochier, Riehle, and Gr\"{u}n}}]{Denker18_1} \bibinfo{author}{\bibfnamefont{M.}~\bibnamefont{Denker}},   \bibinfo{author}{\bibfnamefont{L.}~\bibnamefont{Zehl}},   \bibinfo{author}{\bibfnamefont{B.~E.} \bibnamefont{Kilavik}},   \bibinfo{author}{\bibfnamefont{M.}~\bibnamefont{Diesmann}},   \bibinfo{author}{\bibfnamefont{T.}~\bibnamefont{Brochier}},   \bibinfo{author}{\bibfnamefont{A.}~\bibnamefont{Riehle}}, \bibnamefont{and}   \bibinfo{author}{\bibfnamefont{S.}~\bibnamefont{Gr\"{u}n}},   \bibinfo{journal}{Scientific Reports} \textbf{\bibinfo{volume}{8}},   \bibinfo{pages}{1} (\bibinfo{year}{2018}).
\bibitem[{\citenamefont{Senk et~al.}(2018)\citenamefont{Senk, Korvasov\'{a},   Schuecker, Hagen, Tetzlaff, Diesmann, and Helias}}]{Senk18_arxiv_06046v1} \bibinfo{author}{\bibfnamefont{J.}~\bibnamefont{Senk}},   \bibinfo{author}{\bibfnamefont{K.}~\bibnamefont{Korvasov\'{a}}},   \bibinfo{author}{\bibfnamefont{J.}~\bibnamefont{Schuecker}},   \bibinfo{author}{\bibfnamefont{E.}~\bibnamefont{Hagen}},   \bibinfo{author}{\bibfnamefont{T.}~\bibnamefont{Tetzlaff}},   \bibinfo{author}{\bibfnamefont{M.}~\bibnamefont{Diesmann}}, \bibnamefont{and}   \bibinfo{author}{\bibfnamefont{M.}~\bibnamefont{Helias}},   \bibinfo{journal}{arXiv} p. \bibinfo{pages}{1801.06046}   (\bibinfo{year}{2018}).
\bibitem[{\citenamefont{Barreiro et~al.}(2017)\citenamefont{Barreiro, Kutz, and   Shlizerman}}]{Barreiro17} \bibinfo{author}{\bibfnamefont{A.~K.} \bibnamefont{Barreiro}},   \bibinfo{author}{\bibfnamefont{J.~N.} \bibnamefont{Kutz}}, \bibnamefont{and}   \bibinfo{author}{\bibfnamefont{E.}~\bibnamefont{Shlizerman}},   \bibinfo{journal}{The Journal of Mathematical Neuroscience}   \textbf{\bibinfo{volume}{7}}, \bibinfo{pages}{10} (\bibinfo{year}{2017}),   ISSN \bibinfo{issn}{2190-8567}.
\bibitem[{\citenamefont{Steriade et~al.}(1993)\citenamefont{Steriade,   Nu\~{n}ez, and Amzica}}]{Steriade93} \bibinfo{author}{\bibfnamefont{M.}~\bibnamefont{Steriade}},   \bibinfo{author}{\bibfnamefont{A.}~\bibnamefont{Nu\~{n}ez}},   \bibnamefont{and} \bibinfo{author}{\bibfnamefont{F.}~\bibnamefont{Amzica}},   \bibinfo{journal}{J. Neurosci.} \textbf{\bibinfo{volume}{13}},   \bibinfo{pages}{3252} (\bibinfo{year}{1993}).
\bibitem[{\citenamefont{Destexhe et~al.}(2007)\citenamefont{Destexhe, Hughes,   Rudolph, and Crunelli}}]{Destexhe2007_334} \bibinfo{author}{\bibfnamefont{A.}~\bibnamefont{Destexhe}},   \bibinfo{author}{\bibfnamefont{S.~W.} \bibnamefont{Hughes}},   \bibinfo{author}{\bibfnamefont{M.}~\bibnamefont{Rudolph}}, \bibnamefont{and}   \bibinfo{author}{\bibfnamefont{V.}~\bibnamefont{Crunelli}},   \bibinfo{journal}{TINS} \textbf{\bibinfo{volume}{30}}, \bibinfo{pages}{334}   (\bibinfo{year}{2007}), ISSN \bibinfo{issn}{0166-2236}, \bibinfo{note}{july   INMED/TINS special issue---Physiogenic and pathogenic oscillations: the   beauty and the beast}.
\bibitem[{\citenamefont{Bressloff and Faugeras}(2017)}]{Bressloff17_033206} \bibinfo{author}{\bibfnamefont{P.~C.} \bibnamefont{Bressloff}}   \bibnamefont{and} \bibinfo{author}{\bibfnamefont{O.}~\bibnamefont{Faugeras}},   \bibinfo{journal}{Journal of Statistical Mechanics: Theory and Experiment}   \textbf{\bibinfo{volume}{2017}}, \bibinfo{pages}{033206}   (\bibinfo{year}{2017}).
\bibitem[{\citenamefont{Bravi and Sollich}(2017)}]{Bravi17_045010} \bibinfo{author}{\bibfnamefont{B.}~\bibnamefont{Bravi}} \bibnamefont{and}   \bibinfo{author}{\bibfnamefont{P.}~\bibnamefont{Sollich}},   \bibinfo{journal}{Physical Biology} \textbf{\bibinfo{volume}{14}},   \bibinfo{pages}{045010} (\bibinfo{year}{2017}).
\bibitem[{\citenamefont{Amit and Brunel}(1997{\natexlab{a}})}]{Amit-1997_373} \bibinfo{author}{\bibfnamefont{D.~J.} \bibnamefont{Amit}} \bibnamefont{and}   \bibinfo{author}{\bibfnamefont{N.}~\bibnamefont{Brunel}},   \bibinfo{journal}{Network: Comput. Neural Systems}   \textbf{\bibinfo{volume}{8}}, \bibinfo{pages}{373}   (\bibinfo{year}{1997}{\natexlab{a}}).
\bibitem[{\citenamefont{van Vreeswijk and Sompolinsky}(1998)}]{Vreeswijk98} \bibinfo{author}{\bibfnamefont{C.}~\bibnamefont{van Vreeswijk}}   \bibnamefont{and}   \bibinfo{author}{\bibfnamefont{H.}~\bibnamefont{Sompolinsky}},   \bibinfo{journal}{Neural Comput.} \textbf{\bibinfo{volume}{10}},   \bibinfo{pages}{1321} (\bibinfo{year}{1998}).
\bibitem[{\citenamefont{Husemann and Salmhofer}(2009)}]{Husemann09_195125} \bibinfo{author}{\bibfnamefont{C.}~\bibnamefont{Husemann}} \bibnamefont{and}   \bibinfo{author}{\bibfnamefont{M.}~\bibnamefont{Salmhofer}},   \bibinfo{journal}{Phys. Rev. B} \textbf{\bibinfo{volume}{79}},   \bibinfo{pages}{195125} (\bibinfo{year}{2009}).
\bibitem[{\citenamefont{Wentzell et~al.}(2016)\citenamefont{Wentzell, Li,   Tagliavini, Taranto, Rohringer, Held, Toschi, and   Andergassen}}]{Wentzell16_arxiv} \bibinfo{author}{\bibfnamefont{N.}~\bibnamefont{Wentzell}},   \bibinfo{author}{\bibfnamefont{G.}~\bibnamefont{Li}},   \bibinfo{author}{\bibfnamefont{A.}~\bibnamefont{Tagliavini}},   \bibinfo{author}{\bibfnamefont{C.}~\bibnamefont{Taranto}},   \bibinfo{author}{\bibfnamefont{G.}~\bibnamefont{Rohringer}},   \bibinfo{author}{\bibfnamefont{K.}~\bibnamefont{Held}},   \bibinfo{author}{\bibfnamefont{A.}~\bibnamefont{Toschi}}, \bibnamefont{and}   \bibinfo{author}{\bibfnamefont{S.}~\bibnamefont{Andergassen}},   \bibinfo{journal}{arxiv preprint, arXiv:1610.06520}  (\bibinfo{year}{2016}).
\bibitem[{\citenamefont{Eckhardt et~al.}(2018)\citenamefont{Eckhardt, Schober,   Ehrlich, and Honerkamp}}]{Eckhardt18_075143} \bibinfo{author}{\bibfnamefont{C.~J.} \bibnamefont{Eckhardt}},   \bibinfo{author}{\bibfnamefont{G.~A.~H.} \bibnamefont{Schober}},   \bibinfo{author}{\bibfnamefont{J.}~\bibnamefont{Ehrlich}}, \bibnamefont{and}   \bibinfo{author}{\bibfnamefont{C.}~\bibnamefont{Honerkamp}},   \bibinfo{journal}{Phys. Rev. B} \textbf{\bibinfo{volume}{98}},   \bibinfo{pages}{075143} (\bibinfo{year}{2018}).
\bibitem[{\citenamefont{Delamotte}(2012)}]{Delamotte12} \bibinfo{author}{\bibfnamefont{B.}~\bibnamefont{Delamotte}}, in   \emph{\bibinfo{booktitle}{Renormalization groups and effective field theory   approaches to many-body systems}}, edited by   \bibinfo{editor}{\bibfnamefont{J.~S.} \bibnamefont{Janos~Polonyi}}   (\bibinfo{publisher}{Springer}, \bibinfo{year}{2012}), pp.   \bibinfo{pages}{49--130}.
\bibitem[{\citenamefont{Churchland}(2010)}]{Churchland10} \bibinfo{author}{\bibfnamefont{M.~M. e.~a.} \bibnamefont{Churchland}},   \bibinfo{journal}{Nat. Neurosci.} \textbf{\bibinfo{volume}{13}},   \bibinfo{pages}{369} (\bibinfo{year}{2010}).
\bibitem[{\citenamefont{Litwin-Kumar et~al.}(2012)\citenamefont{Litwin-Kumar,   Chacron, and Doiron}}]{LitvinKumar12_e1002667} \bibinfo{author}{\bibfnamefont{A.}~\bibnamefont{Litwin-Kumar}},   \bibinfo{author}{\bibfnamefont{M.~J.} \bibnamefont{Chacron}},   \bibnamefont{and} \bibinfo{author}{\bibfnamefont{B.}~\bibnamefont{Doiron}},   \bibinfo{journal}{PLOS Comput. Biol.} \textbf{\bibinfo{volume}{8}},   \bibinfo{pages}{e1002667} (\bibinfo{year}{2012}).
\bibitem[{\citenamefont{Ohbayashi et~al.}(2003)\citenamefont{Ohbayashi, Ohki,   and Miyashita}}]{Obayashi03_233} \bibinfo{author}{\bibfnamefont{M.}~\bibnamefont{Ohbayashi}},   \bibinfo{author}{\bibfnamefont{K.}~\bibnamefont{Ohki}}, \bibnamefont{and}   \bibinfo{author}{\bibfnamefont{Y.}~\bibnamefont{Miyashita}},   \bibinfo{journal}{Science} \textbf{\bibinfo{volume}{301}},   \bibinfo{pages}{233} (\bibinfo{year}{2003}), ISSN \bibinfo{issn}{0036-8075},   \eprint{http://science.sciencemag.org/content/301/5630/233.full.pdf}.
\bibitem[{\citenamefont{Henningson and Illes}(2017)}]{Henningson17} \bibinfo{author}{\bibfnamefont{M.}~\bibnamefont{Henningson}} \bibnamefont{and}   \bibinfo{author}{\bibfnamefont{S.}~\bibnamefont{Illes}},   \bibinfo{journal}{Front Comput Neurosci} \textbf{\bibinfo{volume}{11}}   (\bibinfo{year}{2017}).
\bibitem[{\citenamefont{Wiese}(1998)}]{Wiese1998} \bibinfo{author}{\bibfnamefont{K.~J.} \bibnamefont{Wiese}},   \bibinfo{journal}{Journal of Statistical Physics}   \textbf{\bibinfo{volume}{93}}, \bibinfo{pages}{143} (\bibinfo{year}{1998}),   ISSN \bibinfo{issn}{1572-9613}.
\bibitem[{\citenamefont{Priesemann et~al.}(2014)\citenamefont{Priesemann,   Wibral, Valderrama, Pr{\"o}pper, Le~Van~Quyen, Geisel, Triesch, Nikolic, and   Munk}}]{Priesemann14_80} \bibinfo{author}{\bibfnamefont{V.}~\bibnamefont{Priesemann}},   \bibinfo{author}{\bibfnamefont{M.}~\bibnamefont{Wibral}},   \bibinfo{author}{\bibfnamefont{M.}~\bibnamefont{Valderrama}},   \bibinfo{author}{\bibfnamefont{R.}~\bibnamefont{Pr{\"o}pper}},   \bibinfo{author}{\bibfnamefont{M.}~\bibnamefont{Le~Van~Quyen}},   \bibinfo{author}{\bibfnamefont{T.}~\bibnamefont{Geisel}},   \bibinfo{author}{\bibfnamefont{J.}~\bibnamefont{Triesch}},   \bibinfo{author}{\bibfnamefont{D.}~\bibnamefont{Nikolic}}, \bibnamefont{and}   \bibinfo{author}{\bibfnamefont{M.~H.~J.} \bibnamefont{Munk}},   \bibinfo{journal}{Frontiers in Systems Neuroscience}   \textbf{\bibinfo{volume}{8}}, \bibinfo{pages}{80} (\bibinfo{year}{2014}),   ISSN \bibinfo{issn}{1662-5137}.
\bibitem[{\citenamefont{Wilting and Priesemann}(2018)}]{Wilting18_2325} \bibinfo{author}{\bibfnamefont{J.}~\bibnamefont{Wilting}} \bibnamefont{and}   \bibinfo{author}{\bibfnamefont{V.}~\bibnamefont{Priesemann}},   \bibinfo{journal}{Nature Communications} \textbf{\bibinfo{volume}{9}},   \bibinfo{pages}{2325} (\bibinfo{year}{2018}).
\bibitem[{\citenamefont{Bak et~al.}(1987)\citenamefont{Bak, Tang, and   Wiesenfeld}}]{Bak87} \bibinfo{author}{\bibfnamefont{P.}~\bibnamefont{Bak}},   \bibinfo{author}{\bibfnamefont{C.}~\bibnamefont{Tang}}, \bibnamefont{and}   \bibinfo{author}{\bibfnamefont{K.}~\bibnamefont{Wiesenfeld}},   \bibinfo{journal}{Phys. Rev. Lett.} \textbf{\bibinfo{volume}{59}},   \bibinfo{pages}{381} (\bibinfo{year}{1987}).
\bibitem[{\citenamefont{Martignon et~al.}(1995)\citenamefont{Martignon, {von   Hasseln}, Gr\"{u}n, Aertsen, and Palm}}]{Martignon95} \bibinfo{author}{\bibfnamefont{L.}~\bibnamefont{Martignon}},   \bibinfo{author}{\bibfnamefont{H.}~\bibnamefont{{von Hasseln}}},   \bibinfo{author}{\bibfnamefont{S.}~\bibnamefont{Gr\"{u}n}},   \bibinfo{author}{\bibfnamefont{A.}~\bibnamefont{Aertsen}}, \bibnamefont{and}   \bibinfo{author}{\bibfnamefont{G.}~\bibnamefont{Palm}},   \bibinfo{journal}{Biol. Cybern.} \textbf{\bibinfo{volume}{73}},   \bibinfo{pages}{69} (\bibinfo{year}{1995}).
\bibitem[{\citenamefont{Manna}(1991)}]{Manna91_L363} \bibinfo{author}{\bibfnamefont{S.~S.} \bibnamefont{Manna}},   \bibinfo{journal}{Journal of Physics A: Mathematical and General}   \textbf{\bibinfo{volume}{24}}, \bibinfo{pages}{L363} (\bibinfo{year}{1991}).
\bibitem[{\citenamefont{Vespignani et~al.}(1998)\citenamefont{Vespignani,   Dickman, Mu\~noz, and Zapperi}}]{Vespignani98_5676} \bibinfo{author}{\bibfnamefont{A.}~\bibnamefont{Vespignani}},   \bibinfo{author}{\bibfnamefont{R.}~\bibnamefont{Dickman}},   \bibinfo{author}{\bibfnamefont{M.~A.} \bibnamefont{Mu\~noz}},   \bibnamefont{and} \bibinfo{author}{\bibfnamefont{S.}~\bibnamefont{Zapperi}},   \bibinfo{journal}{Phys. Rev. Lett.} \textbf{\bibinfo{volume}{81}},   \bibinfo{pages}{5676} (\bibinfo{year}{1998}).
\bibitem[{\citenamefont{Dickman et~al.}(1998)\citenamefont{Dickman, Vespignani,   and Zapperi}}]{Dickmann98_5095} \bibinfo{author}{\bibfnamefont{R.}~\bibnamefont{Dickman}},   \bibinfo{author}{\bibfnamefont{A.}~\bibnamefont{Vespignani}},   \bibnamefont{and} \bibinfo{author}{\bibfnamefont{S.}~\bibnamefont{Zapperi}},   \bibinfo{journal}{Phys. Rev. E} \textbf{\bibinfo{volume}{57}},   \bibinfo{pages}{5095} (\bibinfo{year}{1998}).
\bibitem[{\citenamefont{Le~Doussal and Wiese}(2015)}]{LeDoussal15_110601} \bibinfo{author}{\bibfnamefont{P.}~\bibnamefont{Le~Doussal}} \bibnamefont{and}   \bibinfo{author}{\bibfnamefont{K.~J.} \bibnamefont{Wiese}},   \bibinfo{journal}{Phys. Rev. Lett.} \textbf{\bibinfo{volume}{114}},   \bibinfo{pages}{110601} (\bibinfo{year}{2015}).
\bibitem[{\citenamefont{Wiese}(2016)}]{Wiese16_042117} \bibinfo{author}{\bibfnamefont{K.~J.} \bibnamefont{Wiese}},   \bibinfo{journal}{Phys. Rev. E} \textbf{\bibinfo{volume}{93}},   \bibinfo{pages}{042117} (\bibinfo{year}{2016}).
\bibitem[{\citenamefont{Harish and Hansel}(2015)}]{Harish15_e1004266} \bibinfo{author}{\bibfnamefont{O.}~\bibnamefont{Harish}} \bibnamefont{and}   \bibinfo{author}{\bibfnamefont{D.}~\bibnamefont{Hansel}},   \bibinfo{journal}{PLOS Comput. Biol.} \textbf{\bibinfo{volume}{11}},   \bibinfo{pages}{e1004266} (\bibinfo{year}{2015}).
\bibitem[{\citenamefont{Crisanti and   Sompolinsky}(2018{\natexlab{b}})}]{Crisanti18_arxiv} \bibinfo{author}{\bibfnamefont{A.}~\bibnamefont{Crisanti}} \bibnamefont{and}   \bibinfo{author}{\bibfnamefont{H.}~\bibnamefont{Sompolinsky}},   \bibinfo{journal}{arxiv preprint, arXiv:1603.08270}   (\bibinfo{year}{2018}{\natexlab{b}}).
\bibitem[{\citenamefont{Hawkes}(1971{\natexlab{a}})}]{Hawkes71_438} \bibinfo{author}{\bibfnamefont{A.}~\bibnamefont{Hawkes}}, \bibinfo{journal}{J.   R. Statist. Soc. Ser. B} \textbf{\bibinfo{volume}{33}}, \bibinfo{pages}{438}   (\bibinfo{year}{1971}{\natexlab{a}}).
\bibitem[{\citenamefont{Hawkes}(1971{\natexlab{b}})}]{Hawkes71_83} \bibinfo{author}{\bibfnamefont{A.}~\bibnamefont{Hawkes}},   \bibinfo{journal}{Biometrika} \textbf{\bibinfo{volume}{58}},   \bibinfo{pages}{83} (\bibinfo{year}{1971}{\natexlab{b}}).
\bibitem[{\citenamefont{Qiu and Chow}(2018)}]{Qiu18_arxiv} \bibinfo{author}{\bibfnamefont{S.}~\bibnamefont{Qiu}} \bibnamefont{and}   \bibinfo{author}{\bibfnamefont{C.}~\bibnamefont{Chow}},   \bibinfo{journal}{arXiv preprint arXiv:1805.03601}  (\bibinfo{year}{2018}).
\bibitem[{\citenamefont{Stein}(1967)}]{Stein67a} \bibinfo{author}{\bibfnamefont{R.~B.} \bibnamefont{Stein}},   \bibinfo{journal}{Biophys. J.} \textbf{\bibinfo{volume}{7}},   \bibinfo{pages}{37} (\bibinfo{year}{1967}).
\bibitem[{\citenamefont{Amit and Brunel}(1997{\natexlab{b}})}]{Amit97} \bibinfo{author}{\bibfnamefont{D.~J.} \bibnamefont{Amit}} \bibnamefont{and}   \bibinfo{author}{\bibfnamefont{N.}~\bibnamefont{Brunel}},   \bibinfo{journal}{Cereb. Cortex} \textbf{\bibinfo{volume}{7}},   \bibinfo{pages}{237} (\bibinfo{year}{1997}{\natexlab{b}}).
\bibitem[{\citenamefont{Sompolinsky and Zippelius}(1982)}]{Sompolinsky82_6860} \bibinfo{author}{\bibfnamefont{H.}~\bibnamefont{Sompolinsky}} \bibnamefont{and}   \bibinfo{author}{\bibfnamefont{A.}~\bibnamefont{Zippelius}},   \bibinfo{journal}{Phys. Rev. B} \textbf{\bibinfo{volume}{25}},   \bibinfo{pages}{6860} (\bibinfo{year}{1982}).
\bibitem[{\citenamefont{Goldenfeld}(1992)}]{Goldenfeld92} \bibinfo{author}{\bibfnamefont{N.}~\bibnamefont{Goldenfeld}},   \emph{\bibinfo{title}{Lectures on phase transitions and the renormalization   group}} (\bibinfo{publisher}{Perseus books}, \bibinfo{address}{Reading,   Massachusetts}, \bibinfo{year}{1992}).
\bibitem[{\citenamefont{Rockafellar}(1970)}]{Rockafellar70} \bibinfo{author}{\bibfnamefont{R.~T.} \bibnamefont{Rockafellar}},   \emph{\bibinfo{title}{Convex Analysis}} (\bibinfo{publisher}{Princeton   University Press}, \bibinfo{year}{1970}), ISBN \bibinfo{isbn}{9780691015866},   \urlprefix\url{https://books.google.de/books?id=GV6YDwAAQBAJ}.
\end{thebibliography}
